\documentclass[reprint,amsmath,amssymb,aps,rmp,nobibnotes]{revtex4-2}

\usepackage{graphicx}
\usepackage{dcolumn}
\usepackage{bm}
\usepackage{upgreek}
\usepackage{textcomp}
\usepackage{hyperref}
\hypersetup{colorlinks,breaklinks,
            urlcolor=[rgb]{0,0,0.64},
            linkcolor=[rgb]{0,0,0.64},
            citecolor=[rgb]{0,0,0.64}}
\usepackage{color}
\usepackage{xcolor}
\definecolor{navy}{rgb}{0,0,0.64}

\setcitestyle{super}
\setcitestyle{comma}

\begin{document}

\title{Time-domain study of coupled collective excitations in quantum materials}

\author{Chenhang~Xu$^{\text{\color{navy}1,2}}$}
\author{Alfred~Zong$^{\text{\color{navy}1,2,3,}}$}
\email[Correspondence to: ]{alfredz@stanford.edu}
\affiliation{$^{\text{\color{navy}1}}$Department of Applied Physics, Stanford University, Stanford, California 94305, USA\looseness=-10}
\affiliation{$^{\text{\color{navy}2}}$Stanford Institute for Materials and Energy Sciences, SLAC National Accelerator Laboratory, Menlo Park, California 94025, USA\looseness=-10}
\affiliation{$^{\text{\color{navy}3}}$Department of Physics, Stanford University, Stanford, California 94305, USA\looseness=-10}

\date{\today}

\begin{abstract}

Quantum materials hold immense promises for future applications due to their intriguing electronic, magnetic, thermal, and mechanical properties that often arise from a complex interplay between microscopic degrees of freedom. Important insights of such interactions come from studying the collective excitations of electrons, spins, orbitals, and lattice, whose cooperative motions play a crucial role in determining the novel behavior of these systems and offer us a key tuning knob to modify material properties on-demand through external perturbations. In this regard, ultrafast light-matter interaction has shown great potential in controlling the couplings of collective excitations, and rapid progress in a plethora of time-resolved techniques down to the attosecond regime has significantly advanced our understanding of the coupling mechanisms and guided us in manipulating the dynamical properties of quantum materials. This review aims to highlight recent experiments on visualizing collective excitations in the time domain, focusing on the coupling mechanisms between different collective modes such as phonon-phonon, phonon-magnon, phonon-exciton, magnon-magnon, magnon-exciton, and various polaritons. We introduce how these collective modes are excited by an ultrashort laser pulse and probed by different ultrafast techniques, and we explain how the coupling between collective excitations governs the ensuing nonequilibrium dynamics. We also provide some perspectives on future studies that can lead to discoveries of the emergent properties of quantum materials both in and out of equilibrium.
\end{abstract}

\maketitle

\tableofcontents

\section{Introduction}\label{sec:intro}

Quantum materials have garnered considerable interest due to their remarkable properties associated with, for instance, unconventional superconductivity \cite{lee2006doping,stewart2011supercond,keimer2015fromquantum,fernandes2022iron}, multiferroicity \cite{cheong2007multiferroics}, colossal magnetoresistance \cite{salamon2001thephy}, and topological quantum states \cite{wang2017topological}. These properties hold great promise for future applications \cite{tokura2017emergent} such as topological quantum computing \cite{nayak2008nonabelian} and spintronic devices \cite{vzutic2004spintronics}, yet a precise understanding of the underlying mechanisms remains elusive in many cases. At the microscopic level, these properties emerge as a result of complex interactions among the charge, spin, orbital, and lattice degrees of freedom. Take superconductivity as an example: the formation of Cooper pairs in the ground state is either aided by electron-phonon interaction in conventional superconductors \cite{cooper1956bound,bardeen1957theory} or other types of electron-boson coupling, if any, in unconventional superconductors. Unraveling the interplay among these degrees of freedom is crucial for advancing our knowledge of the rich phase diagrams of quantum systems. However, one key challenge has been the coexistence of multi-way couplings of a myriad of quasiparticles and excitations, which makes it difficult to isolate a specific type of interaction we are interested in.

To tackle this challenge, recent advances in time-resolved techniques based on ultrafast light-matter interaction have identified \emph{time} as a new dimension to investigate the couplings among the microscopic degrees of freedom. Leveraging the time resolution of ultrashort pulses down to the attosecond level, experiments based on the pump-probe protocol offer a unique way to disentangle the intertwined couplings at the intrinsic timescales of microscopic interactions \cite{petek1997femtosecond,bovensiepen2012elementary}. Aside from studying the dynamics of individual quasiparticles, a cooperative and wave-like motion of different degrees of freedom --- referred to as \emph{collective excitations} --- can be directly observed in the time domain. One of the most notable examples is photoinduced coherent phonons \cite{zeiger1992theory,stevens2002coherent}, where a macroscopic number of atoms move in perfect synchronization. Such lattice dynamics show up in a variety of time-resolved probes, such as the periodic oscillations of diffraction peak intensity in time-resolved scattering \cite{bugayev2011coherent, chatelain2014coherent, gerber2017femtosecond, zong2018ultrafast, zong2023spin, huang2023ultrafastmeasurement} and the oscillations of optical reflectivity and transmissivity in transient optical spectroscopy \cite{zong2019dynamicalslowing,bartram2022ultrafast,zong2023emerging}. Oftentimes, more than one type of collective excitation emerges after a photoexcitation event, and their coupling is governed by the underlying interaction among the constituent degrees of freedom. Investigating these collective modes and their couplings provides an important perspective to study the microscopic interactions in quantum materials, which in turn offers an opportunity to use light to modify dynamical material properties.

This review focuses on recent experimental progress in understanding the couplings between collective excitations thanks to the development of ultrafast techniques with ever better spatiotemporal resolution and the capability to generate intense laser pulses with highly tailored properties. We begin in Sec.~\ref{sec:collective} by explaining why we are interested in studying collective excitations and how time-domain probes of these excitations and their couplings bring out new insights into material properties that are otherwise missing in equilibrium probes. Given that a single probe cannot capture all the degrees of freedom and their collective modes, we subsequently introduce some select ultrafast techniques in Sec.~\ref{sec:techniques}. In Sec.~\ref{sec:couplings}, we delve deeper into the couplings of collective excitations by introducing specific cases such as photon-phonon, phonon-phonon, phonon-magnon, phonon-exciton, and other coupling mechanisms. An outlook for future studies in this field is included in Sec.~\ref{sec:outlook}. 

While this review does not aim to cover the more general area of ultrafast science in condensed matter systems, such understanding of the broader field is helpful for contextualizing the present article. Therefore, we refer readers to a more general survey of theoretical advances and experimental techniques in ref.~\onlinecite{delatorre2021nonthermal} as well as a few other topical reviews. For techniques, see the following review articles on ultrafast optical spectroscopy \cite{fausti2011time,orenstein2012ultrafast,woerner2013ultrafast,giannetti2016ultrafastoptical, llloyd2021the2021ultrafast, zong2023emerging, dong2023recentdevelop}, ultrafast electron diffraction (UED) and microscopy \cite{zong2021unconventional,filippetto2022ultrafastelectron, lee2024structural}, time-resolved X-ray scattering \cite{buzzi2018probing,cao2019ultrafast, mitrano2020probing,mitrano2024exploring,lee2024structural}, and time- and angle-resolved photoemission spectroscopy (tr-ARPES) \cite{boven2012elementary, smallwood2016ultrafast, zhou2018newdevelopments, huang2022highresilution, boschini2024timeresolvedarpes}. Regarding precision engineering of material properties through ultrafast light pulses, we refer readers to excellent reviews on nonequilibrium engineering such as nonlinear phononics \cite{mankowsky2016nonequilibrium, nicoletti2016nonlinear,disa2021engineering} and Floquet engineering \cite{oka2019floquet,rudner2020band}. From the perspective of different material classes, see reviews on two-dimensional materials \cite{bao2022lightinduced}, magnetic materials \cite{kirilyuk2010ultrafastoptical}, high-temperature superconductors \cite{giannetti2016ultrafastoptical,nicoletti2016nonlinear}, complex oxides \cite{zhang2014dynamicalcontrol}, and topological insulators \cite{rudner2020band}. These lists are by no means exhaustive but provide sufficient conceptual background for appreciating the experiments highlighted in this review.

\begin{figure*}[!t]
\centering
\includegraphics[width=1\textwidth]{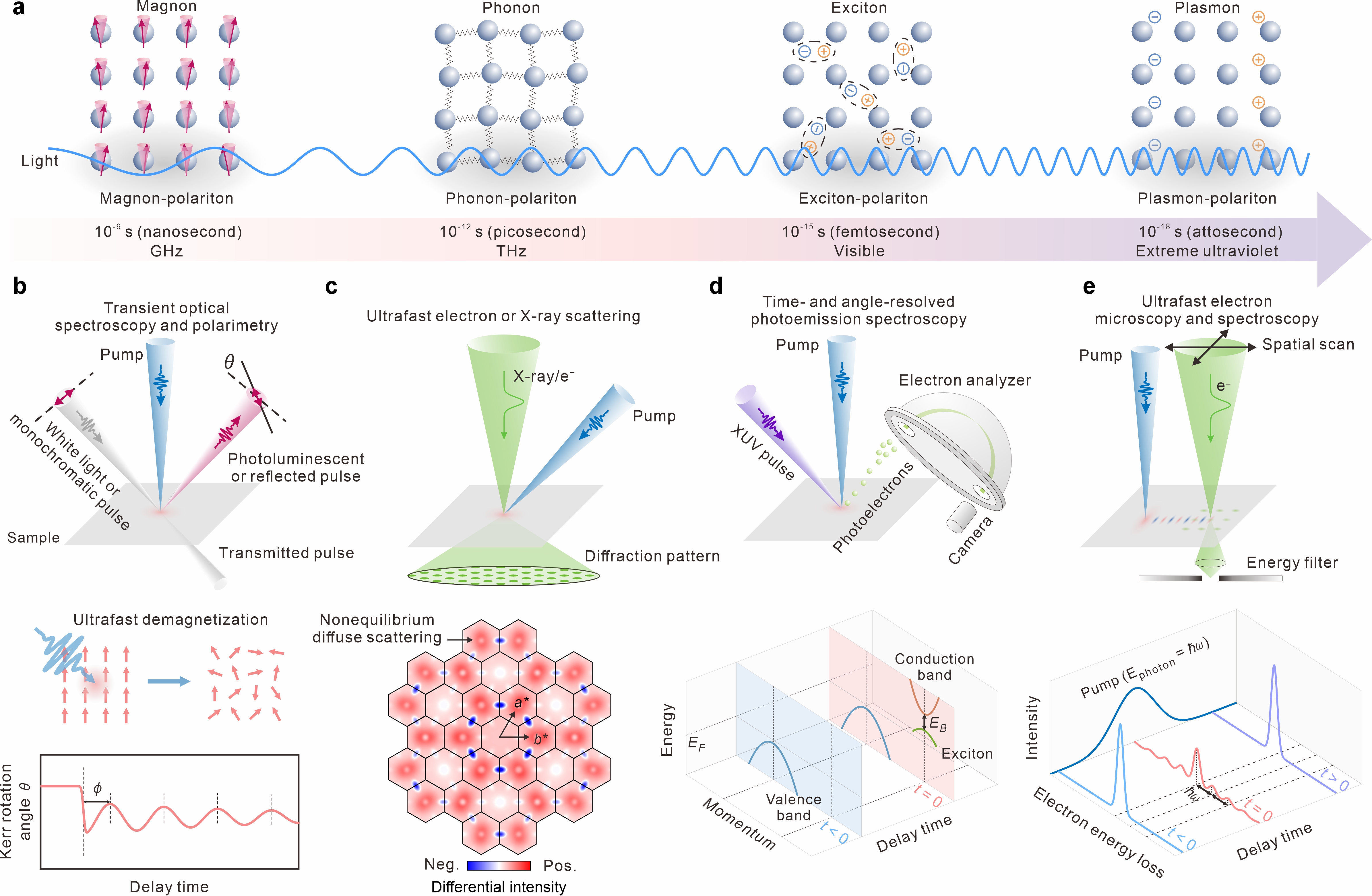}
\caption{\textbf{Overview of select collective excitations in quantum materials and corresponding probing techniques.}  \textbf{a},~Schematic of four different collective excitations in solids and their respective coupling with light arranged by their approximate timescale in order of magnitude. The blue sinusoidal wave represents a light wave. The gray shadow represents hybridization between light and each collective excitation to form different polaritons. \textbf{b},~\textit{Top}:~Schematic setup of ultrafast optical experiments. \textit{Middle}:~Schematic of optically-induced ultrafast demagnetization of a ferromagnet. \textit{Bottom}:~Representative data for the oscillatory component of time-resolved optical polarimetry, showing a sudden decrease of the probe laser polarization angle $\theta$ due to ultrafast demagnetization followed by a periodic modulation due to a coherent magnon mode. The phase $\phi$ of the waveform is labeled. \textbf{c},~\textit{Top}:~Schematic setup of time-resolved electron or X-ray diffraction. \textit{Bottom}:~Representative data for electron diffuse scattering, showing a specific momentum space distribution of diffuse intensity due to nonthermal population of certain phonon branches. \textbf{d},~\textit{Top}:~Schematic setup of time- and angle-resolved photoemission spectroscopy. \textit{Bottom}:~Representative transient photoemission data for probing unoccupied states in a semiconductor, showing new features in the measured band dispersion such as the conduction band (red curve) and excitons (green curve). \textbf{e},~\textit{Top}:~Schematic setup for ultrafast near-field electron microscopy and spectroscopy. Alternating red and blue regions represent the photoinduced field distribution due to a plasmon-polariton. Green ovals on the sample represent the electron scanning positions. \textit{Bottom}:~Representative data for probing the interaction between free electrons and photons. When the probe electron packet overlaps spatiotemporally with the photoinduced plasmon-polariton, electrons undergo quantized gain or loss of the quasiparticle energy ($\hbar\omega$), resulting in discrete energy peaks in the electron energy loss spectrum. Panel~\textbf{c} adapted from ref.~\onlinecite{cheng2024ultrafast_TiSe2},~Springer Nature Ltd.}
\label{fig:intro}
\end{figure*}

\section{Collective excitations viewed from an ultrafast lens} \label{sec:collective}

\subsection{Collective excitations in and out of equilibrium}

Collective excitations, as the name suggests, manifest as  cooperative motions of individual degrees of freedom in solids \cite{baldini2018nonequilibrium}. As illustrated in Fig.~\ref{fig:intro}, some common examples include magnons, phonons, excitons, and plasmons. While these collective excitations are intrinsic to a material, others emerge due to hybridization with an external field. For instance, photons become ``dressed'' when they propagate inside a solid, forming various types of polaritons \cite{basov2016polaritons, basov2020polariton}. These distinct collective excitations are characterized by different timescales in their underlying periodic dynamics, where the associated energy scale ranges from $\upmu$eV (GHz frequency) to tens of eV (extreme ultraviolet frequency). 

In equilibrium, collective excitations can reveal crucial information about the ground state orders \cite{baldini2020discovery} and the associated phase transitions due to a symmetry change. For example, structural or magnetic transitions are often accompanied by soft phonons or magnons, respectively, which signify an instability in the system. In addition, these collective modes are highly responsive to external stimuli, making it possible to achieve on-demand properties of quantum materials as one applies electric or magnetic fields, pressure, strain, or ultrafast photoexcitation to modify the collective dynamics \cite{basov2017towards}. While collective excitations are readily detected in an equilibrium setting and have hallmarks in a variety of measurements ranging from transport to scattering experiments, the underlying microscopic motions are incoherent and hence only time-averaged dynamics are obtained. 

This situation changed with the development of ultrafast techniques that are capable of visualizing microscopic motion with a time resolution commensurate with the characteristic period of collective excitations. Under the paradigm of pump-probe measurements, coherent dynamics can be excited by an ultrashort laser pulse, offering a general framework for inducing and controlling coherent dynamics in solids. For example, coherent phonons or magnons are often observed following a photoexcitation event, where ionic displacement or spin precession occurs in complete synchrony with the same temporal phase over a macroscopic sample volume. Such coherent motions are in stark contrast to the random movements that account for incoherent phonons and magnons in thermal equilibrium.

\subsection{Photoinduced phase transitions}

The ability to observe these coherent dynamics yields critical information about light-induced phase transitions. One extensively studied example is transitions that involves a symmetry change under the Ginzburg-Landau framework, such as those found in material systems hosting a charge density wave (CDW). One type of collective excitations of the CDW ground state is the amplitude mode, which has been widely observed following an above-gap photoexcitation that transiently suppresses the order parameter amplitude and hence launches a coherent amplitude oscillation \cite{demsar1999singleparticle,demsar2002femtosecond,schmitt2008transient}. Importantly, the photo-driven amplitude mode, in particular its magnitude and frequency, can inform us how the phase transition unfolds after the laser pulse incidence, such as whether the phase change proceeds in a spatially homogeneous fashion or involves heterogeneity like the formation of topological defects\cite{yusupov2010coherent,mihailovic2019importance}. As another example, in a prototypical CDW material LaTe$_3$, the CDW order parameter can be suddenly quenched within a time set by the characteristic time of the amplitude mode oscillations \cite{zong2019dynamicalslowing}. With varying photoexcitation densities, the suppression of the CDW order takes the longest time at the threshold photoexcitation density at which the order parameter transiently vanishes, demonstrating dynamical slowing-down in a nonequilibrium transition. Using more than one pump pulses, one can further harness the coherence in photoinduced collective excitations to precisely control the extent of the nonequilibrium transition. This coherent control was recently demonstrated in 1D Si(111)-In nanowires \cite{horstmann2020coherent} and quasi-2D 1$T$-TaS$_2$ (ref.~\onlinecite{maklar2023coherent}), both of which are archetypical CDW systems.

While most collective modes studied in the context of photoinduced phase transitions are driven by above-gap photoexcitation with ultraviolet to near-infrared light, they can also be induced by an intense mid-infrared or terahertz field. Instead of photo-doping and injecting mobile carriers into the solids, long-wavelength light can directly excite (i)~dipole-active modes through the oscillating electric field, (ii)~Raman-active modes through two-photon processes \cite{maehrlein2017terahertz}, or (iii)~other strong-field effects \cite{shi2023intrinsic} to initiate the collective mode that subsequently drives a transition. Novel metastable states and phenomenology can emerge, such as dynamic multiferroicity \cite{juraschek2017dynamicalmuti, basini2024terahertz, davies2024phononic, romao2024light}, a change in electronic topology \cite{sie2019ultrafast,tang2023light}, and magnetism induced by light-driven chiral phonons \cite{juraschek2022gianteffective, kahana2024lightinduced, luo2023large} or as a result of strong spin-lattice coupling \cite{ilyas_terahertz_2024}. We again use a CDW material as an example. In metallic 3$R$-Ta$_{1+x}$Se$_2$, when its CDW amplitude mode is excited by an intense terahertz pulse, an energy gap appears in the optical conductivity spectrum and executes a dynamical evolution in sync with the amplitude mode oscillation, suggesting a transition into a transient insulating state that has no equilibrium counterpart \cite{yoshikawa2021ultrafastswitch}.

\subsection{Dressed states}

Collective modes, aside from their role during the course of a nonequilibrium phase transition, can also help induce novel dynamical states of matter via coherent light-matter interactions. For instance, whereas Floquet-Bloch states have been conventionally realized by hybridization between electronic states and an external periodic light field \cite{wang2013obsercation, mahood2016selective,zhou2023pseudospin,zhou2023floquet,ito2023buildup,choi2024direct,merboldt2024observation}, there are a number of theoretical proposals for achieving the Floquet states by replacing photons with phonons \cite{hubener2018phonon,chaudhary2020phonon} or excitons \cite{chan2023giant}. One signature of Floquet-Bloch states is the appearance of electronic band replicas detected in tr-ARPES experiments, where these replica bands follow specific intensity relations and polarization selection rules given by the exciting photons. Importantly, these bands are spectrally spaced apart by $\hbar\omega_B$, where $\omega_B$ is the angular frequency of the bosonic field, be it a photon, phonon, or exciton. In a recent photoemission experiment using a momentum microscope specifically designed to study extremely small samples, exciton-induced Floquet effects were reported in monolayer WS$_2$ in the presence of a high density of resonantly-driven photoexcited excitons \cite{paraeek2024driving}, which also bear the hallmark of an excitonic insulator phase.

\subsection{Phase resolution of coherent collective excitations}

When coherent dynamics are launched in solids upon photoexcitation, the motion of collective excitations manifests as wave-like oscillations. The waveform of these oscillations can be directly visualized via an ultrafast probe, an example of which is shown in Fig.~\ref{fig:intro}b that depicts an oscillating Kerr rotation signal caused by a coherent magnon. Therefore, the \emph{phase} of the collective excitation oscillations can be determined (labeled as $\phi$ in Fig.~\ref{fig:intro}b, bottom panel), which equilibrium experiments cannot measure directly.
 
Phase-sensitivity can be used to isolate a single collective mode that affects multiple degrees of freedom, providing a way to measure the coupling strength between them \cite{gerber2017femtosecond, huang2023ultrafastmeasurement}. More specifically, by selectively inducing a particular oscillatory response using a tailored pump, and by leveraging phase-locked coherent features from different observables for the corresponding degrees of freedom, one can obtain the coupling strength directly, independent of the theoretical model. For example, following the excitation of a coherent phonon, tr-ARPES can probe spectral shifts in the electronic band energy while time-resolved X-ray diffraction (tr-XRD) can quantify the shifts in each atomic position for the phonon of interest. By combining these two measurements, electron-phonon coupling of a specific electronic band to a specific phonon mode can be calculated. This procedure has been applied to an iron-based superconductor FeSe (ref.~\onlinecite{gerber2017femtosecond}) and three-dimensional topological insulators Bi$_2$Se$_3$ and Bi$_2$Te$_3$ (ref.~\onlinecite{huang2023ultrafastmeasurement}). The extracted electron-phonon coupling strength in Bi$_2$Se$_3$ and Bi$_2$Te$_3$ is generally consistent with the expectation from density functional theory (DFT) computations while that in FeSe shows a significant deviation from DFT, which underestimates the strength. Instead, after electron-electron correlation effects were included in a self-consistent DFT-dynamical mean field theory, a quantitative agreement emerged, indicating an enhancement of the coupling strength due to strong electron correlations. Besides electron-phonon coupling, this multi-probe experimental protocol has also been deployed to measure the dynamical magnetoelectric coupling in multiferroic materials. For instance, in NiI$_2$, a giant coupling strength was obtained \cite{gao2024giant} where a single electromagnon mode induces periodic modulations in the dipolar and magnetic orders, revealing an interesting $\pi$/2 phase shift and lock-in behavior between electric polarization and magnetization oscillations that are at terahertz frequencies.

Even with a single experimental probe, a change in the oscillation phase with respect to some tuning parameter can be used to reveal material properties \cite{golias2016observation,ron2020ultrafast, ergeccen2023coherent} and identify microscopic processes underlying the collective excitation \cite{su2023delamination}. For instance, far above the Curie temperature, the $A^3_g$ optical phonon mode in a layered ferromagnetic insulator CrSiTe$_3$ can be excited due to the conventional displacive mechanism \cite{zeiger1992theory}, but the coherent phonon phase experiences a $\pi$ shift when the material cools below a temperature scale where in-plane ferromagnetic correlations start to develop \cite{ron2020ultrafast}. Such a $\pi$ phase shift is only observed when the excitation photon energy is sufficiently large to induce charge transfer from the Te$^{2-}$ to Cr$^{3+}$ ions, a process that also increases the superexchange energy. The $\pi$ phase shift is interpreted as evidence for a coherent phonon mechanism due to superexchange enhancement via magneto-elastic coupling, which only applies in the presence of short-range ferromagnetic correlation in this material. Notably, the $\pi$ phase shift of an $A^1_g$ mode across the N\'{e}el temperature also appears in an antiferromagnetic insulator FePS$_3$, providing evidence for selective coupling between the magnetic order and the trigonal distortions in this layered system \cite{ergeccen2023coherent}. Another example that highlights the importance of the oscillation phase comes from a coherent acoustic phonon in a freestanding film of La$_{2/3}$Ca$_{1/3}$MnO$_3$. Three different observables in time-resolved diffraction (Bragg peak intensity, width, and position) were observed to synchronously oscillate with a particular phase relation --- either in-phase or $\pi$ out-of-phase --- after photoexcitation \cite{su2023delamination}. By theoretically modeling the phase relation across the distinct observables from this coherent acoustic mode, Su and coworkers interpreted the phase relation as features of ultrafast wrinkle formation that results from delamination between the freestanding film and the substrate at its boundary, offering unique insights into the oft-neglected film–substrate interaction that governs the properties of freestanding structures.

\section{Time-domain probes of collective excitations\label{sec:techniques}} 

To explore collective excitations and their couplings in the time domain, a variety of advanced ultrafast tools have been developed, each with distinct advantages. Here, we briefly introduce a few selected techniques that serve as the workhorses in the experiments we reviewed in this article. A schematic for each category of techniques is depicted in Fig.~\ref{fig:intro}b--e. For a typical table-top time-resolved probe introduced below, a femtosecond laser pulse is split into two beams, one of which is used to generate a tailored pump pulse to drive the nonequilibrium dynamics while the other is used to produce a probe pulse, which is either a photon or an electron pulse. The frequency of the pump pulse can range from sub-terahertz up to ultraviolet and its polarization can be tuned anywhere between linear and circular, depending on the target excitation.

\subsection{Time-resolved optical spectroscopy and polarimetry}

The most commonly used ultrafast pump-probe technique is time-resolved optical spectroscopy and polarimetry, depicted in Fig.~\ref{fig:intro}b. In a typical experiment, white light or monochromatic pulses are used as probes, and experimental observables can be transmission, reflectivity, photoluminescence, birefringence, and dichroism, all of which can be measured with varying energies and polarization states of the probe beam. These measurements are not confined to the fundamental frequency of the probe but also apply to its second or higher-order harmonics, which can be more sensitive to order parameter symmetry hidden in the linear order measurement. Based on these observables, different variants of the techniques have been developed, such as transient optical transmittance or reflectance spectroscopy \cite{orenstein2012ultrafast, woerner2013ultrafast, giannetti2016ultrafastoptical,geneaus2019transient,li2020attosecond,llloyd2021the2021ultrafast,dong2023recentdevelop,zong2023emerging}, time-resolved magneto-optical Kerr effect \cite{zhang2009paradigm}, time-resolved second harmonic generation (SHG) \cite{nawakowsko2015timeresolved, zhang2021probingultrafast}, and time-resolved Raman scattering \cite{Yoshizawa1999femtosecond,sachu2009two,fausti2011time,versteeg2018tunable,yang2020ultrafast,Buchenau2023optical}. These measurements allow us to extract the complex optical constants of a material \cite{dong2023recentdevelop}, which make it possible to infer collective excitations associated with carriers \cite{ breusing2009ultrafastcarrier,ulbricht2011carrierdynamics, lagarde2014carrier, robert2016exciton}, spins \cite{kirilyuk2010ultrafastoptical,bartram2022ultrafast}, and polarizations \cite{zhang2021probingultrafast} in quantum materials.

\subsection{Ultrafast electron and X-ray scattering}

In contrast to optical spectroscopy in the infrared to ultraviolet regime, high energy electrons and X-rays have much shorter wavelengths that are comparable to the length scale of lattice parameters of common crystals, making time-resolved electron and X-ray diffraction suitable for tracking phonon dynamics \cite{buzzi2018probing,qi2020breaking,filippetto2022ultrafastelectron,lee2024structural} and charge or orbital order evolutions that are modulated by the lattice \cite{beaud2009ultrafast,li2016dichotomy}. In a typical diffraction experiment, a tailored pump is used to launch the nonequilibrium dynamics of nuclear ions. It should be noted that the pump pulse duration should always be shorter than the period of the collective excitations involved. The ensuing periodic motion of ions modulates the Bragg condition and can lead to a coherent oscillation of the intensity or position of Bragg peaks --- the former is commonly associated with an optical phonon while the latter with an acoustic mode \cite{kirkland2020advanced}. 

Besides coherent phonons, incoherent phonons can also be populated following an excitation event, where \emph{incoherent} means there is no net displacement of ionic positions, namely, the expectation value of phonon amplitude at wave vector $\mathbf{q}$ is zero at all time $t$, $\langle u_\mathbf{q}(t)\rangle = 0$. Inelastic scattering of a specific phonon branch can lead to a particular distribution of diffuse scattering \cite{wall2018ultrafast,durr2021revealing,zong2021role,pan2023vibrational, britt2023ultrafast,cheng2024ultrafast_TiSe2} depending on how this phonon modulates the structure factor. The bottom panel of Fig.~\ref{fig:intro}c shows a simulated nonequilibrium electron diffuse scattering pattern resulting from a nonthermal population of phonons induced by light \cite{cheng2024ultrafast_TiSe2}. 

It is worth noting that the absence of coherent phonons does not imply the absence of time-oscillatory features in a diffraction measurement. Without a net displacement of ions, the pump light can couple to the lattice to the second order in the phonon amplitude, where pairs of phonons at $\pm\mathbf{q}$ are generated to conserve the total momentum, leading to oscillations of the variance in the atomic displacement at twice the frequency of the phonon mode itself \cite{trigo2018ultrafast}. This second-order effect induces intensity oscillations in either Bragg peaks \cite{johnson2009directly} or diffuse scattering \cite{trigo2013fourier}, and is also known as squeezed phonons \cite{hu1996squeezed,garrett1997vacuum,trigo2013fourier,henighan2016control}. 

Another rapidly developing area that rides on recent progress in the X-ray free-electron laser technology is time-resolved resonant inelastic X-ray scattering (RIXS), which not only reveals phonon dynamics but can also access charge, orbital, and spin excitations as well as high-order, multi-particle correlation functions that describe coupled spin/charge/orbital excitations \cite{dean2016ultrafast,mitrano2020probing,mazzone2021laser,mitrano2024exploring}. With ongoing effort towards entering the sub-10-meV energy resolution in RIXS \cite{kim2018quartz}, this emerging technique provides a unique window into the nonequilibrium states in correlated materials through dynamical signatures across time, momentum, and energy axes.

\subsection{Time-resolved photoemission}

Time- and angle-resolved photoemission is a complementary method to ultrafast diffraction, enabling us to directly measure the evolution of electronic structure \cite{boven2012elementary,smallwood2016ultrafast,zhou2018newdevelopments,huang2022highresilution,boschini2024timeresolvedarpes} (see Fig.~\ref{fig:intro}d). In an equilibrium ARPES experiment, extreme ultraviolet to soft X-ray photons are illuminated onto a material to eject photoelectrons, whose energy and momentum inform us about the single-particle spectral function of the corresponding electrons inside a material \cite{damascelli2003angle,sobota2021angle}. By introducing a pump pulse, tr-ARPES yields further insights into nonequilibrium carrier dynamics and unoccupied states, reveals the electronic origin of ultrafast phase transitions \cite{zong2019evidence, duan2023ultrafast}, and unambiguously demonstrates how band structure can be engineered by light \cite{wang2013obsercation,mahood2016selective,oka2019floquet,rudner2020band,zhou2023floquet,zhou2023pseudospin,paraeek2024driving}. 

The advent of momentum microscopes in tr-ARPES experiments has also recently made it feasible to investigate micrometer-sized samples that were challenging to access before. For example, tr-ARPES experiments \cite{madeo2020directly,karni2022structure,paraeek2024driving} have demonstrated the ability to probe dark excitons and moir\'e excitons in monolayer and heterobilayer transition metal dichalcogenides, respectively, whereas they were hard to probe by traditional photon-absorption or photon-emission spectroscopy. As illustrated in the bottom panel of Fig.~\ref{fig:intro}d, only the occupied states at the valence band top can be visualized before the optical pump. With resonant photoexcitation at an exciton energy, photoemission signals from excitons can be observed upon the arrival of the pump pulses \cite{madeo2020directly}. The exciton photoemission signals feature an energy-momentum distribution centered below the conduction band bottom with a gap of binding energy $E_{\rm{B}}$ (green dispersion line), in agreement with theoretical expectations \cite{prefetto2016first, steinhoff2017exciton, rustagi2018photoemission, rustagi2019coherent, christiansen2019theory}.

\subsection{Time- and space-resolved probes}

Time-resolved spectroscopic and diffraction methods introduced thus far can only provide spatially-averaged information inside the probed region, whereas in general ultrafast dynamics may vary in different regions due to different local environments. Therefore, techniques with both spatial and temporal resolutions play an important role in studying inhomogeneous effects and near-field distribution during ultrafast light-matter interactions. Techniques such as ultrafast electron microscopy \cite{zewail20094d, zwail2010fourdimentional}, photoinduced near-field electron microscopy (PINEM) \cite{barwick2009photon, liu2016infrared}, coherent correlated X-ray imaging \cite{buttner2021observation,klose2023coherent,johnson2023ultrafast}, and lightwave-driven terahertz scanning tunneling microscopy \cite{jelic2024atomic,sheng2024terahertz} have all shown great potential in studying heterogeneity and spatially-varying dynamics present in quantum materials. 

As an example to illustrate the relation to collective excitations discussed in this article, ultrafast PINEM exhibits exceptional capabilities in studying the evolution of light-induced plasmonic field distribution \cite{barwick2015photonics..pinem} (see Fig.~\ref{fig:intro}e), and we will return to specific examples in Sec.~\ref{sec:other-couplings}. PINEM is based on ultrafast electron microscopy, which typically uses elastically scattered electrons for imaging. In the PINEM mode, inelastically scattered electrons are used instead, where electrons can absorb and emit spatiotemporally overlapped photons and produce discrete peaks with a constant energy interval in the electron energy loss spectrum. Using the energy-filtered electrons for imaging, one can obtain information about dressed photons such as the near-field distribution of surface plasmon-polaritons \cite{piazza2015simultaneous} and propagation of phonon-polaritons \cite{kurman2021spatiotemporal}. By focusing pump and probe pulses at distinct points and spatially scanning the probe electron pulse, the near-field distributions with space, time, and energy resolution can also be obtained \cite{yurtsever2012subparticle}.\\

The techniques introduced in this section are by no means exhaustive, and here we refer readers to other excellent topical reviews on different classes of ultrafast probes \cite{orenstein2012ultrafast,boven2012elementary,woerner2013ultrafast,giannetti2016ultrafastoptical, smallwood2016ultrafast, zhou2018newdevelopments,buzzi2018probing,cao2019ultrafast, mitrano2020probing,mitrano2024exploring, llloyd2021the2021ultrafast, zong2021unconventional, huang2022highresilution,filippetto2022ultrafastelectron, zong2023emerging, dong2023recentdevelop, lee2024structural, boschini2024timeresolvedarpes}. 

\begin{figure*}[!bt]
\centering
\includegraphics[width=\textwidth]{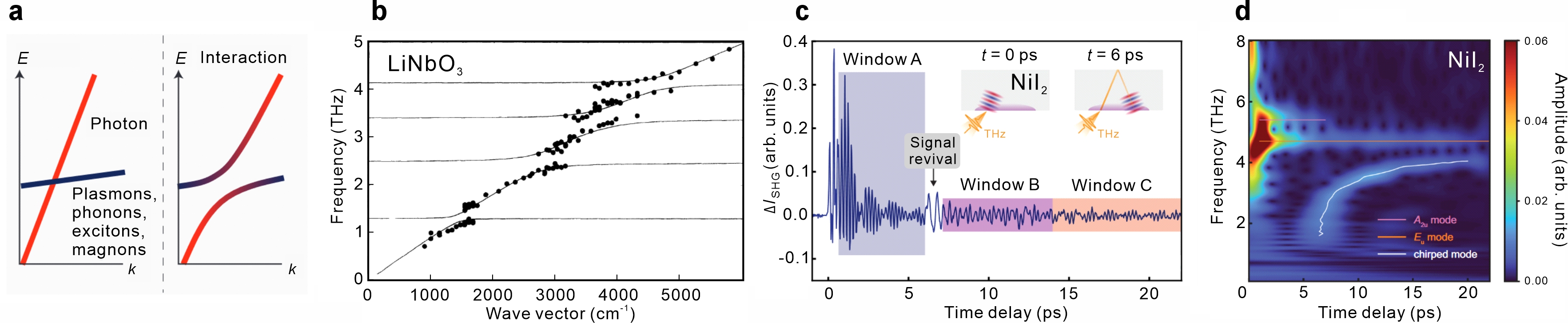}
\caption{\textbf{Measurements of phonon-polariton dispersions.} \textbf{a},~Sketch of the energy-momentum dispersion when photons are dressed with magnons, phonons, excitons, and plasmons, resulting in an avoided crossing. \textbf{b},~The phonon-polariton dispersion of ferroelectric LiNbO$_3$ at 300~K measured via four-wave mixing. Circular markers are measured data while curves are the calculated dispersion. \textbf{c},~Evolution of THz field-induced SHG signal of 800-nm probe pulses in NiI$_2$; see text for a discussion of the dynamics in the three windows A to C. The insets illustrate the schematics of the dominant signal contribution before and after the phonon-polaritons are reflected by the back sample surface. The gray rectangle represents the sample, the orange pulse corresponds to the THz pump, and the purple-shaded region denotes the probing area. The periodic waves in red and blue represent the excited phonon-polaritons. \textbf{d},~Wavelet transform spectrum of the SHG signal in \textbf{c}. Infrared-active $E_u$ (4.7~THz) and $A_{2u}$ (5.3~THz) modes are observed, which are highlighted by the orange and pink lines. A chirped mode appears after 6~ps, indicated by the white curve, and it was assigned to the phonon-polariton wavepacket. Panel~\textbf{a} adapted from ref.~\onlinecite{basov2017towards},~Springer Nature Ltd. Panel~\textbf{b} adapted from ref.~\onlinecite{bakker1994investigation},~American Physical Society. Panels~\textbf{c} and \textbf{d} adapted from ref.~\onlinecite{luo2024time},~\href{https://creativecommons.org/licenses/by/4.0/}{CC~BY~4.0}.} 
\label{fig:pola_spect}
\end{figure*}

\section{Couplings between collective excitations at nonequilibrium\label{sec:couplings}}

\subsection{Dressing photons with phonons: phonon-polariton}\label{sec:phonon-polariton}

When solids are exposed to external electromagnetic waves, many dressed states or hybridized quasiparticles emerge. Polaritons are those effects that arise from a strong light-matter interaction between photons and dipole-active excitations. Polaritons can be of various types \cite{basov2016polaritons,basov2020polariton} depending on which excitations they hybridize with. For instance, light can hybridize with magnons, phonons, excitons, and plasmons, forming magnon-polaritons, phonon-polaritons, exciton-polaritons, and plasmon-polaritons (Fig.~\ref{fig:intro}). One hallmark feature of these interactions is the appearance of an avoided crossing between the dispersions of two quasiparticles \cite{basov2017towards}, as illustrated in Fig.~\ref{fig:pola_spect}a. The size of the gap is a measure of the interaction strength, where a larger gap suggests a stronger interaction.

Here we focus on recent ultrafast studies of phonon-polaritons, which arise from the interaction between lattice vibrations and electromagnetic waves. When light passes through a crystal, it can drive a transverse optical phonon mode if the frequency of the electromagnetic wave resonates with the phonon, which leads to dipole oscillations and consequent re-radiation of the electromagnetic wave. The re-radiated waves can further excite the dipolar vibrations, resulting in the propagation of hybridized light and phonons, namely, phonon-polaritons. As the hybridized quasiparticles inherit the attributes from both light and phonons, their frequencies lie in the frequency range between gigahertz and tens of terahertz, showing great potential for applications in THz spectroscopy, THz imaging, and high-bandwidth signal processing \cite{Feurer2007Polaritonics}.

Many early ultrafast experiments of phonon-polaritons focused on the study of ferroelectrics \cite{bakker1998coherent}, in which photoexcited coherent phonon-polaritons were used as a probe for low-frequency anharmonic phonons that play a crucial role in driving structural and ferroelectric phase transition near the critical point. Raman spectroscopy was first used to measure the frequency of phonon-polaritons \cite{henry1965raman} but lacked phase information of their wavepacket. With the development of femtosecond lasers, ultrafast optical spectroscopy with spatially distinct pump and probe techniques has enabled us to directly measure the phonon-polaritons waveform in the time domain. In one experimental scheme, broadband polar phonon-polariton modes are produced by illuminating the surface of a ferroelectric material with a femtosecond laser pulse. These modes are generated from impulsive stimulated Raman scattering (ISRS) through mixing different frequency components contained within the pulse bandwidth \cite{cheung1985excitation,yan1985impulsive}. However, this method does not generate phonon-polariton modes with a well-defined frequency. To solve this problem, a four-wave mixing method has been developed \cite{planken1992femtosecond, bakker1994investigation,koehl1999direct,koehl1999real}. In this configuration, two identical, spatiotemporally overlapped femtosecond pulses are used to launch the phonon-polaritons in a nonlinear crystal. The two pump pulses form a transient grating due to interference, and they selectively excite phonon-polaritons with a well-defined scattering wavevector that  can be tuned by varying the incidence angle between the pump pulses. As the photoexcited phonon-polariton field modulates the index of refraction that in turn affects the scattering of a separate probe pulse, the periodic probe signal can reveal the presence of phonon-polaritons in a phase-resolved manner. Figure~\ref{fig:pola_spect}b shows the measured dispersion relation of phonon-polaritons in LiNbO$_3$ using this four-wave mixing scheme \cite{bakker1994investigation}, where five branches and four avoided crossings were observed, showing strong couplings between the photons and phonons in this material.

Experiments using four-wave mixing have some limitations: ISRS can only excite Raman-active modes \cite{yan1985impulsive}, while to the lowest order only IR-active phonon modes can couple with electromagnetic waves to form phonon-polaritons. Therefore, this scheme is only feasible for noncentrosymmetric materials in which IR phonons are also Raman-active, such as in LiNbO$_3$ and LiTaO$_3$. In addition, these experiments need careful adjustment of the geometry of the two pump beams in order to excite phonon-polaritons with a selected wavevector. An alternative method that overcomes these limitations is time-domain THz spectroscopy \cite{kojima2003far}, where phonon-polaritons are excited by a single beam of THz radiation instead of two beams of visible or near-infrared light pulses. However, in a typical time-domain THz experiment that measures phonon-polariton dispersion, a reference crystal is required for accurate electro-optic sampling to retrieve the phase information across the spectrum of the THz pulse, hence complicating the data analysis and interpretation.

Recently, Luo and coworkers developed a time-of-flight detection method that overcame these challenges and was applied to probing phonon-polaritons in van der Waals magnetic materials \cite{luo2024time}, including NiI$_2$ and MnPS$_3$. In this scheme, a pulse centered around 3~THz was used to excite phonon-polaritons while the SHG signal of a 800-nm laser pulse was used as a probe to capture phonon-polaritons before or after they were reflected from the back surface of the sample. More specifically, when the THz pump pulse arrived at the sample surface, multiple oscillatory responses due to phonons and magnons were observed (Fig.~\ref{fig:pola_spect}c, window~A). After the initial decay of the oscillations, there was a sudden amplitude revival at around 6~ps, suggesting a reflection of the propagating phonon-polariton wavepacket from the sample back surface (Fig.~\ref{fig:pola_spect}c, insets). After this revival, the signal frequency was also observed to increase towards longer time delay (Fig.~\ref{fig:pola_spect}c, windows~B and C). A clearer view of this phenomenon can be seen in the wavelet transform spectrum in Fig.~\ref{fig:pola_spect}d, which shows the time-dependent signal frequencies. Around time zero, as defined by the THz pulse incidence, the excitation spectrum and the second harmonic frequencies of the THz pump can be seen, including the photoexcited $E_u$ and $A_{2u}$ phonon modes. After 6~ps, a chirped mode emerges, whose frequency increases from 2 to 4~THz over, which corresponds to the phonon-polariton wavepacket. Given the known sample thickness, the time-dependent frequency spectrum in Fig.~\ref{fig:pola_spect}d can yield the group velocity at different frequencies, which in turn helps reconstruct the energy-momentum dispersion of the phonon-polariton. Furthermore, the time-dependent attenuation of the phonon-polariton during its propagation allows one to compute the imaginary part of the dispersion, whose divergence near the phonon resonance conforms to the expectation that light quickly dissipates its energy through the phonon mode.

\begin{figure*}[!htbp]
\centering
\includegraphics[width=1\textwidth]{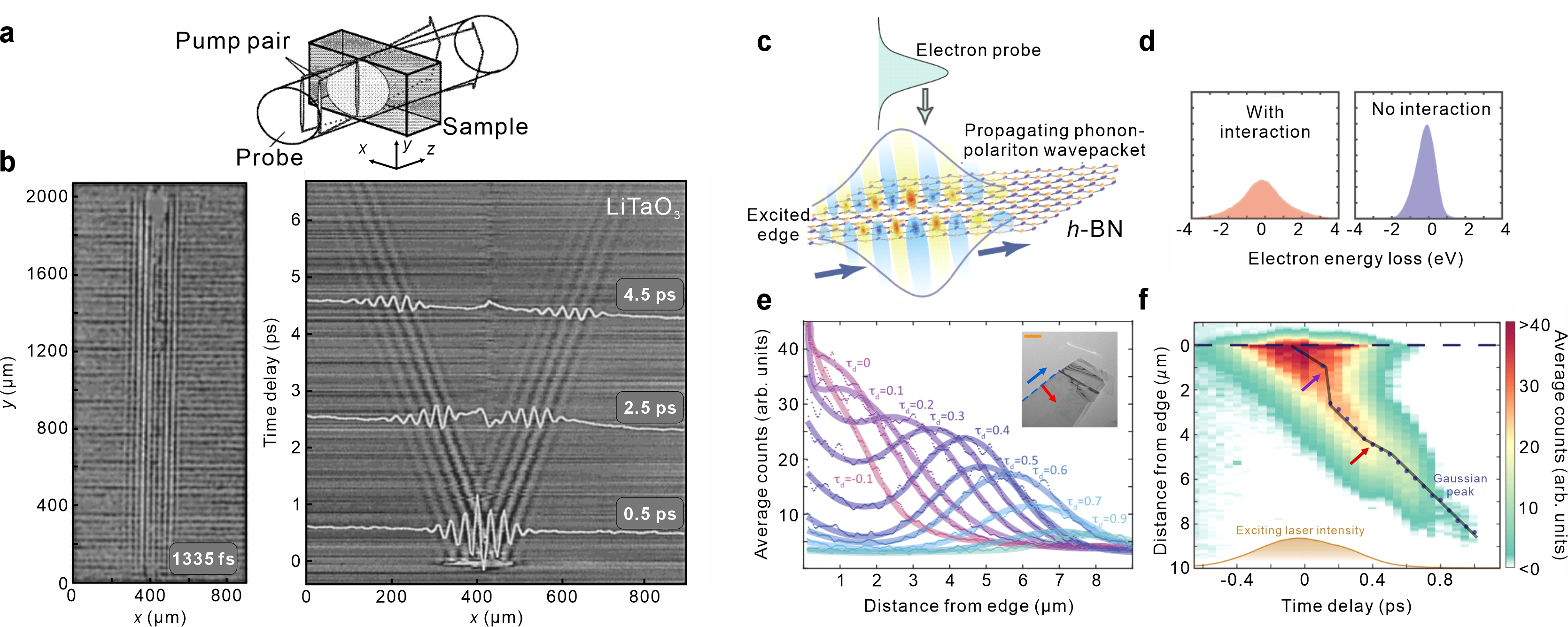}
\caption{\textbf{Phonon-polariton imaging using ultrafast optical or electron microscopy.} \textbf{a},~Experimental geometry for dynamical phonon-polariton imaging involving a pair of anisotropically-shaped pump pulses and a isotropically-shaped probe pulse. \textbf{b},~\textit{Left~panel}:~Snapshot of an optical micrograph of phonon-polaritons in LiTaO$_3$. \textit{Right~panel}:~2D spatiotemporal mapping of phonon-polariton propagation after photoexcitation. \textbf{c},~Schematic of phonon-polariton generation and free electron probing of its propagation inside a $h$-BN flake. \textbf{d},~Electron energy loss spectra with (left) or without (right) interactions with the phonon-polariton wavepacket. \textbf{e},~Measurement of the energy-filtered electrons for different time delays $\tau_d$ expressed in picoseconds between the laser pump and the electron probe. The experimental signals (dots) are averaged along the edge direction (indicated by the blue arrow in the inset). Solid curves are fits to a Gaussian profile plus an exponential decay. \textit{Inset}: A bright-field image of the $h$-BN flake; scale bar, 5~$\upmu$m. \textbf{f},~2D mapping of phonon-polariton wavepacket propagation as a function of time and distance from the edge. Blue dots indicate the positions of Gaussian wavepacket peaks, and the orange curve represents the temporal profile of the laser excitation. The gray translucent line is a guide to the eye. The phonon-polariton group velocity undergoes a sudden acceleration followed by deceleration, which are indicated by the purple and red arrows, respectively. The measurements in \textbf{e} and \textbf{f} used a 55-nm-thick $h$-BN flake that was excited by a 6470-nm laser with a bandwidth of 175~nm. Panels~\textbf{a} and \textbf{b} adapted from ref.~\onlinecite{koehl1999real},~AIP Publishing. Panels~\textbf{c}--\textbf{f} adapted from ref.~\onlinecite{kurman2021spatiotemporal},~AAAS.}  
\label{fig:polariton_microscopy}
\end{figure*}

While the aforementioned experiments can detect the presence and dispersion of phonon-polariton wavepackets in the probed volume, they do not reveal the spatial structure of these hybridized quasiparticles. To image the spatiotemporal propagation of phonon-polaritons, ultrafast optical or electron microscopy can be used. Nelson and colleagues first demonstrated direct imaging of optical phonon-polariton wavepackets in LiTaO$_3$ using ultrafast optical microscopy that leveraged four-wave mixing \cite{koehl1999direct,koehl1999real}. In one of the experiments, a pump pulse with an anisotropic spatial profile was applied (see Fig.~\ref{fig:polariton_microscopy}a), which excited phonon-polaritons that propagated mainly along the horizontal direction of the sample \cite{koehl1999direct}. The spot size of the probe pulse was significantly larger than the excitation region, so spatial information of the propagation dynamics was readily acquired. The left panel of Fig.~\ref{fig:polariton_microscopy}b shows a representative snapshot of the wavepacket distribution at 1335~fs after the pump incidence. After averaging over the vertical direction in each image captured at different time delays, the spatiotemporal propagation of phonon-polaritons in LaTiO$_3$ was obtained in the right panel of Fig.~\ref{fig:polariton_microscopy}b. Under this experimental framework, the phonon-polariton response can be coherently controlled via spatiotemporal shaping of the optical pump pulses \cite{Feurer2007Polaritonics}. For a more comprehensive survey on the coherent control of phonon-polaritons and THz polaritonics, we refer readers to ref.~\onlinecite{Feurer2007Polaritonics}.

Apart from optical probes, electrons have been successfully deployed to image phonon-polariton propagation dynamics in van der Waals materials \cite{kurman2021spatiotemporal}, leveraging the spatial and spectral information of PINEM. As shown in Fig.~\ref{fig:polariton_microscopy}c, an optical pump pulse first illuminates the edge of a hexagonal boron nitride ($h$-BN) and launches multi-branch phonon-polariton wavepackets. The probing free electrons interact with the propagating phonon-polaritons, broadening the electron energy loss profile due to inelastic scattering (Fig.~\ref{fig:polariton_microscopy}d). In this experiment by Kurman \textit{et al.}, it should be noted that the energy of each phonon-polariton is smaller than the spectral resolution of the instrument. Hence, instead of showing discrete peaks drawn in the schematic in Fig.~\ref{fig:intro}e, the electron energy loss spectrum profile is broadened. Figure~\ref{fig:polariton_microscopy}e shows the averaged energy-filtered electron counts along the edge direction (blue arrow in the inset of Fig.~\ref{fig:polariton_microscopy}e) as a function of distance from the edge (red arrow in the inset of Fig.~\ref{fig:polariton_microscopy}e), indicating the field profile of phonon-polaritons along the propagation direction. By studying the distance-time relation of the traveling wavepackets (Fig.~\ref{fig:polariton_microscopy}f), one can observe both acceleration (purple arrow) and deceleration (red arrow) during the propagation of the photoexcited phonon-polaritons. The observed acceleration and deceleration result from the excitation of different spectral components at different times during the pump-$h$-BN interaction. Interestingly, Kurman \textit{et al.} also showed that the accelerating and decelerating propagation can be modified by changing the exciting photon wavelength and sample thickness \cite{kurman2021spatiotemporal}.

Note that the two featured examples of spatiotemporal imaging differ in the launching mechanism of the phonon-polaritons. While ultrafast optical microscopy used two near-infrared pulses to form a transient grating to excite phonon-polaritons based on the ISRS mechanism \cite{koehl1999direct}, the PINEM experiment directly used a mid-IR pulse to coherently excite them \cite{kurman2021spatiotemporal}. Hence, the former methodology was deployed to noncentrosymmetric materials while the latter was used to directly excite dipole-active IR modes to launch the phonon-polaritons. In both cases, the additional spatial information has deepened our understanding of how phonon-polaritons propagate in materials, which will be especially helpful when one considers spatially heterogeneous samples in a real device setting \cite{Feurer2007Polaritonics,alfaro2019deeply,fotei2019phonon}.

\subsection{Phonon-phonon coupling}\label{sec:phonon-phonon}

\begin{figure}[!tbp]
\centering
\includegraphics[width=\columnwidth]{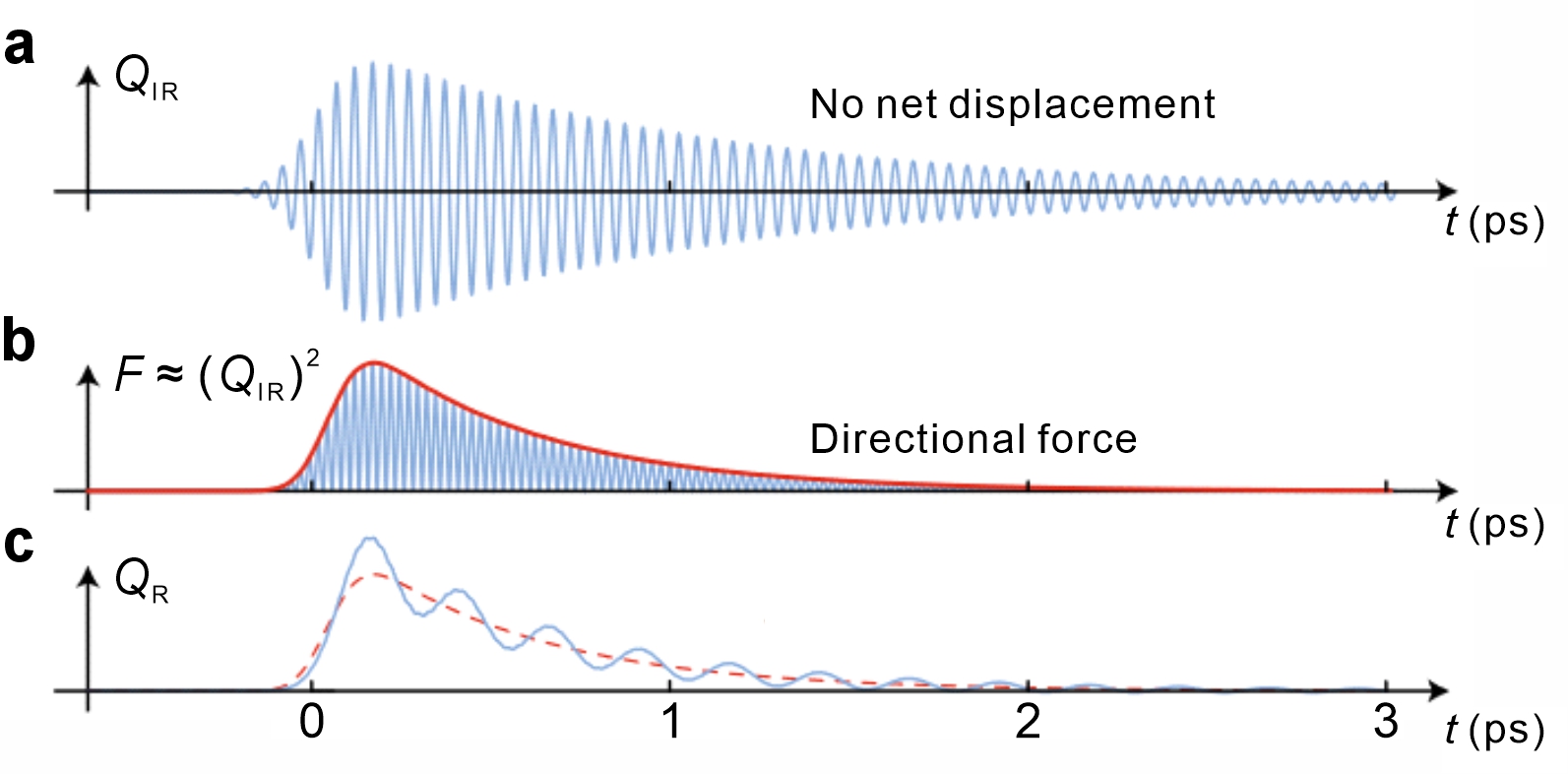}
\caption{\textbf{Schematic of a nonlinear phononic process.} \textbf{a},~Time-dependent displacement of a resonantly-driven IR phonon excited by an ultrashort, intense laser pulse in the mid-infrared to terahertz frequencies. \textbf{b},~The resulting force on the crystal lattice induced by the IR phonon driven to a large amplitude. \textbf{c},~The displacement of a coupled Raman mode that results in a net structural distortion as a function of time. Adapted from ref.~\onlinecite{disa2021engineering},~Springer Nature Ltd.} 
\label{fig:nonlinear_phonon_schematic}
\end{figure}

The lattice structure often dictates the electronic, optical, and mechanical properties of quantum materials. The collective excitations of atomic vibrations, or phonons, hence play an important role in determining these physical attributes as well. Under equilibrium conditions, phonons and their couplings can be characterized by techniques such as IR spectroscopy \cite{stuart2004infrared}, Raman spectroscopy \cite{gardiner1989introduction}, and inelastic neutron, X-ray, or electron scatterings \cite{burkel2000phonon}. Though these static methods are powerful, a lack of precise control over particular phonon modes impedes us from revealing how individual phonon modes couple to other quasiparticles in solids. In this regard, ultrafast techniques allow us to tackle this challenge by extending the measurement into the time domain. For example, advances in generating intense mid-IR and THz pulses with either broadband or tunable narrow-band frequencies have been instrumental in the emergence of the field of \emph{nonlinear phononics} \cite{forst2011nonlinearphononics, disa2021engineering}. By selectively exciting particular phonon modes using tailored low-photon-energy pulses, we can study phonon-phonon couplings and engineer crystal structure in a nonequilibrium state.

\begin{figure*}[!htbp]
\centering
\includegraphics[width=\textwidth]{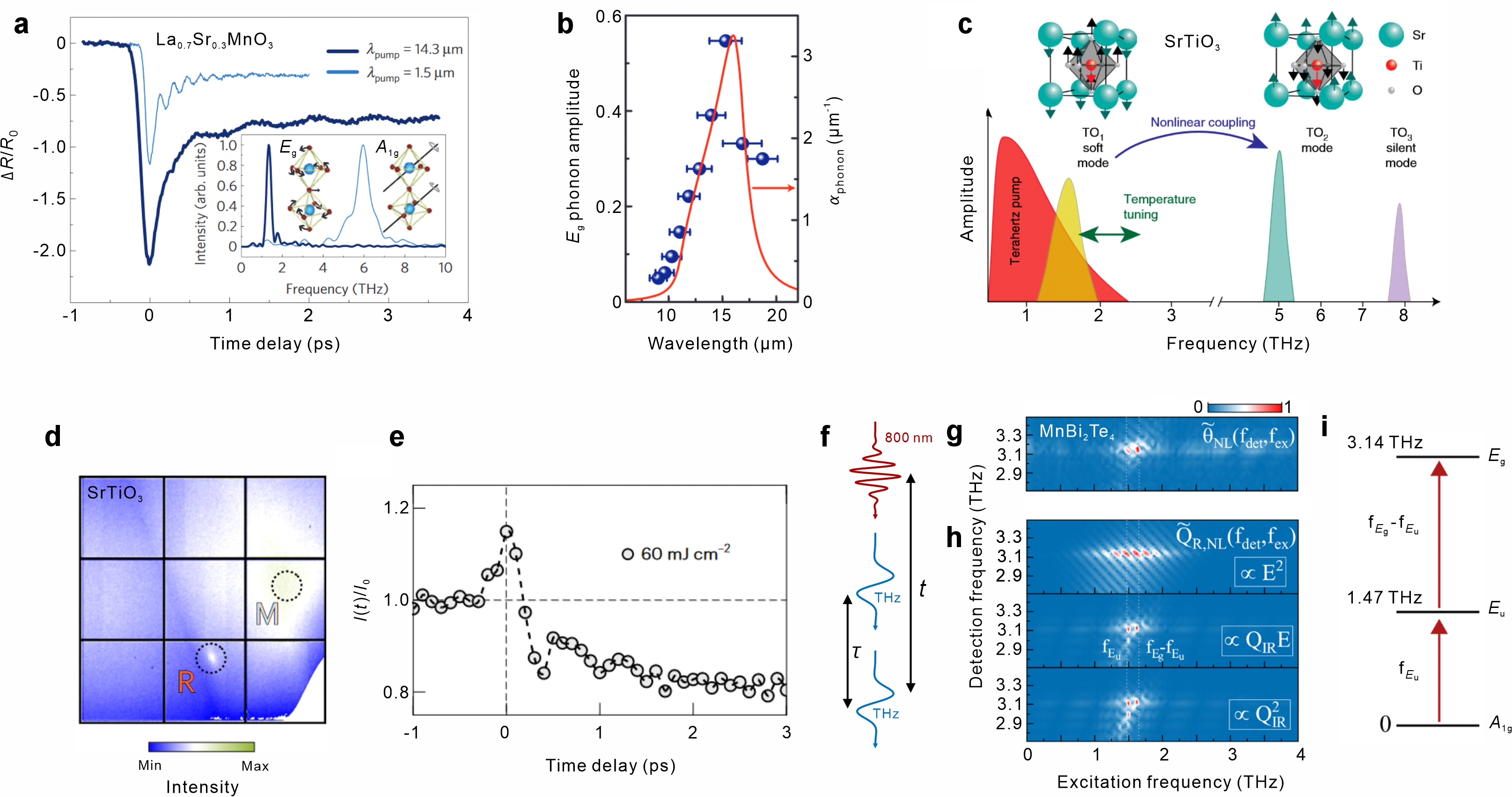}
\caption{\textbf{Nonlinear phononics demonstrated via time-resolved optical spectroscopy, X-ray scattering, and 2D terahertz spectroscopy.} \textbf{a},~Transient reflectivity changes of La$_{0.7}$Sr$_{0.3}$MnO$_3$ at 800~nm for two different pump wavelengths, 1.5~$\upmu$m and 14.3~$\upmu$m. \textit{Inset}:~Fourier transform spectra of the oscillatory component and schematics of the corresponding phonons. \textbf{b},~Amplitude of the coherently-driven $E_g$ phonon at different pump wavelengths, indicating a resonant enhancement at the mode frequency in \textbf{a}. Horizontal error bars indicate the bandwidths of the mid-IR pump pulses. The red curve is the calculated linear absorption due to the IR-active $E_\textrm{u}$ phonon. \textbf{c},~Terahertz-driven phonon up-conversion in SrTiO$_3$. When the soft transverse optical (TO) phonon (yellow, TO$_1$) is resonantly driven by a strong terahertz pulse (red), energy is exchanged with higher-frequency phonon modes through nonlinear phononic couplings, resulting in the TO$_2$ (turquoise) and  TO$_3$ (purple) mode. A schematic of the two lowest-frequency zone-center TO modes are indicated at the top. \textbf{d},~Equilibrium X-ray intensity of SrTiO$_3$ at 135~K, where selected high-symmetry points are labeled, including $R$ (1/2~1/2~1/2) and $M$ (1/2~1/2~0). The $R$ point hosts the antiferrodistortive fluctuations of the cubic-to-tetragonal phase transition at 110~K. \textbf{e},~Time-resolved changes in the X-ray diffuse scattering intensity at the $R$ point after photoexcitation by a mid-IR pulse with a fluence of 60~mJ/cm$^2$. \textbf{f}--\textbf{i},~2D terahertz spectroscopy of nonlinear phononics in MnBi$_2$Te$_4$. \textbf{f},~Schematic diagram of the measurement. \textbf{g},~Normalized 2D fast Fourier transform (FFT) of the nonlinear response, $\widetilde{\theta}_{\rm{NL}}(f_{\rm{ex}}, f_{\rm{det}})$; $f_{\rm{det}}$ and $f_{\rm{ex}}$ denote the detection frequency and excitation frequency, respectively. \textbf{h},~Normalized 2D~FFT of the simulated nonlinear response, $\widetilde{\theta}_{\rm{R, NL}}(f_{\rm{ex}}, f_{\rm{det}})$, modeled by three different mechanisms: photonic ($\propto E^2$), combined photophononic ($\propto Q_{\rm{IR}}E$), and phononic ($\propto Q_{\rm{IR}}^2$), where $E$ is the electric field of the terahertz excitation pulse, $Q_{\rm{IR}}$ is the normal coordinate of the IR-active mode $E_u$, and $Q_{\rm{R}}$ is the normal coordinate of the Raman-active mode $E_g$. The photophononic scenario shows the best agreement, which confirms that the excitation of the Raman-active $E_g$ phonon is mediated by the IR-active $E_u$ phonon via the photophononic mechanism. $f_{E_g}$ and $f_{E_u}$ are 3.14~THz and 1.47~THz, respectively. \textbf{i},~Schematic illustration of the excitation mechanism in a model with two oscillators, corresponding to the $E_u$ and $E_g$ phonons. The arrows indicate the stimulated transitions. Panels~\textbf{a} and \textbf{b} adapted from ref.~\onlinecite{forst2011nonlinearphononics},~Springer Nature Ltd. Panel~\textbf{c} adapted from ref.~\onlinecite{kozina2019terahertz},~Springer Nature Ltd. Panels~\textbf{d} and \textbf{e} adapted from ref.~\onlinecite{fechner2024quenchedlattice},~\href{https://creativecommons.org/licenses/by/4.0/}{CC~BY~4.0}. Panels~\textbf{g}--\textbf{i} adapted from ref.~\onlinecite{blank2023twodimensionaltera},~American Physical Society.}
\label{fig:nonlinear_phonon_example}
\end{figure*}

The first experimental demonstration of nonlinear phononics was reported in 2011 by F\"orst and coworkers \cite{forst2011nonlinearphononics}. After the $E_u$ IR mode in a bulk single crystal of La$_{0.7}$Sr$_{0.3}$MnO$_3$ was excited by 14.3-$\upmu$m light pulses, the reflectivity exhibited oscillations at 1.2~THz, as shown by the dark curve in Fig.~\ref{fig:nonlinear_phonon_example}a. The oscillations were attributed to the $E_g$ Raman mode that is associated with oxygen octahedral rotation in this perovskite. Moreover, the amplitude of the $E_g$ mode showed a strong pump wavelength dependence and reached a maximum when it was resonant with the $E_u$ mode (see Fig.~\ref{fig:nonlinear_phonon_example}b). This intriguing phenomenon can be described by an anharmonic coupling between the IR-active $E_u$ mode and the Raman-active $E_g$ mode, as illustrated in Fig.~\ref{fig:nonlinear_phonon_schematic}. During this process, a mid-IR light pulse resonantly couples to the dipole of the IR-active phonon and sets off coherent oscillations of atoms around their equilibrium positions (Fig.~\ref{fig:nonlinear_phonon_schematic}a). The anharmonic coupling term ($V_\text{anh} \propto Q^2_{\rm{IR}}Q_{\rm{R}}$) applies a directional force (Fig.~\ref{fig:nonlinear_phonon_schematic}b) on the Raman mode (Fig.~\ref{fig:nonlinear_phonon_schematic}c), thus driving its oscillations. For a more detailed description of this mechanism such as symmetry considerations during anharmonic interactions, we refer readers to refs.~\cite{radaelli2018breaking, disa2021engineering}. In contrast to the resonant excitation condition, 5.8-THz oscillations were observed when La$_{0.7}$Sr$_{0.3}$MnO$_3$ was photoexcited by a 1.5-$\upmu$m pulse (Fig.~\ref{fig:nonlinear_phonon_example}a, light curve). This phonon mode was attributed to the $A_{1g}$ mode, which was induced by the more conventional mechanism involving a displacive excitation of coherent phonons \cite{zeiger1992theory}. 

This work by F\"orst \textit{et al.} inspired ideas to engineer crystals using resonant mid-IR or terahertz excitations. This nonlinear phononic mechanism has since been utilized to manipulate properties in many perovskite systems, such as controlling an anisotropic strain wave in LaAlO$_3$ (ref.~\cite{hortensius2020ultrafast}), as well as changing ferroelectric polarizations in systems such as LiNbO$_3$ (refs.~\cite{mankowsky2017ultrafast,henstridge2022nonlocal}), SrTiO$_3$ (refs.~\cite{li2019terahertz,nova2019metastable,fechner2024quenchedlattice}), and BiFeO$_3$ (ref.~\cite{lopez2023ultrafast}). Besides the effect on the lattice and hence coupled polar orders, nonlinear phononics has also been shown to influence electronic and magnetic properties \cite{disa2021engineering}. A particularly intriguing case concerns the apparent resemblance to a transient superconducting state when IR-active phonons of specific wavelengths are driven to a large amplitude by resonant excitation \cite{fausti2011light,mankowsky2014nonlinear,mankowsky2017optically,kaiser2017light,liu2020pump,fava2024magnetic}. However, other works have suggested that certain superconducting-like behavior may be a manifestation of the excitation of quasiparticles with a low scattering rate \cite{zhang2020photoinduced, zhang2024light}. Therefore, care needs to be taken in processing the transient optical data for drawing precise conclusions about the nonequilibrium state \cite{dodge2023optical,dodge2023status,buzzi2023comment}. This potential of light-induced superconductivity remains an intensely studied area with many interesting, open questions to be answered.

Compared with ultrafast optical spectroscopy, time-resolved X-ray diffraction can directly quantify the changes in a crystal structure, thus offering a powerful alternative to study phonon-phonon couplings \cite{buzzi2018probing,lee2024structural}. Here, we highlight several experimental works on SrTiO$_3$, a quantum paraelectric material where the competition between ferroelectric and antiferrodistortive instabilities prevents it from entering a ferroelectric phase at low temperature. Using THz pump pulses, the direct control of its ferroelectric soft phonon mode has been intensely studied and several nonlinear phononic effects have been observed \cite{li2019terahertz,nova2019metastable,fechner2024quenchedlattice}. For example, phonon up-conversion in SrTiO$_3$ was demonstrated in a THz pump, X-ray diffraction probe experiment \cite{kozina2019terahertz}. Kozina \textit{et al.} used an intense THz pump pulse with a spectral range of 0.2--2.5~THz (red region in Fig.~\ref{fig:nonlinear_phonon_example}c), which overlapped with the IR-active TO$_1$ soft mode in SrTiO$_3$ at 100~K (the yellow peak). Interestingly, the oscillatory Bragg peak intensity observed in X-ray diffraction not only showed the frequency of the driving THz field but also exhibited several high frequencies that correspond to the TO$_2$ and TO$_3$ modes (turquoise and purple peaks in Fig.~\ref{fig:nonlinear_phonon_example}c). When temperature increased such that the hardened TO$_1$ mode fell outside the THz excitation spectrum, no signatures of the TO$_2$ and TO$_3$ modes were seen, providing compelling evidence for phonon frequency up-conversion from the soft TO$_1$ mode to these high-frequency modes. 

Beyond the proof-of-principle demonstration by Kozina \textit{et al.}, nonlinear phononic effects have also been used to induce ferroelectricity in unstrained SrTiO$_3$ under ambient pressure. In 2019, Li and coworkers reported that SrTiO$_3$ can be driven to a transient ferroelectric phase via intense THz pumping as evidenced by the second harmonic signal of a 800-nm probe pulse \cite{li2019terahertz}, although later experiments performed on a cousin compound KTaO$_3$ suggested that the THz-induced transient second-harmonic signal may come from defect-induced local polar structures without any long-range ferroelectric ordering \cite{cheng2023terahertz}. A contemporary work in 2019 reported a metastable ferroelectric phase of SrTiO$_3$, which was induced by a mid-IR pump pulse instead of a THz pulse \cite{nova2019metastable}. In this experiment, Nova and coworkers resonantly drove the high-frequency IR-active $A_{2u}$ mode in SrTiO$_3$ at 4~K, and detected a metastable noncentrosymmetric phase via second-harmonic spectroscopy. The results were interpreted via anharmonic phonon-phonon interactions between the pulse-induced IR-active mode and acoustic strain, the latter of which suppressed the antiferrodistortive distortion and hence stabilized the ferroelectric phase. This mechanism was further clarified by a subsequent work using time-resolved X-ray diffusing scattering \cite{fechner2024quenchedlattice}. The antiferrodistortive fluctuations can lead to a significant diffuse scattering intensity at the $R$ point in the Brillouin zone, as shown in the equilibrium X-ray detector image in Fig.~\ref{fig:nonlinear_phonon_example}d. At 135~K, upon photoexcitation with a 17-THz pulse that resonantly drives the IR-active mode, the X-ray scattering intensity at $R$ shows a sudden increase followed by a decay (Fig.~\ref{fig:nonlinear_phonon_example}e). The sudden intensity increase is a result of the enhancement of antiferrodistortive fluctuations induced by their anharmonic coupling with the resonantly excited high-frequency IR-active mode. The same IR-active mode can also couple to low-frequency acoustic modes, which induce a strain that dominates at long time delays ($> 0.5$~ps) and suppresses the antiferrodistortive fluctuations, resulting in the decrease of X-ray intensity at the $R$ point. Though the X-ray experiment was performed at 135~K, which is above the antiferrodistortive transition, the authors expect the same mechanism would remain valid to explain the metastable ferroelectric property of SrTiO$_3$ at 4~K, where a suppressed antiferro-distortion at long times leads to the emergence of a ferroelectric phase.

\begin{figure}[!htbp]
\centering
\includegraphics[width=0.3\textwidth]{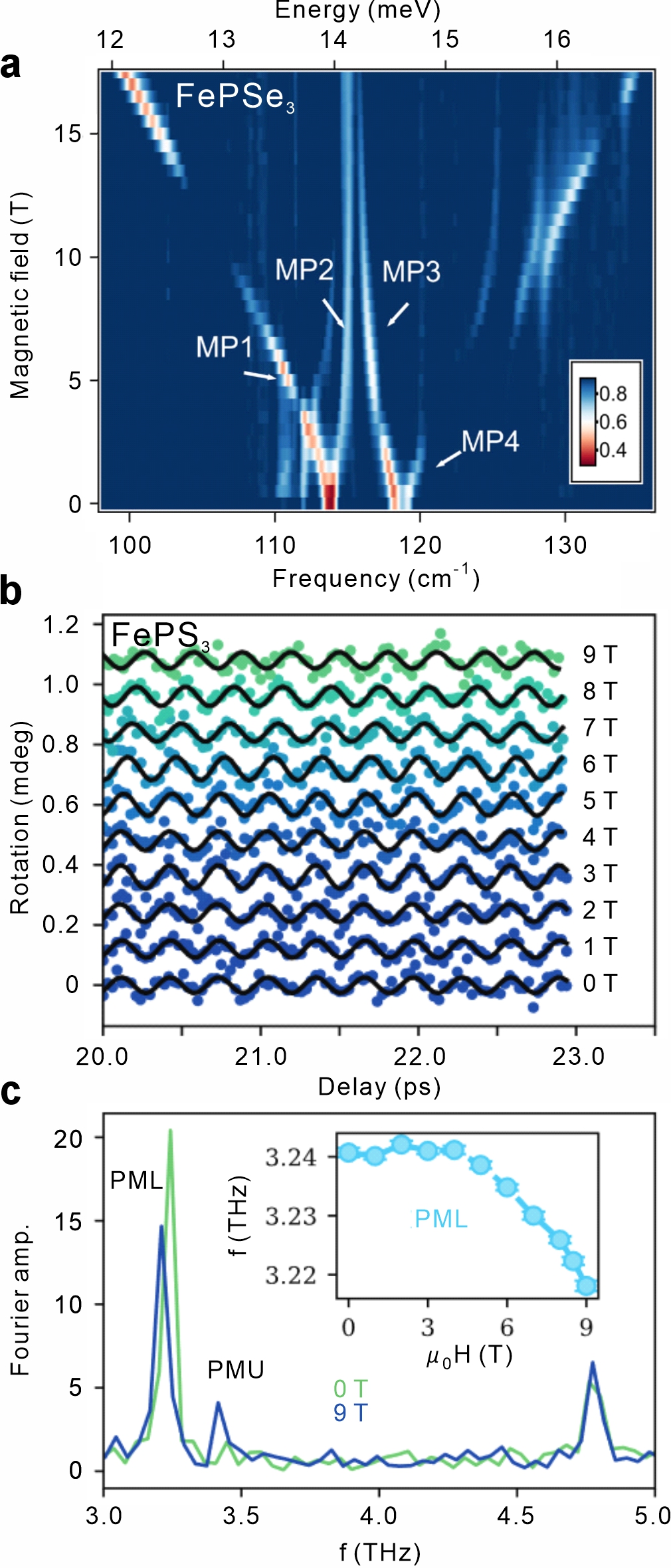}
\caption{\textbf{Phonon-magnon couplings in van der Waals magnets.} \textbf{a},~Normalized far-infrared transmission spectra of bulk FePSe$_3$ taken at 4.2~K, which is well below the antiferromagnetic transition temperature. Four magnon polaron (MP) branches are identified as MP1 to MP4. \textbf{b},~Magnetic field-dependent time evolution of the polarization rotation of a 1.45-eV probe pulse after photoexciting a FePS$_3$ sample in its antiferromagnetic state by a 1.03~eV pulse. \textbf{c},~Fast Fourier transform of the time traces in the absence of an external field (green) and at 9~T (blue). PML labels the lower branch of the phonon–magnon mode, which corresponds to the 3.2~THz phonon in the absence of an applied magnetic field. PMU refers to the upper branch of the phonon–magnon mode, which only appears under the 9~T applied field. \textit{Inset}:~Field-dependent frequency shift of the PML mode due to phonon-magnon hybridization. Panel~\textbf{a} adapted from ref.~\onlinecite{cui2023chirality},~\href{https://creativecommons.org/licenses/by/4.0/}{CC~BY~4.0}. Panels~\textbf{b} and \textbf{c} adapted from ref.~\onlinecite{mertens2023ultrafastcoherent},~\href{https://creativecommons.org/licenses/by/4.0/}{CC~BY~4.0}.}
\label{fig:magnon_phonon_2Dmagnet}
\end{figure}

Aside from the standard pump-probe scheme employed in transient optical spectroscopy and X-ray scattering, 2D~THz spectroscopy offers another route to determine the interaction pathway between different quasiparticles, showing particular relevance in revealing phonon-phonon coupling \cite{woerner2013ultrafast}. In typical 2D THz spectroscopy setups, two intense THz pump pulses are used to induce nonlinear responses with a time delay $\tau$ between the two pulses, as shown in Fig.~\ref{fig:nonlinear_phonon_example}f. A third optical or near-infrared probe pulse with time delay $t$ is then applied to record the optical response of the sample. By performing a 2D Fourier transform on the time-domain data with varying $\tau$ and $t$, spectra as a function of THz excitation frequency and detection frequency can be generated, yielding insights into the interaction process between different modes. As a recent example, 2D~THz spectroscopy was employed to reveal the energy-flow pathway in a topological antiferromagnet MnBi$_2$Te$_4$ (ref.~\cite{blank2023twodimensionaltera}). The normalized 2D Fourier transform spectrum of the nonlinear signal is shown in Fig.~\ref{fig:nonlinear_phonon_example}g, while Fig.~\ref{fig:nonlinear_phonon_example}h shows three simulated spectra based on the different coupling pathways of exciting the Raman-active phonon: two-photon absorption (``photonic''), nonlinear phononics (``phononic''), and a combination of the two (``photophononic'') . Careful comparison with the experimental observation in Fig.~\ref{fig:nonlinear_phonon_example}g pinpoints a photophononic pathway as the most probable scenario, where the THz electric field first excites the IR-active $E_u$ mode and then interacts with this mode to further excite the Raman-active $E_g$ mode, as summarized in Fig.~\ref{fig:nonlinear_phonon_example}i.

\begin{figure*}[!htbp]
\centering
\includegraphics[width=\textwidth]{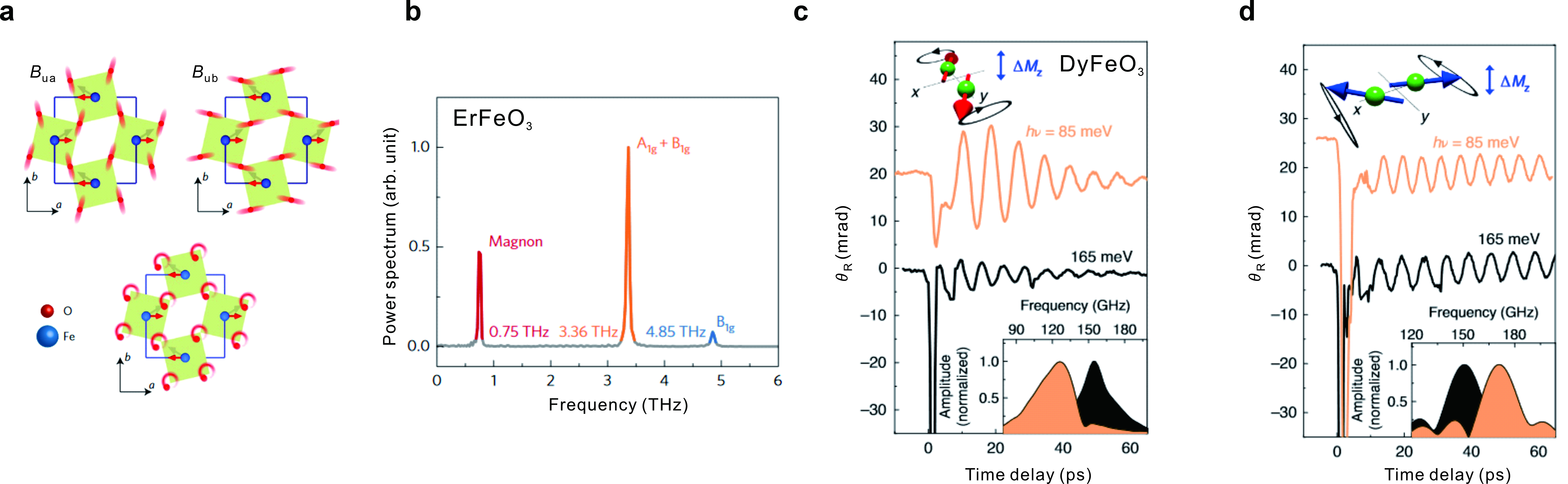}
\caption{\textbf{Time-domain probe of phonon-magnon couplings via phonon excitation.} \textbf{a},~\textit{Top}:~Schematic of the two in-plane phonon modes ($B_{ua}$ and $B_{ub}$) excited by a 20-THz pump pulse in ErFeO$_3$. \textit{Bottom}:~Schematic of the resulting elliptically polarized phononic field induced by mixing the two non-degenerate $B_{ua}$ and $B_{ub}$ modes. \textbf{b},~Power spectrum of the oscillatory component of the Faraday rotation signal, where the pump was a 20-THz mid-infrared pulse and the probe was a 800-nm near-infrared pulse. The ErFeO$_3$ sample was kept at 100~K. \textbf{c},\textbf{d},~Time-resolved polarization rotation $\theta_{\rm{R}}$ of the 800-nm probe pulse after photoexciting DyFeO$_3$ with a pump photon energy of 85~meV (orange) and 165~meV (black). The DyFeO$_3$ sample was in the antiferromagnetic phase (\textbf{c}) and weakly ferromagnetic (\textbf{d}) phase, respectively. \textit{Top insets}:~Schematics of the corresponding spin precessions and the resulting oscillating magnetic component $\Delta M_z$. \textit{Bottom insets}:~Fourier spectra of the oscillations that share the same color codes as the main panel. Panels~\textbf{a} and \textbf{b} adapted from ref.~\onlinecite{nova2017aneffective},~Springer Nature Ltd. Panels~\textbf{c} and \textbf{d} adapted from ref.~\onlinecite{afanasiev2021ultrafastcontrol},~Springer Nature Ltd.}
\label{fig:phonon_magnon_ultrafast}
\end{figure*}

\subsection{Phonon-magnon coupling}\label{sec:phonon-magnon}

Novel properties in quantum materials often arise from the coupling between the spin and lattice degrees of freedom, for example, the Einstein-de~Haas effect \cite{frenkel1979history}, the Barnett effect \cite{barnett1915magnetization,barnett1935gyromagnetic}, magnetostriction \cite{lee1955magnetostriction}, and piezomagnetism \cite{dzialoshinskii1958problem, borovik1994piezomagnetism}. In certain contexts, their strong coupling implies a unique spin configuration dictated by the crystalline lattice, and one prototypical example is the realization of quantum spin liquid predicted in a frustrated geometry such as in a triangular or kagome lattice \cite{broholm2020quantum}.

While effects due to strong spin-lattice coupling have been extensively investigated in thermal equilibrium \cite{hu2023spin}, ultrafast techniques allow us to probe the exact energy-flow pathway and reveal how angular momentum is transferred between the electron spin and the crystalline lattice. For instance, ferromagnetic systems are known to lose their net magnetization well within a picosecond when they are subjected to an intense ultrafast laser pulse \cite{beaurepaire1996ultrafast}, but how the conservation of total angular momentum is maintained at this timescale has been a long-standing puzzle. Recent works using time-resolved X-ray and electron diffraction offered some clues to this question \cite{dornes2019ultrafast,tauchert2022polarized}, where circularly-polarized high-frequency phonons were first excited to absorb the angular momentum, which was then transferred to low-frequency shear acoustic modes and a macroscopic rotation of the sample. Interestingly, acoustic shearing has been similarly reported during ultrafast demagnetization of a van der Waals antiferromagnet whose net spin angular momentum is zero \cite{zhou2022dynamical,zong2023spin,zhou2023ultrafast}, and a detailed description of the angular momentum flow in the antiferromagnetic case remains an open question.

Strong spin-lattice couplings can sometimes manifest in the form of magnon-phonon coupling, where hybridized magnon-phonon modes, magnon polarons, can readily be detected in equilibrium. Similar to phonon-polaritons discussed in Sec.~\ref{sec:phonon-polariton}, one characteristic feature of magnon polarons is the avoided crossing of energy levels of the constituent magnon and phonon modes. An example of avoided crossings is shown in Fig.~\ref{fig:magnon_phonon_2Dmagnet}a for a layered antiferromagnet FePSe$_3$, measured by magneto-infrared transmission spectroscopy \cite{cui2023chirality}. In the detected spectral window of FePSe$_3$, while the constituent phonons are individually chiral with opposite chiralities, their near degeneracy leads to no net chirality. However, the coupling to chiral magnons lifts such degeneracy, resulting in chiral magnon polarons that were demonstrated by magneto-Raman spectroscopy. The hybridization of magnon and phonon can also be directly measured in the time domain using ultrafast techniques. For instance, in a cousin van der Waals antiferromagnet FePS$_3$, two Raman-active optical phonons at 3.2 and 4.8~THz can be excited by a near-infrared pump pulse, evidenced by the coherent oscillations in the polarization rotation signal of the probe pulse (Fig.~\ref{fig:magnon_phonon_2Dmagnet}b) \cite{mertens2023ultrafastcoherent}. As the applied magnetic field increases, the 3.2~THz phonon splits into two phonon-magnon modes due to phonon-magnon coupling (Fig.~\ref{fig:magnon_phonon_2Dmagnet}c), which is reminiscent of the hybridization-induced peak splittings detected in equilibrium \cite{liu2021direct,vaclavkova2021magnon,zhang2021coherent}.

\begin{figure*}[!t]
\centering
\includegraphics[width=\textwidth]{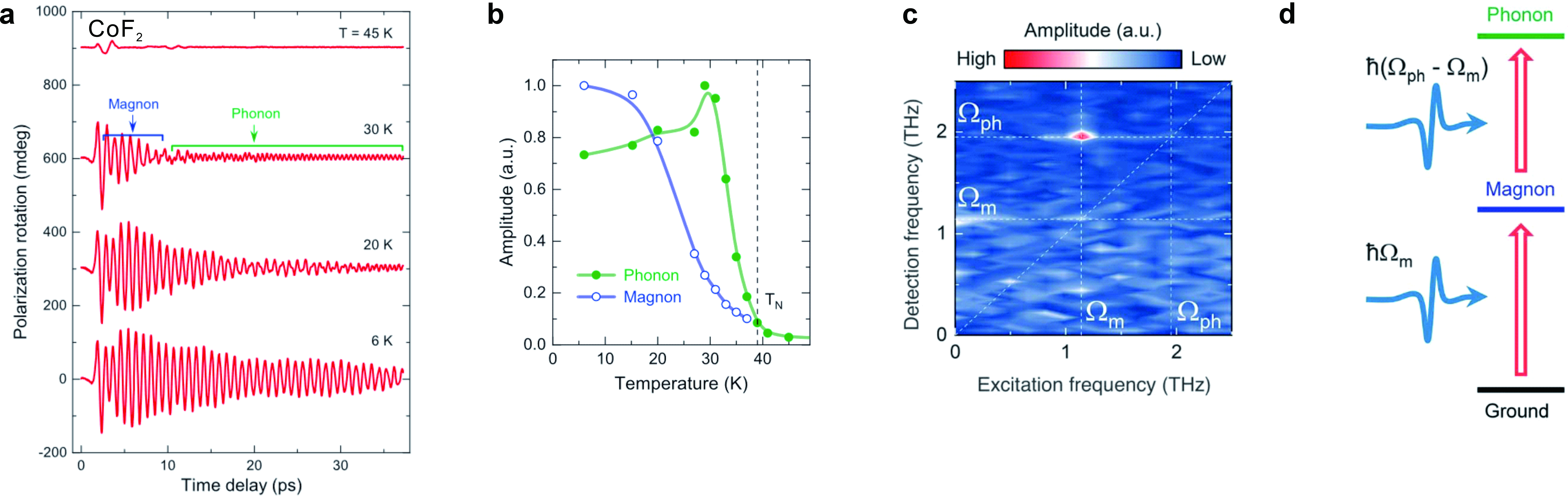}
\caption{\textbf{Time-domain probe of phonon-magnon couplings via magnon excitation.} \textbf{a},~THz pump–induced polarization rotation of the 800-nm probe pulse measured in CoF$_2$ at different temperatures, whose antiferromagnetic transition temperature is at 39~K.  \textbf{b},~Temperature-dependent Fourier amplitudes of the magnon mode (blue open circles) and phonon mode (green filled circles), both vanishing above the N\'eel temperature, $T_\text{N}$. \textbf{c},~Fourier spectrum of the nonlinear amplitude measured in 2D terahertz spectroscopy on CoF$_2$ at 10~K. \textbf{d},~Schematic illustration of magnon-mediated excitation of the $B_{1g}$ phonon by the THz magnetic field. $\hbar$ is the reduced Planck's constant; $\Omega_\text{m}$ and $\Omega_\text{ph}$ are the magnon and phonon frequency, respectively. Adapted from ref.~\cite{mashkovich2021terahertz}, AAAS.}
\label{fig:magnon_phonon_ultrafast}
\end{figure*}

In the study of phonon-magnon coupling, what sets nonequilibrium experiments apart from their equilibrium counterpart is the coherence of the dynamical motion of both atoms and spins, leading to a number of interesting phenomena. For example, circularly polarized phonons can be coherently excited, leading to an effective magnetic field that can in turn excite coherent spin precession. In an early demonstration of this effect, Nova and coworkers used a 20-THz mid-infrared pump to resonantly drive the in-plane $B_{ua}$ and $B_{ub}$ phonons in an antiferromagnetic insulator ErFeO$_3$ (ref.~\cite{nova2017aneffective}), whose phonon eigenvectors are sketched in Fig.~\ref{fig:phonon_magnon_ultrafast}a. Due to the orthorhombic distortion in ErFeO$_3$, these two phonons exhibit different eigenfrequencies and oscillator strengths. When the modes are excited simultaneously, the two modes begin with a nonzero relative phase and evolve at different rates, resulting in an elliptical motion of oxygen atoms, as illustrated in the bottom panel of Fig.~\ref{fig:phonon_magnon_ultrafast}a. The elliptically polarized phononic field mimics the application of a magnetic field and results in a spin precession, leading to the excitation of a coherent magnon at 0.75~THz frequency, which is readily detected in the Fourier spectrum of the pump-induced Faraday rotation signal (Fig.~\ref{fig:phonon_magnon_ultrafast}b). The elliptically polarized phonons are otherwise known as chiral phonons, which have attracted both theoretical \cite{zhang2015chiral,suri2021chiral} and experimental \cite{zhu2018observation,luo2023large,davies2024phononic,basini2024terahertz} interests because of their ability to create an effective magnetic field that paves the way for ``chiral phonomagnetism'' in magnetization-based devices.

Coherent magnons can also be excited by phonon-induced lattice distortions that modify the exchange interaction between neighboring atoms, a mechanism recently demonstrated in a closely related antiferromagnet DyFeO$_3$ by Afanasiev and colleagues \cite{afanasiev2021ultrafastcontrol}. In this material, the highest-frequency IR-active transverse optical phonon mode ($B_u$) is associated with a periodic stretching of the Fe-O bond. Using a tailored terahertz pump pulse, this phonon can be resonantly driven to a large amplitude and couple to a finite lattice distortion along the coordinate of a Raman-active $A_g$ mode via the nonlinear phononics pathway discussed in Sec.~\ref{sec:phonon-phonon} (see Fig.~\ref{fig:nonlinear_phonon_schematic}). This lattice distortion involves an antipolar motion of the Dy$^{3+}$ ions, resulting in a rectification of the Fe-Dy exchange integral that can reach approximately 1--2~$\upmu$eV per unit cell according to DFT calculations, thus setting off coherent magnon oscillations. Depending on whether DyFeO$_3$ is in the antiferromagnetic state ($<50$~K) or the weakly ferromagnetic state ($>50$~K), distinct magnon modes are induced by the resonant driving of the same optical phonon (Fig.~\ref{fig:phonon_magnon_ultrafast}c,d). Importantly, the magnon mode frequencies under the resonant driving condition (orange peak in the insets) are clearly distinct from those under off-resonant pumping (black peak in the insets), the latter of which are excited via impulsive stimulated Raman scattering \cite{kalashnikova2007impulsive}. Leveraging the modification of exchange interaction under a resonant pumping condition, Afanasiev and colleagues further showed that with a very large fluence of the resonant pump, an ultrafast magnetic transition from an antiferromagnetic state to a weakly ferromagnetic state can be induced coherently, a transition that proceeded much faster compared to an above-bandgap excitation scheme.

While the preceding examples illustrated how coherent magnons can be induced by resonant excitation of coherent phonons, the reverse pathway from magnons to phonons has also been demonstrated. In an experiment conducted on antiferromagnetic CoF$_2$, Mashkovich \textit{et al.} used an intense, nearly single-cycle terahertz pulse to pump the crystal below its N\'eel temperature, where the magnetic field of the pulse resonantly excited a magnon mode (Fig.~\ref{fig:magnon_phonon_ultrafast}a) \cite{mashkovich2021terahertz}. After approximately 10~ps, the amplitude of the magnon decreased, followed by a persistent $B^1_g$ phonon oscillation over the next tens of picoseconds. By contrast, the magnon and phonon oscillations both disappeared above the N\'eel temperature (Fig.~\ref{fig:magnon_phonon_ultrafast}b), strongly suggesting that the phonons are induced by the magnons and not vice versa. The researchers further performed 2D terahertz spectroscopy to demonstrate the causal relation that a coherent magnon mediated the nonlinear excitation of phonons, as shown by the Fourier spectrum of the nonlinear amplitude in Fig.~\ref{fig:magnon_phonon_ultrafast}c. Specifically, a peak was observed only when the excitation frequency coincided with the magnon mode and the detection frequency coincided with the phonon mode; on the other hand, no peak was seen above the noise level when one swapped the excitation and detection frequency. The magnon-mediated excitation pathway for the coherent phonon is summarized in Fig.~\ref{fig:magnon_phonon_ultrafast}d. In the first step, an intense terahertz field resonantly populates the coherent magnonic state at frequency $\Omega_\text{m}$, creating an intermediate state. In the second step, another terahertz photon at frequency $(\Omega_\text{ph} - \Omega_\text{m})$ interacts with this intermediate state and coherently excites the $B_{1g}$ phonon.

\subsection{Phonon-exciton coupling}\label{sec:phonon-exciton}

When a coherent phonon mode is driven out of equilibrium following an optical pump, the resulting atomic motions can in turn modulate the optical properties in a coherent manner. In materials whose optical response is strongly modified by the presence of excitons --- or bound electron-hole pairs --- the exciton binding energy \cite{baldini2019exciton,thouin2019phonon,trovatello2020strongly}, lifetime \cite{robert2016exciton,Trovatello2020theultrafast}, and transport behavior \cite{peng2022long} can also be changed. Such phonon-exciton coupling has been widely studied in quantum dots and quantum wells \cite{morello2007temperature,sagar2008size,Nazir2016modelling}, and more recently, in complex bulk and atomically-thin materials \cite{baldini2019exciton,thouin2019phonon,trovatello2020strongly}. By comparing the exciton lineshape in broadband absorption or photoluminescence spectra before and after the photoexcitation event, one can observe how excitons respond to phonons in the time domain, yielding insights for tuning exciton properties via ultrafast phonon engineering.

For example, in anatase TiO$_2$, a bulk semiconductor, Baldini \textit{et~al.} used photoinduced coherent acoustic phonons to control its excitons at room temperature, observing an exciton energy shift up to 30~meV, which is one of the largest reported in solids under a time-dependent perturbation \cite{baldini2019exciton}. In a prototypical two-dimensional lead iodide perovskite, (PEA)$_2$PbI$_4$ (PEA~=~phenylethylammonium), it was further discovered that different excitons have varying coupling strengths with different coherent phonons, suggesting distinct polaronic characters of each exciton in this system and possibly in a broader class of hybrid semiconductors \cite{thouin2019phonon}. In the atomically thin limit, Trovatello \textit{et~al.} also showed a pronounced phonon-exciton coupling effect in a monolayer MoS$_2$, whose $C$ excitons are strongly affected by a photoinduced coherent phonon \cite{trovatello2020strongly}. Using transient absorption spectroscopy, the authors studied the interaction between out-of-plane phonon modes and excitons, noting a marked enhancement in the phonon oscillation amplitude at resonant exciton frequencies. 

\begin{figure}[t!]
\centering
\includegraphics[width=0.48\textwidth]{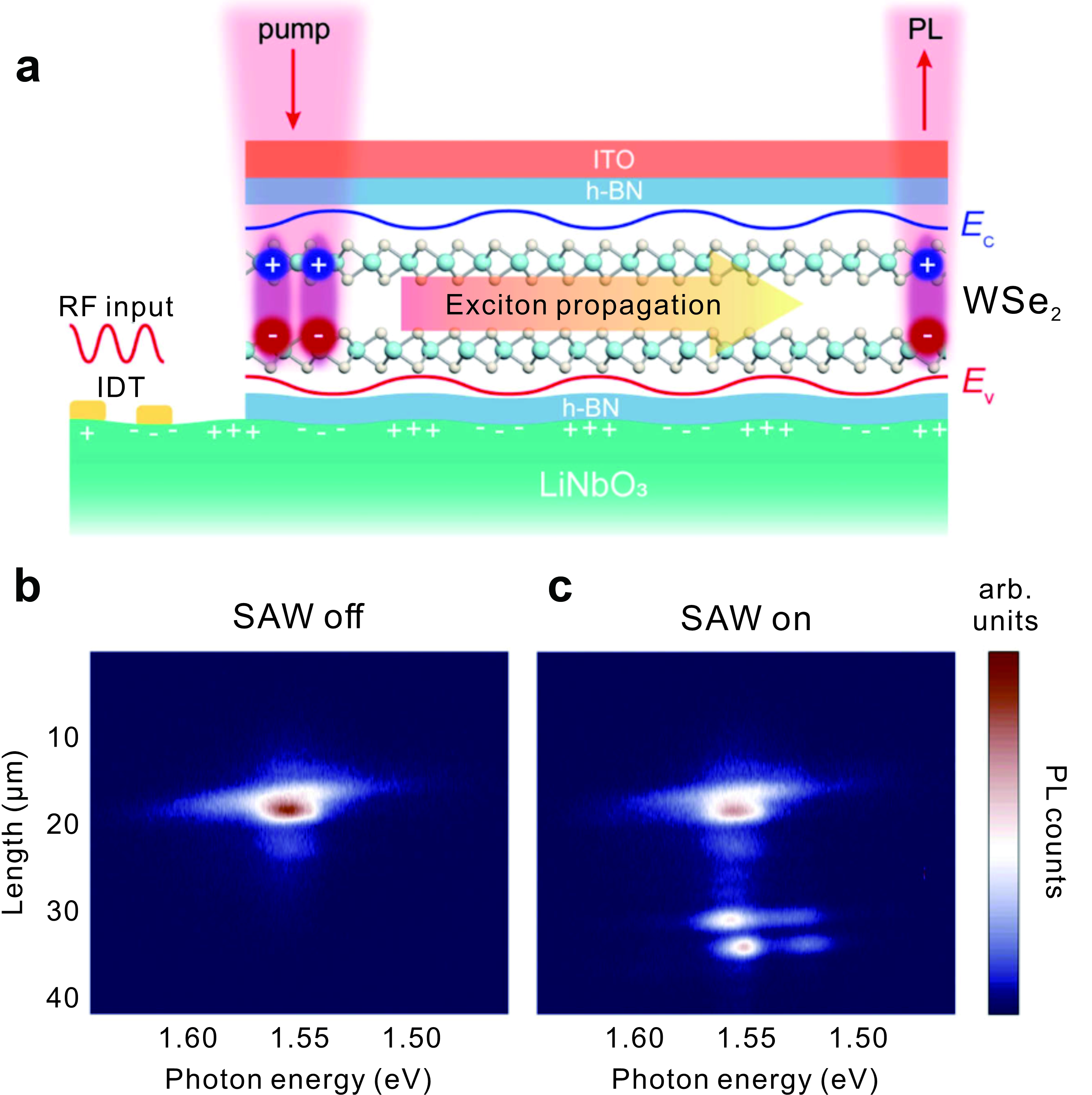}
\caption{\textbf{Exciton transport assisted by surface acoustic waves.} \textbf{a},~Schematic of the experimental setup. Interlayer excitons in bilayer WSe$_2$ are launched on the left via a pump laser and propagate to the right by riding on a surface acoustic wave (SAW) generated by an interdigital transducer (IDT). The excitons are detected by the photoluminescence (PL) signal upon electron-hole recombination. \textbf{b},\textbf{c}~Spectrally-resolved photoluminescence image with and without the surface acoustic wave. The length refers to the spatial distance in the exciton propagation direction. $h$-BN:~Hexagonal boron nitride; ITO:~Indium-tin-oxide. Adapted from ref.~\onlinecite{peng2022long},~\href{https://creativecommons.org/licenses/by/4.0/}{CC~BY~4.0}.}
\label{fig:phonon_exciton_ultrafast}
\end{figure}

The aforementioned studies of phonon-exciton coupling, however, lack spatial information, which is important for learning the transport behavior of excitons under the influence of phonons. Recently, Peng and coworkers demonstrated microscopic imaging of directional transport of interlayer excitons in bilayer WSe$_2$ with the assistance of surface acoustic waves \cite{peng2022long}, and their experimental scheme is summarized in Fig.~\ref{fig:phonon_exciton_ultrafast}a. In this geometry, excitons were optically excited by a continuous-wave laser on the left edge of the sample and probed at different locations via photoluminescence imaging, giving rise to a spatially-resolved mapping of exciton propagation. As shown in Fig.~\ref{fig:phonon_exciton_ultrafast}b, the excitons were launched near the position at 20~$\upmu$m, where the strongest photoluminescence intensity appears. In the absence of a surface acoustic wave, the exciton is largely localized (Fig.~\ref{fig:phonon_exciton_ultrafast}b). However, by launching a surface acoustic wave using an interdigital transducer, excitons can travel  a long distance ($\geq 20~\upmu$m, Fig.~\ref{fig:phonon_exciton_ultrafast}c), which is much longer than their typical diffusion length. Despite the lack of temporal resolution in this measurement, one can imagine a spatiotemporal experimental scheme by using a pulsed laser to launch the excitons. Crucially, this experiment introduces a new way to control exciton transport via electrically-excited or photoinduced acoustic waves, a versatile approach that can be adapted to many other 2D~semiconductors and heterostructures.

\begin{figure*}[!htbp]
\centering
\includegraphics[width=\textwidth]{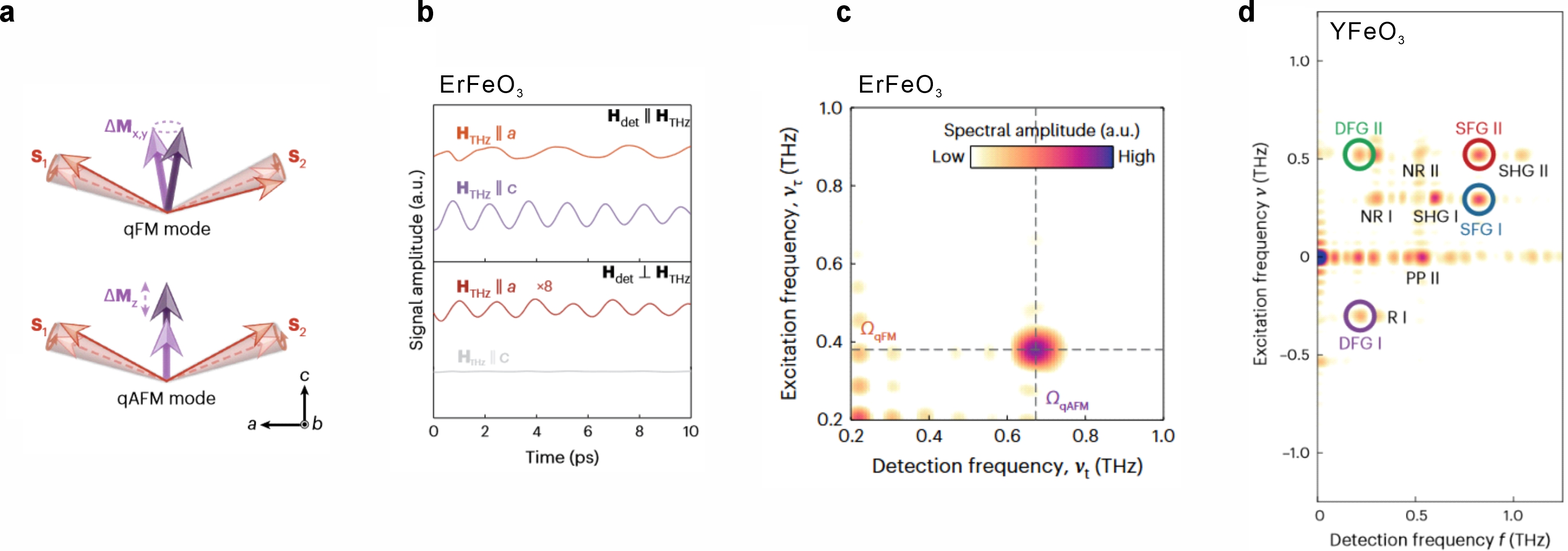}
\caption{\textbf{Magnon-magnon couplings in rare-earth ferrites.} \textbf{a},~Schematic of two magnon modes in canted antiferromagnet YFeO$_3$ and ErFeO$_3$: quasi-ferromagnetic (qFM) mode, which corresponds to a precession of the magnetization orientation, and the quasi-antiferromagnetic (qAFM) mode, which corresponds to an oscillation of the magnetization amplitude. Microscopically, spin dynamics in the qFM (or qAFM) mode correspond to the out-of-phase (or in-phase) precession of the sublattice spins $\mathbf{S}_1$ and $\mathbf{S}_2$. \textbf{b},~Terahertz field-induced free-induction decay signals of ErFeO$_3$, corresponding to excitation of either the $a$-axis qFM mode or the $c$-axis qAFM mode for both $\mathbf{H}_{\rm{det}}\parallel\mathbf{H}_{\rm{THz}}$ and $\mathbf{H}_{\rm{det}}\perp\mathbf{H}_{\rm{THz}}$ detection configurations. Here, $\mathbf{H}_\text{THz}$ is the magnetic field in the terahertz pulse and $\mathbf{H}_\text{det}$ is the detected free-induction decay signals. \textbf{c}~Room-temperature nonlinear 2D THz spectrum of ErFeO$_3$ collected in the $\mathbf{H}_{\rm{det}}\perp\mathbf{H}_{\rm{THz}}$ detection geometry, with $\mathbf{H}_{\rm{THz}}\parallel{c}$ showing the strong off-diagonal qFM-to-qAFM magnon up-conversion peak. \textbf{d},~Nonlinear 2D THz spectrum of YFeO$_3$ for the geometry where $\mathbf{H}_{\rm{THz}}\parallel{ac}$ is the bisector direction. Labeled peaks correspond to pump-probe (PP), rephasing or photon echo (R), non-rephasing (NR), second-harmonic-generation (SHG), sum-frequency-generation (SFG), and difference-frequency-generation (DFG) signals; I and II refer to the qFM and qAFM modes, respectively. For the sum and difference frequency signals, the assignment refers to the excitation frequency and indicates which magnon mode was excited by the first THz field. Panels~\textbf{a} and \textbf{d} adapted from ref.~\onlinecite{zhang2024terahertzDFG}, and Panels~\textbf{b} and \textbf{c} adapted from ref.~\onlinecite{zhang2024terahertzup},~Springer Nature Ltd.}
\label{fig:magnon_magnon}
\end{figure*}

These studies underscore the diversity and tunability of phonon-exciton interactions across material systems, providing insights into phononic control over excitonic properties, which has far-reaching implications for the next-generation design of optoelectronic devices.

\subsection{Other collective excitation couplings}\label{sec:other-couplings}

Despite the special focus on phonons as one of the coupled modes in the preceding sections, the full combinatorics of collective excitation coupling has been applied to many other scenarios such as magnon-magnon \cite{Blank2023empowering,zhang2024terahertzup, zhang2024terahertzDFG,leenders2024canted,zhang2024terahertzstimu}, exciton-magnon \cite{bae2022exciton}, plasmon-polariton \cite{zayats2005nano,piazza2015simultaneous} among other interactions. In this section, we discuss a few examples that showcase the diversity of collective excitation couplings and their role in determining the ultrafast dynamics of quantum materials.

Similar to nonlinear phononics, which opens a new pathway for the coherent manipulation of the crystalline structure, magnon-magnon coupling provides an avenue for controlling the spin dynamics. Utilizing 2D THz spectroscopy, Blank and coworkers revealed the down-conversion coupling of two magnon modes in the canted antiferromagnet FeBO$_3$ (ref.~\onlinecite{Blank2023empowering}). In another canted antiferromagnet, ErFeO$_3$, Zhang and coworkers also demonstrated a unidirectional up-conversion coupling between distinct magnon modes \cite{zhang2024terahertzup}. The canted spins in these materials result from the competition between antiferromagnetic interaction and Dzyaloshinskii-Moriya interaction, yielding a small net magnetic moment $\mathbf{M}$. As a result, there are two primary cooperative motions of the sublattice spins, corresponding to two distinct magnon modes sketched in Fig.~\ref{fig:magnon_magnon}a: the quasi-ferromagnetic (qFM) mode where $\mathbf{M}$ precesses around the $c$-axis, and the quasi-antiferromagnetic (qAFM) mode where $|\mathbf{M}|$ oscillates. Given the different contributions to $\Delta \textbf{M}$ from the two modes, they can be selectively excited by tuning the direction of the magnetic field $\mathbf{H}_\text{THz}$ in the incident THz pump pulse, and their dynamics can be detected in the radiated free-induction decay signals. For example, when the magnetic field of the emitted THz pulse is parallel to that of the excitation field ($\mathbf{H}_\text{det} \parallel \mathbf{H}_\text{THz}$), qFM mode and qAFM mode are selectively excited and observed when $\mathbf{H}_\text{THz} \parallel a$ and $\mathbf{H}_\text{THz} \parallel c$, respectively (upper panel in Fig.~\ref{fig:magnon_magnon}b). This observation is expected as qFM involves a precession in the $a$-$b$ plane while qAFM involves an amplitude oscillation along $c$. Surprisingly, when $\mathbf{H}_\text{det} \perp \mathbf{H}_\text{THz}$, qAFM mode is still observed with $\mathbf{H}_\text{THz} \parallel a$ (lower panel in Fig.~\ref{fig:magnon_magnon}b), which is not expected if there were only linear interactions between the driving THz magnetic field and magnons. To elicit the nonlinear interactions, 2D THz spectroscopy measurements were employed, where the nonlinear spectrum reveals a unidirectional up-conversion process from qFM to qAFM. As shown in Fig.~\ref{fig:magnon_magnon}c, the unidirectionality is evidenced by a single peak at the qFM excitation frequency and qAFM detection frequency, and importantly, a peak is absent when the corresponding excitation and detection frequencies are interchanged. Besides magnon up-conversion that echoes the phonon up-conversion seen in SrTiO$_3$ (Fig.~\ref{fig:nonlinear_phonon_example}c, ref.~\onlinecite{kozina2019terahertz}), the sum and difference frequencies between these two magnon modes have been demonstrated in a cousin rare-earth ferrite, YFeO$_3$, which are featured by a plethora of off-diagonal peaks in the nonlinear 2D terahertz spectrum in Fig.~\ref{fig:magnon_magnon}d (ref.~\onlinecite{zhang2024terahertzDFG}). Like nonlinear phononics, these nonlinear effects of magnons demonstrated in the time domain help usher in new ways to coherently manipulate the spin degree of freedom in magnetic materials.

The analogy between phonons and magnons goes beyond anharmonic couplings. Like phonons, a magnon wavepacket can propagate in materials, which forms the bedrock for spintronic and magnonic applications. When coupled with other collective excitations, magnon propagation can be modified. For instance, for magnons that are coupled with acoustic phonons, theoretical \cite{shen2018theory} and experimental \cite{ogawa2015photodrive,wang2023long} works have shown that magnon polarons travel over long distances with a faster group velocity than magnons alone. In a study of a van der Waals antiferromagnetic semiconductor CrSBr, magnons were observed to coherently travel beyond several micrometers with a very fast group velocity, an effect that Bae and coworkers interpreted as the coupling between the magnon and acoustic phonon branches \cite{bae2022exciton,bae2024transient}. What is interesting about this experiment is that coherent magnons were probed by laser pulses tuned to the exciton resonances in CrSBr, demonstrating efficient optical access to spintronic information due to strong exciton-magnon couplings in this material. A subsequent experiment suggested that optically launched magnon propagation in CrSBr and likely other van der Waals magnets are instead mediated by long-range magnetic-dipole interactions \cite{sun2024dipolar}, calling for additional spatiotemporally-resolved studies to understand magnon dynamics in this important class of materials. 

\begin{figure}[!t]
\centering
\includegraphics[width=0.48\textwidth]{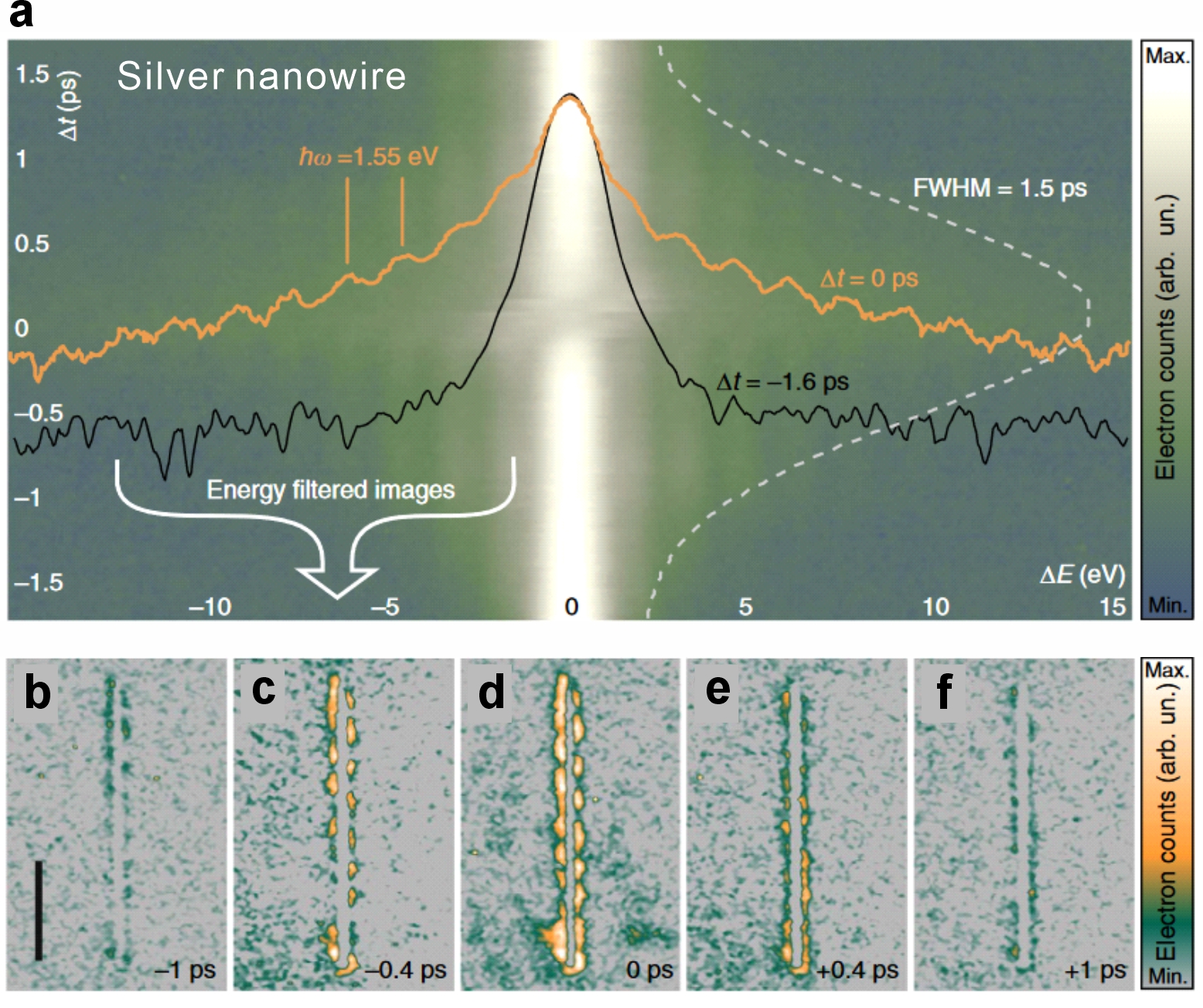}
\caption{\textbf{Surface plasmon-polaritons in a silver nanowire.} \textbf{a},~Intensity map of electron energy loss ($\Delta E$) vs. time delays ($\Delta t$) between the 800-nm optical pump and electron probe, taken on a single silver nanowire (5.7~mm length, $\simeq$ 67~nm radius). The azimuthal angle between the pump polarization and the nanowire long axis was 45$^\circ$. Representative energy spectra at $\Delta t  = -1.6$~ps (black solid curve) and $\Delta t = 0$~ps (orange solid curve) are superimposed. The intensity in both the map and the two representative spectra is plotted on a logarithmic scale. The dashed white curve shows the temporal cross-correlation of the optical pump and electron probe. \textbf{b}--\textbf{f},~Snapshots of an isolated nanowire at different time delays obtained using energy-filtered electrons as indicated by the white arrow in \textbf{a}. Electron counts are on a linear scale. Scale bar, 2~$\upmu$m. Adapted from ref.~\onlinecite{piazza2015simultaneous},~\href{https://creativecommons.org/licenses/by/4.0/}{CC~BY~4.0}.}
\label{fig:plasmon_polariton}
\end{figure}

Non-local propagation effects mediated by the coupling between collective excitations are not restricted to magnons. Other examples include exciton propagation with the aid of surface acoustic waves in a GaAs quantum well \cite{violante2014dynamics,grasselli2018classical} or in a transition metal dichalcogenide \cite{peng2022long}, and surface-plasmon-polariton propagation at a metal-dielectric interface \cite{zhang2012surface}. Similar to a phonon-polariton, a plasmon-polariton is a hybridized excitation between charge oscillations and light. More specifically, when light illuminates a metal-dielectric interface, the oscillating electric field of light drives a collective oscillation of the surface charge, which in turn radiates electromagnetic waves. As a result, a hybridized wave, namely surface-plasmon-polariton that involves both the charge motion on the metal surface and electromagnetic field in the dielectric, propagates along the interface. To observe and control its propagation dynamics, PINEM is one of the leading techniques \cite{yurtsever2012subparticle, piazza2015simultaneous}. Figure~\ref{fig:plasmon_polariton}a shows the spatiotemporally-resolved imaging of surface-plasmon-polaritons induced by near-infrared laser pulses on an isolated silver nanowire suspended on a graphene film \cite{piazza2015simultaneous}, where surface-plasmon-polaritons are evidenced by the additional peaks that appear in the electron energy loss spectra when the light and electron pulses are temporally overlapped. These peaks result from the quantized gain and loss of energy when electrons interact with the surface-plasmon-polariton field. Using energy-filtered electrons, the temporal dynamics of the surface-plasmon-polariton field was imaged with high spatial resolution as it formed an evolving interferometric standing wave pattern (Fig.~\ref{fig:plasmon_polariton}b--f). Moreover, with different light polarizations, different surface-plasmon-polariton field distributions were obtained, demonstrating the potential to control surface-plasmon-polaritons using light.

\begin{figure*}[!t]
\centering
\includegraphics[width=1.0\textwidth]{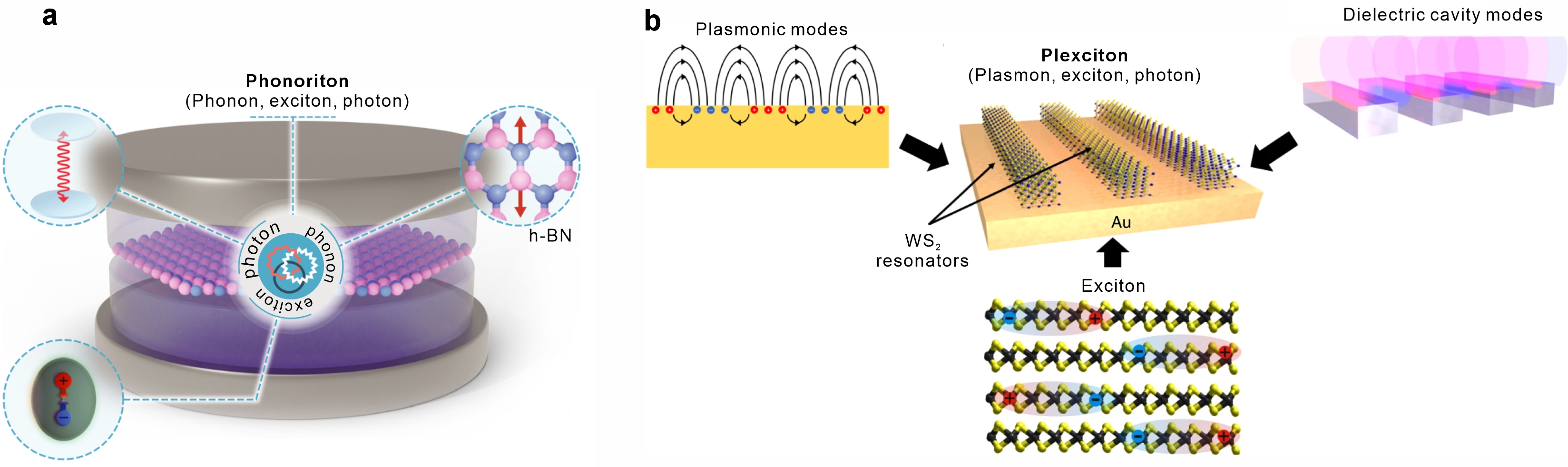}
\caption{\textbf{Theoretical prediction and experimental realization of multipartite couplings of collective modes.} \textbf{a},~Schematic of a theoretically proposed ``phonoriton'' in a monolayer $h$-BN embedded in an optical cavity, where phonoritons are hybridized quasiparticles consisting of excitons, phonons, and photons. \textbf{b},~Schematic of a ``plexciton'', which is a hybrid plasmon-exciton-photon, experimentally realized in nano-structured, multilayer WS$_2$ on gold. Panel~\textbf{a} adapted from ref.~\onlinecite{latini2021phonoritons},~\href{https://creativecommons.org/licenses/by/4.0/}{CC~BY~4.0}. Panel~\textbf{b} adapted from ref.~\cite{zhang2020hybrid},~\href{https://creativecommons.org/licenses/by/4.0/}{CC~BY~4.0}.}
\label{fig:phonoriton}
\end{figure*}

\section{Summary and outlook}\label{sec:outlook}

The study and manipulation of coupled collective excitations by ultrafast light pulses offer a new perspective on disentangling intertwined degrees of freedom that are difficult to isolate in equilibrium. In this review, we have highlighted several couplings that both yield an in-depth understanding of the equilibrium properties and enable the coherent control of the nonequilibrium dynamics in the host material.

While examples mentioned so far only include two participating collective excitations, multipartite couplings have been theoretically pondered, such as a hybrid exciton-phonon-photon that gives rise to a ``phonoriton'' in an optical cavity (Fig.~\ref{fig:phonoriton}a) \cite{keldysh2024coulomb,latini2021phonoritons}, and experimentally realized, such as the hybrid plasmon-exciton-photon --- known as a ``plexciton'' --- in WS$_2$ (Fig.~\ref{fig:phonoriton}b), where cavity effects were achieved through nano-patterning \cite{zhang2020hybrid,tang2022interacting}. These studies fall into a larger body of works that aims to manipulate quantum materials via cavity control, in which the light-matter coupling strength can be greatly enhanced by replacing classical laser fields with quantum-mechanical photon modes. One of the most celebrated examples is microcavity exciton–polaritons in GaAs quantum wells, which arise from the strong coupling between photons and excitons and exhibit remarkable effects such as Bose-Einstein condensation \cite{deng2002condensation,deng2010exciton,luo2023nanophotonics}. More recently, a multitude of theoretical proposals have been put forward for tailoring the unique properties of quantum materials via cavity design \cite{ashida2020quantum,schlawin2022cavity}, with experimental progress in the study of topological states \cite{peng2024topological}, superconductivity \cite{garcia-vidal2021manipulating}, and phase transitions \cite{jarc2023cavity} in different material platforms.

Besides the potential in the experimental realization of such multi-way couplings in the context of cavity quantum materials, we comment on a few other directions in the study of coupled excitations from the viewpoint of ultrafast science.

First, advances in engineering the waveform of excitation light pulses --- such as their polarization, carrier envelope phase, and orbital angular momentum  --- can greatly expand the toolkits for resonant excitation of collective modes and hence enable a deterministic control over material properties. A recent example is the synthesis of an intense, circularly-polarized THz pulse used to induce circular motions of atoms, otherwise known as chiral phonons, which can in turn induce a large effective magnetic field \cite{luo2023large,romao2024light,davies2024phononic,basini2024terahertz}. In particular, Davies \textit{et al.} demonstrated that the induced field can significantly alter the magnetic order even at considerable distances from the phonon creation point \cite{davies2024phononic}, underscoring the need for deeper insight into chiral phonon transport. Equipped with the capability to deterministically excite particular phonon and magnon modes using the mechanisms introduced in this review, one can envision future experiments where chiral phonons and magnons are created at distinct sample locations, and their mutual interaction over space and time can be used to gain a mechanistic understanding of the propagation mechanism for realizing ultrafast non-local control of magnetic ordering.

Second, from a technical point of view, great care needs to be taken when comparing the dynamics of different collective excitations probed by distinct experimental setups, making a quantitative study of their couplings challenging. An example of this issue arises when one extracts the electron-phonon coupling strength by comparing the electronic band change and atomic displacement, which are probed by time-resolved photoemission and X-ray diffraction respectively \cite{gerber2017femtosecond,huang2023ultrafastmeasurement}. For the same incident pump fluence, the effective excitation fluence can differ substantially because photoemission is a surface-sensitive probe (sub- to few nanometers), whereas X-rays can probe much deeper into the bulk (tens of nanometers to micrometers). Therefore, careful fluence calibrations are needed when comparing these probes unless monolayer samples are used to eliminate the variance. An alternative route is to look for complementary probes with a similar probing depth, such as high-energy electron diffraction and extreme ultraviolet absorption spectroscopy, both operated under a transmission geometry. While the former is sensitive to the lattice degree of freedom, the latter excels in capturing carrier, magnetic, and excitonic excitations with exceptional time resolution \cite{zong2023emerging,zong2024core}. One can envision a chamber design that integrates both probes in a single setup, making it possible to concurrently access lattice and electronic information following the same excitation pulse.

Third, while the review has largely focused on coherent collective dynamics with a well-defined phase, not all light-induced collective excitations are coherent or fully reproducible over multiple pump pulses, particularly in systems with spatial inhomogeneities where defects and disorders are prevalent \cite{o2015inhomogeneity,cremons2016femtosecond,perez2022multi,carbin2023evidence}. In this aspect, techniques with both spatial and temporal resolutions that operate in the single-shot mode are important for revealing the local, incoherent, and stochastic dynamics that vary from pulse to pulse, as recently demonstrated in the study of magnetic fluctuations in magnetic multilayer thin films \cite{buttner2021observation,klose2023coherent} and the heterogeneous insulator-to-metal transition in VO$_2$ (ref.~\onlinecite{johnson2023ultrafast}), both of which utilized a coherent X-ray source in large-scale user facilities. It is worth noting that stochastic dynamics can also be accessed without relying on single-shot detection \cite{mcclellan2024hidden}, which makes it possible to study fluctuating systems with the stroboscopic scheme employed in the vast majority of ultrafast measurements. Along with rapid progress in facility-based ultrashort coherent X-ray sources, we also envision two key developments in table-top experiments for probing local and incoherent collective modes. At the extreme spatial scale, innovations in scanning tunneling microscopy have enabled new ways to image collective excitations at the atomic scale \cite{jelic2024atomic,sheng2024terahertz}, paving the way for controlling these local dynamics through creative engineering of nanoscopic defects. At the extreme temporal scale, ultrafast electron diffraction and microscopy, which have witnessed a rapid improvement of their temporal resolution \cite{kim2019towards,qi2020breaking,nabben2023attosecond,hui2024attosecond,2024arXiv241114518B}, would provide a useful platform to study the initial emergence of incoherent phonons following photoexcitation \cite{rene2019time, britt2022direct}, which play an important role in determining the ensuing defect structure \cite{cheng2024ultrafast_TiSe2} and energy flow in the host material system \cite{sidiropoulos2021probing}.

As we conclude this review that has touched a diverse set of experimental techniques and observables, it is apt to be reminded that the overarching goal of studying coupled collective excitations is to uncover a set of unifying principles that govern the interaction among the different degrees of freedom in solids. While studying collective excitations is certainly not the only way to unravel the intricate microscopic interactions, they offer us a way to describe a many-body system without the need to keep track of individual dynamics of an astronomical number of particles, making it tractable to study and engineer material properties both in and out of equilibrium.

\begin{acknowledgments}
We thank helpful input from Edoardo Baldini, Henry Bell, Thomas Blank, Yu-Che Chien, Shaofeng Duan, Nuh Gedik, Patrick Liu, Megan Loh, Tianchuang Luo, Enrico Perfetto, Ruixin Qiu, Gianluca Stefanucci, Mariano Trigo, and Xunqing Yin during the preparation of this manuscript. 
\end{acknowledgments}


\begin{thebibliography}{262}%
\makeatletter
\providecommand \@ifxundefined [1]{%
 \@ifx{#1\undefined}
}%
\providecommand \@ifnum [1]{%
 \ifnum #1\expandafter \@firstoftwo
 \else \expandafter \@secondoftwo
 \fi
}%
\providecommand \@ifx [1]{%
 \ifx #1\expandafter \@firstoftwo
 \else \expandafter \@secondoftwo
 \fi
}%
\providecommand \natexlab [1]{#1}%
\providecommand \enquote  [1]{``#1''}%
\providecommand \bibnamefont  [1]{#1}%
\providecommand \bibfnamefont [1]{#1}%
\providecommand \citenamefont [1]{#1}%
\providecommand \href@noop [0]{\@secondoftwo}%
\providecommand \href [0]{\begingroup \@sanitize@url \@href}%
\providecommand \@href[1]{\@@startlink{#1}\@@href}%
\providecommand \@@href[1]{\endgroup#1\@@endlink}%
\providecommand \@sanitize@url [0]{\catcode `\\12\catcode `\$12\catcode `\&12\catcode `\#12\catcode `\^12\catcode `\_12\catcode `\%12\relax}%
\providecommand \@@startlink[1]{}%
\providecommand \@@endlink[0]{}%
\providecommand \url  [0]{\begingroup\@sanitize@url \@url }%
\providecommand \@url [1]{\endgroup\@href {#1}{\urlprefix }}%
\providecommand \urlprefix  [0]{URL }%
\providecommand \Eprint [0]{\href }%
\providecommand \doibase [0]{https://doi.org/}%
\providecommand \selectlanguage [0]{\@gobble}%
\providecommand \bibinfo  [0]{\@secondoftwo}%
\providecommand \bibfield  [0]{\@secondoftwo}%
\providecommand \translation [1]{[#1]}%
\providecommand \BibitemOpen [0]{}%
\providecommand \bibitemStop [0]{}%
\providecommand \bibitemNoStop [0]{.\EOS\space}%
\providecommand \EOS [0]{\spacefactor3000\relax}%
\providecommand \BibitemShut  [1]{\csname bibitem#1\endcsname}%
\let\auto@bib@innerbib\@empty
\bibitem [{\citenamefont {{Lee}}\ \emph {et~al.}(2006)\citenamefont {{Lee}}, \citenamefont {{Nagaosa}},\ and\ \citenamefont {{Wen}}}]{lee2006doping}%
  \BibitemOpen
  \bibfield  {author} {\bibinfo {author} {\bibfnamefont {P.~A.}\ \bibnamefont {{Lee}}}, \bibinfo {author} {\bibfnamefont {N.}~\bibnamefont {{Nagaosa}}},\ and\ \bibinfo {author} {\bibfnamefont {X.-G.}\ \bibnamefont {{Wen}}},\ }\bibfield  {title} {\emph {\bibinfo {title} {{Doping a Mott insulator: Physics of high-temperature superconductivity}}},\ }\href {https://doi.org/10.1103/RevModPhys.78.17} {\bibfield  {journal} {\bibinfo  {journal} {Rev. Mod. Phys.}\ }\textbf {\bibinfo {volume} {78}},\ \bibinfo {pages} {17} (\bibinfo {year} {2006})}\BibitemShut {NoStop}%
\bibitem [{\citenamefont {{Stewart}}(2011)}]{stewart2011supercond}%
  \BibitemOpen
  \bibfield  {author} {\bibinfo {author} {\bibfnamefont {G.~R.}\ \bibnamefont {{Stewart}}},\ }\bibfield  {title} {\emph {\bibinfo {title} {{Superconductivity in iron compounds}}},\ }\href {https://doi.org/10.1103/RevModPhys.83.1589} {\bibfield  {journal} {\bibinfo  {journal} {Rev. Mod. Phys.}\ }\textbf {\bibinfo {volume} {83}},\ \bibinfo {pages} {1589} (\bibinfo {year} {2011})}\BibitemShut {NoStop}%
\bibitem [{\citenamefont {{Keimer}}\ \emph {et~al.}(2015)\citenamefont {{Keimer}}, \citenamefont {{Kivelson}}, \citenamefont {{Norman}}, \citenamefont {{Uchida}},\ and\ \citenamefont {{Zaanen}}}]{keimer2015fromquantum}%
  \BibitemOpen
  \bibfield  {author} {\bibinfo {author} {\bibfnamefont {B.}~\bibnamefont {{Keimer}}}, \bibinfo {author} {\bibfnamefont {S.~A.}\ \bibnamefont {{Kivelson}}}, \bibinfo {author} {\bibfnamefont {M.~R.}\ \bibnamefont {{Norman}}}, \bibinfo {author} {\bibfnamefont {S.}~\bibnamefont {{Uchida}}},\ and\ \bibinfo {author} {\bibfnamefont {J.}~\bibnamefont {{Zaanen}}},\ }\bibfield  {title} {\emph {\bibinfo {title} {{From quantum matter to high-temperature superconductivity in copper oxides}}},\ }\href {https://doi.org/10.1038/nature14165} {\bibfield  {journal} {\bibinfo  {journal} {Nature}\ }\textbf {\bibinfo {volume} {518}},\ \bibinfo {pages} {179} (\bibinfo {year} {2015})}\BibitemShut {NoStop}%
\bibitem [{\citenamefont {{Fernandes}}\ \emph {et~al.}(2022)\citenamefont {{Fernandes}}, \citenamefont {{Coldea}}, \citenamefont {{Ding}}, \citenamefont {{Fisher}}, \citenamefont {{Hirschfeld}},\ and\ \citenamefont {{Kotliar}}}]{fernandes2022iron}%
  \BibitemOpen
  \bibfield  {author} {\bibinfo {author} {\bibfnamefont {R.~M.}\ \bibnamefont {{Fernandes}}}, \bibinfo {author} {\bibfnamefont {A.~I.}\ \bibnamefont {{Coldea}}}, \bibinfo {author} {\bibfnamefont {H.}~\bibnamefont {{Ding}}}, \bibinfo {author} {\bibfnamefont {I.~R.}\ \bibnamefont {{Fisher}}}, \bibinfo {author} {\bibfnamefont {P.~J.}\ \bibnamefont {{Hirschfeld}}},\ and\ \bibinfo {author} {\bibfnamefont {G.}~\bibnamefont {{Kotliar}}},\ }\bibfield  {title} {\emph {\bibinfo {title} {{Iron pnictides and chalcogenides: a new paradigm for superconductivity}}},\ }\href {https://doi.org/10.1038/s41586-021-04073-2} {\bibfield  {journal} {\bibinfo  {journal} {Nature}\ }\textbf {\bibinfo {volume} {601}},\ \bibinfo {pages} {35} (\bibinfo {year} {2022})}\BibitemShut {NoStop}%
\bibitem [{\citenamefont {{Cheong}}\ and\ \citenamefont {{Mostovoy}}(2007)}]{cheong2007multiferroics}%
  \BibitemOpen
  \bibfield  {author} {\bibinfo {author} {\bibfnamefont {S.-W.}\ \bibnamefont {{Cheong}}}\ and\ \bibinfo {author} {\bibfnamefont {M.}~\bibnamefont {{Mostovoy}}},\ }\bibfield  {title} {\emph {\bibinfo {title} {{Multiferroics: a magnetic twist for ferroelectricity}}},\ }\href {https://doi.org/10.1038/nmat1804} {\bibfield  {journal} {\bibinfo  {journal} {Nat. Mater.}\ }\textbf {\bibinfo {volume} {6}},\ \bibinfo {pages} {13} (\bibinfo {year} {2007})}\BibitemShut {NoStop}%
\bibitem [{\citenamefont {{Salamon}}\ and\ \citenamefont {{Jaime}}(2001)}]{salamon2001thephy}%
  \BibitemOpen
  \bibfield  {author} {\bibinfo {author} {\bibfnamefont {M.~B.}\ \bibnamefont {{Salamon}}}\ and\ \bibinfo {author} {\bibfnamefont {M.}~\bibnamefont {{Jaime}}},\ }\bibfield  {title} {\emph {\bibinfo {title} {{The physics of manganites: Structure and transport}}},\ }\href {https://doi.org/10.1103/RevModPhys.73.583} {\bibfield  {journal} {\bibinfo  {journal} {Rev. Mod. Phys.}\ }\textbf {\bibinfo {volume} {73}},\ \bibinfo {pages} {583} (\bibinfo {year} {2001})}\BibitemShut {NoStop}%
\bibitem [{\citenamefont {Wang}\ and\ \citenamefont {Zhang}(2017)}]{wang2017topological}%
  \BibitemOpen
  \bibfield  {author} {\bibinfo {author} {\bibfnamefont {J.}~\bibnamefont {Wang}}\ and\ \bibinfo {author} {\bibfnamefont {S.-C.}\ \bibnamefont {Zhang}},\ }\bibfield  {title} {\emph {\bibinfo {title} {{Topological states of condensed matter}}},\ }\href {https://doi.org/10.1038/nmat5012} {\bibfield  {journal} {\bibinfo  {journal} {Nat. Mater.}\ }\textbf {\bibinfo {volume} {16}},\ \bibinfo {pages} {1062} (\bibinfo {year} {2017})}\BibitemShut {NoStop}%
\bibitem [{\citenamefont {{Tokura}}\ \emph {et~al.}(2017)\citenamefont {{Tokura}}, \citenamefont {{Kawasaki}},\ and\ \citenamefont {{Nagaosa}}}]{tokura2017emergent}%
  \BibitemOpen
  \bibfield  {author} {\bibinfo {author} {\bibfnamefont {Y.}~\bibnamefont {{Tokura}}}, \bibinfo {author} {\bibfnamefont {M.}~\bibnamefont {{Kawasaki}}},\ and\ \bibinfo {author} {\bibfnamefont {N.}~\bibnamefont {{Nagaosa}}},\ }\bibfield  {title} {\emph {\bibinfo {title} {{Emergent functions of quantum materials}}},\ }\href {https://doi.org/10.1038/nphys4274} {\bibfield  {journal} {\bibinfo  {journal} {Nat. Phys.}\ }\textbf {\bibinfo {volume} {13}},\ \bibinfo {pages} {1056} (\bibinfo {year} {2017})}\BibitemShut {NoStop}%
\bibitem [{\citenamefont {{Nayak}}\ \emph {et~al.}(2008)\citenamefont {{Nayak}}, \citenamefont {{Simon}}, \citenamefont {{Stern}}, \citenamefont {{Freedman}},\ and\ \citenamefont {{Das Sarma}}}]{nayak2008nonabelian}%
  \BibitemOpen
  \bibfield  {author} {\bibinfo {author} {\bibfnamefont {C.}~\bibnamefont {{Nayak}}}, \bibinfo {author} {\bibfnamefont {S.~H.}\ \bibnamefont {{Simon}}}, \bibinfo {author} {\bibfnamefont {A.}~\bibnamefont {{Stern}}}, \bibinfo {author} {\bibfnamefont {M.}~\bibnamefont {{Freedman}}},\ and\ \bibinfo {author} {\bibfnamefont {S.}~\bibnamefont {{Das Sarma}}},\ }\bibfield  {title} {\emph {\bibinfo {title} {{Non-Abelian anyons and topological quantum computation}}},\ }\href {https://doi.org/10.1103/RevModPhys.80.1083} {\bibfield  {journal} {\bibinfo  {journal} {Rev. Mod. Phys.}\ }\textbf {\bibinfo {volume} {80}},\ \bibinfo {pages} {1083} (\bibinfo {year} {2008})}\BibitemShut {NoStop}%
\bibitem [{\citenamefont {{{\v{Z}}uti{\'c}}}\ \emph {et~al.}(2004)\citenamefont {{{\v{Z}}uti{\'c}}}, \citenamefont {{Fabian}},\ and\ \citenamefont {{Das Sarma}}}]{vzutic2004spintronics}%
  \BibitemOpen
  \bibfield  {author} {\bibinfo {author} {\bibfnamefont {I.}~\bibnamefont {{{\v{Z}}uti{\'c}}}}, \bibinfo {author} {\bibfnamefont {J.}~\bibnamefont {{Fabian}}},\ and\ \bibinfo {author} {\bibfnamefont {S.}~\bibnamefont {{Das Sarma}}},\ }\bibfield  {title} {\emph {\bibinfo {title} {{Spintronics: Fundamentals and applications}}},\ }\href {https://doi.org/10.1103/RevModPhys.76.323} {\bibfield  {journal} {\bibinfo  {journal} {Rev. Mod. Phys.}\ }\textbf {\bibinfo {volume} {76}},\ \bibinfo {pages} {323} (\bibinfo {year} {2004})}\BibitemShut {NoStop}%
\bibitem [{\citenamefont {{Cooper}}(1956)}]{cooper1956bound}%
  \BibitemOpen
  \bibfield  {author} {\bibinfo {author} {\bibfnamefont {L.~N.}\ \bibnamefont {{Cooper}}},\ }\bibfield  {title} {\emph {\bibinfo {title} {{Bound electron pairs in a degenerate Fermi gas}}},\ }\href {https://doi.org/10.1103/PhysRev.104.1189} {\bibfield  {journal} {\bibinfo  {journal} {Phys. Rev.}\ }\textbf {\bibinfo {volume} {104}},\ \bibinfo {pages} {1189} (\bibinfo {year} {1956})}\BibitemShut {NoStop}%
\bibitem [{\citenamefont {Bardeen}\ \emph {et~al.}(1957)\citenamefont {Bardeen}, \citenamefont {Cooper},\ and\ \citenamefont {Schrieffer}}]{bardeen1957theory}%
  \BibitemOpen
  \bibfield  {author} {\bibinfo {author} {\bibfnamefont {J.}~\bibnamefont {Bardeen}}, \bibinfo {author} {\bibfnamefont {L.~N.}\ \bibnamefont {Cooper}},\ and\ \bibinfo {author} {\bibfnamefont {J.~R.}\ \bibnamefont {Schrieffer}},\ }\bibfield  {title} {\emph {\bibinfo {title} {{Theory of superconductivity}}},\ }\href {https://doi.org/10.1103/PhysRev.108.1175} {\bibfield  {journal} {\bibinfo  {journal} {Phys. Rev.}\ }\textbf {\bibinfo {volume} {108}},\ \bibinfo {pages} {1175} (\bibinfo {year} {1957})}\BibitemShut {NoStop}%
\bibitem [{\citenamefont {Petek}\ and\ \citenamefont {Ogawa}(1997)}]{petek1997femtosecond}%
  \BibitemOpen
  \bibfield  {author} {\bibinfo {author} {\bibfnamefont {H.}~\bibnamefont {Petek}}\ and\ \bibinfo {author} {\bibfnamefont {S.}~\bibnamefont {Ogawa}},\ }\bibfield  {title} {\emph {\bibinfo {title} {{Femtosecond time-resolved two-photon photoemission studies of electron dynamics in metals}}},\ }\href {https://doi.org/10.1016/S0079-6816(98)00002-1} {\bibfield  {journal} {\bibinfo  {journal} {Prog. Surf. Sci.}\ }\textbf {\bibinfo {volume} {56}},\ \bibinfo {pages} {239--310} (\bibinfo {year} {1997})}\BibitemShut {NoStop}%
\bibitem [{\citenamefont {Bovensiepen}\ and\ \citenamefont {Kirchmann}(2012)}]{bovensiepen2012elementary}%
  \BibitemOpen
  \bibfield  {author} {\bibinfo {author} {\bibfnamefont {U.}~\bibnamefont {Bovensiepen}}\ and\ \bibinfo {author} {\bibfnamefont {P.~S.}\ \bibnamefont {Kirchmann}},\ }\bibfield  {title} {\emph {\bibinfo {title} {{Elementary relaxation processes investigated by femtosecond photoelectron spectroscopy of two-dimensional materials}}},\ }\href {https://doi.org/10.1002/lpor.201000035} {\bibfield  {journal} {\bibinfo  {journal} {Laser Photonics Rev.}\ }\textbf {\bibinfo {volume} {6}},\ \bibinfo {pages} {589--606} (\bibinfo {year} {2012})}\BibitemShut {NoStop}%
\bibitem [{\citenamefont {{Zeiger}}\ \emph {et~al.}(1992)\citenamefont {{Zeiger}}, \citenamefont {{Vidal}}, \citenamefont {{Cheng}}, \citenamefont {{Ippen}}, \citenamefont {{Dresselhaus}},\ and\ \citenamefont {{Dresselhaus}}}]{zeiger1992theory}%
  \BibitemOpen
  \bibfield  {author} {\bibinfo {author} {\bibfnamefont {H.~J.}\ \bibnamefont {{Zeiger}}}, \bibinfo {author} {\bibfnamefont {J.}~\bibnamefont {{Vidal}}}, \bibinfo {author} {\bibfnamefont {T.~K.}\ \bibnamefont {{Cheng}}}, \bibinfo {author} {\bibfnamefont {E.~P.}\ \bibnamefont {{Ippen}}}, \bibinfo {author} {\bibfnamefont {G.}~\bibnamefont {{Dresselhaus}}},\ and\ \bibinfo {author} {\bibfnamefont {M.~S.}\ \bibnamefont {{Dresselhaus}}},\ }\bibfield  {title} {\emph {\bibinfo {title} {{Theory for displacive excitation of coherent phonons}}},\ }\href {https://doi.org/10.1103/PhysRevB.45.768} {\bibfield  {journal} {\bibinfo  {journal} {Phys. Rev. B}\ }\textbf {\bibinfo {volume} {45}},\ \bibinfo {pages} {768} (\bibinfo {year} {1992})}\BibitemShut {NoStop}%
\bibitem [{\citenamefont {{Stevens}}\ \emph {et~al.}(2002)\citenamefont {{Stevens}}, \citenamefont {{Kuhl}},\ and\ \citenamefont {{Merlin}}}]{stevens2002coherent}%
  \BibitemOpen
  \bibfield  {author} {\bibinfo {author} {\bibfnamefont {T.~E.}\ \bibnamefont {{Stevens}}}, \bibinfo {author} {\bibfnamefont {J.}~\bibnamefont {{Kuhl}}},\ and\ \bibinfo {author} {\bibfnamefont {R.}~\bibnamefont {{Merlin}}},\ }\bibfield  {title} {\emph {\bibinfo {title} {{Coherent phonon generation and the two stimulated Raman tensors}}},\ }\href {https://doi.org/10.1103/PhysRevB.65.144304} {\bibfield  {journal} {\bibinfo  {journal} {Phys. Rev. B}\ }\textbf {\bibinfo {volume} {65}},\ \bibinfo {pages} {144304} (\bibinfo {year} {2002})}\BibitemShut {NoStop}%
\bibitem [{\citenamefont {{Bugayev}}\ \emph {et~al.}(2011)\citenamefont {{Bugayev}}, \citenamefont {{Esmail}}, \citenamefont {{Abdel-Fattah}},\ and\ \citenamefont {{Elsayed-Ali}}}]{bugayev2011coherent}%
  \BibitemOpen
  \bibfield  {author} {\bibinfo {author} {\bibfnamefont {A.}~\bibnamefont {{Bugayev}}}, \bibinfo {author} {\bibfnamefont {A.}~\bibnamefont {{Esmail}}}, \bibinfo {author} {\bibfnamefont {M.}~\bibnamefont {{Abdel-Fattah}}},\ and\ \bibinfo {author} {\bibfnamefont {H.~E.}\ \bibnamefont {{Elsayed-Ali}}},\ }\bibfield  {title} {\emph {\bibinfo {title} {{Coherent phonons in bismuth film observed by ultrafast electron diffraction}}},\ }\href {https://doi.org/10.1063/1.3574888} {\bibfield  {journal} {\bibinfo  {journal} {AIP Adv.}\ }\textbf {\bibinfo {volume} {1}},\ \bibinfo {pages} {012117} (\bibinfo {year} {2011})}\BibitemShut {NoStop}%
\bibitem [{\citenamefont {{Chatelain}}\ \emph {et~al.}(2014)\citenamefont {{Chatelain}}, \citenamefont {{Morrison}}, \citenamefont {{Klarenaar}},\ and\ \citenamefont {{Siwick}}}]{chatelain2014coherent}%
  \BibitemOpen
  \bibfield  {author} {\bibinfo {author} {\bibfnamefont {R.~P.}\ \bibnamefont {{Chatelain}}}, \bibinfo {author} {\bibfnamefont {V.~R.}\ \bibnamefont {{Morrison}}}, \bibinfo {author} {\bibfnamefont {B.~L.~M.}\ \bibnamefont {{Klarenaar}}},\ and\ \bibinfo {author} {\bibfnamefont {B.~J.}\ \bibnamefont {{Siwick}}},\ }\bibfield  {title} {\emph {\bibinfo {title} {{Coherent and incoherent electron-phonon coupling in graphite observed with radio-frequency compressed ultrafast electron diffraction}}},\ }\href {https://doi.org/10.1103/PhysRevLett.113.235502} {\bibfield  {journal} {\bibinfo  {journal} {Phys. Rev. Lett.}\ }\textbf {\bibinfo {volume} {113}},\ \bibinfo {pages} {235502} (\bibinfo {year} {2014})}\BibitemShut {NoStop}%
\bibitem [{\citenamefont {{Gerber}}\ \emph {et~al.}(2017)\citenamefont {{Gerber}}, \citenamefont {{Yang}}, \citenamefont {{Zhu}}, \citenamefont {{Soifer}}, \citenamefont {{Sobota}}, \citenamefont {{Rebec}}, \citenamefont {{Lee}}, \citenamefont {{Jia}}, \citenamefont {{Moritz}}, \citenamefont {{Jia}}, \citenamefont {{Gauthier}}, \citenamefont {{Li}}, \citenamefont {{Leuenberger}}, \citenamefont {{Zhang}}, \citenamefont {{Chaix}}, \citenamefont {{Li}}, \citenamefont {{Jang}}, \citenamefont {{Lee}}, \citenamefont {{Yi}}, \citenamefont {{Dakovski}}, \citenamefont {{Song}}, \citenamefont {{Glownia}}, \citenamefont {{Nelson}}, \citenamefont {{Kim}}, \citenamefont {{Chuang}}, \citenamefont {{Hussain}}, \citenamefont {{Moore}}, \citenamefont {{Devereaux}}, \citenamefont {{Lee}}, \citenamefont {{Kirchmann}},\ and\ \citenamefont {{Shen}}}]{gerber2017femtosecond}%
  \BibitemOpen
  \bibfield  {author} {\bibinfo {author} {\bibfnamefont {S.}~\bibnamefont {{Gerber}}}, \bibinfo {author} {\bibfnamefont {S.~L.}\ \bibnamefont {{Yang}}}, \bibinfo {author} {\bibfnamefont {D.}~\bibnamefont {{Zhu}}}, \bibinfo {author} {\bibfnamefont {H.}~\bibnamefont {{Soifer}}}, \bibinfo {author} {\bibfnamefont {J.~A.}\ \bibnamefont {{Sobota}}}, \bibinfo {author} {\bibfnamefont {S.}~\bibnamefont {{Rebec}}}, \bibinfo {author} {\bibfnamefont {J.~J.}\ \bibnamefont {{Lee}}}, \bibinfo {author} {\bibfnamefont {T.}~\bibnamefont {{Jia}}}, \bibinfo {author} {\bibfnamefont {B.}~\bibnamefont {{Moritz}}}, \bibinfo {author} {\bibfnamefont {C.}~\bibnamefont {{Jia}}}, \bibinfo {author} {\bibfnamefont {A.}~\bibnamefont {{Gauthier}}}, \bibinfo {author} {\bibfnamefont {Y.}~\bibnamefont {{Li}}}, \bibinfo {author} {\bibfnamefont {D.}~\bibnamefont {{Leuenberger}}}, \bibinfo {author} {\bibfnamefont {Y.}~\bibnamefont {{Zhang}}}, \bibinfo {author} {\bibfnamefont {L.}~\bibnamefont {{Chaix}}}, \bibinfo {author} {\bibfnamefont
  {W.}~\bibnamefont {{Li}}}, \bibinfo {author} {\bibfnamefont {H.}~\bibnamefont {{Jang}}}, \bibinfo {author} {\bibfnamefont {J.~S.}\ \bibnamefont {{Lee}}}, \bibinfo {author} {\bibfnamefont {M.}~\bibnamefont {{Yi}}}, \bibinfo {author} {\bibfnamefont {G.~L.}\ \bibnamefont {{Dakovski}}}, \bibinfo {author} {\bibfnamefont {S.}~\bibnamefont {{Song}}}, \bibinfo {author} {\bibfnamefont {J.~M.}\ \bibnamefont {{Glownia}}}, \bibinfo {author} {\bibfnamefont {S.}~\bibnamefont {{Nelson}}}, \bibinfo {author} {\bibfnamefont {K.~W.}\ \bibnamefont {{Kim}}}, \bibinfo {author} {\bibfnamefont {Y.~D.}\ \bibnamefont {{Chuang}}}, \bibinfo {author} {\bibfnamefont {Z.}~\bibnamefont {{Hussain}}}, \bibinfo {author} {\bibfnamefont {R.~G.}\ \bibnamefont {{Moore}}}, \bibinfo {author} {\bibfnamefont {T.~P.}\ \bibnamefont {{Devereaux}}}, \bibinfo {author} {\bibfnamefont {W.~S.}\ \bibnamefont {{Lee}}}, \bibinfo {author} {\bibfnamefont {P.~S.}\ \bibnamefont {{Kirchmann}}},\ and\ \bibinfo {author} {\bibfnamefont {Z.~X.}\ \bibnamefont
  {{Shen}}},\ }\bibfield  {title} {\emph {\bibinfo {title} {{Femtosecond electron-phonon lock-in by photoemission and x-ray free-electron laser}}},\ }\href {https://doi.org/10.1126/science.aak9946} {\bibfield  {journal} {\bibinfo  {journal} {Science}\ }\textbf {\bibinfo {volume} {357}},\ \bibinfo {pages} {71} (\bibinfo {year} {2017})}\BibitemShut {NoStop}%
\bibitem [{\citenamefont {{Zong}}\ \emph {et~al.}(2018)\citenamefont {{Zong}}, \citenamefont {{Shen}}, \citenamefont {{Kogar}}, \citenamefont {{Ye}}, \citenamefont {{Marks}}, \citenamefont {{Chowdhury}}, \citenamefont {{Rohwer}}, \citenamefont {{Freelon}}, \citenamefont {{Weathersby}}, \citenamefont {{Li}}, \citenamefont {{Yang}}, \citenamefont {{Checkelsky}}, \citenamefont {{Wang}},\ and\ \citenamefont {{Gedik}}}]{zong2018ultrafast}%
  \BibitemOpen
  \bibfield  {author} {\bibinfo {author} {\bibfnamefont {A.}~\bibnamefont {{Zong}}}, \bibinfo {author} {\bibfnamefont {X.}~\bibnamefont {{Shen}}}, \bibinfo {author} {\bibfnamefont {A.}~\bibnamefont {{Kogar}}}, \bibinfo {author} {\bibfnamefont {L.}~\bibnamefont {{Ye}}}, \bibinfo {author} {\bibfnamefont {C.}~\bibnamefont {{Marks}}}, \bibinfo {author} {\bibfnamefont {D.}~\bibnamefont {{Chowdhury}}}, \bibinfo {author} {\bibfnamefont {T.}~\bibnamefont {{Rohwer}}}, \bibinfo {author} {\bibfnamefont {B.}~\bibnamefont {{Freelon}}}, \bibinfo {author} {\bibfnamefont {S.}~\bibnamefont {{Weathersby}}}, \bibinfo {author} {\bibfnamefont {R.}~\bibnamefont {{Li}}}, \bibinfo {author} {\bibfnamefont {J.}~\bibnamefont {{Yang}}}, \bibinfo {author} {\bibfnamefont {J.}~\bibnamefont {{Checkelsky}}}, \bibinfo {author} {\bibfnamefont {X.}~\bibnamefont {{Wang}}},\ and\ \bibinfo {author} {\bibfnamefont {N.}~\bibnamefont {{Gedik}}},\ }\bibfield  {title} {\emph {\bibinfo {title} {{Ultrafast manipulation of mirror domain walls in a charge
  density wave}}},\ }\href {https://doi.org/10.1126/sciadv.aau5501} {\bibfield  {journal} {\bibinfo  {journal} {Sci. Adv.}\ }\textbf {\bibinfo {volume} {4}},\ \bibinfo {pages} {eaau5501} (\bibinfo {year} {2018})}\BibitemShut {NoStop}%
\bibitem [{\citenamefont {{Zong}}\ \emph {et~al.}(2023{\natexlab{a}})\citenamefont {{Zong}}, \citenamefont {{Zhang}}, \citenamefont {{Zhou}}, \citenamefont {{Su}}, \citenamefont {{Hwangbo}}, \citenamefont {{Shen}}, \citenamefont {{Jiang}}, \citenamefont {{Liu}}, \citenamefont {{Gage}}, \citenamefont {{Walko}}, \citenamefont {{Kozina}}, \citenamefont {{Luo}}, \citenamefont {{Reid}}, \citenamefont {{Yang}}, \citenamefont {{Park}}, \citenamefont {{Lapidus}}, \citenamefont {{Chu}}, \citenamefont {{Arslan}}, \citenamefont {{Wang}}, \citenamefont {{Xiao}}, \citenamefont {{Xu}}, \citenamefont {{Gedik}},\ and\ \citenamefont {{Wen}}}]{zong2023spin}%
  \BibitemOpen
  \bibfield  {author} {\bibinfo {author} {\bibfnamefont {A.}~\bibnamefont {{Zong}}}, \bibinfo {author} {\bibfnamefont {Q.}~\bibnamefont {{Zhang}}}, \bibinfo {author} {\bibfnamefont {F.}~\bibnamefont {{Zhou}}}, \bibinfo {author} {\bibfnamefont {Y.}~\bibnamefont {{Su}}}, \bibinfo {author} {\bibfnamefont {K.}~\bibnamefont {{Hwangbo}}}, \bibinfo {author} {\bibfnamefont {X.}~\bibnamefont {{Shen}}}, \bibinfo {author} {\bibfnamefont {Q.}~\bibnamefont {{Jiang}}}, \bibinfo {author} {\bibfnamefont {H.}~\bibnamefont {{Liu}}}, \bibinfo {author} {\bibfnamefont {T.~E.}\ \bibnamefont {{Gage}}}, \bibinfo {author} {\bibfnamefont {D.~A.}\ \bibnamefont {{Walko}}}, \bibinfo {author} {\bibfnamefont {M.~E.}\ \bibnamefont {{Kozina}}}, \bibinfo {author} {\bibfnamefont {D.}~\bibnamefont {{Luo}}}, \bibinfo {author} {\bibfnamefont {A.~H.}\ \bibnamefont {{Reid}}}, \bibinfo {author} {\bibfnamefont {J.}~\bibnamefont {{Yang}}}, \bibinfo {author} {\bibfnamefont {S.}~\bibnamefont {{Park}}}, \bibinfo {author} {\bibfnamefont {S.~H.}\
  \bibnamefont {{Lapidus}}}, \bibinfo {author} {\bibfnamefont {J.-H.}\ \bibnamefont {{Chu}}}, \bibinfo {author} {\bibfnamefont {I.}~\bibnamefont {{Arslan}}}, \bibinfo {author} {\bibfnamefont {X.}~\bibnamefont {{Wang}}}, \bibinfo {author} {\bibfnamefont {D.}~\bibnamefont {{Xiao}}}, \bibinfo {author} {\bibfnamefont {X.}~\bibnamefont {{Xu}}}, \bibinfo {author} {\bibfnamefont {N.}~\bibnamefont {{Gedik}}},\ and\ \bibinfo {author} {\bibfnamefont {H.}~\bibnamefont {{Wen}}},\ }\bibfield  {title} {\emph {\bibinfo {title} {{Spin-mediated shear oscillators in a van der Waals antiferromagnet}}},\ }\href {https://doi.org/10.1038/s41586-023-06279-y} {\bibfield  {journal} {\bibinfo  {journal} {Nature}\ }\textbf {\bibinfo {volume} {620}},\ \bibinfo {pages} {988} (\bibinfo {year} {2023}{\natexlab{a}})}\BibitemShut {NoStop}%
\bibitem [{\citenamefont {{Huang}}\ \emph {et~al.}(2023)\citenamefont {{Huang}}, \citenamefont {{Querales-Flores}}, \citenamefont {{Teitelbaum}}, \citenamefont {{Cao}}, \citenamefont {{Henighan}}, \citenamefont {{Liu}}, \citenamefont {{Jiang}}, \citenamefont {{De la Pe{\~n}a}}, \citenamefont {{Krapivin}}, \citenamefont {{Haber}}, \citenamefont {{Sato}}, \citenamefont {{Chollet}}, \citenamefont {{Zhu}}, \citenamefont {{Katayama}}, \citenamefont {{Power}}, \citenamefont {{Allen}}, \citenamefont {{Rotundu}}, \citenamefont {{Bailey}}, \citenamefont {{Uher}}, \citenamefont {{Trigo}}, \citenamefont {{Kirchmann}}, \citenamefont {{Murray}}, \citenamefont {{Shen}}, \citenamefont {{Savi{\'c}}}, \citenamefont {{Fahy}}, \citenamefont {{Sobota}},\ and\ \citenamefont {{Reis}}}]{huang2023ultrafastmeasurement}%
  \BibitemOpen
  \bibfield  {author} {\bibinfo {author} {\bibfnamefont {Y.}~\bibnamefont {{Huang}}}, \bibinfo {author} {\bibfnamefont {J.~D.}\ \bibnamefont {{Querales-Flores}}}, \bibinfo {author} {\bibfnamefont {S.~W.}\ \bibnamefont {{Teitelbaum}}}, \bibinfo {author} {\bibfnamefont {J.}~\bibnamefont {{Cao}}}, \bibinfo {author} {\bibfnamefont {T.}~\bibnamefont {{Henighan}}}, \bibinfo {author} {\bibfnamefont {H.}~\bibnamefont {{Liu}}}, \bibinfo {author} {\bibfnamefont {M.}~\bibnamefont {{Jiang}}}, \bibinfo {author} {\bibfnamefont {G.}~\bibnamefont {{De la Pe{\~n}a}}}, \bibinfo {author} {\bibfnamefont {V.}~\bibnamefont {{Krapivin}}}, \bibinfo {author} {\bibfnamefont {J.}~\bibnamefont {{Haber}}}, \bibinfo {author} {\bibfnamefont {T.}~\bibnamefont {{Sato}}}, \bibinfo {author} {\bibfnamefont {M.}~\bibnamefont {{Chollet}}}, \bibinfo {author} {\bibfnamefont {D.}~\bibnamefont {{Zhu}}}, \bibinfo {author} {\bibfnamefont {T.}~\bibnamefont {{Katayama}}}, \bibinfo {author} {\bibfnamefont {R.}~\bibnamefont {{Power}}}, \bibinfo {author}
  {\bibfnamefont {M.}~\bibnamefont {{Allen}}}, \bibinfo {author} {\bibfnamefont {C.~R.}\ \bibnamefont {{Rotundu}}}, \bibinfo {author} {\bibfnamefont {T.~P.}\ \bibnamefont {{Bailey}}}, \bibinfo {author} {\bibfnamefont {C.}~\bibnamefont {{Uher}}}, \bibinfo {author} {\bibfnamefont {M.}~\bibnamefont {{Trigo}}}, \bibinfo {author} {\bibfnamefont {P.~S.}\ \bibnamefont {{Kirchmann}}}, \bibinfo {author} {\bibfnamefont {{\'E}.~D.}\ \bibnamefont {{Murray}}}, \bibinfo {author} {\bibfnamefont {Z.-X.}\ \bibnamefont {{Shen}}}, \bibinfo {author} {\bibfnamefont {I.}~\bibnamefont {{Savi{\'c}}}}, \bibinfo {author} {\bibfnamefont {S.}~\bibnamefont {{Fahy}}}, \bibinfo {author} {\bibfnamefont {J.~A.}\ \bibnamefont {{Sobota}}},\ and\ \bibinfo {author} {\bibfnamefont {D.~A.}\ \bibnamefont {{Reis}}},\ }\bibfield  {title} {\emph {\bibinfo {title} {{Ultrafast measurements of mode-specific deformation potentials of Bi$_{2}$Te$_{3}$ and Bi$_{2}$Se$_{3}$}}},\ }\href {https://doi.org/10.1103/PhysRevX.13.041050} {\bibfield  {journal}
  {\bibinfo  {journal} {Phys. Rev. X}\ }\textbf {\bibinfo {volume} {13}},\ \bibinfo {pages} {041050} (\bibinfo {year} {2023})}\BibitemShut {NoStop}%
\bibitem [{\citenamefont {{Zong}}\ \emph {et~al.}(2019{\natexlab{a}})\citenamefont {{Zong}}, \citenamefont {{Dolgirev}}, \citenamefont {{Kogar}}, \citenamefont {{Erge{\c{c}}en}}, \citenamefont {{Yilmaz}}, \citenamefont {{Bie}}, \citenamefont {{Rohwer}}, \citenamefont {{Tung}}, \citenamefont {{Straquadine}}, \citenamefont {{Wang}}, \citenamefont {{Yang}}, \citenamefont {{Shen}}, \citenamefont {{Li}}, \citenamefont {{Yang}}, \citenamefont {{Park}}, \citenamefont {{Hoffmann}}, \citenamefont {{Ofori-Okai}}, \citenamefont {{Kozina}}, \citenamefont {{Wen}}, \citenamefont {{Wang}}, \citenamefont {{Fisher}}, \citenamefont {{Jarillo-Herrero}},\ and\ \citenamefont {{Gedik}}}]{zong2019dynamicalslowing}%
  \BibitemOpen
  \bibfield  {author} {\bibinfo {author} {\bibfnamefont {A.}~\bibnamefont {{Zong}}}, \bibinfo {author} {\bibfnamefont {P.~E.}\ \bibnamefont {{Dolgirev}}}, \bibinfo {author} {\bibfnamefont {A.}~\bibnamefont {{Kogar}}}, \bibinfo {author} {\bibfnamefont {E.}~\bibnamefont {{Erge{\c{c}}en}}}, \bibinfo {author} {\bibfnamefont {M.~B.}\ \bibnamefont {{Yilmaz}}}, \bibinfo {author} {\bibfnamefont {Y.-Q.}\ \bibnamefont {{Bie}}}, \bibinfo {author} {\bibfnamefont {T.}~\bibnamefont {{Rohwer}}}, \bibinfo {author} {\bibfnamefont {I.~C.}\ \bibnamefont {{Tung}}}, \bibinfo {author} {\bibfnamefont {J.}~\bibnamefont {{Straquadine}}}, \bibinfo {author} {\bibfnamefont {X.}~\bibnamefont {{Wang}}}, \bibinfo {author} {\bibfnamefont {Y.}~\bibnamefont {{Yang}}}, \bibinfo {author} {\bibfnamefont {X.}~\bibnamefont {{Shen}}}, \bibinfo {author} {\bibfnamefont {R.}~\bibnamefont {{Li}}}, \bibinfo {author} {\bibfnamefont {J.}~\bibnamefont {{Yang}}}, \bibinfo {author} {\bibfnamefont {S.}~\bibnamefont {{Park}}}, \bibinfo {author} {\bibfnamefont
  {M.~C.}\ \bibnamefont {{Hoffmann}}}, \bibinfo {author} {\bibfnamefont {B.~K.}\ \bibnamefont {{Ofori-Okai}}}, \bibinfo {author} {\bibfnamefont {M.~E.}\ \bibnamefont {{Kozina}}}, \bibinfo {author} {\bibfnamefont {H.}~\bibnamefont {{Wen}}}, \bibinfo {author} {\bibfnamefont {X.}~\bibnamefont {{Wang}}}, \bibinfo {author} {\bibfnamefont {I.~R.}\ \bibnamefont {{Fisher}}}, \bibinfo {author} {\bibfnamefont {P.}~\bibnamefont {{Jarillo-Herrero}}},\ and\ \bibinfo {author} {\bibfnamefont {N.}~\bibnamefont {{Gedik}}},\ }\bibfield  {title} {\emph {\bibinfo {title} {{Dynamical slowing-down in an ultrafast photoinduced phase transition}}},\ }\href {https://doi.org/10.1103/PhysRevLett.123.097601} {\bibfield  {journal} {\bibinfo  {journal} {Phys. Rev. Lett.}\ }\textbf {\bibinfo {volume} {123}},\ \bibinfo {pages} {097601} (\bibinfo {year} {2019}{\natexlab{a}})}\BibitemShut {NoStop}%
\bibitem [{\citenamefont {{Bartram}}\ \emph {et~al.}(2022)\citenamefont {{Bartram}}, \citenamefont {{Leng}}, \citenamefont {{Wang}}, \citenamefont {{Liu}}, \citenamefont {{Chen}}, \citenamefont {{Peng}}, \citenamefont {{Li}}, \citenamefont {{Yu}}, \citenamefont {{Wu}}, \citenamefont {{Lin}}, \citenamefont {{Zhang}}, \citenamefont {{Tan}},\ and\ \citenamefont {{Yang}}}]{bartram2022ultrafast}%
  \BibitemOpen
  \bibfield  {author} {\bibinfo {author} {\bibfnamefont {F.~M.}\ \bibnamefont {{Bartram}}}, \bibinfo {author} {\bibfnamefont {Y.-C.}\ \bibnamefont {{Leng}}}, \bibinfo {author} {\bibfnamefont {Y.}~\bibnamefont {{Wang}}}, \bibinfo {author} {\bibfnamefont {L.}~\bibnamefont {{Liu}}}, \bibinfo {author} {\bibfnamefont {X.}~\bibnamefont {{Chen}}}, \bibinfo {author} {\bibfnamefont {H.}~\bibnamefont {{Peng}}}, \bibinfo {author} {\bibfnamefont {H.}~\bibnamefont {{Li}}}, \bibinfo {author} {\bibfnamefont {P.}~\bibnamefont {{Yu}}}, \bibinfo {author} {\bibfnamefont {Y.}~\bibnamefont {{Wu}}}, \bibinfo {author} {\bibfnamefont {M.-L.}\ \bibnamefont {{Lin}}}, \bibinfo {author} {\bibfnamefont {J.}~\bibnamefont {{Zhang}}}, \bibinfo {author} {\bibfnamefont {P.-H.}\ \bibnamefont {{Tan}}},\ and\ \bibinfo {author} {\bibfnamefont {L.}~\bibnamefont {{Yang}}},\ }\bibfield  {title} {\emph {\bibinfo {title} {{Ultrafast coherent interlayer phonon dynamics in atomically thin layers of MnBi$_{2}$Te$_{4}$}}},\ }\href
  {https://doi.org/10.1038/s41535-022-00495-x} {\bibfield  {journal} {\bibinfo  {journal} {npj Quantum Mater.}\ }\textbf {\bibinfo {volume} {7}},\ \bibinfo {pages} {84} (\bibinfo {year} {2022})}\BibitemShut {NoStop}%
\bibitem [{\citenamefont {{Zong}}\ \emph {et~al.}(2023{\natexlab{b}})\citenamefont {{Zong}}, \citenamefont {{Nebgen}}, \citenamefont {{Lin}}, \citenamefont {{Spies}},\ and\ \citenamefont {{Zuerch}}}]{zong2023emerging}%
  \BibitemOpen
  \bibfield  {author} {\bibinfo {author} {\bibfnamefont {A.}~\bibnamefont {{Zong}}}, \bibinfo {author} {\bibfnamefont {B.~R.}\ \bibnamefont {{Nebgen}}}, \bibinfo {author} {\bibfnamefont {S.-C.}\ \bibnamefont {{Lin}}}, \bibinfo {author} {\bibfnamefont {J.~A.}\ \bibnamefont {{Spies}}},\ and\ \bibinfo {author} {\bibfnamefont {M.}~\bibnamefont {{Zuerch}}},\ }\bibfield  {title} {\emph {\bibinfo {title} {{Emerging ultrafast techniques for studying quantum materials}}},\ }\href {https://doi.org/10.1038/s41578-022-00530-0} {\bibfield  {journal} {\bibinfo  {journal} {Nat. Rev. Mater.}\ }\textbf {\bibinfo {volume} {8}},\ \bibinfo {pages} {224} (\bibinfo {year} {2023}{\natexlab{b}})}\BibitemShut {NoStop}%
\bibitem [{\citenamefont {{de la Torre}}\ \emph {et~al.}(2021)\citenamefont {{de la Torre}}, \citenamefont {{Kennes}}, \citenamefont {{Claassen}}, \citenamefont {{Gerber}}, \citenamefont {{McIver}},\ and\ \citenamefont {{Sentef}}}]{delatorre2021nonthermal}%
  \BibitemOpen
  \bibfield  {author} {\bibinfo {author} {\bibfnamefont {A.}~\bibnamefont {{de la Torre}}}, \bibinfo {author} {\bibfnamefont {D.~M.}\ \bibnamefont {{Kennes}}}, \bibinfo {author} {\bibfnamefont {M.}~\bibnamefont {{Claassen}}}, \bibinfo {author} {\bibfnamefont {S.}~\bibnamefont {{Gerber}}}, \bibinfo {author} {\bibfnamefont {J.~W.}\ \bibnamefont {{McIver}}},\ and\ \bibinfo {author} {\bibfnamefont {M.~A.}\ \bibnamefont {{Sentef}}},\ }\bibfield  {title} {\emph {\bibinfo {title} {{Colloquium: Nonthermal pathways to ultrafast control in quantum materials}}},\ }\href {https://doi.org/10.1103/RevModPhys.93.041002} {\bibfield  {journal} {\bibinfo  {journal} {Rev. Mod. Phys.}\ }\textbf {\bibinfo {volume} {93}},\ \bibinfo {pages} {041002} (\bibinfo {year} {2021})}\BibitemShut {NoStop}%
\bibitem [{\citenamefont {{Fausti}}\ and\ \citenamefont {{van Loosdrecht}}(2011)}]{fausti2011time}%
  \BibitemOpen
  \bibfield  {author} {\bibinfo {author} {\bibfnamefont {D.}~\bibnamefont {{Fausti}}}\ and\ \bibinfo {author} {\bibfnamefont {P.~H.~M.}\ \bibnamefont {{van Loosdrecht}}},\ }\href {https://doi.org/10.1201/b11040-20} {\emph {\bibinfo {title} {Optical techniques for solid-state materials characterization}}}\ (\bibinfo  {publisher} {CRC Press Boca Raton},\ \bibinfo {year} {2011})\BibitemShut {NoStop}%
\bibitem [{\citenamefont {{Orenstein}}(2012)}]{orenstein2012ultrafast}%
  \BibitemOpen
  \bibfield  {author} {\bibinfo {author} {\bibfnamefont {J.}~\bibnamefont {{Orenstein}}},\ }\bibfield  {title} {\emph {\bibinfo {title} {{Ultrafast spectroscopy of quantum materials}}},\ }\href {https://doi.org/10.1063/PT.3.1717} {\bibfield  {journal} {\bibinfo  {journal} {Phys. Today}\ }\textbf {\bibinfo {volume} {65}},\ \bibinfo {pages} {44} (\bibinfo {year} {2012})}\BibitemShut {NoStop}%
\bibitem [{\citenamefont {{Woerner}}\ \emph {et~al.}(2013)\citenamefont {{Woerner}}, \citenamefont {{Kuehn}}, \citenamefont {{Bowlan}}, \citenamefont {{Reimann}},\ and\ \citenamefont {{Elsaesser}}}]{woerner2013ultrafast}%
  \BibitemOpen
  \bibfield  {author} {\bibinfo {author} {\bibfnamefont {M.}~\bibnamefont {{Woerner}}}, \bibinfo {author} {\bibfnamefont {W.}~\bibnamefont {{Kuehn}}}, \bibinfo {author} {\bibfnamefont {P.}~\bibnamefont {{Bowlan}}}, \bibinfo {author} {\bibfnamefont {K.}~\bibnamefont {{Reimann}}},\ and\ \bibinfo {author} {\bibfnamefont {T.}~\bibnamefont {{Elsaesser}}},\ }\bibfield  {title} {\emph {\bibinfo {title} {{Ultrafast two-dimensional terahertz spectroscopy of elementary excitations in solids}}},\ }\href {https://doi.org/10.1088/1367-2630/15/2/025039} {\bibfield  {journal} {\bibinfo  {journal} {New J. Phys.}\ }\textbf {\bibinfo {volume} {15}},\ \bibinfo {pages} {025039} (\bibinfo {year} {2013})}\BibitemShut {NoStop}%
\bibitem [{\citenamefont {{Giannetti}}\ \emph {et~al.}(2016)\citenamefont {{Giannetti}}, \citenamefont {{Capone}}, \citenamefont {{Fausti}}, \citenamefont {{Fabrizio}}, \citenamefont {{Parmigiani}},\ and\ \citenamefont {{Mihailovic}}}]{giannetti2016ultrafastoptical}%
  \BibitemOpen
  \bibfield  {author} {\bibinfo {author} {\bibfnamefont {C.}~\bibnamefont {{Giannetti}}}, \bibinfo {author} {\bibfnamefont {M.}~\bibnamefont {{Capone}}}, \bibinfo {author} {\bibfnamefont {D.}~\bibnamefont {{Fausti}}}, \bibinfo {author} {\bibfnamefont {M.}~\bibnamefont {{Fabrizio}}}, \bibinfo {author} {\bibfnamefont {F.}~\bibnamefont {{Parmigiani}}},\ and\ \bibinfo {author} {\bibfnamefont {D.}~\bibnamefont {{Mihailovic}}},\ }\bibfield  {title} {\emph {\bibinfo {title} {{Ultrafast optical spectroscopy of strongly correlated materials and high-temperature superconductors: a non-equilibrium approach}}},\ }\href {https://doi.org/10.1080/00018732.2016.1194044} {\bibfield  {journal} {\bibinfo  {journal} {Adv. Phys.}\ }\textbf {\bibinfo {volume} {65}},\ \bibinfo {pages} {58} (\bibinfo {year} {2016})}\BibitemShut {NoStop}%
\bibitem [{\citenamefont {{Lloyd-Hughes}}\ \emph {et~al.}(2021)\citenamefont {{Lloyd-Hughes}}, \citenamefont {{Oppeneer}}, \citenamefont {{Pereira dos Santos}}, \citenamefont {{Schleife}}, \citenamefont {{Meng}}, \citenamefont {{Sentef}}, \citenamefont {{Ruggenthaler}}, \citenamefont {{Rubio}}, \citenamefont {{Radu}}, \citenamefont {{Murnane}}, \citenamefont {{Shi}}, \citenamefont {{Kapteyn}}, \citenamefont {{Stadtm{\"u}ller}}, \citenamefont {{Dani}}, \citenamefont {{da Jornada}}, \citenamefont {{Prinz}}, \citenamefont {{Aeschlimann}}, \citenamefont {{Milot}}, \citenamefont {{Burdanova}}, \citenamefont {{Boland}}, \citenamefont {{Cocker}},\ and\ \citenamefont {{Hegmann}}}]{llloyd2021the2021ultrafast}%
  \BibitemOpen
  \bibfield  {author} {\bibinfo {author} {\bibfnamefont {J.}~\bibnamefont {{Lloyd-Hughes}}}, \bibinfo {author} {\bibfnamefont {P.~M.}\ \bibnamefont {{Oppeneer}}}, \bibinfo {author} {\bibfnamefont {T.}~\bibnamefont {{Pereira dos Santos}}}, \bibinfo {author} {\bibfnamefont {A.}~\bibnamefont {{Schleife}}}, \bibinfo {author} {\bibfnamefont {S.}~\bibnamefont {{Meng}}}, \bibinfo {author} {\bibfnamefont {M.~A.}\ \bibnamefont {{Sentef}}}, \bibinfo {author} {\bibfnamefont {M.}~\bibnamefont {{Ruggenthaler}}}, \bibinfo {author} {\bibfnamefont {A.}~\bibnamefont {{Rubio}}}, \bibinfo {author} {\bibfnamefont {I.}~\bibnamefont {{Radu}}}, \bibinfo {author} {\bibfnamefont {M.}~\bibnamefont {{Murnane}}}, \bibinfo {author} {\bibfnamefont {X.}~\bibnamefont {{Shi}}}, \bibinfo {author} {\bibfnamefont {H.}~\bibnamefont {{Kapteyn}}}, \bibinfo {author} {\bibfnamefont {B.}~\bibnamefont {{Stadtm{\"u}ller}}}, \bibinfo {author} {\bibfnamefont {K.~M.}\ \bibnamefont {{Dani}}}, \bibinfo {author} {\bibfnamefont {F.~H.}\ \bibnamefont {{da
  Jornada}}}, \bibinfo {author} {\bibfnamefont {E.}~\bibnamefont {{Prinz}}}, \bibinfo {author} {\bibfnamefont {M.}~\bibnamefont {{Aeschlimann}}}, \bibinfo {author} {\bibfnamefont {R.~L.}\ \bibnamefont {{Milot}}}, \bibinfo {author} {\bibfnamefont {M.}~\bibnamefont {{Burdanova}}}, \bibinfo {author} {\bibfnamefont {J.}~\bibnamefont {{Boland}}}, \bibinfo {author} {\bibfnamefont {T.}~\bibnamefont {{Cocker}}},\ and\ \bibinfo {author} {\bibfnamefont {F.}~\bibnamefont {{Hegmann}}},\ }\bibfield  {title} {\emph {\bibinfo {title} {{The 2021 ultrafast spectroscopic probes of condensed matter roadmap}}},\ }\href {https://doi.org/10.1088/1361-648X/abfe21} {\bibfield  {journal} {\bibinfo  {journal} {J. Phys. Condens. Matter}\ }\textbf {\bibinfo {volume} {33}},\ \bibinfo {pages} {353001} (\bibinfo {year} {2021})}\BibitemShut {NoStop}%
\bibitem [{\citenamefont {{Dong}}\ \emph {et~al.}(2023)\citenamefont {{Dong}}, \citenamefont {{Zhang}},\ and\ \citenamefont {{Wang}}}]{dong2023recentdevelop}%
  \BibitemOpen
  \bibfield  {author} {\bibinfo {author} {\bibfnamefont {T.}~\bibnamefont {{Dong}}}, \bibinfo {author} {\bibfnamefont {S.-J.}\ \bibnamefont {{Zhang}}},\ and\ \bibinfo {author} {\bibfnamefont {N.-L.}\ \bibnamefont {{Wang}}},\ }\bibfield  {title} {\emph {\bibinfo {title} {{Recent development of ultrafast optical characterizations for quantum materials}}},\ }\href {https://doi.org/10.1002/adma.202110068} {\bibfield  {journal} {\bibinfo  {journal} {Adv. Mater.}\ }\textbf {\bibinfo {volume} {35}},\ \bibinfo {pages} {2110068} (\bibinfo {year} {2023})}\BibitemShut {NoStop}%
\bibitem [{\citenamefont {{Zong}}\ \emph {et~al.}(2021)\citenamefont {{Zong}}, \citenamefont {{Kogar}},\ and\ \citenamefont {{Gedik}}}]{zong2021unconventional}%
  \BibitemOpen
  \bibfield  {author} {\bibinfo {author} {\bibfnamefont {A.}~\bibnamefont {{Zong}}}, \bibinfo {author} {\bibfnamefont {A.}~\bibnamefont {{Kogar}}},\ and\ \bibinfo {author} {\bibfnamefont {N.}~\bibnamefont {{Gedik}}},\ }\bibfield  {title} {\emph {\bibinfo {title} {{Unconventional light-induced states visualized by ultrafast electron diffraction and microscopy}}},\ }\href {https://doi.org/10.1557/s43577-021-00163-8} {\bibfield  {journal} {\bibinfo  {journal} {MRS Bull.}\ }\textbf {\bibinfo {volume} {46}},\ \bibinfo {pages} {720} (\bibinfo {year} {2021})}\BibitemShut {NoStop}%
\bibitem [{\citenamefont {{Filippetto}}\ \emph {et~al.}(2022)\citenamefont {{Filippetto}}, \citenamefont {{Musumeci}}, \citenamefont {{Li}}, \citenamefont {{Siwick}}, \citenamefont {{Otto}}, \citenamefont {{Centurion}},\ and\ \citenamefont {{Nunes}}}]{filippetto2022ultrafastelectron}%
  \BibitemOpen
  \bibfield  {author} {\bibinfo {author} {\bibfnamefont {D.}~\bibnamefont {{Filippetto}}}, \bibinfo {author} {\bibfnamefont {P.}~\bibnamefont {{Musumeci}}}, \bibinfo {author} {\bibfnamefont {R.~K.}\ \bibnamefont {{Li}}}, \bibinfo {author} {\bibfnamefont {B.~J.}\ \bibnamefont {{Siwick}}}, \bibinfo {author} {\bibfnamefont {M.~R.}\ \bibnamefont {{Otto}}}, \bibinfo {author} {\bibfnamefont {M.}~\bibnamefont {{Centurion}}},\ and\ \bibinfo {author} {\bibfnamefont {J.~P.~F.}\ \bibnamefont {{Nunes}}},\ }\bibfield  {title} {\emph {\bibinfo {title} {{Ultrafast electron diffraction: Visualizing dynamic states of matter}}},\ }\href {https://doi.org/10.1103/RevModPhys.94.045004} {\bibfield  {journal} {\bibinfo  {journal} {Rev. Mod. Phys.}\ }\textbf {\bibinfo {volume} {94}},\ \bibinfo {pages} {045004} (\bibinfo {year} {2022})}\BibitemShut {NoStop}%
\bibitem [{\citenamefont {Lee}\ \emph {et~al.}(2024)\citenamefont {Lee}, \citenamefont {Oang}, \citenamefont {Kim},\ and\ \citenamefont {Ihee}}]{lee2024structural}%
  \BibitemOpen
  \bibfield  {author} {\bibinfo {author} {\bibfnamefont {Y.}~\bibnamefont {Lee}}, \bibinfo {author} {\bibfnamefont {K.~Y.}\ \bibnamefont {Oang}}, \bibinfo {author} {\bibfnamefont {D.}~\bibnamefont {Kim}},\ and\ \bibinfo {author} {\bibfnamefont {H.}~\bibnamefont {Ihee}},\ }\bibfield  {title} {\emph {\bibinfo {title} {{A comparative review of time-resolved x-ray and electron scattering to probe structural dynamics}}},\ }\href {https://doi.org/10.1063/4.0000249} {\bibfield  {journal} {\bibinfo  {journal} {Struct. Dyn.}\ }\textbf {\bibinfo {volume} {11}},\ \bibinfo {pages} {031301} (\bibinfo {year} {2024})}\BibitemShut {NoStop}%
\bibitem [{\citenamefont {{Buzzi}}\ \emph {et~al.}(2018)\citenamefont {{Buzzi}}, \citenamefont {{F{\"o}rst}}, \citenamefont {{Mankowsky}},\ and\ \citenamefont {{Cavalleri}}}]{buzzi2018probing}%
  \BibitemOpen
  \bibfield  {author} {\bibinfo {author} {\bibfnamefont {M.}~\bibnamefont {{Buzzi}}}, \bibinfo {author} {\bibfnamefont {M.}~\bibnamefont {{F{\"o}rst}}}, \bibinfo {author} {\bibfnamefont {R.}~\bibnamefont {{Mankowsky}}},\ and\ \bibinfo {author} {\bibfnamefont {A.}~\bibnamefont {{Cavalleri}}},\ }\bibfield  {title} {\emph {\bibinfo {title} {{Probing dynamics in quantum materials with femtosecond X-rays}}},\ }\href {https://doi.org/10.1038/s41578-018-0024-9} {\bibfield  {journal} {\bibinfo  {journal} {Nat. Rev. Mater.}\ }\textbf {\bibinfo {volume} {3}},\ \bibinfo {pages} {299} (\bibinfo {year} {2018})}\BibitemShut {NoStop}%
\bibitem [{\citenamefont {{Cao}}\ \emph {et~al.}(2019)\citenamefont {{Cao}}, \citenamefont {{Mazzone}}, \citenamefont {{Meyers}}, \citenamefont {{Hill}}, \citenamefont {{Liu}}, \citenamefont {{Wall}},\ and\ \citenamefont {{Dean}}}]{cao2019ultrafast}%
  \BibitemOpen
  \bibfield  {author} {\bibinfo {author} {\bibfnamefont {Y.}~\bibnamefont {{Cao}}}, \bibinfo {author} {\bibfnamefont {D.~G.}\ \bibnamefont {{Mazzone}}}, \bibinfo {author} {\bibfnamefont {D.}~\bibnamefont {{Meyers}}}, \bibinfo {author} {\bibfnamefont {J.~P.}\ \bibnamefont {{Hill}}}, \bibinfo {author} {\bibfnamefont {X.}~\bibnamefont {{Liu}}}, \bibinfo {author} {\bibfnamefont {S.}~\bibnamefont {{Wall}}},\ and\ \bibinfo {author} {\bibfnamefont {M.~P.~M.}\ \bibnamefont {{Dean}}},\ }\bibfield  {title} {\emph {\bibinfo {title} {{Ultrafast dynamics of spin and orbital correlations in quantum materials: an energy- and momentum-resolved perspective}}},\ }\href {https://doi.org/10.1098/rsta.2017.0480} {\bibfield  {journal} {\bibinfo  {journal} {Philos. Trans. R. Soc. A}\ }\textbf {\bibinfo {volume} {377}},\ \bibinfo {pages} {20170480} (\bibinfo {year} {2019})}\BibitemShut {NoStop}%
\bibitem [{\citenamefont {{Mitrano}}\ and\ \citenamefont {{Wang}}(2020)}]{mitrano2020probing}%
  \BibitemOpen
  \bibfield  {author} {\bibinfo {author} {\bibfnamefont {M.}~\bibnamefont {{Mitrano}}}\ and\ \bibinfo {author} {\bibfnamefont {Y.}~\bibnamefont {{Wang}}},\ }\bibfield  {title} {\emph {\bibinfo {title} {{Probing light-driven quantum materials with ultrafast resonant inelastic X-ray scattering}}},\ }\href {https://doi.org/10.1038/s42005-020-00447-6} {\bibfield  {journal} {\bibinfo  {journal} {Commun. Phys.}\ }\textbf {\bibinfo {volume} {3}},\ \bibinfo {pages} {184} (\bibinfo {year} {2020})}\BibitemShut {NoStop}%
\bibitem [{\citenamefont {Mitrano}\ \emph {et~al.}(2024)\citenamefont {Mitrano}, \citenamefont {Johnston}, \citenamefont {Kim},\ and\ \citenamefont {Dean}}]{mitrano2024exploring}%
  \BibitemOpen
  \bibfield  {author} {\bibinfo {author} {\bibfnamefont {M.}~\bibnamefont {Mitrano}}, \bibinfo {author} {\bibfnamefont {S.}~\bibnamefont {Johnston}}, \bibinfo {author} {\bibfnamefont {Y.-J.}\ \bibnamefont {Kim}},\ and\ \bibinfo {author} {\bibfnamefont {M.~P.~M.}\ \bibnamefont {Dean}},\ }\bibfield  {title} {\emph {\bibinfo {title} {Exploring quantum materials with resonant inelastic x-ray scattering}},\ }\href {https://doi.org/10.1103/PhysRevX.14.040501} {\bibfield  {journal} {\bibinfo  {journal} {Phys. Rev. X}\ }\textbf {\bibinfo {volume} {14}},\ \bibinfo {pages} {040501} (\bibinfo {year} {2024})}\BibitemShut {NoStop}%
\bibitem [{\citenamefont {{Bovensiepen}}\ and\ \citenamefont {{Kirchmann}}(2012)}]{boven2012elementary}%
  \BibitemOpen
  \bibfield  {author} {\bibinfo {author} {\bibfnamefont {U.}~\bibnamefont {{Bovensiepen}}}\ and\ \bibinfo {author} {\bibfnamefont {P.~S.}\ \bibnamefont {{Kirchmann}}},\ }\bibfield  {title} {\emph {\bibinfo {title} {{Elementary relaxation processes investigated by femtosecond photoelectron spectroscopy of two-dimensional materials}}},\ }\href {https://doi.org/10.1002/lpor.201000035} {\bibfield  {journal} {\bibinfo  {journal} {Laser Photonics Rev.}\ }\textbf {\bibinfo {volume} {6}},\ \bibinfo {pages} {589} (\bibinfo {year} {2012})}\BibitemShut {NoStop}%
\bibitem [{\citenamefont {{Smallwood}}\ \emph {et~al.}(2016)\citenamefont {{Smallwood}}, \citenamefont {{Kaindl}},\ and\ \citenamefont {{Lanzara}}}]{smallwood2016ultrafast}%
  \BibitemOpen
  \bibfield  {author} {\bibinfo {author} {\bibfnamefont {C.~L.}\ \bibnamefont {{Smallwood}}}, \bibinfo {author} {\bibfnamefont {R.~A.}\ \bibnamefont {{Kaindl}}},\ and\ \bibinfo {author} {\bibfnamefont {A.}~\bibnamefont {{Lanzara}}},\ }\bibfield  {title} {\emph {\bibinfo {title} {{Ultrafast angle-resolved photoemission spectroscopy of quantum materials}}},\ }\href {https://doi.org/10.1209/0295-5075/115/27001} {\bibfield  {journal} {\bibinfo  {journal} {EPL}\ }\textbf {\bibinfo {volume} {115}},\ \bibinfo {pages} {27001} (\bibinfo {year} {2016})}\BibitemShut {NoStop}%
\bibitem [{\citenamefont {{Zhou}}\ \emph {et~al.}(2018)\citenamefont {{Zhou}}, \citenamefont {{He}}, \citenamefont {{Liu}}, \citenamefont {{Zhao}}, \citenamefont {{Yu}},\ and\ \citenamefont {{Zhang}}}]{zhou2018newdevelopments}%
  \BibitemOpen
  \bibfield  {author} {\bibinfo {author} {\bibfnamefont {X.}~\bibnamefont {{Zhou}}}, \bibinfo {author} {\bibfnamefont {S.}~\bibnamefont {{He}}}, \bibinfo {author} {\bibfnamefont {G.}~\bibnamefont {{Liu}}}, \bibinfo {author} {\bibfnamefont {L.}~\bibnamefont {{Zhao}}}, \bibinfo {author} {\bibfnamefont {L.}~\bibnamefont {{Yu}}},\ and\ \bibinfo {author} {\bibfnamefont {W.}~\bibnamefont {{Zhang}}},\ }\bibfield  {title} {\emph {\bibinfo {title} {{New developments in laser-based photoemission spectroscopy and its scientific applications: a key issues review}}},\ }\href {https://doi.org/10.1088/1361-6633/aab0cc} {\bibfield  {journal} {\bibinfo  {journal} {Rep. Prog. Phys.}\ }\textbf {\bibinfo {volume} {81}},\ \bibinfo {pages} {062101} (\bibinfo {year} {2018})}\BibitemShut {NoStop}%
\bibitem [{\citenamefont {Huang}\ \emph {et~al.}(2022)\citenamefont {Huang}, \citenamefont {Duan},\ and\ \citenamefont {Zhang}}]{huang2022highresilution}%
  \BibitemOpen
  \bibfield  {author} {\bibinfo {author} {\bibfnamefont {C.}~\bibnamefont {Huang}}, \bibinfo {author} {\bibfnamefont {S.}~\bibnamefont {Duan}},\ and\ \bibinfo {author} {\bibfnamefont {W.}~\bibnamefont {Zhang}},\ }\bibfield  {title} {\emph {\bibinfo {title} {{High-resolution time-and angle-resolved photoemission studies on quantum materials}}},\ }\href {https://doi.org/10.1007/s44214-022-00013-x} {\bibfield  {journal} {\bibinfo  {journal} {Quantum Frontiers}\ }\textbf {\bibinfo {volume} {1}},\ \bibinfo {pages} {15} (\bibinfo {year} {2022})}\BibitemShut {NoStop}%
\bibitem [{\citenamefont {{Boschini}}\ \emph {et~al.}(2024)\citenamefont {{Boschini}}, \citenamefont {{Zonno}},\ and\ \citenamefont {{Damascelli}}}]{boschini2024timeresolvedarpes}%
  \BibitemOpen
  \bibfield  {author} {\bibinfo {author} {\bibfnamefont {F.}~\bibnamefont {{Boschini}}}, \bibinfo {author} {\bibfnamefont {M.}~\bibnamefont {{Zonno}}},\ and\ \bibinfo {author} {\bibfnamefont {A.}~\bibnamefont {{Damascelli}}},\ }\bibfield  {title} {\emph {\bibinfo {title} {{Time-resolved ARPES studies of quantum materials}}},\ }\href {https://doi.org/10.1103/RevModPhys.96.015003} {\bibfield  {journal} {\bibinfo  {journal} {Rev. Mod. Phys.}\ }\textbf {\bibinfo {volume} {96}},\ \bibinfo {pages} {015003} (\bibinfo {year} {2024})}\BibitemShut {NoStop}%
\bibitem [{\citenamefont {{Mankowsky}}\ \emph {et~al.}(2016)\citenamefont {{Mankowsky}}, \citenamefont {{F{\"o}rst}},\ and\ \citenamefont {{Cavalleri}}}]{mankowsky2016nonequilibrium}%
  \BibitemOpen
  \bibfield  {author} {\bibinfo {author} {\bibfnamefont {R.}~\bibnamefont {{Mankowsky}}}, \bibinfo {author} {\bibfnamefont {M.}~\bibnamefont {{F{\"o}rst}}},\ and\ \bibinfo {author} {\bibfnamefont {A.}~\bibnamefont {{Cavalleri}}},\ }\bibfield  {title} {\emph {\bibinfo {title} {{Non-equilibrium control of complex solids by nonlinear phononics}}},\ }\href {https://doi.org/10.1088/0034-4885/79/6/064503} {\bibfield  {journal} {\bibinfo  {journal} {Rep. Prog. Phys.}\ }\textbf {\bibinfo {volume} {79}},\ \bibinfo {pages} {064503} (\bibinfo {year} {2016})}\BibitemShut {NoStop}%
\bibitem [{\citenamefont {{Nicoletti}}\ and\ \citenamefont {{Cavalleri}}(2016)}]{nicoletti2016nonlinear}%
  \BibitemOpen
  \bibfield  {author} {\bibinfo {author} {\bibfnamefont {D.}~\bibnamefont {{Nicoletti}}}\ and\ \bibinfo {author} {\bibfnamefont {A.}~\bibnamefont {{Cavalleri}}},\ }\bibfield  {title} {\emph {\bibinfo {title} {{Nonlinear light{\textendash}matter interaction at terahertz frequencies}}},\ }\href {https://doi.org/10.1364/AOP.8.000401} {\bibfield  {journal} {\bibinfo  {journal} {Adv. Opt. Photonics}\ }\textbf {\bibinfo {volume} {8}},\ \bibinfo {pages} {401} (\bibinfo {year} {2016})}\BibitemShut {NoStop}%
\bibitem [{\citenamefont {{Disa}}\ \emph {et~al.}(2021)\citenamefont {{Disa}}, \citenamefont {{Nova}},\ and\ \citenamefont {{Cavalleri}}}]{disa2021engineering}%
  \BibitemOpen
  \bibfield  {author} {\bibinfo {author} {\bibfnamefont {A.~S.}\ \bibnamefont {{Disa}}}, \bibinfo {author} {\bibfnamefont {T.~F.}\ \bibnamefont {{Nova}}},\ and\ \bibinfo {author} {\bibfnamefont {A.}~\bibnamefont {{Cavalleri}}},\ }\bibfield  {title} {\emph {\bibinfo {title} {{Engineering crystal structures with light}}},\ }\href {https://doi.org/10.1038/s41567-021-01366-1} {\bibfield  {journal} {\bibinfo  {journal} {Nat. Phys.}\ }\textbf {\bibinfo {volume} {17}},\ \bibinfo {pages} {1087} (\bibinfo {year} {2021})}\BibitemShut {NoStop}%
\bibitem [{\citenamefont {{Oka}}\ and\ \citenamefont {{Kitamura}}(2019)}]{oka2019floquet}%
  \BibitemOpen
  \bibfield  {author} {\bibinfo {author} {\bibfnamefont {T.}~\bibnamefont {{Oka}}}\ and\ \bibinfo {author} {\bibfnamefont {S.}~\bibnamefont {{Kitamura}}},\ }\bibfield  {title} {\emph {\bibinfo {title} {{Floquet engineering of quantum materials}}},\ }\href {https://doi.org/10.1146/annurev-conmatphys-031218-013423} {\bibfield  {journal} {\bibinfo  {journal} {Annu. Rev. Condens. Matter Phys}\ }\textbf {\bibinfo {volume} {10}},\ \bibinfo {pages} {387} (\bibinfo {year} {2019})}\BibitemShut {NoStop}%
\bibitem [{\citenamefont {{Rudner}}\ and\ \citenamefont {{Lindner}}(2020)}]{rudner2020band}%
  \BibitemOpen
  \bibfield  {author} {\bibinfo {author} {\bibfnamefont {M.~S.}\ \bibnamefont {{Rudner}}}\ and\ \bibinfo {author} {\bibfnamefont {N.~H.}\ \bibnamefont {{Lindner}}},\ }\bibfield  {title} {\emph {\bibinfo {title} {{Band structure engineering and non-equilibrium dynamics in Floquet topological insulators}}},\ }\href {https://doi.org/10.1038/s42254-020-0170-z} {\bibfield  {journal} {\bibinfo  {journal} {Nat. Rev. Phys.}\ }\textbf {\bibinfo {volume} {2}},\ \bibinfo {pages} {229} (\bibinfo {year} {2020})}\BibitemShut {NoStop}%
\bibitem [{\citenamefont {{Bao}}\ \emph {et~al.}(2022)\citenamefont {{Bao}}, \citenamefont {{Tang}}, \citenamefont {{Sun}},\ and\ \citenamefont {{Zhou}}}]{bao2022lightinduced}%
  \BibitemOpen
  \bibfield  {author} {\bibinfo {author} {\bibfnamefont {C.}~\bibnamefont {{Bao}}}, \bibinfo {author} {\bibfnamefont {P.}~\bibnamefont {{Tang}}}, \bibinfo {author} {\bibfnamefont {D.}~\bibnamefont {{Sun}}},\ and\ \bibinfo {author} {\bibfnamefont {S.}~\bibnamefont {{Zhou}}},\ }\bibfield  {title} {\emph {\bibinfo {title} {{Light-induced emergent phenomena in 2D materials and topological materials}}},\ }\href {https://doi.org/10.1038/s42254-021-00388-1} {\bibfield  {journal} {\bibinfo  {journal} {Nat. Rev. Phys.}\ }\textbf {\bibinfo {volume} {4}},\ \bibinfo {pages} {33} (\bibinfo {year} {2022})}\BibitemShut {NoStop}%
\bibitem [{\citenamefont {{Kirilyuk}}\ \emph {et~al.}(2010)\citenamefont {{Kirilyuk}}, \citenamefont {{Kimel}},\ and\ \citenamefont {{Rasing}}}]{kirilyuk2010ultrafastoptical}%
  \BibitemOpen
  \bibfield  {author} {\bibinfo {author} {\bibfnamefont {A.}~\bibnamefont {{Kirilyuk}}}, \bibinfo {author} {\bibfnamefont {A.~V.}\ \bibnamefont {{Kimel}}},\ and\ \bibinfo {author} {\bibfnamefont {T.}~\bibnamefont {{Rasing}}},\ }\bibfield  {title} {\emph {\bibinfo {title} {{Ultrafast optical manipulation of magnetic order}}},\ }\href {https://doi.org/10.1103/RevModPhys.82.2731} {\bibfield  {journal} {\bibinfo  {journal} {Rev. Mod. Phys.}\ }\textbf {\bibinfo {volume} {82}},\ \bibinfo {pages} {2731} (\bibinfo {year} {2010})}\BibitemShut {NoStop}%
\bibitem [{\citenamefont {{Zhang}}\ and\ \citenamefont {{Averitt}}(2014)}]{zhang2014dynamicalcontrol}%
  \BibitemOpen
  \bibfield  {author} {\bibinfo {author} {\bibfnamefont {J.}~\bibnamefont {{Zhang}}}\ and\ \bibinfo {author} {\bibfnamefont {R.~D.}\ \bibnamefont {{Averitt}}},\ }\bibfield  {title} {\emph {\bibinfo {title} {{Dynamics and control in complex transition metal oxides}}},\ }\href {https://doi.org/10.1146/annurev-matsci-070813-113258} {\bibfield  {journal} {\bibinfo  {journal} {Annu. Rev. Mater. Res.}\ }\textbf {\bibinfo {volume} {44}},\ \bibinfo {pages} {19} (\bibinfo {year} {2014})}\BibitemShut {NoStop}%
\bibitem [{\citenamefont {{Cheng}}\ \emph {et~al.}(2024)\citenamefont {{Cheng}}, \citenamefont {{Zong}}, \citenamefont {{Wu}}, \citenamefont {{Meng}}, \citenamefont {{Xia}}, \citenamefont {{Qi}}, \citenamefont {{Zhu}}, \citenamefont {{Zou}}, \citenamefont {{Jiang}}, \citenamefont {{Guo}}, \citenamefont {{van Wezel}}, \citenamefont {{Kogar}}, \citenamefont {{Zuerch}}, \citenamefont {{Zhang}}, \citenamefont {{Zhu}},\ and\ \citenamefont {{Xiang}}}]{cheng2024ultrafast_TiSe2}%
  \BibitemOpen
  \bibfield  {author} {\bibinfo {author} {\bibfnamefont {Y.}~\bibnamefont {{Cheng}}}, \bibinfo {author} {\bibfnamefont {A.}~\bibnamefont {{Zong}}}, \bibinfo {author} {\bibfnamefont {L.}~\bibnamefont {{Wu}}}, \bibinfo {author} {\bibfnamefont {Q.}~\bibnamefont {{Meng}}}, \bibinfo {author} {\bibfnamefont {W.}~\bibnamefont {{Xia}}}, \bibinfo {author} {\bibfnamefont {F.}~\bibnamefont {{Qi}}}, \bibinfo {author} {\bibfnamefont {P.}~\bibnamefont {{Zhu}}}, \bibinfo {author} {\bibfnamefont {X.}~\bibnamefont {{Zou}}}, \bibinfo {author} {\bibfnamefont {T.}~\bibnamefont {{Jiang}}}, \bibinfo {author} {\bibfnamefont {Y.}~\bibnamefont {{Guo}}}, \bibinfo {author} {\bibfnamefont {J.}~\bibnamefont {{van Wezel}}}, \bibinfo {author} {\bibfnamefont {A.}~\bibnamefont {{Kogar}}}, \bibinfo {author} {\bibfnamefont {M.~W.}\ \bibnamefont {{Zuerch}}}, \bibinfo {author} {\bibfnamefont {J.}~\bibnamefont {{Zhang}}}, \bibinfo {author} {\bibfnamefont {Y.}~\bibnamefont {{Zhu}}},\ and\ \bibinfo {author} {\bibfnamefont {D.}~\bibnamefont
  {{Xiang}}},\ }\bibfield  {title} {\emph {\bibinfo {title} {{Ultrafast formation of topological defects in a two-dimensional charge density wave}}},\ }\href {https://doi.org/10.1038/s41567-023-02279-x} {\bibfield  {journal} {\bibinfo  {journal} {Nat. Phys.}\ }\textbf {\bibinfo {volume} {20}},\ \bibinfo {pages} {54} (\bibinfo {year} {2024})}\BibitemShut {NoStop}%
\bibitem [{\citenamefont {Baldini}(2018)}]{baldini2018nonequilibrium}%
  \BibitemOpen
  \bibfield  {author} {\bibinfo {author} {\bibfnamefont {E.}~\bibnamefont {Baldini}},\ }\href {https://doi.org/10.1007/978-3-319-77498-5} {\emph {\bibinfo {title} {{Nonequilibrium dynamics of collective excitations in quantum materials}}}}\ (\bibinfo  {publisher} {Springer},\ \bibinfo {year} {2018})\BibitemShut {NoStop}%
\bibitem [{\citenamefont {{Basov}}\ \emph {et~al.}(2016)\citenamefont {{Basov}}, \citenamefont {{Fogler}},\ and\ \citenamefont {Garc{\'\i}a~de Abajo}}]{basov2016polaritons}%
  \BibitemOpen
  \bibfield  {author} {\bibinfo {author} {\bibfnamefont {D.~N.}\ \bibnamefont {{Basov}}}, \bibinfo {author} {\bibfnamefont {M.~M.}\ \bibnamefont {{Fogler}}},\ and\ \bibinfo {author} {\bibfnamefont {F.~J.}\ \bibnamefont {Garc{\'\i}a~de Abajo}},\ }\bibfield  {title} {\emph {\bibinfo {title} {{Polaritons in van der Waals materials}}},\ }\href {https://doi.org/10.1126/science.aag1992} {\bibfield  {journal} {\bibinfo  {journal} {Science}\ }\textbf {\bibinfo {volume} {354}},\ \bibinfo {pages} {aag1992} (\bibinfo {year} {2016})}\BibitemShut {NoStop}%
\bibitem [{\citenamefont {{Basov}}\ \emph {et~al.}(2020)\citenamefont {{Basov}}, \citenamefont {{Asenjo-Garcia}}, \citenamefont {{Schuck}}, \citenamefont {{Zhu}},\ and\ \citenamefont {{Rubio}}}]{basov2020polariton}%
  \BibitemOpen
  \bibfield  {author} {\bibinfo {author} {\bibfnamefont {D.~N.}\ \bibnamefont {{Basov}}}, \bibinfo {author} {\bibfnamefont {A.}~\bibnamefont {{Asenjo-Garcia}}}, \bibinfo {author} {\bibfnamefont {P.~J.}\ \bibnamefont {{Schuck}}}, \bibinfo {author} {\bibfnamefont {X.}~\bibnamefont {{Zhu}}},\ and\ \bibinfo {author} {\bibfnamefont {A.}~\bibnamefont {{Rubio}}},\ }\bibfield  {title} {\emph {\bibinfo {title} {{Polariton panorama}}},\ }\href {https://doi.org/10.1515/nanoph-2020-0449} {\bibfield  {journal} {\bibinfo  {journal} {Nanophotonics}\ }\textbf {\bibinfo {volume} {10}},\ \bibinfo {pages} {449} (\bibinfo {year} {2020})}\BibitemShut {NoStop}%
\bibitem [{\citenamefont {Baldini}\ \emph {et~al.}(2020)\citenamefont {Baldini}, \citenamefont {Belvin}, \citenamefont {Rodriguez-Vega}, \citenamefont {Ozel}, \citenamefont {Legut}, \citenamefont {Kozłowski}, \citenamefont {Oleś}, \citenamefont {Parlinski}, \citenamefont {Piekarz}, \citenamefont {Lorenzana}, \citenamefont {Fiete},\ and\ \citenamefont {Gedik}}]{baldini2020discovery}%
  \BibitemOpen
  \bibfield  {author} {\bibinfo {author} {\bibfnamefont {E.}~\bibnamefont {Baldini}}, \bibinfo {author} {\bibfnamefont {C.~A.}\ \bibnamefont {Belvin}}, \bibinfo {author} {\bibfnamefont {M.}~\bibnamefont {Rodriguez-Vega}}, \bibinfo {author} {\bibfnamefont {I.~O.}\ \bibnamefont {Ozel}}, \bibinfo {author} {\bibfnamefont {D.}~\bibnamefont {Legut}}, \bibinfo {author} {\bibfnamefont {A.}~\bibnamefont {Kozłowski}}, \bibinfo {author} {\bibfnamefont {A.~M.}\ \bibnamefont {Oleś}}, \bibinfo {author} {\bibfnamefont {K.}~\bibnamefont {Parlinski}}, \bibinfo {author} {\bibfnamefont {P.}~\bibnamefont {Piekarz}}, \bibinfo {author} {\bibfnamefont {J.}~\bibnamefont {Lorenzana}}, \bibinfo {author} {\bibfnamefont {G.~A.}\ \bibnamefont {Fiete}},\ and\ \bibinfo {author} {\bibfnamefont {N.}~\bibnamefont {Gedik}},\ }\bibfield  {title} {\emph {\bibinfo {title} {Discovery of the soft electronic modes of the trimeron order in magnetite}},\ }\href {https://doi.org/10.1038/s41567-020-0823-y} {\bibfield  {journal} {\bibinfo  {journal}
  {Nat. Phys.}\ }\textbf {\bibinfo {volume} {16}},\ \bibinfo {pages} {541--545} (\bibinfo {year} {2020})}\BibitemShut {NoStop}%
\bibitem [{\citenamefont {{Basov}}\ \emph {et~al.}(2017)\citenamefont {{Basov}}, \citenamefont {{Averitt}},\ and\ \citenamefont {{Hsieh}}}]{basov2017towards}%
  \BibitemOpen
  \bibfield  {author} {\bibinfo {author} {\bibfnamefont {D.~N.}\ \bibnamefont {{Basov}}}, \bibinfo {author} {\bibfnamefont {R.~D.}\ \bibnamefont {{Averitt}}},\ and\ \bibinfo {author} {\bibfnamefont {D.}~\bibnamefont {{Hsieh}}},\ }\bibfield  {title} {\emph {\bibinfo {title} {{Towards properties on demand in quantum materials}}},\ }\href {https://doi.org/10.1038/nmat5017} {\bibfield  {journal} {\bibinfo  {journal} {Nat. Mater.}\ }\textbf {\bibinfo {volume} {16}},\ \bibinfo {pages} {1077} (\bibinfo {year} {2017})}\BibitemShut {NoStop}%
\bibitem [{\citenamefont {{Demsar}}\ \emph {et~al.}(1999)\citenamefont {{Demsar}}, \citenamefont {{Biljakovi{\'c} }},\ and\ \citenamefont {{Mihailovic}}}]{demsar1999singleparticle}%
  \BibitemOpen
  \bibfield  {author} {\bibinfo {author} {\bibfnamefont {J.}~\bibnamefont {{Demsar}}}, \bibinfo {author} {\bibfnamefont {K.}~\bibnamefont {{Biljakovi{\'c} }}},\ and\ \bibinfo {author} {\bibfnamefont {D.}~\bibnamefont {{Mihailovic}}},\ }\bibfield  {title} {\emph {\bibinfo {title} {{Single particle and collective excitations in the one-dimensional charge density wave solid K$_{0.3}$MoO$_{3}$ probed in real time by femtosecond spectroscopy}}},\ }\href {https://doi.org/10.1103/PhysRevLett.83.800} {\bibfield  {journal} {\bibinfo  {journal} {Phys. Rev. Lett.}\ }\textbf {\bibinfo {volume} {83}},\ \bibinfo {pages} {800} (\bibinfo {year} {1999})}\BibitemShut {NoStop}%
\bibitem [{\citenamefont {{Demsar}}\ \emph {et~al.}(2002)\citenamefont {{Demsar}}, \citenamefont {{Forr{\'o}}}, \citenamefont {{Berger}},\ and\ \citenamefont {{Mihailovic}}}]{demsar2002femtosecond}%
  \BibitemOpen
  \bibfield  {author} {\bibinfo {author} {\bibfnamefont {J.}~\bibnamefont {{Demsar}}}, \bibinfo {author} {\bibfnamefont {L.}~\bibnamefont {{Forr{\'o}}}}, \bibinfo {author} {\bibfnamefont {H.}~\bibnamefont {{Berger}}},\ and\ \bibinfo {author} {\bibfnamefont {D.}~\bibnamefont {{Mihailovic}}},\ }\bibfield  {title} {\emph {\bibinfo {title} {{Femtosecond snapshots of gap-forming charge-density-wave correlations in quasi-two-dimensional dichalcogenides 1T-TaS$_{2}$ and 2H-TaSe$_{2}$}}},\ }\href {https://doi.org/10.1103/PhysRevB.66.041101} {\bibfield  {journal} {\bibinfo  {journal} {Phys. Rev. B}\ }\textbf {\bibinfo {volume} {66}},\ \bibinfo {pages} {041101} (\bibinfo {year} {2002})}\BibitemShut {NoStop}%
\bibitem [{\citenamefont {{Schmitt}}\ \emph {et~al.}(2008)\citenamefont {{Schmitt}}, \citenamefont {{Kirchmann}}, \citenamefont {{Bovensiepen}}, \citenamefont {{Moore}}, \citenamefont {{Rettig}}, \citenamefont {{Krenz}}, \citenamefont {{Chu}}, \citenamefont {{Ru}}, \citenamefont {{Perfetti}}, \citenamefont {{Lu}}, \citenamefont {{Wolf}}, \citenamefont {{Fisher}},\ and\ \citenamefont {{Shen}}}]{schmitt2008transient}%
  \BibitemOpen
  \bibfield  {author} {\bibinfo {author} {\bibfnamefont {F.}~\bibnamefont {{Schmitt}}}, \bibinfo {author} {\bibfnamefont {P.~S.}\ \bibnamefont {{Kirchmann}}}, \bibinfo {author} {\bibfnamefont {U.}~\bibnamefont {{Bovensiepen}}}, \bibinfo {author} {\bibfnamefont {R.~G.}\ \bibnamefont {{Moore}}}, \bibinfo {author} {\bibfnamefont {L.}~\bibnamefont {{Rettig}}}, \bibinfo {author} {\bibfnamefont {M.}~\bibnamefont {{Krenz}}}, \bibinfo {author} {\bibfnamefont {J.~H.}\ \bibnamefont {{Chu}}}, \bibinfo {author} {\bibfnamefont {N.}~\bibnamefont {{Ru}}}, \bibinfo {author} {\bibfnamefont {L.}~\bibnamefont {{Perfetti}}}, \bibinfo {author} {\bibfnamefont {D.~H.}\ \bibnamefont {{Lu}}}, \bibinfo {author} {\bibfnamefont {M.}~\bibnamefont {{Wolf}}}, \bibinfo {author} {\bibfnamefont {I.~R.}\ \bibnamefont {{Fisher}}},\ and\ \bibinfo {author} {\bibfnamefont {Z.~X.}\ \bibnamefont {{Shen}}},\ }\bibfield  {title} {\emph {\bibinfo {title} {{Transient electronic structure and melting of a charge density wave in TbTe$_{3}$}}},\ }\href
  {https://doi.org/10.1126/science.1160778} {\bibfield  {journal} {\bibinfo  {journal} {Science}\ }\textbf {\bibinfo {volume} {321}},\ \bibinfo {pages} {1649} (\bibinfo {year} {2008})}\BibitemShut {NoStop}%
\bibitem [{\citenamefont {Yusupov}\ \emph {et~al.}(2010)\citenamefont {Yusupov}, \citenamefont {Mertelj}, \citenamefont {Kabanov}, \citenamefont {Brazovskii}, \citenamefont {Kusar}, \citenamefont {Chu}, \citenamefont {Fisher},\ and\ \citenamefont {Mihailovic}}]{yusupov2010coherent}%
  \BibitemOpen
  \bibfield  {author} {\bibinfo {author} {\bibfnamefont {R.}~\bibnamefont {Yusupov}}, \bibinfo {author} {\bibfnamefont {T.}~\bibnamefont {Mertelj}}, \bibinfo {author} {\bibfnamefont {V.~V.}\ \bibnamefont {Kabanov}}, \bibinfo {author} {\bibfnamefont {S.}~\bibnamefont {Brazovskii}}, \bibinfo {author} {\bibfnamefont {P.}~\bibnamefont {Kusar}}, \bibinfo {author} {\bibfnamefont {J.-H.}\ \bibnamefont {Chu}}, \bibinfo {author} {\bibfnamefont {I.~R.}\ \bibnamefont {Fisher}},\ and\ \bibinfo {author} {\bibfnamefont {D.}~\bibnamefont {Mihailovic}},\ }\bibfield  {title} {\emph {\bibinfo {title} {{Coherent dynamics of macroscopic electronic order through a symmetry breaking transition}}},\ }\href {https://doi.org/10.1038/nphys1738} {\bibfield  {journal} {\bibinfo  {journal} {Nat. Phys.}\ }\textbf {\bibinfo {volume} {6}},\ \bibinfo {pages} {681--684} (\bibinfo {year} {2010})}\BibitemShut {NoStop}%
\bibitem [{\citenamefont {Mihailovic}(2019)}]{mihailovic2019importance}%
  \BibitemOpen
  \bibfield  {author} {\bibinfo {author} {\bibfnamefont {D.}~\bibnamefont {Mihailovic}},\ }\bibfield  {title} {\emph {\bibinfo {title} {{The importance of topological defects in photoexcited phase transitions including memory applications}}},\ }\href {https://doi.org/10.3390/app9050890} {\bibfield  {journal} {\bibinfo  {journal} {Appl. Sci.}\ }\textbf {\bibinfo {volume} {9}},\ \bibinfo {pages} {890} (\bibinfo {year} {2019})}\BibitemShut {NoStop}%
\bibitem [{\citenamefont {Horstmann}\ \emph {et~al.}(2020)\citenamefont {Horstmann}, \citenamefont {Böckmann}, \citenamefont {Wit}, \citenamefont {Kurtz}, \citenamefont {Storeck},\ and\ \citenamefont {Ropers}}]{horstmann2020coherent}%
  \BibitemOpen
  \bibfield  {author} {\bibinfo {author} {\bibfnamefont {J.~G.}\ \bibnamefont {Horstmann}}, \bibinfo {author} {\bibfnamefont {H.}~\bibnamefont {Böckmann}}, \bibinfo {author} {\bibfnamefont {B.}~\bibnamefont {Wit}}, \bibinfo {author} {\bibfnamefont {F.}~\bibnamefont {Kurtz}}, \bibinfo {author} {\bibfnamefont {G.}~\bibnamefont {Storeck}},\ and\ \bibinfo {author} {\bibfnamefont {C.}~\bibnamefont {Ropers}},\ }\bibfield  {title} {\emph {\bibinfo {title} {{Coherent control of a surface structural phase transition}}},\ }\href {https://doi.org/10.1038/s41586-020-2440-4} {\bibfield  {journal} {\bibinfo  {journal} {Nature}\ }\textbf {\bibinfo {volume} {583}},\ \bibinfo {pages} {232--236} (\bibinfo {year} {2020})}\BibitemShut {NoStop}%
\bibitem [{\citenamefont {Maklar}\ \emph {et~al.}(2023)\citenamefont {Maklar}, \citenamefont {Sarkar}, \citenamefont {Dong}, \citenamefont {Gerasimenko}, \citenamefont {Pincelli}, \citenamefont {Beaulieu}, \citenamefont {Kirchmann}, \citenamefont {Sobota}, \citenamefont {Yang}, \citenamefont {Leuenberger}, \citenamefont {Moore}, \citenamefont {Shen}, \citenamefont {Wolf}, \citenamefont {Mihailovic}, \citenamefont {Ernstorfer},\ and\ \citenamefont {Rettig}}]{maklar2023coherent}%
  \BibitemOpen
  \bibfield  {author} {\bibinfo {author} {\bibfnamefont {J.}~\bibnamefont {Maklar}}, \bibinfo {author} {\bibfnamefont {J.}~\bibnamefont {Sarkar}}, \bibinfo {author} {\bibfnamefont {S.}~\bibnamefont {Dong}}, \bibinfo {author} {\bibfnamefont {Y.~A.}\ \bibnamefont {Gerasimenko}}, \bibinfo {author} {\bibfnamefont {T.}~\bibnamefont {Pincelli}}, \bibinfo {author} {\bibfnamefont {S.}~\bibnamefont {Beaulieu}}, \bibinfo {author} {\bibfnamefont {P.~S.}\ \bibnamefont {Kirchmann}}, \bibinfo {author} {\bibfnamefont {J.~A.}\ \bibnamefont {Sobota}}, \bibinfo {author} {\bibfnamefont {S.}~\bibnamefont {Yang}}, \bibinfo {author} {\bibfnamefont {D.}~\bibnamefont {Leuenberger}}, \bibinfo {author} {\bibfnamefont {R.~G.}\ \bibnamefont {Moore}}, \bibinfo {author} {\bibfnamefont {Z.-X.}\ \bibnamefont {Shen}}, \bibinfo {author} {\bibfnamefont {M.}~\bibnamefont {Wolf}}, \bibinfo {author} {\bibfnamefont {D.}~\bibnamefont {Mihailovic}}, \bibinfo {author} {\bibfnamefont {R.}~\bibnamefont {Ernstorfer}},\ and\ \bibinfo {author}
  {\bibfnamefont {L.}~\bibnamefont {Rettig}},\ }\bibfield  {title} {\emph {\bibinfo {title} {{Coherent light control of a metastable hidden state}}},\ }\href {https://doi.org/10.1126/sciadv.adi4661} {\bibfield  {journal} {\bibinfo  {journal} {Sci. Adv.}\ }\textbf {\bibinfo {volume} {9}},\ \bibinfo {pages} {eadi4661} (\bibinfo {year} {2023})}\BibitemShut {NoStop}%
\bibitem [{\citenamefont {{Maehrlein}}\ \emph {et~al.}(2017)\citenamefont {{Maehrlein}}, \citenamefont {{Paarmann}}, \citenamefont {{Wolf}},\ and\ \citenamefont {{Kampfrath}}}]{maehrlein2017terahertz}%
  \BibitemOpen
  \bibfield  {author} {\bibinfo {author} {\bibfnamefont {S.}~\bibnamefont {{Maehrlein}}}, \bibinfo {author} {\bibfnamefont {A.}~\bibnamefont {{Paarmann}}}, \bibinfo {author} {\bibfnamefont {M.}~\bibnamefont {{Wolf}}},\ and\ \bibinfo {author} {\bibfnamefont {T.}~\bibnamefont {{Kampfrath}}},\ }\bibfield  {title} {\emph {\bibinfo {title} {{Terahertz sum-frequency excitation of a Raman-active phonon}}},\ }\href {https://doi.org/10.1103/PhysRevLett.119.127402} {\bibfield  {journal} {\bibinfo  {journal} {Phys. Rev. Lett.}\ }\textbf {\bibinfo {volume} {119}},\ \bibinfo {pages} {127402} (\bibinfo {year} {2017})}\BibitemShut {NoStop}%
\bibitem [{\citenamefont {Shi}\ \emph {et~al.}(2023)\citenamefont {Shi}, \citenamefont {Bie}, \citenamefont {Zong}, \citenamefont {Fang}, \citenamefont {Chen}, \citenamefont {Han}, \citenamefont {Cao}, \citenamefont {Zhang}, \citenamefont {Taniguchi}, \citenamefont {Watanabe}, \citenamefont {Fu}, \citenamefont {Bulović}, \citenamefont {Kaxiras}, \citenamefont {Baldini}, \citenamefont {Jarillo-Herrero},\ and\ \citenamefont {Nelson}}]{shi2023intrinsic}%
  \BibitemOpen
  \bibfield  {author} {\bibinfo {author} {\bibfnamefont {J.}~\bibnamefont {Shi}}, \bibinfo {author} {\bibfnamefont {Y.-Q.}\ \bibnamefont {Bie}}, \bibinfo {author} {\bibfnamefont {A.}~\bibnamefont {Zong}}, \bibinfo {author} {\bibfnamefont {S.}~\bibnamefont {Fang}}, \bibinfo {author} {\bibfnamefont {W.}~\bibnamefont {Chen}}, \bibinfo {author} {\bibfnamefont {J.}~\bibnamefont {Han}}, \bibinfo {author} {\bibfnamefont {Z.}~\bibnamefont {Cao}}, \bibinfo {author} {\bibfnamefont {Y.}~\bibnamefont {Zhang}}, \bibinfo {author} {\bibfnamefont {T.}~\bibnamefont {Taniguchi}}, \bibinfo {author} {\bibfnamefont {K.}~\bibnamefont {Watanabe}}, \bibinfo {author} {\bibfnamefont {X.}~\bibnamefont {Fu}}, \bibinfo {author} {\bibfnamefont {V.}~\bibnamefont {Bulović}}, \bibinfo {author} {\bibfnamefont {E.}~\bibnamefont {Kaxiras}}, \bibinfo {author} {\bibfnamefont {E.}~\bibnamefont {Baldini}}, \bibinfo {author} {\bibfnamefont {P.}~\bibnamefont {Jarillo-Herrero}},\ and\ \bibinfo {author} {\bibfnamefont {K.~A.}\ \bibnamefont {Nelson}},\
  }\bibfield  {title} {\emph {\bibinfo {title} {{Intrinsic $1T'$ phase induced in atomically thin 2$H$-MoTe$_2$ by a single terahertz pulse}}},\ }\href {https://doi.org/10.1038/s41467-023-41291-w} {\bibfield  {journal} {\bibinfo  {journal} {Nat. Commun.}\ }\textbf {\bibinfo {volume} {14}},\ \bibinfo {pages} {5905} (\bibinfo {year} {2023})}\BibitemShut {NoStop}%
\bibitem [{\citenamefont {{Juraschek}}\ \emph {et~al.}(2017)\citenamefont {{Juraschek}}, \citenamefont {{Fechner}}, \citenamefont {{Balatsky}},\ and\ \citenamefont {{Spaldin}}}]{juraschek2017dynamicalmuti}%
  \BibitemOpen
  \bibfield  {author} {\bibinfo {author} {\bibfnamefont {D.~M.}\ \bibnamefont {{Juraschek}}}, \bibinfo {author} {\bibfnamefont {M.}~\bibnamefont {{Fechner}}}, \bibinfo {author} {\bibfnamefont {A.~V.}\ \bibnamefont {{Balatsky}}},\ and\ \bibinfo {author} {\bibfnamefont {N.~A.}\ \bibnamefont {{Spaldin}}},\ }\bibfield  {title} {\emph {\bibinfo {title} {{Dynamical multiferroicity}}},\ }\href {https://doi.org/10.1103/PhysRevMaterials.1.014401} {\bibfield  {journal} {\bibinfo  {journal} {Phys. Rev. Mater.}\ }\textbf {\bibinfo {volume} {1}},\ \bibinfo {pages} {014401} (\bibinfo {year} {2017})}\BibitemShut {NoStop}%
\bibitem [{\citenamefont {{Basini}}\ \emph {et~al.}(2024)\citenamefont {{Basini}}, \citenamefont {{Pancaldi}}, \citenamefont {{Wehinger}}, \citenamefont {{Udina}}, \citenamefont {{Unikandanunni}}, \citenamefont {{Tadano}}, \citenamefont {{Hoffmann}}, \citenamefont {{Balatsky}},\ and\ \citenamefont {{Bonetti}}}]{basini2024terahertz}%
  \BibitemOpen
  \bibfield  {author} {\bibinfo {author} {\bibfnamefont {M.}~\bibnamefont {{Basini}}}, \bibinfo {author} {\bibfnamefont {M.}~\bibnamefont {{Pancaldi}}}, \bibinfo {author} {\bibfnamefont {B.}~\bibnamefont {{Wehinger}}}, \bibinfo {author} {\bibfnamefont {M.}~\bibnamefont {{Udina}}}, \bibinfo {author} {\bibfnamefont {V.}~\bibnamefont {{Unikandanunni}}}, \bibinfo {author} {\bibfnamefont {T.}~\bibnamefont {{Tadano}}}, \bibinfo {author} {\bibfnamefont {M.~C.}\ \bibnamefont {{Hoffmann}}}, \bibinfo {author} {\bibfnamefont {A.~V.}\ \bibnamefont {{Balatsky}}},\ and\ \bibinfo {author} {\bibfnamefont {S.}~\bibnamefont {{Bonetti}}},\ }\bibfield  {title} {\emph {\bibinfo {title} {{Terahertz electric-field-driven dynamical multiferroicity in SrTiO$_{3}$}}},\ }\href {https://doi.org/10.1038/s41586-024-07175-9} {\bibfield  {journal} {\bibinfo  {journal} {Nature}\ }\textbf {\bibinfo {volume} {628}},\ \bibinfo {pages} {534} (\bibinfo {year} {2024})}\BibitemShut {NoStop}%
\bibitem [{\citenamefont {{Davies}}\ \emph {et~al.}(2024)\citenamefont {{Davies}}, \citenamefont {{Fennema}}, \citenamefont {{Tsukamoto}}, \citenamefont {{Razdolski}}, \citenamefont {{Kimel}},\ and\ \citenamefont {{Kirilyuk}}}]{davies2024phononic}%
  \BibitemOpen
  \bibfield  {author} {\bibinfo {author} {\bibfnamefont {C.~S.}\ \bibnamefont {{Davies}}}, \bibinfo {author} {\bibfnamefont {F.~G.~N.}\ \bibnamefont {{Fennema}}}, \bibinfo {author} {\bibfnamefont {A.}~\bibnamefont {{Tsukamoto}}}, \bibinfo {author} {\bibfnamefont {I.}~\bibnamefont {{Razdolski}}}, \bibinfo {author} {\bibfnamefont {A.~V.}\ \bibnamefont {{Kimel}}},\ and\ \bibinfo {author} {\bibfnamefont {A.}~\bibnamefont {{Kirilyuk}}},\ }\bibfield  {title} {\emph {\bibinfo {title} {{Phononic switching of magnetization by the ultrafast Barnett effect}}},\ }\href {https://doi.org/10.1038/s41586-024-07200-x} {\bibfield  {journal} {\bibinfo  {journal} {Nature}\ }\textbf {\bibinfo {volume} {628}},\ \bibinfo {pages} {540} (\bibinfo {year} {2024})}\BibitemShut {NoStop}%
\bibitem [{\citenamefont {{Romao}}\ and\ \citenamefont {{Juraschek}}(2024)}]{romao2024light}%
  \BibitemOpen
  \bibfield  {author} {\bibinfo {author} {\bibfnamefont {C.~P.}\ \bibnamefont {{Romao}}}\ and\ \bibinfo {author} {\bibfnamefont {D.~M.}\ \bibnamefont {{Juraschek}}},\ }\bibfield  {title} {\emph {\bibinfo {title} {{Light makes atoms behave like electromagnetic coils}}},\ }\href {https://doi.org/10.1038/d41586-024-00889-w} {\bibfield  {journal} {\bibinfo  {journal} {Nature}\ }\textbf {\bibinfo {volume} {628}},\ \bibinfo {pages} {505} (\bibinfo {year} {2024})}\BibitemShut {NoStop}%
\bibitem [{\citenamefont {Sie}\ \emph {et~al.}(2019)\citenamefont {Sie}, \citenamefont {Nyby}, \citenamefont {Pemmaraju}, \citenamefont {Park}, \citenamefont {Shen}, \citenamefont {Yang}, \citenamefont {Hoffmann}, \citenamefont {Ofori-Okai}, \citenamefont {Li}, \citenamefont {Reid}, \citenamefont {Weathersby}, \citenamefont {Mannebach}, \citenamefont {Finney}, \citenamefont {Rhodes}, \citenamefont {Chenet}, \citenamefont {Antony}, \citenamefont {Balicas}, \citenamefont {Hone}, \citenamefont {Devereaux}, \citenamefont {Heinz}, \citenamefont {Wang},\ and\ \citenamefont {Lindenberg}}]{sie2019ultrafast}%
  \BibitemOpen
  \bibfield  {author} {\bibinfo {author} {\bibfnamefont {E.~J.}\ \bibnamefont {Sie}}, \bibinfo {author} {\bibfnamefont {C.~M.}\ \bibnamefont {Nyby}}, \bibinfo {author} {\bibfnamefont {C.~D.}\ \bibnamefont {Pemmaraju}}, \bibinfo {author} {\bibfnamefont {S.~J.}\ \bibnamefont {Park}}, \bibinfo {author} {\bibfnamefont {X.}~\bibnamefont {Shen}}, \bibinfo {author} {\bibfnamefont {J.}~\bibnamefont {Yang}}, \bibinfo {author} {\bibfnamefont {M.~C.}\ \bibnamefont {Hoffmann}}, \bibinfo {author} {\bibfnamefont {B.~K.}\ \bibnamefont {Ofori-Okai}}, \bibinfo {author} {\bibfnamefont {R.}~\bibnamefont {Li}}, \bibinfo {author} {\bibfnamefont {A.~H.}\ \bibnamefont {Reid}}, \bibinfo {author} {\bibfnamefont {S.}~\bibnamefont {Weathersby}}, \bibinfo {author} {\bibfnamefont {E.}~\bibnamefont {Mannebach}}, \bibinfo {author} {\bibfnamefont {N.}~\bibnamefont {Finney}}, \bibinfo {author} {\bibfnamefont {D.}~\bibnamefont {Rhodes}}, \bibinfo {author} {\bibfnamefont {D.}~\bibnamefont {Chenet}}, \bibinfo {author} {\bibfnamefont
  {A.}~\bibnamefont {Antony}}, \bibinfo {author} {\bibfnamefont {L.}~\bibnamefont {Balicas}}, \bibinfo {author} {\bibfnamefont {J.}~\bibnamefont {Hone}}, \bibinfo {author} {\bibfnamefont {T.~P.}\ \bibnamefont {Devereaux}}, \bibinfo {author} {\bibfnamefont {T.~F.}\ \bibnamefont {Heinz}}, \bibinfo {author} {\bibfnamefont {X.}~\bibnamefont {Wang}},\ and\ \bibinfo {author} {\bibfnamefont {A.~M.}\ \bibnamefont {Lindenberg}},\ }\bibfield  {title} {\emph {\bibinfo {title} {An ultrafast symmetry switch in a {Weyl} semimetal}},\ }\href {https://doi.org/10.1038/s41586-018-0809-4} {\bibfield  {journal} {\bibinfo  {journal} {Nature}\ }\textbf {\bibinfo {volume} {565}},\ \bibinfo {pages} {61--66} (\bibinfo {year} {2019})}\BibitemShut {NoStop}%
\bibitem [{\citenamefont {{Tang}}\ \emph {et~al.}(2023)\citenamefont {{Tang}}, \citenamefont {{Boi}},\ and\ \citenamefont {{Cheng}}}]{tang2023light}%
  \BibitemOpen
  \bibfield  {author} {\bibinfo {author} {\bibfnamefont {R.}~\bibnamefont {{Tang}}}, \bibinfo {author} {\bibfnamefont {F.}~\bibnamefont {{Boi}}},\ and\ \bibinfo {author} {\bibfnamefont {Y.-H.}\ \bibnamefont {{Cheng}}},\ }\bibfield  {title} {\emph {\bibinfo {title} {{Light-induced topological phase transition via nonlinear phononics in superconductor CsV$_{3}$Sb$_{5}$}}},\ }\href {https://doi.org/10.1038/s41535-023-00609-z} {\bibfield  {journal} {\bibinfo  {journal} {npj Quantum Mater.}\ }\textbf {\bibinfo {volume} {8}},\ \bibinfo {pages} {78} (\bibinfo {year} {2023})}\BibitemShut {NoStop}%
\bibitem [{\citenamefont {{Juraschek}}\ \emph {et~al.}(2022)\citenamefont {{Juraschek}}, \citenamefont {{Neuman}},\ and\ \citenamefont {{Narang}}}]{juraschek2022gianteffective}%
  \BibitemOpen
  \bibfield  {author} {\bibinfo {author} {\bibfnamefont {D.~M.}\ \bibnamefont {{Juraschek}}}, \bibinfo {author} {\bibfnamefont {T.}~\bibnamefont {{Neuman}}},\ and\ \bibinfo {author} {\bibfnamefont {P.}~\bibnamefont {{Narang}}},\ }\bibfield  {title} {\emph {\bibinfo {title} {{{Giant effective magnetic fields from optically driven chiral phonons in 4$f$ paramagnets}}}},\ }\href {https://doi.org/10.1103/PhysRevResearch.4.013129} {\bibfield  {journal} {\bibinfo  {journal} {Phys. Rev. Res.}\ }\textbf {\bibinfo {volume} {4}},\ \bibinfo {pages} {013129} (\bibinfo {year} {2022})}\BibitemShut {NoStop}%
\bibitem [{\citenamefont {Kahana}\ \emph {et~al.}(2024)\citenamefont {Kahana}, \citenamefont {Bustamante~Lopez},\ and\ \citenamefont {Juraschek}}]{kahana2024lightinduced}%
  \BibitemOpen
  \bibfield  {author} {\bibinfo {author} {\bibfnamefont {T.}~\bibnamefont {Kahana}}, \bibinfo {author} {\bibfnamefont {D.~A.}\ \bibnamefont {Bustamante~Lopez}},\ and\ \bibinfo {author} {\bibfnamefont {D.~M.}\ \bibnamefont {Juraschek}},\ }\bibfield  {title} {\emph {\bibinfo {title} {Light-induced magnetization from magnonic rectification}},\ }\href {https://doi.org/10.1126/sciadv.ado0722} {\bibfield  {journal} {\bibinfo  {journal} {Sci. Adv.}\ }\textbf {\bibinfo {volume} {10}},\ \bibinfo {pages} {eado0722} (\bibinfo {year} {2024})}\BibitemShut {NoStop}%
\bibitem [{\citenamefont {{Luo}}\ \emph {et~al.}(2023{\natexlab{a}})\citenamefont {{Luo}}, \citenamefont {{Lin}}, \citenamefont {{Zhang}}, \citenamefont {{Chen}}, \citenamefont {{Blackert}}, \citenamefont {{Xu}}, \citenamefont {{Yakobson}},\ and\ \citenamefont {{Zhu}}}]{luo2023large}%
  \BibitemOpen
  \bibfield  {author} {\bibinfo {author} {\bibfnamefont {J.}~\bibnamefont {{Luo}}}, \bibinfo {author} {\bibfnamefont {T.}~\bibnamefont {{Lin}}}, \bibinfo {author} {\bibfnamefont {J.}~\bibnamefont {{Zhang}}}, \bibinfo {author} {\bibfnamefont {X.}~\bibnamefont {{Chen}}}, \bibinfo {author} {\bibfnamefont {E.~R.}\ \bibnamefont {{Blackert}}}, \bibinfo {author} {\bibfnamefont {R.}~\bibnamefont {{Xu}}}, \bibinfo {author} {\bibfnamefont {B.~I.}\ \bibnamefont {{Yakobson}}},\ and\ \bibinfo {author} {\bibfnamefont {H.}~\bibnamefont {{Zhu}}},\ }\bibfield  {title} {\emph {\bibinfo {title} {{Large effective magnetic fields from chiral phonons in rare-earth halides}}},\ }\href {https://doi.org/10.1126/science.adi9601} {\bibfield  {journal} {\bibinfo  {journal} {Science}\ }\textbf {\bibinfo {volume} {382}},\ \bibinfo {pages} {698} (\bibinfo {year} {2023}{\natexlab{a}})}\BibitemShut {NoStop}%
\bibitem [{\citenamefont {Ilyas}\ \emph {et~al.}(2024)\citenamefont {Ilyas}, \citenamefont {Luo}, \citenamefont {von Hoegen}, \citenamefont {Viñas~Boström}, \citenamefont {Zhang}, \citenamefont {Park}, \citenamefont {Kim}, \citenamefont {Park}, \citenamefont {Nelson}, \citenamefont {Rubio},\ and\ \citenamefont {Gedik}}]{ilyas_terahertz_2024}%
  \BibitemOpen
  \bibfield  {author} {\bibinfo {author} {\bibfnamefont {B.}~\bibnamefont {Ilyas}}, \bibinfo {author} {\bibfnamefont {T.}~\bibnamefont {Luo}}, \bibinfo {author} {\bibfnamefont {A.}~\bibnamefont {von Hoegen}}, \bibinfo {author} {\bibfnamefont {E.}~\bibnamefont {Viñas~Boström}}, \bibinfo {author} {\bibfnamefont {Z.}~\bibnamefont {Zhang}}, \bibinfo {author} {\bibfnamefont {J.}~\bibnamefont {Park}}, \bibinfo {author} {\bibfnamefont {J.}~\bibnamefont {Kim}}, \bibinfo {author} {\bibfnamefont {J.-G.}\ \bibnamefont {Park}}, \bibinfo {author} {\bibfnamefont {K.~A.}\ \bibnamefont {Nelson}}, \bibinfo {author} {\bibfnamefont {A.}~\bibnamefont {Rubio}},\ and\ \bibinfo {author} {\bibfnamefont {N.}~\bibnamefont {Gedik}},\ }\bibfield  {title} {\emph {\bibinfo {title} {Terahertz field-induced metastable magnetization near criticality in {FePS$_3$}}},\ }\href {https://doi.org/10.1038/s41586-024-08226-x} {\bibfield  {journal} {\bibinfo  {journal} {Nature}\ }\textbf {\bibinfo {volume} {636}},\ \bibinfo {pages} {609--614}
  (\bibinfo {year} {2024})}\BibitemShut {NoStop}%
\bibitem [{\citenamefont {{Yoshikawa}}\ \emph {et~al.}(2021)\citenamefont {{Yoshikawa}}, \citenamefont {{Suganuma}}, \citenamefont {{Matsuoka}}, \citenamefont {{Tanaka}}, \citenamefont {{Hemme}}, \citenamefont {{Cazayous}}, \citenamefont {{Gallais}}, \citenamefont {{Nakano}}, \citenamefont {{Iwasa}},\ and\ \citenamefont {{Shimano}}}]{yoshikawa2021ultrafastswitch}%
  \BibitemOpen
  \bibfield  {author} {\bibinfo {author} {\bibfnamefont {N.}~\bibnamefont {{Yoshikawa}}}, \bibinfo {author} {\bibfnamefont {H.}~\bibnamefont {{Suganuma}}}, \bibinfo {author} {\bibfnamefont {H.}~\bibnamefont {{Matsuoka}}}, \bibinfo {author} {\bibfnamefont {Y.}~\bibnamefont {{Tanaka}}}, \bibinfo {author} {\bibfnamefont {P.}~\bibnamefont {{Hemme}}}, \bibinfo {author} {\bibfnamefont {M.}~\bibnamefont {{Cazayous}}}, \bibinfo {author} {\bibfnamefont {Y.}~\bibnamefont {{Gallais}}}, \bibinfo {author} {\bibfnamefont {M.}~\bibnamefont {{Nakano}}}, \bibinfo {author} {\bibfnamefont {Y.}~\bibnamefont {{Iwasa}}},\ and\ \bibinfo {author} {\bibfnamefont {R.}~\bibnamefont {{Shimano}}},\ }\bibfield  {title} {\emph {\bibinfo {title} {{Ultrafast switching to an insulating-like metastable state by amplitudon excitation of a charge density wave}}},\ }\href {https://doi.org/10.1038/s41567-021-01267-3} {\bibfield  {journal} {\bibinfo  {journal} {Nat. Phys.}\ }\textbf {\bibinfo {volume} {17}},\ \bibinfo {pages} {909} (\bibinfo {year}
  {2021})}\BibitemShut {NoStop}%
\bibitem [{\citenamefont {{Wang}}\ \emph {et~al.}(2013)\citenamefont {{Wang}}, \citenamefont {{Steinberg}}, \citenamefont {{Jarillo-Herrero}},\ and\ \citenamefont {{Gedik}}}]{wang2013obsercation}%
  \BibitemOpen
  \bibfield  {author} {\bibinfo {author} {\bibfnamefont {Y.~H.}\ \bibnamefont {{Wang}}}, \bibinfo {author} {\bibfnamefont {H.}~\bibnamefont {{Steinberg}}}, \bibinfo {author} {\bibfnamefont {P.}~\bibnamefont {{Jarillo-Herrero}}},\ and\ \bibinfo {author} {\bibfnamefont {N.}~\bibnamefont {{Gedik}}},\ }\bibfield  {title} {\emph {\bibinfo {title} {{{Observation of Floquet-Bloch states on the surface of a topological insulator}}}},\ }\href {https://doi.org/10.1126/science.1239834} {\bibfield  {journal} {\bibinfo  {journal} {Science}\ }\textbf {\bibinfo {volume} {342}},\ \bibinfo {pages} {453} (\bibinfo {year} {2013})}\BibitemShut {NoStop}%
\bibitem [{\citenamefont {{Mahmood}}\ \emph {et~al.}(2016)\citenamefont {{Mahmood}}, \citenamefont {{Chan}}, \citenamefont {{Alpichshev}}, \citenamefont {{Gardner}}, \citenamefont {{Lee}}, \citenamefont {{Lee}},\ and\ \citenamefont {{Gedik}}}]{mahood2016selective}%
  \BibitemOpen
  \bibfield  {author} {\bibinfo {author} {\bibfnamefont {F.}~\bibnamefont {{Mahmood}}}, \bibinfo {author} {\bibfnamefont {C.-K.}\ \bibnamefont {{Chan}}}, \bibinfo {author} {\bibfnamefont {Z.}~\bibnamefont {{Alpichshev}}}, \bibinfo {author} {\bibfnamefont {D.}~\bibnamefont {{Gardner}}}, \bibinfo {author} {\bibfnamefont {Y.}~\bibnamefont {{Lee}}}, \bibinfo {author} {\bibfnamefont {P.~A.}\ \bibnamefont {{Lee}}},\ and\ \bibinfo {author} {\bibfnamefont {N.}~\bibnamefont {{Gedik}}},\ }\bibfield  {title} {\emph {\bibinfo {title} {{Selective scattering between Floquet-Bloch and Volkov states in a topological insulator}}},\ }\href {https://doi.org/10.1038/nphys3609} {\bibfield  {journal} {\bibinfo  {journal} {Nat. Phys.}\ }\textbf {\bibinfo {volume} {12}},\ \bibinfo {pages} {306} (\bibinfo {year} {2016})}\BibitemShut {NoStop}%
\bibitem [{\citenamefont {{Zhou}}\ \emph {et~al.}(2023{\natexlab{a}})\citenamefont {{Zhou}}, \citenamefont {{Bao}}, \citenamefont {{Fan}}, \citenamefont {{Zhou}}, \citenamefont {{Gao}}, \citenamefont {{Zhong}}, \citenamefont {{Lin}}, \citenamefont {{Liu}}, \citenamefont {{Yu}}, \citenamefont {{Tang}}, \citenamefont {{Meng}}, \citenamefont {{Duan}},\ and\ \citenamefont {{Zhou}}}]{zhou2023pseudospin}%
  \BibitemOpen
  \bibfield  {author} {\bibinfo {author} {\bibfnamefont {S.}~\bibnamefont {{Zhou}}}, \bibinfo {author} {\bibfnamefont {C.}~\bibnamefont {{Bao}}}, \bibinfo {author} {\bibfnamefont {B.}~\bibnamefont {{Fan}}}, \bibinfo {author} {\bibfnamefont {H.}~\bibnamefont {{Zhou}}}, \bibinfo {author} {\bibfnamefont {Q.}~\bibnamefont {{Gao}}}, \bibinfo {author} {\bibfnamefont {H.}~\bibnamefont {{Zhong}}}, \bibinfo {author} {\bibfnamefont {T.}~\bibnamefont {{Lin}}}, \bibinfo {author} {\bibfnamefont {H.}~\bibnamefont {{Liu}}}, \bibinfo {author} {\bibfnamefont {P.}~\bibnamefont {{Yu}}}, \bibinfo {author} {\bibfnamefont {P.}~\bibnamefont {{Tang}}}, \bibinfo {author} {\bibfnamefont {S.}~\bibnamefont {{Meng}}}, \bibinfo {author} {\bibfnamefont {W.}~\bibnamefont {{Duan}}},\ and\ \bibinfo {author} {\bibfnamefont {S.}~\bibnamefont {{Zhou}}},\ }\bibfield  {title} {\emph {\bibinfo {title} {{Pseudospin-selective Floquet band engineering in black phosphorus}}},\ }\href {https://doi.org/10.1038/s41586-022-05610-3} {\bibfield  {journal}
  {\bibinfo  {journal} {Nature}\ }\textbf {\bibinfo {volume} {614}},\ \bibinfo {pages} {75} (\bibinfo {year} {2023}{\natexlab{a}})}\BibitemShut {NoStop}%
\bibitem [{\citenamefont {{Zhou}}\ \emph {et~al.}(2023{\natexlab{b}})\citenamefont {{Zhou}}, \citenamefont {{Bao}}, \citenamefont {{Fan}}, \citenamefont {{Wang}}, \citenamefont {{Zhong}}, \citenamefont {{Zhang}}, \citenamefont {{Tang}}, \citenamefont {{Duan}},\ and\ \citenamefont {{Zhou}}}]{zhou2023floquet}%
  \BibitemOpen
  \bibfield  {author} {\bibinfo {author} {\bibfnamefont {S.}~\bibnamefont {{Zhou}}}, \bibinfo {author} {\bibfnamefont {C.}~\bibnamefont {{Bao}}}, \bibinfo {author} {\bibfnamefont {B.}~\bibnamefont {{Fan}}}, \bibinfo {author} {\bibfnamefont {F.}~\bibnamefont {{Wang}}}, \bibinfo {author} {\bibfnamefont {H.}~\bibnamefont {{Zhong}}}, \bibinfo {author} {\bibfnamefont {H.}~\bibnamefont {{Zhang}}}, \bibinfo {author} {\bibfnamefont {P.}~\bibnamefont {{Tang}}}, \bibinfo {author} {\bibfnamefont {W.}~\bibnamefont {{Duan}}},\ and\ \bibinfo {author} {\bibfnamefont {S.}~\bibnamefont {{Zhou}}},\ }\bibfield  {title} {\emph {\bibinfo {title} {{Floquet engineering of black phosphorus upon below-gap pumping}}},\ }\href {https://doi.org/10.1103/PhysRevLett.131.116401} {\bibfield  {journal} {\bibinfo  {journal} {Phys. Rev. Lett.}\ }\textbf {\bibinfo {volume} {131}},\ \bibinfo {pages} {116401} (\bibinfo {year} {2023}{\natexlab{b}})}\BibitemShut {NoStop}%
\bibitem [{\citenamefont {Ito}\ \emph {et~al.}(2023)\citenamefont {Ito}, \citenamefont {Schüler}, \citenamefont {Meierhofer}, \citenamefont {Schlauderer}, \citenamefont {Freudenstein}, \citenamefont {Reimann}, \citenamefont {Afanasiev}, \citenamefont {Kokh}, \citenamefont {Tereshchenko}, \citenamefont {Güdde}, \citenamefont {Sentef}, \citenamefont {Höfer},\ and\ \citenamefont {Huber}}]{ito2023buildup}%
  \BibitemOpen
  \bibfield  {author} {\bibinfo {author} {\bibfnamefont {S.}~\bibnamefont {Ito}}, \bibinfo {author} {\bibfnamefont {M.}~\bibnamefont {Schüler}}, \bibinfo {author} {\bibfnamefont {M.}~\bibnamefont {Meierhofer}}, \bibinfo {author} {\bibfnamefont {S.}~\bibnamefont {Schlauderer}}, \bibinfo {author} {\bibfnamefont {J.}~\bibnamefont {Freudenstein}}, \bibinfo {author} {\bibfnamefont {J.}~\bibnamefont {Reimann}}, \bibinfo {author} {\bibfnamefont {D.}~\bibnamefont {Afanasiev}}, \bibinfo {author} {\bibfnamefont {K.~A.}\ \bibnamefont {Kokh}}, \bibinfo {author} {\bibfnamefont {O.~E.}\ \bibnamefont {Tereshchenko}}, \bibinfo {author} {\bibfnamefont {J.}~\bibnamefont {Güdde}}, \bibinfo {author} {\bibfnamefont {M.~A.}\ \bibnamefont {Sentef}}, \bibinfo {author} {\bibfnamefont {U.}~\bibnamefont {Höfer}},\ and\ \bibinfo {author} {\bibfnamefont {R.}~\bibnamefont {Huber}},\ }\bibfield  {title} {\emph {\bibinfo {title} {Build-up and dephasing of {Floquet}–{Bloch} bands on subcycle timescales}},\ }\href
  {https://doi.org/10.1038/s41586-023-05850-x} {\bibfield  {journal} {\bibinfo  {journal} {Nature}\ }\textbf {\bibinfo {volume} {616}},\ \bibinfo {pages} {696--701} (\bibinfo {year} {2023})}\BibitemShut {NoStop}%
\bibitem [{\citenamefont {{Choi}}\ \emph {et~al.}(2024)\citenamefont {{Choi}}, \citenamefont {{Mogi}}, \citenamefont {{De Giovannini}}, \citenamefont {{Azoury}}, \citenamefont {{Lv}}, \citenamefont {{Su}}, \citenamefont {{H{\"u}bener}}, \citenamefont {{Rubio}},\ and\ \citenamefont {{Gedik}}}]{choi2024direct}%
  \BibitemOpen
  \bibfield  {author} {\bibinfo {author} {\bibfnamefont {D.}~\bibnamefont {{Choi}}}, \bibinfo {author} {\bibfnamefont {M.}~\bibnamefont {{Mogi}}}, \bibinfo {author} {\bibfnamefont {U.}~\bibnamefont {{De Giovannini}}}, \bibinfo {author} {\bibfnamefont {D.}~\bibnamefont {{Azoury}}}, \bibinfo {author} {\bibfnamefont {B.}~\bibnamefont {{Lv}}}, \bibinfo {author} {\bibfnamefont {Y.}~\bibnamefont {{Su}}}, \bibinfo {author} {\bibfnamefont {H.}~\bibnamefont {{H{\"u}bener}}}, \bibinfo {author} {\bibfnamefont {A.}~\bibnamefont {{Rubio}}},\ and\ \bibinfo {author} {\bibfnamefont {N.}~\bibnamefont {{Gedik}}},\ }\href {https://doi.org/10.48550/arXiv.2404.14392} {\bibinfo {title} {{\textit{Direct observation of Floquet-Bloch states in monolayer graphene}}}} (\bibinfo {year} {2024}),\ \Eprint {https://arxiv.org/abs/2404.14392} {arXiv:2404.14392} \BibitemShut {NoStop}%
\bibitem [{\citenamefont {{Merboldt}}\ \emph {et~al.}(2024)\citenamefont {{Merboldt}}, \citenamefont {{Sch{\"u}ler}}, \citenamefont {{Schmitt}}, \citenamefont {{Bange}}, \citenamefont {{Bennecke}}, \citenamefont {{Gadge}}, \citenamefont {{Pierz}}, \citenamefont {{Schumacher}}, \citenamefont {{Momeni}}, \citenamefont {{Steil}}, \citenamefont {{Manmana}}, \citenamefont {{Sentef}}, \citenamefont {{Reutzel}},\ and\ \citenamefont {{Mathias}}}]{merboldt2024observation}%
  \BibitemOpen
  \bibfield  {author} {\bibinfo {author} {\bibfnamefont {M.}~\bibnamefont {{Merboldt}}}, \bibinfo {author} {\bibfnamefont {M.}~\bibnamefont {{Sch{\"u}ler}}}, \bibinfo {author} {\bibfnamefont {D.}~\bibnamefont {{Schmitt}}}, \bibinfo {author} {\bibfnamefont {J.~P.}\ \bibnamefont {{Bange}}}, \bibinfo {author} {\bibfnamefont {W.}~\bibnamefont {{Bennecke}}}, \bibinfo {author} {\bibfnamefont {K.}~\bibnamefont {{Gadge}}}, \bibinfo {author} {\bibfnamefont {K.}~\bibnamefont {{Pierz}}}, \bibinfo {author} {\bibfnamefont {H.~W.}\ \bibnamefont {{Schumacher}}}, \bibinfo {author} {\bibfnamefont {D.}~\bibnamefont {{Momeni}}}, \bibinfo {author} {\bibfnamefont {D.}~\bibnamefont {{Steil}}}, \bibinfo {author} {\bibfnamefont {S.~R.}\ \bibnamefont {{Manmana}}}, \bibinfo {author} {\bibfnamefont {M.}~\bibnamefont {{Sentef}}}, \bibinfo {author} {\bibfnamefont {M.}~\bibnamefont {{Reutzel}}},\ and\ \bibinfo {author} {\bibfnamefont {S.}~\bibnamefont {{Mathias}}},\ }\href {https://doi.org/10.48550/arXiv.2404.12791} {\bibinfo {title}
  {{\textit{Observation of Floquet states in graphene}}}} (\bibinfo {year} {2024}),\ \Eprint {https://arxiv.org/abs/2404.12791} {arXiv:2404.12791} \BibitemShut {NoStop}%
\bibitem [{\citenamefont {Hübener}\ \emph {et~al.}(2018)\citenamefont {Hübener}, \citenamefont {De~Giovannini},\ and\ \citenamefont {Rubio}}]{hubener2018phonon}%
  \BibitemOpen
  \bibfield  {author} {\bibinfo {author} {\bibfnamefont {H.}~\bibnamefont {Hübener}}, \bibinfo {author} {\bibfnamefont {U.}~\bibnamefont {De~Giovannini}},\ and\ \bibinfo {author} {\bibfnamefont {A.}~\bibnamefont {Rubio}},\ }\bibfield  {title} {\emph {\bibinfo {title} {{Phonon driven Floquet matter}}},\ }\href {https://doi.org/10.1021/acs.nanolett.7b05391} {\bibfield  {journal} {\bibinfo  {journal} {Nano Lett.}\ }\textbf {\bibinfo {volume} {18}},\ \bibinfo {pages} {1535--1542} (\bibinfo {year} {2018})}\BibitemShut {NoStop}%
\bibitem [{\citenamefont {Chaudhary}\ \emph {et~al.}(2020)\citenamefont {Chaudhary}, \citenamefont {Haim}, \citenamefont {Peng},\ and\ \citenamefont {Refael}}]{chaudhary2020phonon}%
  \BibitemOpen
  \bibfield  {author} {\bibinfo {author} {\bibfnamefont {S.}~\bibnamefont {Chaudhary}}, \bibinfo {author} {\bibfnamefont {A.}~\bibnamefont {Haim}}, \bibinfo {author} {\bibfnamefont {Y.}~\bibnamefont {Peng}},\ and\ \bibinfo {author} {\bibfnamefont {G.}~\bibnamefont {Refael}},\ }\bibfield  {title} {\emph {\bibinfo {title} {{Phonon-induced Floquet topological phases protected by space-time symmetries}}},\ }\href {https://doi.org/10.1103/PhysRevResearch.2.043431} {\bibfield  {journal} {\bibinfo  {journal} {Phys. Rev. Res.}\ }\textbf {\bibinfo {volume} {2}},\ \bibinfo {pages} {043431} (\bibinfo {year} {2020})}\BibitemShut {NoStop}%
\bibitem [{\citenamefont {Chan}\ \emph {et~al.}(2023)\citenamefont {Chan}, \citenamefont {Qiu}, \citenamefont {da~Jornada},\ and\ \citenamefont {Louie}}]{chan2023giant}%
  \BibitemOpen
  \bibfield  {author} {\bibinfo {author} {\bibfnamefont {Y.-H.}\ \bibnamefont {Chan}}, \bibinfo {author} {\bibfnamefont {D.~Y.}\ \bibnamefont {Qiu}}, \bibinfo {author} {\bibfnamefont {F.~H.}\ \bibnamefont {da~Jornada}},\ and\ \bibinfo {author} {\bibfnamefont {S.~G.}\ \bibnamefont {Louie}},\ }\bibfield  {title} {\emph {\bibinfo {title} {Giant self-driven exciton-{Floquet} signatures in time-resolved photoemission spectroscopy of {MoS$_2$} from time-dependent {GW} approach}},\ }\href {https://doi.org/10.1073/pnas.2301957120} {\bibfield  {journal} {\bibinfo  {journal} {Proc. Natl. Acad. Sci. U.S.A.}\ }\textbf {\bibinfo {volume} {120}},\ \bibinfo {pages} {e2301957120} (\bibinfo {year} {2023})}\BibitemShut {NoStop}%
\bibitem [{\citenamefont {{Pareek}}\ \emph {et~al.}(2024)\citenamefont {{Pareek}}, \citenamefont {{Bacon}}, \citenamefont {{Zhu}}, \citenamefont {{Chan}}, \citenamefont {{Bussolotti}}, \citenamefont {{Chan}}, \citenamefont {{P{\'e}rez Urquizo}}, \citenamefont {{Watanabe}}, \citenamefont {{Taniguchi}}, \citenamefont {{Man}}, \citenamefont {{Mad{\'e}o}}, \citenamefont {{Qiu}}, \citenamefont {{Eng Johnson Goh}}, \citenamefont {{da Jornada}},\ and\ \citenamefont {{Dani}}}]{paraeek2024driving}%
  \BibitemOpen
  \bibfield  {author} {\bibinfo {author} {\bibfnamefont {V.}~\bibnamefont {{Pareek}}}, \bibinfo {author} {\bibfnamefont {D.~R.}\ \bibnamefont {{Bacon}}}, \bibinfo {author} {\bibfnamefont {X.}~\bibnamefont {{Zhu}}}, \bibinfo {author} {\bibfnamefont {Y.-H.}\ \bibnamefont {{Chan}}}, \bibinfo {author} {\bibfnamefont {F.}~\bibnamefont {{Bussolotti}}}, \bibinfo {author} {\bibfnamefont {N.~S.}\ \bibnamefont {{Chan}}}, \bibinfo {author} {\bibfnamefont {J.}~\bibnamefont {{P{\'e}rez Urquizo}}}, \bibinfo {author} {\bibfnamefont {K.}~\bibnamefont {{Watanabe}}}, \bibinfo {author} {\bibfnamefont {T.}~\bibnamefont {{Taniguchi}}}, \bibinfo {author} {\bibfnamefont {M.~K.~L.}\ \bibnamefont {{Man}}}, \bibinfo {author} {\bibfnamefont {J.}~\bibnamefont {{Mad{\'e}o}}}, \bibinfo {author} {\bibfnamefont {D.~Y.}\ \bibnamefont {{Qiu}}}, \bibinfo {author} {\bibfnamefont {K.}~\bibnamefont {{Eng Johnson Goh}}}, \bibinfo {author} {\bibfnamefont {F.~H.}\ \bibnamefont {{da Jornada}}},\ and\ \bibinfo {author} {\bibfnamefont {K.~M.}\
  \bibnamefont {{Dani}}},\ }\href {https://doi.org/10.48550/arXiv.2403.08725} {\bibinfo {title} {{\textit{Driving non-trivial quantum phases in conventional semiconductors with intense excitonic fields}}}} (\bibinfo {year} {2024}),\ \Eprint {https://arxiv.org/abs/2403.08725} {arXiv:2403.08725} \BibitemShut {NoStop}%
\bibitem [{\citenamefont {{Gao}}\ \emph {et~al.}(2024)\citenamefont {{Gao}}, \citenamefont {{Peng}}, \citenamefont {{Cheng}}, \citenamefont {{Vi{\~n}as Bostr{\"o}m}}, \citenamefont {{Kim}}, \citenamefont {{Jain}}, \citenamefont {{Vishnu}}, \citenamefont {{Raju}}, \citenamefont {{Sankar}}, \citenamefont {{Lee}}, \citenamefont {{Sentef}}, \citenamefont {{Kurumaji}}, \citenamefont {{Li}}, \citenamefont {{Tang}}, \citenamefont {{Rubio}},\ and\ \citenamefont {{Baldini}}}]{gao2024giant}%
  \BibitemOpen
  \bibfield  {author} {\bibinfo {author} {\bibfnamefont {F.~Y.}\ \bibnamefont {{Gao}}}, \bibinfo {author} {\bibfnamefont {X.}~\bibnamefont {{Peng}}}, \bibinfo {author} {\bibfnamefont {X.}~\bibnamefont {{Cheng}}}, \bibinfo {author} {\bibfnamefont {E.}~\bibnamefont {{Vi{\~n}as Bostr{\"o}m}}}, \bibinfo {author} {\bibfnamefont {D.~S.}\ \bibnamefont {{Kim}}}, \bibinfo {author} {\bibfnamefont {R.~K.}\ \bibnamefont {{Jain}}}, \bibinfo {author} {\bibfnamefont {D.}~\bibnamefont {{Vishnu}}}, \bibinfo {author} {\bibfnamefont {K.}~\bibnamefont {{Raju}}}, \bibinfo {author} {\bibfnamefont {R.}~\bibnamefont {{Sankar}}}, \bibinfo {author} {\bibfnamefont {S.-F.}\ \bibnamefont {{Lee}}}, \bibinfo {author} {\bibfnamefont {M.~A.}\ \bibnamefont {{Sentef}}}, \bibinfo {author} {\bibfnamefont {T.}~\bibnamefont {{Kurumaji}}}, \bibinfo {author} {\bibfnamefont {X.}~\bibnamefont {{Li}}}, \bibinfo {author} {\bibfnamefont {P.}~\bibnamefont {{Tang}}}, \bibinfo {author} {\bibfnamefont {A.}~\bibnamefont {{Rubio}}},\ and\ \bibinfo {author}
  {\bibfnamefont {E.}~\bibnamefont {{Baldini}}},\ }\bibfield  {title} {\emph {\bibinfo {title} {{Giant chiral magnetoelectric oscillations in a van der Waals multiferroic}}},\ }\href {https://doi.org/10.1038/s41586-024-07678-5} {\bibfield  {journal} {\bibinfo  {journal} {Nature}\ }\textbf {\bibinfo {volume} {632}},\ \bibinfo {pages} {273} (\bibinfo {year} {2024})}\BibitemShut {NoStop}%
\bibitem [{\citenamefont {{Golias}}\ and\ \citenamefont {{S{\'a}nchez-Barriga}}(2016)}]{golias2016observation}%
  \BibitemOpen
  \bibfield  {author} {\bibinfo {author} {\bibfnamefont {E.}~\bibnamefont {{Golias}}}\ and\ \bibinfo {author} {\bibfnamefont {J.}~\bibnamefont {{S{\'a}nchez-Barriga}}},\ }\bibfield  {title} {\emph {\bibinfo {title} {{Observation of antiphase coherent phonons in the warped Dirac cone of Bi$_{2}$Te$_{3}$}}},\ }\href {https://doi.org/10.1103/PhysRevB.94.161113} {\bibfield  {journal} {\bibinfo  {journal} {Phys. Rev. B}\ }\textbf {\bibinfo {volume} {94}},\ \bibinfo {pages} {161113} (\bibinfo {year} {2016})}\BibitemShut {NoStop}%
\bibitem [{\citenamefont {{Ron}}\ \emph {et~al.}(2020)\citenamefont {{Ron}}, \citenamefont {{Chaudhary}}, \citenamefont {{Zhang}}, \citenamefont {{Ning}}, \citenamefont {{Zoghlin}}, \citenamefont {{Wilson}}, \citenamefont {{Averitt}}, \citenamefont {{Refael}},\ and\ \citenamefont {{Hsieh}}}]{ron2020ultrafast}%
  \BibitemOpen
  \bibfield  {author} {\bibinfo {author} {\bibfnamefont {A.}~\bibnamefont {{Ron}}}, \bibinfo {author} {\bibfnamefont {S.}~\bibnamefont {{Chaudhary}}}, \bibinfo {author} {\bibfnamefont {G.}~\bibnamefont {{Zhang}}}, \bibinfo {author} {\bibfnamefont {H.}~\bibnamefont {{Ning}}}, \bibinfo {author} {\bibfnamefont {E.}~\bibnamefont {{Zoghlin}}}, \bibinfo {author} {\bibfnamefont {S.~D.}\ \bibnamefont {{Wilson}}}, \bibinfo {author} {\bibfnamefont {R.~D.}\ \bibnamefont {{Averitt}}}, \bibinfo {author} {\bibfnamefont {G.}~\bibnamefont {{Refael}}},\ and\ \bibinfo {author} {\bibfnamefont {D.}~\bibnamefont {{Hsieh}}},\ }\bibfield  {title} {\emph {\bibinfo {title} {{{Ultrafast enhancement of ferromagnetic spin exchange induced by ligand-to-metal charge transfer}}}},\ }\href {https://doi.org/10.1103/PhysRevLett.125.197203} {\bibfield  {journal} {\bibinfo  {journal} {Phys. Rev. Lett.}\ }\textbf {\bibinfo {volume} {125}},\ \bibinfo {pages} {197203} (\bibinfo {year} {2020})}\BibitemShut {NoStop}%
\bibitem [{\citenamefont {Erge{\c{c}}en}\ \emph {et~al.}(2023)\citenamefont {Erge{\c{c}}en}, \citenamefont {Ilyas}, \citenamefont {Kim}, \citenamefont {Park}, \citenamefont {Yilmaz}, \citenamefont {Luo}, \citenamefont {Xiao}, \citenamefont {Okamoto}, \citenamefont {Park},\ and\ \citenamefont {Gedik}}]{ergeccen2023coherent}%
  \BibitemOpen
  \bibfield  {author} {\bibinfo {author} {\bibfnamefont {E.}~\bibnamefont {Erge{\c{c}}en}}, \bibinfo {author} {\bibfnamefont {B.}~\bibnamefont {Ilyas}}, \bibinfo {author} {\bibfnamefont {J.}~\bibnamefont {Kim}}, \bibinfo {author} {\bibfnamefont {J.}~\bibnamefont {Park}}, \bibinfo {author} {\bibfnamefont {M.~B.}\ \bibnamefont {Yilmaz}}, \bibinfo {author} {\bibfnamefont {T.}~\bibnamefont {Luo}}, \bibinfo {author} {\bibfnamefont {D.}~\bibnamefont {Xiao}}, \bibinfo {author} {\bibfnamefont {S.}~\bibnamefont {Okamoto}}, \bibinfo {author} {\bibfnamefont {J.-G.}\ \bibnamefont {Park}},\ and\ \bibinfo {author} {\bibfnamefont {N.}~\bibnamefont {Gedik}},\ }\bibfield  {title} {\emph {\bibinfo {title} {{Coherent detection of hidden spin-lattice coupling in a van der Waals antiferromagnet}}},\ }\href {https://doi.org/10.1073/pnas.2208968120} {\bibfield  {journal} {\bibinfo  {journal} {Proc. Natl. Acad. Sci. U.S.A.}\ }\textbf {\bibinfo {volume} {120}},\ \bibinfo {pages} {e2208968120} (\bibinfo {year} {2023})}\BibitemShut
  {NoStop}%
\bibitem [{\citenamefont {{Su}}\ \emph {et~al.}(2023)\citenamefont {{Su}}, \citenamefont {{Zong}}, \citenamefont {{Kogar}}, \citenamefont {{Lu}}, \citenamefont {{Hong}}, \citenamefont {{Freelon}}, \citenamefont {{Rohwer}}, \citenamefont {{Wang}}, \citenamefont {{Hwang}},\ and\ \citenamefont {{Gedik}}}]{su2023delamination}%
  \BibitemOpen
  \bibfield  {author} {\bibinfo {author} {\bibfnamefont {Y.}~\bibnamefont {{Su}}}, \bibinfo {author} {\bibfnamefont {A.}~\bibnamefont {{Zong}}}, \bibinfo {author} {\bibfnamefont {A.}~\bibnamefont {{Kogar}}}, \bibinfo {author} {\bibfnamefont {D.}~\bibnamefont {{Lu}}}, \bibinfo {author} {\bibfnamefont {S.~S.}\ \bibnamefont {{Hong}}}, \bibinfo {author} {\bibfnamefont {B.}~\bibnamefont {{Freelon}}}, \bibinfo {author} {\bibfnamefont {T.}~\bibnamefont {{Rohwer}}}, \bibinfo {author} {\bibfnamefont {B.~Y.}\ \bibnamefont {{Wang}}}, \bibinfo {author} {\bibfnamefont {H.~Y.}\ \bibnamefont {{Hwang}}},\ and\ \bibinfo {author} {\bibfnamefont {N.}~\bibnamefont {{Gedik}}},\ }\bibfield  {title} {\emph {\bibinfo {title} {{Delamination-assisted ultrafast wrinkle formation in a freestanding film}}},\ }\href {https://doi.org/10.1021/acs.nanolett.3c02898} {\bibfield  {journal} {\bibinfo  {journal} {Nano Lett.}\ }\textbf {\bibinfo {volume} {23}},\ \bibinfo {pages} {10772} (\bibinfo {year} {2023})}\BibitemShut {NoStop}%
\bibitem [{\citenamefont {{Geneaux}}\ \emph {et~al.}(2019)\citenamefont {{Geneaux}}, \citenamefont {{Marroux}}, \citenamefont {{Guggenmos}}, \citenamefont {{Neumark}},\ and\ \citenamefont {{Leone}}}]{geneaus2019transient}%
  \BibitemOpen
  \bibfield  {author} {\bibinfo {author} {\bibfnamefont {R.}~\bibnamefont {{Geneaux}}}, \bibinfo {author} {\bibfnamefont {H.~J.~B.}\ \bibnamefont {{Marroux}}}, \bibinfo {author} {\bibfnamefont {A.}~\bibnamefont {{Guggenmos}}}, \bibinfo {author} {\bibfnamefont {D.~M.}\ \bibnamefont {{Neumark}}},\ and\ \bibinfo {author} {\bibfnamefont {S.~R.}\ \bibnamefont {{Leone}}},\ }\bibfield  {title} {\emph {\bibinfo {title} {{Transient absorption spectroscopy using high harmonic generation: a review of ultrafast X-ray dynamics in molecules and solids}}},\ }\href {https://doi.org/10.1098/rsta.2017.0463} {\bibfield  {journal} {\bibinfo  {journal} {Philos. Trans. R. Soc. A}\ }\textbf {\bibinfo {volume} {377}},\ \bibinfo {pages} {20170463} (\bibinfo {year} {2019})}\BibitemShut {NoStop}%
\bibitem [{\citenamefont {{Li}}\ \emph {et~al.}(2020)\citenamefont {{Li}}, \citenamefont {{Lu}}, \citenamefont {{Chew}}, \citenamefont {{Han}}, \citenamefont {{Li}}, \citenamefont {{Wu}}, \citenamefont {{Wang}}, \citenamefont {{Ghimire}},\ and\ \citenamefont {{Chang}}}]{li2020attosecond}%
  \BibitemOpen
  \bibfield  {author} {\bibinfo {author} {\bibfnamefont {J.}~\bibnamefont {{Li}}}, \bibinfo {author} {\bibfnamefont {J.}~\bibnamefont {{Lu}}}, \bibinfo {author} {\bibfnamefont {A.}~\bibnamefont {{Chew}}}, \bibinfo {author} {\bibfnamefont {S.}~\bibnamefont {{Han}}}, \bibinfo {author} {\bibfnamefont {J.}~\bibnamefont {{Li}}}, \bibinfo {author} {\bibfnamefont {Y.}~\bibnamefont {{Wu}}}, \bibinfo {author} {\bibfnamefont {H.}~\bibnamefont {{Wang}}}, \bibinfo {author} {\bibfnamefont {S.}~\bibnamefont {{Ghimire}}},\ and\ \bibinfo {author} {\bibfnamefont {Z.}~\bibnamefont {{Chang}}},\ }\bibfield  {title} {\emph {\bibinfo {title} {{Attosecond science based on high harmonic generation from gases and solids}}},\ }\href {https://doi.org/10.1038/s41467-020-16480-6} {\bibfield  {journal} {\bibinfo  {journal} {Nat. Commun.}\ }\textbf {\bibinfo {volume} {11}},\ \bibinfo {pages} {2748} (\bibinfo {year} {2020})}\BibitemShut {NoStop}%
\bibitem [{\citenamefont {{Zhang}}\ \emph {et~al.}(2009)\citenamefont {{Zhang}}, \citenamefont {{H{\"u}bner}}, \citenamefont {{Lefkidis}}, \citenamefont {{Bai}},\ and\ \citenamefont {{George}}}]{zhang2009paradigm}%
  \BibitemOpen
  \bibfield  {author} {\bibinfo {author} {\bibfnamefont {G.~P.}\ \bibnamefont {{Zhang}}}, \bibinfo {author} {\bibfnamefont {W.}~\bibnamefont {{H{\"u}bner}}}, \bibinfo {author} {\bibfnamefont {G.}~\bibnamefont {{Lefkidis}}}, \bibinfo {author} {\bibfnamefont {Y.}~\bibnamefont {{Bai}}},\ and\ \bibinfo {author} {\bibfnamefont {T.~F.}\ \bibnamefont {{George}}},\ }\bibfield  {title} {\emph {\bibinfo {title} {{Paradigm of the time-resolved magneto-optical Kerr effect for femtosecond magnetism}}},\ }\href {https://doi.org/10.1038/nphys1315} {\bibfield  {journal} {\bibinfo  {journal} {Nat. Phys.}\ }\textbf {\bibinfo {volume} {5}},\ \bibinfo {pages} {499} (\bibinfo {year} {2009})}\BibitemShut {NoStop}%
\bibitem [{\citenamefont {{Nowakowski}}\ \emph {et~al.}(2015)\citenamefont {{Nowakowski}}, \citenamefont {{Woods}}, \citenamefont {{Bain}},\ and\ \citenamefont {{Verlet}}}]{nawakowsko2015timeresolved}%
  \BibitemOpen
  \bibfield  {author} {\bibinfo {author} {\bibfnamefont {P.~J.}\ \bibnamefont {{Nowakowski}}}, \bibinfo {author} {\bibfnamefont {D.~A.}\ \bibnamefont {{Woods}}}, \bibinfo {author} {\bibfnamefont {C.~D.}\ \bibnamefont {{Bain}}},\ and\ \bibinfo {author} {\bibfnamefont {J.~R.~R.}\ \bibnamefont {{Verlet}}},\ }\bibfield  {title} {\emph {\bibinfo {title} {{Time-resolved phase-sensitive second harmonic generation spectroscopy}}},\ }\href {https://doi.org/10.1063/1.4909522} {\bibfield  {journal} {\bibinfo  {journal} {J. Chem. Phys.}\ }\textbf {\bibinfo {volume} {142}},\ \bibinfo {pages} {084201} (\bibinfo {year} {2015})}\BibitemShut {NoStop}%
\bibitem [{\citenamefont {Zhang}\ \emph {et~al.}(2021{\natexlab{a}})\citenamefont {Zhang}, \citenamefont {Dai}, \citenamefont {Zhong}, \citenamefont {Zhang}, \citenamefont {Zhong},\ and\ \citenamefont {Li}}]{zhang2021probingultrafast}%
  \BibitemOpen
  \bibfield  {author} {\bibinfo {author} {\bibfnamefont {Y.}~\bibnamefont {Zhang}}, \bibinfo {author} {\bibfnamefont {J.}~\bibnamefont {Dai}}, \bibinfo {author} {\bibfnamefont {X.}~\bibnamefont {Zhong}}, \bibinfo {author} {\bibfnamefont {D.}~\bibnamefont {Zhang}}, \bibinfo {author} {\bibfnamefont {G.}~\bibnamefont {Zhong}},\ and\ \bibinfo {author} {\bibfnamefont {J.}~\bibnamefont {Li}},\ }\bibfield  {title} {\emph {\bibinfo {title} {{Probing ultrafast dynamics of ferroelectrics by time-resolved pump-probe spectroscopy}}},\ }\href {https://doi.org/10.1002/advs.202102488} {\bibfield  {journal} {\bibinfo  {journal} {Adv. Sci.}\ }\textbf {\bibinfo {volume} {8}},\ \bibinfo {pages} {2102488} (\bibinfo {year} {2021}{\natexlab{a}})}\BibitemShut {NoStop}%
\bibitem [{\citenamefont {{Yoshizawa}}\ and\ \citenamefont {{Kurosawa}}(1999)}]{Yoshizawa1999femtosecond}%
  \BibitemOpen
  \bibfield  {author} {\bibinfo {author} {\bibfnamefont {M.}~\bibnamefont {{Yoshizawa}}}\ and\ \bibinfo {author} {\bibfnamefont {M.}~\bibnamefont {{Kurosawa}}},\ }\bibfield  {title} {\emph {\bibinfo {title} {{Femtosecond time-resolved Raman spectroscopy using stimulated Raman scattering}}},\ }\href {https://doi.org/10.1103/PhysRevA.61.013808} {\bibfield  {journal} {\bibinfo  {journal} {\pra}\ }\textbf {\bibinfo {volume} {61}},\ \bibinfo {pages} {013808} (\bibinfo {year} {1999})}\BibitemShut {NoStop}%
\bibitem [{\citenamefont {{Saichu}}\ \emph {et~al.}(2009)\citenamefont {{Saichu}}, \citenamefont {{Mahns}}, \citenamefont {{Goos}}, \citenamefont {{Binder}}, \citenamefont {{May}}, \citenamefont {{Singer}}, \citenamefont {{Schulz}}, \citenamefont {{Rusydi}}, \citenamefont {{Unterhinninghofen}}, \citenamefont {{Manske}}, \citenamefont {{Guptasarma}}, \citenamefont {{Williamsen}},\ and\ \citenamefont {{R{\"u}bhausen}}}]{sachu2009two}%
  \BibitemOpen
  \bibfield  {author} {\bibinfo {author} {\bibfnamefont {R.~P.}\ \bibnamefont {{Saichu}}}, \bibinfo {author} {\bibfnamefont {I.}~\bibnamefont {{Mahns}}}, \bibinfo {author} {\bibfnamefont {A.}~\bibnamefont {{Goos}}}, \bibinfo {author} {\bibfnamefont {S.}~\bibnamefont {{Binder}}}, \bibinfo {author} {\bibfnamefont {P.}~\bibnamefont {{May}}}, \bibinfo {author} {\bibfnamefont {S.~G.}\ \bibnamefont {{Singer}}}, \bibinfo {author} {\bibfnamefont {B.}~\bibnamefont {{Schulz}}}, \bibinfo {author} {\bibfnamefont {A.}~\bibnamefont {{Rusydi}}}, \bibinfo {author} {\bibfnamefont {J.}~\bibnamefont {{Unterhinninghofen}}}, \bibinfo {author} {\bibfnamefont {D.}~\bibnamefont {{Manske}}}, \bibinfo {author} {\bibfnamefont {P.}~\bibnamefont {{Guptasarma}}}, \bibinfo {author} {\bibfnamefont {M.~S.}\ \bibnamefont {{Williamsen}}},\ and\ \bibinfo {author} {\bibfnamefont {M.}~\bibnamefont {{R{\"u}bhausen}}},\ }\bibfield  {title} {\emph {\bibinfo {title} {{Two-component dynamics of the order parameter of high temperature
  Bi$_{2}$Sr$_{2}$CaCu$_{2}$O$_{8+{\ensuremath{\delta}}}$ superconductors revealed by time-resolved Raman scattering}}},\ }\href {https://doi.org/10.1103/PhysRevLett.102.177004} {\bibfield  {journal} {\bibinfo  {journal} {\prl}\ }\textbf {\bibinfo {volume} {102}},\ \bibinfo {pages} {177004} (\bibinfo {year} {2009})}\BibitemShut {NoStop}%
\bibitem [{\citenamefont {{Versteeg}}\ \emph {et~al.}(2018)\citenamefont {{Versteeg}}, \citenamefont {{Zhu}}, \citenamefont {{Padmanabhan}}, \citenamefont {{Boguschewski}}, \citenamefont {{German}}, \citenamefont {{Goedecke}}, \citenamefont {{Becker}},\ and\ \citenamefont {{van Loosdrecht}}}]{versteeg2018tunable}%
  \BibitemOpen
  \bibfield  {author} {\bibinfo {author} {\bibfnamefont {R.~B.}\ \bibnamefont {{Versteeg}}}, \bibinfo {author} {\bibfnamefont {J.}~\bibnamefont {{Zhu}}}, \bibinfo {author} {\bibfnamefont {P.}~\bibnamefont {{Padmanabhan}}}, \bibinfo {author} {\bibfnamefont {C.}~\bibnamefont {{Boguschewski}}}, \bibinfo {author} {\bibfnamefont {R.}~\bibnamefont {{German}}}, \bibinfo {author} {\bibfnamefont {M.}~\bibnamefont {{Goedecke}}}, \bibinfo {author} {\bibfnamefont {P.}~\bibnamefont {{Becker}}},\ and\ \bibinfo {author} {\bibfnamefont {P.~H.~M.}\ \bibnamefont {{van Loosdrecht}}},\ }\bibfield  {title} {\emph {\bibinfo {title} {{A tunable time-resolved spontaneous Raman spectroscopy setup for probing ultrafast collective excitation and quasiparticle dynamics in quantum materials}}},\ }\href {https://doi.org/10.1063/1.5037784} {\bibfield  {journal} {\bibinfo  {journal} {Struct. Dyn.}\ }\textbf {\bibinfo {volume} {5}},\ \bibinfo {pages} {044301} (\bibinfo {year} {2018})}\BibitemShut {NoStop}%
\bibitem [{\citenamefont {{Yang}}\ \emph {et~al.}(2020)\citenamefont {{Yang}}, \citenamefont {{Pellatz}}, \citenamefont {{Wolf}}, \citenamefont {{Nandkishore}},\ and\ \citenamefont {{Reznik}}}]{yang2020ultrafast}%
  \BibitemOpen
  \bibfield  {author} {\bibinfo {author} {\bibfnamefont {J.-A.}\ \bibnamefont {{Yang}}}, \bibinfo {author} {\bibfnamefont {N.}~\bibnamefont {{Pellatz}}}, \bibinfo {author} {\bibfnamefont {T.}~\bibnamefont {{Wolf}}}, \bibinfo {author} {\bibfnamefont {R.}~\bibnamefont {{Nandkishore}}},\ and\ \bibinfo {author} {\bibfnamefont {D.}~\bibnamefont {{Reznik}}},\ }\bibfield  {title} {\emph {\bibinfo {title} {{Ultrafast magnetic dynamics in insulating YBa$_{2}$Cu$_{3}$O$_{6.1}$ revealed by time resolved two-magnon Raman scattering}}},\ }\href {https://doi.org/10.1038/s41467-020-16275-9} {\bibfield  {journal} {\bibinfo  {journal} {Nat. Commun.}\ }\textbf {\bibinfo {volume} {11}},\ \bibinfo {pages} {2548} (\bibinfo {year} {2020})}\BibitemShut {NoStop}%
\bibitem [{\citenamefont {{Buchenau}}\ \emph {et~al.}(2023)\citenamefont {{Buchenau}}, \citenamefont {{Grimm-Lebsanft}}, \citenamefont {{Biebl}}, \citenamefont {{Glier}}, \citenamefont {{Westphal}}, \citenamefont {{Reichstetter}}, \citenamefont {{Manske}}, \citenamefont {{Fechner}}, \citenamefont {{Cavalleri}}, \citenamefont {{Herres-Pawlis}},\ and\ \citenamefont {{R{\"u}bhausen}}}]{Buchenau2023optical}%
  \BibitemOpen
  \bibfield  {author} {\bibinfo {author} {\bibfnamefont {S.}~\bibnamefont {{Buchenau}}}, \bibinfo {author} {\bibfnamefont {B.}~\bibnamefont {{Grimm-Lebsanft}}}, \bibinfo {author} {\bibfnamefont {F.}~\bibnamefont {{Biebl}}}, \bibinfo {author} {\bibfnamefont {T.}~\bibnamefont {{Glier}}}, \bibinfo {author} {\bibfnamefont {L.}~\bibnamefont {{Westphal}}}, \bibinfo {author} {\bibfnamefont {J.}~\bibnamefont {{Reichstetter}}}, \bibinfo {author} {\bibfnamefont {D.}~\bibnamefont {{Manske}}}, \bibinfo {author} {\bibfnamefont {M.}~\bibnamefont {{Fechner}}}, \bibinfo {author} {\bibfnamefont {A.}~\bibnamefont {{Cavalleri}}}, \bibinfo {author} {\bibfnamefont {S.}~\bibnamefont {{Herres-Pawlis}}},\ and\ \bibinfo {author} {\bibfnamefont {M.}~\bibnamefont {{R{\"u}bhausen}}},\ }\bibfield  {title} {\emph {\bibinfo {title} {{Optically induced avoided crossing in graphene}}},\ }\href {https://doi.org/10.1103/PhysRevB.108.075419} {\bibfield  {journal} {\bibinfo  {journal} {Phys. Rev. B}\ }\textbf {\bibinfo {volume} {108}},\ \bibinfo
  {pages} {075419} (\bibinfo {year} {2023})}\BibitemShut {NoStop}%
\bibitem [{\citenamefont {{Breusing}}\ \emph {et~al.}(2009)\citenamefont {{Breusing}}, \citenamefont {{Ropers}},\ and\ \citenamefont {{Elsaesser}}}]{breusing2009ultrafastcarrier}%
  \BibitemOpen
  \bibfield  {author} {\bibinfo {author} {\bibfnamefont {M.}~\bibnamefont {{Breusing}}}, \bibinfo {author} {\bibfnamefont {C.}~\bibnamefont {{Ropers}}},\ and\ \bibinfo {author} {\bibfnamefont {T.}~\bibnamefont {{Elsaesser}}},\ }\bibfield  {title} {\emph {\bibinfo {title} {{Ultrafast carrier dynamics in graphite}}},\ }\href {https://doi.org/10.1103/PhysRevLett.102.086809} {\bibfield  {journal} {\bibinfo  {journal} {Phys. Rev. Lett.}\ }\textbf {\bibinfo {volume} {102}},\ \bibinfo {pages} {086809} (\bibinfo {year} {2009})}\BibitemShut {NoStop}%
\bibitem [{\citenamefont {{Ulbricht}}\ \emph {et~al.}(2011)\citenamefont {{Ulbricht}}, \citenamefont {{Hendry}}, \citenamefont {{Shan}}, \citenamefont {{Heinz}},\ and\ \citenamefont {{Bonn}}}]{ulbricht2011carrierdynamics}%
  \BibitemOpen
  \bibfield  {author} {\bibinfo {author} {\bibfnamefont {R.}~\bibnamefont {{Ulbricht}}}, \bibinfo {author} {\bibfnamefont {E.}~\bibnamefont {{Hendry}}}, \bibinfo {author} {\bibfnamefont {J.}~\bibnamefont {{Shan}}}, \bibinfo {author} {\bibfnamefont {T.~F.}\ \bibnamefont {{Heinz}}},\ and\ \bibinfo {author} {\bibfnamefont {M.}~\bibnamefont {{Bonn}}},\ }\bibfield  {title} {\emph {\bibinfo {title} {{Carrier dynamics in semiconductors studied with time-resolved terahertz spectroscopy}}},\ }\href {https://doi.org/10.1103/RevModPhys.83.543} {\bibfield  {journal} {\bibinfo  {journal} {Rev. Mod. Phys.}\ }\textbf {\bibinfo {volume} {83}},\ \bibinfo {pages} {543} (\bibinfo {year} {2011})}\BibitemShut {NoStop}%
\bibitem [{\citenamefont {{Lagarde}}\ \emph {et~al.}(2014)\citenamefont {{Lagarde}}, \citenamefont {{Bouet}}, \citenamefont {{Marie}}, \citenamefont {{Zhu}}, \citenamefont {{Liu}}, \citenamefont {{Amand}}, \citenamefont {{Tan}},\ and\ \citenamefont {{Urbaszek}}}]{lagarde2014carrier}%
  \BibitemOpen
  \bibfield  {author} {\bibinfo {author} {\bibfnamefont {D.}~\bibnamefont {{Lagarde}}}, \bibinfo {author} {\bibfnamefont {L.}~\bibnamefont {{Bouet}}}, \bibinfo {author} {\bibfnamefont {X.}~\bibnamefont {{Marie}}}, \bibinfo {author} {\bibfnamefont {C.~R.}\ \bibnamefont {{Zhu}}}, \bibinfo {author} {\bibfnamefont {B.~L.}\ \bibnamefont {{Liu}}}, \bibinfo {author} {\bibfnamefont {T.}~\bibnamefont {{Amand}}}, \bibinfo {author} {\bibfnamefont {P.~H.}\ \bibnamefont {{Tan}}},\ and\ \bibinfo {author} {\bibfnamefont {B.}~\bibnamefont {{Urbaszek}}},\ }\bibfield  {title} {\emph {\bibinfo {title} {{Carrier and polarization dynamics in monolayer MoS$_{2}$}}},\ }\href {https://doi.org/10.1103/PhysRevLett.112.047401} {\bibfield  {journal} {\bibinfo  {journal} {Phys. Rev. Lett.}\ }\textbf {\bibinfo {volume} {112}},\ \bibinfo {pages} {047401} (\bibinfo {year} {2014})}\BibitemShut {NoStop}%
\bibitem [{\citenamefont {{Robert}}\ \emph {et~al.}(2016)\citenamefont {{Robert}}, \citenamefont {{Lagarde}}, \citenamefont {{Cadiz}}, \citenamefont {{Wang}}, \citenamefont {{Lassagne}}, \citenamefont {{Amand}}, \citenamefont {{Balocchi}}, \citenamefont {{Renucci}}, \citenamefont {{Tongay}}, \citenamefont {{Urbaszek}},\ and\ \citenamefont {{Marie}}}]{robert2016exciton}%
  \BibitemOpen
  \bibfield  {author} {\bibinfo {author} {\bibfnamefont {C.}~\bibnamefont {{Robert}}}, \bibinfo {author} {\bibfnamefont {D.}~\bibnamefont {{Lagarde}}}, \bibinfo {author} {\bibfnamefont {F.}~\bibnamefont {{Cadiz}}}, \bibinfo {author} {\bibfnamefont {G.}~\bibnamefont {{Wang}}}, \bibinfo {author} {\bibfnamefont {B.}~\bibnamefont {{Lassagne}}}, \bibinfo {author} {\bibfnamefont {T.}~\bibnamefont {{Amand}}}, \bibinfo {author} {\bibfnamefont {A.}~\bibnamefont {{Balocchi}}}, \bibinfo {author} {\bibfnamefont {P.}~\bibnamefont {{Renucci}}}, \bibinfo {author} {\bibfnamefont {S.}~\bibnamefont {{Tongay}}}, \bibinfo {author} {\bibfnamefont {B.}~\bibnamefont {{Urbaszek}}},\ and\ \bibinfo {author} {\bibfnamefont {X.}~\bibnamefont {{Marie}}},\ }\bibfield  {title} {\emph {\bibinfo {title} {{Exciton radiative lifetime in transition metal dichalcogenide monolayers}}},\ }\href {https://doi.org/10.1103/PhysRevB.93.205423} {\bibfield  {journal} {\bibinfo  {journal} {Phys. Rev. B}\ }\textbf {\bibinfo {volume} {93}},\ \bibinfo
  {pages} {205423} (\bibinfo {year} {2016})}\BibitemShut {NoStop}%
\bibitem [{\citenamefont {{Qi}}\ \emph {et~al.}(2020)\citenamefont {{Qi}}, \citenamefont {{Ma}}, \citenamefont {{Zhao}}, \citenamefont {{Cheng}}, \citenamefont {{Jiang}}, \citenamefont {{Lu}}, \citenamefont {{Jiang}}, \citenamefont {{Qian}}, \citenamefont {{Wang}}, \citenamefont {{Zhang}}, \citenamefont {{Zhu}}, \citenamefont {{Zou}}, \citenamefont {{Wan}}, \citenamefont {{Xiang}},\ and\ \citenamefont {{Zhang}}}]{qi2020breaking}%
  \BibitemOpen
  \bibfield  {author} {\bibinfo {author} {\bibfnamefont {F.}~\bibnamefont {{Qi}}}, \bibinfo {author} {\bibfnamefont {Z.}~\bibnamefont {{Ma}}}, \bibinfo {author} {\bibfnamefont {L.}~\bibnamefont {{Zhao}}}, \bibinfo {author} {\bibfnamefont {Y.}~\bibnamefont {{Cheng}}}, \bibinfo {author} {\bibfnamefont {W.}~\bibnamefont {{Jiang}}}, \bibinfo {author} {\bibfnamefont {C.}~\bibnamefont {{Lu}}}, \bibinfo {author} {\bibfnamefont {T.}~\bibnamefont {{Jiang}}}, \bibinfo {author} {\bibfnamefont {D.}~\bibnamefont {{Qian}}}, \bibinfo {author} {\bibfnamefont {Z.}~\bibnamefont {{Wang}}}, \bibinfo {author} {\bibfnamefont {W.}~\bibnamefont {{Zhang}}}, \bibinfo {author} {\bibfnamefont {P.}~\bibnamefont {{Zhu}}}, \bibinfo {author} {\bibfnamefont {X.}~\bibnamefont {{Zou}}}, \bibinfo {author} {\bibfnamefont {W.}~\bibnamefont {{Wan}}}, \bibinfo {author} {\bibfnamefont {D.}~\bibnamefont {{Xiang}}},\ and\ \bibinfo {author} {\bibfnamefont {J.}~\bibnamefont {{Zhang}}},\ }\bibfield  {title} {\emph {\bibinfo {title} {{Breaking 50
  femtosecond resolution barrier in MeV ultrafast electron diffraction with a double bend achromat compressor}}},\ }\href {https://doi.org/10.1103/PhysRevLett.124.134803} {\bibfield  {journal} {\bibinfo  {journal} {Phys. Rev. Lett.}\ }\textbf {\bibinfo {volume} {124}},\ \bibinfo {pages} {134803} (\bibinfo {year} {2020})}\BibitemShut {NoStop}%
\bibitem [{\citenamefont {{Beaud}}\ \emph {et~al.}(2009)\citenamefont {{Beaud}}, \citenamefont {{Johnson}}, \citenamefont {{Vorobeva}}, \citenamefont {{Staub}}, \citenamefont {{Souza}}, \citenamefont {{Milne}}, \citenamefont {{Jia}},\ and\ \citenamefont {{Ingold}}}]{beaud2009ultrafast}%
  \BibitemOpen
  \bibfield  {author} {\bibinfo {author} {\bibfnamefont {P.}~\bibnamefont {{Beaud}}}, \bibinfo {author} {\bibfnamefont {S.~L.}\ \bibnamefont {{Johnson}}}, \bibinfo {author} {\bibfnamefont {E.}~\bibnamefont {{Vorobeva}}}, \bibinfo {author} {\bibfnamefont {U.}~\bibnamefont {{Staub}}}, \bibinfo {author} {\bibfnamefont {R.~A.~D.}\ \bibnamefont {{Souza}}}, \bibinfo {author} {\bibfnamefont {C.~J.}\ \bibnamefont {{Milne}}}, \bibinfo {author} {\bibfnamefont {Q.~X.}\ \bibnamefont {{Jia}}},\ and\ \bibinfo {author} {\bibfnamefont {G.}~\bibnamefont {{Ingold}}},\ }\bibfield  {title} {\emph {\bibinfo {title} {{Ultrafast structural phase transition driven by photoinduced melting of charge and orbital order}}},\ }\href {https://doi.org/10.1103/PhysRevLett.103.155702} {\bibfield  {journal} {\bibinfo  {journal} {Phys. Rev. Lett.}\ }\textbf {\bibinfo {volume} {103}},\ \bibinfo {pages} {155702} (\bibinfo {year} {2009})}\BibitemShut {NoStop}%
\bibitem [{\citenamefont {{Li}}\ \emph {et~al.}(2016)\citenamefont {{Li}}, \citenamefont {{Yin}}, \citenamefont {{Wu}}, \citenamefont {{Zhu}}, \citenamefont {{Konstantinova}}, \citenamefont {{Tao}}, \citenamefont {{Yang}}, \citenamefont {{Cheong}}, \citenamefont {{Carbone}}, \citenamefont {{Misewich}}, \citenamefont {{Hill}}, \citenamefont {{Wang}}, \citenamefont {{Cava}},\ and\ \citenamefont {{Zhu}}}]{li2016dichotomy}%
  \BibitemOpen
  \bibfield  {author} {\bibinfo {author} {\bibfnamefont {J.}~\bibnamefont {{Li}}}, \bibinfo {author} {\bibfnamefont {W.-G.}\ \bibnamefont {{Yin}}}, \bibinfo {author} {\bibfnamefont {L.}~\bibnamefont {{Wu}}}, \bibinfo {author} {\bibfnamefont {P.}~\bibnamefont {{Zhu}}}, \bibinfo {author} {\bibfnamefont {T.}~\bibnamefont {{Konstantinova}}}, \bibinfo {author} {\bibfnamefont {J.}~\bibnamefont {{Tao}}}, \bibinfo {author} {\bibfnamefont {J.}~\bibnamefont {{Yang}}}, \bibinfo {author} {\bibfnamefont {S.-W.}\ \bibnamefont {{Cheong}}}, \bibinfo {author} {\bibfnamefont {F.}~\bibnamefont {{Carbone}}}, \bibinfo {author} {\bibfnamefont {J.~A.}\ \bibnamefont {{Misewich}}}, \bibinfo {author} {\bibfnamefont {J.~P.}\ \bibnamefont {{Hill}}}, \bibinfo {author} {\bibfnamefont {X.}~\bibnamefont {{Wang}}}, \bibinfo {author} {\bibfnamefont {R.~J.}\ \bibnamefont {{Cava}}},\ and\ \bibinfo {author} {\bibfnamefont {Y.}~\bibnamefont {{Zhu}}},\ }\bibfield  {title} {\emph {\bibinfo {title} {{Dichotomy in ultrafast atomic dynamics as direct
  evidence of polaron formation in manganites}}},\ }\href {https://doi.org/10.1038/npjquantmats.2016.26} {\bibfield  {journal} {\bibinfo  {journal} {npj Quantum Mater.}\ }\textbf {\bibinfo {volume} {1}},\ \bibinfo {pages} {16026} (\bibinfo {year} {2016})}\BibitemShut {NoStop}%
\bibitem [{\citenamefont {Kirkland}(2020)}]{kirkland2020advanced}%
  \BibitemOpen
  \bibfield  {author} {\bibinfo {author} {\bibfnamefont {E.~J.}\ \bibnamefont {Kirkland}},\ }\href {https://doi.org/10.1007/978-3-030-33260-0} {\emph {\bibinfo {title} {{Advanced computing in electron microscopy}}}},\ Vol.~\bibinfo {volume} {12}\ (\bibinfo  {publisher} {Springer},\ \bibinfo {year} {2020})\BibitemShut {NoStop}%
\bibitem [{\citenamefont {{Wall}}\ \emph {et~al.}(2018)\citenamefont {{Wall}}, \citenamefont {{Yang}}, \citenamefont {{Vidas}}, \citenamefont {{Chollet}}, \citenamefont {{Glownia}}, \citenamefont {{Kozina}}, \citenamefont {{Katayama}}, \citenamefont {{Henighan}}, \citenamefont {{Jiang}}, \citenamefont {{Miller}}, \citenamefont {{Reis}}, \citenamefont {{Boatner}}, \citenamefont {{Delaire}},\ and\ \citenamefont {{Trigo}}}]{wall2018ultrafast}%
  \BibitemOpen
  \bibfield  {author} {\bibinfo {author} {\bibfnamefont {S.}~\bibnamefont {{Wall}}}, \bibinfo {author} {\bibfnamefont {S.}~\bibnamefont {{Yang}}}, \bibinfo {author} {\bibfnamefont {L.}~\bibnamefont {{Vidas}}}, \bibinfo {author} {\bibfnamefont {M.}~\bibnamefont {{Chollet}}}, \bibinfo {author} {\bibfnamefont {J.~M.}\ \bibnamefont {{Glownia}}}, \bibinfo {author} {\bibfnamefont {M.}~\bibnamefont {{Kozina}}}, \bibinfo {author} {\bibfnamefont {T.}~\bibnamefont {{Katayama}}}, \bibinfo {author} {\bibfnamefont {T.}~\bibnamefont {{Henighan}}}, \bibinfo {author} {\bibfnamefont {M.}~\bibnamefont {{Jiang}}}, \bibinfo {author} {\bibfnamefont {T.~A.}\ \bibnamefont {{Miller}}}, \bibinfo {author} {\bibfnamefont {D.~A.}\ \bibnamefont {{Reis}}}, \bibinfo {author} {\bibfnamefont {L.~A.}\ \bibnamefont {{Boatner}}}, \bibinfo {author} {\bibfnamefont {O.}~\bibnamefont {{Delaire}}},\ and\ \bibinfo {author} {\bibfnamefont {M.}~\bibnamefont {{Trigo}}},\ }\bibfield  {title} {\emph {\bibinfo {title} {{Ultrafast disordering of vanadium
  dimers in photoexcited VO$_{2}$}}},\ }\href {https://doi.org/10.1126/science.aau3873} {\bibfield  {journal} {\bibinfo  {journal} {Science}\ }\textbf {\bibinfo {volume} {362}},\ \bibinfo {pages} {572} (\bibinfo {year} {2018})}\BibitemShut {NoStop}%
\bibitem [{\citenamefont {{D{\"u}rr}}\ \emph {et~al.}(2021)\citenamefont {{D{\"u}rr}}, \citenamefont {{Ernstorfer}},\ and\ \citenamefont {{Siwick}}}]{durr2021revealing}%
  \BibitemOpen
  \bibfield  {author} {\bibinfo {author} {\bibfnamefont {H.~A.}\ \bibnamefont {{D{\"u}rr}}}, \bibinfo {author} {\bibfnamefont {R.}~\bibnamefont {{Ernstorfer}}},\ and\ \bibinfo {author} {\bibfnamefont {B.~J.}\ \bibnamefont {{Siwick}}},\ }\bibfield  {title} {\emph {\bibinfo {title} {{Revealing momentum-dependent electron-phonon and phonon-phonon coupling in complex materials with ultrafast electron diffuse scattering}}},\ }\href {https://doi.org/10.1557/s43577-021-00156-7} {\bibfield  {journal} {\bibinfo  {journal} {MRS Bull.}\ }\textbf {\bibinfo {volume} {46}},\ \bibinfo {pages} {731} (\bibinfo {year} {2021})}\BibitemShut {NoStop}%
\bibitem [{\citenamefont {Zong}\ \emph {et~al.}(2021)\citenamefont {Zong}, \citenamefont {Dolgirev}, \citenamefont {Kogar}, \citenamefont {Su}, \citenamefont {Shen}, \citenamefont {Straquadine}, \citenamefont {Wang}, \citenamefont {Luo}, \citenamefont {Kozina}, \citenamefont {Reid}, \citenamefont {Li}, \citenamefont {Yang}, \citenamefont {Weathersby}, \citenamefont {Park}, \citenamefont {Sie}, \citenamefont {Jarillo-Herrero}, \citenamefont {Fisher}, \citenamefont {Wang}, \citenamefont {Demler},\ and\ \citenamefont {Gedik}}]{zong2021role}%
  \BibitemOpen
  \bibfield  {author} {\bibinfo {author} {\bibfnamefont {A.}~\bibnamefont {Zong}}, \bibinfo {author} {\bibfnamefont {P.~E.}\ \bibnamefont {Dolgirev}}, \bibinfo {author} {\bibfnamefont {A.}~\bibnamefont {Kogar}}, \bibinfo {author} {\bibfnamefont {Y.}~\bibnamefont {Su}}, \bibinfo {author} {\bibfnamefont {X.}~\bibnamefont {Shen}}, \bibinfo {author} {\bibfnamefont {J.~A.~W.}\ \bibnamefont {Straquadine}}, \bibinfo {author} {\bibfnamefont {X.}~\bibnamefont {Wang}}, \bibinfo {author} {\bibfnamefont {D.}~\bibnamefont {Luo}}, \bibinfo {author} {\bibfnamefont {M.~E.}\ \bibnamefont {Kozina}}, \bibinfo {author} {\bibfnamefont {A.~H.}\ \bibnamefont {Reid}}, \bibinfo {author} {\bibfnamefont {R.}~\bibnamefont {Li}}, \bibinfo {author} {\bibfnamefont {J.}~\bibnamefont {Yang}}, \bibinfo {author} {\bibfnamefont {S.~P.}\ \bibnamefont {Weathersby}}, \bibinfo {author} {\bibfnamefont {S.}~\bibnamefont {Park}}, \bibinfo {author} {\bibfnamefont {E.~J.}\ \bibnamefont {Sie}}, \bibinfo {author} {\bibfnamefont {P.}~\bibnamefont
  {Jarillo-Herrero}}, \bibinfo {author} {\bibfnamefont {I.~R.}\ \bibnamefont {Fisher}}, \bibinfo {author} {\bibfnamefont {X.}~\bibnamefont {Wang}}, \bibinfo {author} {\bibfnamefont {E.}~\bibnamefont {Demler}},\ and\ \bibinfo {author} {\bibfnamefont {N.}~\bibnamefont {Gedik}},\ }\bibfield  {title} {\emph {\bibinfo {title} {Role of equilibrium fluctuations in light-induced order}},\ }\href {https://doi.org/10.1103/PhysRevLett.127.227401} {\bibfield  {journal} {\bibinfo  {journal} {Phys. Rev. Lett.}\ }\textbf {\bibinfo {volume} {127}},\ \bibinfo {pages} {227401} (\bibinfo {year} {2021})}\BibitemShut {NoStop}%
\bibitem [{\citenamefont {{Pan}}\ and\ \citenamefont {{Caruso}}(2023)}]{pan2023vibrational}%
  \BibitemOpen
  \bibfield  {author} {\bibinfo {author} {\bibfnamefont {Y.}~\bibnamefont {{Pan}}}\ and\ \bibinfo {author} {\bibfnamefont {F.}~\bibnamefont {{Caruso}}},\ }\bibfield  {title} {\emph {\bibinfo {title} {{Vibrational dichroism of chiral valley phonons}}},\ }\href {https://doi.org/10.1021/acs.nanolett.3c01904} {\bibfield  {journal} {\bibinfo  {journal} {Nano Lett.}\ }\textbf {\bibinfo {volume} {23}},\ \bibinfo {pages} {7463} (\bibinfo {year} {2023})}\BibitemShut {NoStop}%
\bibitem [{\citenamefont {{Britt}}\ and\ \citenamefont {{Siwick}}(2023)}]{britt2023ultrafast}%
  \BibitemOpen
  \bibfield  {author} {\bibinfo {author} {\bibfnamefont {T.~L.}\ \bibnamefont {{Britt}}}\ and\ \bibinfo {author} {\bibfnamefont {B.~J.}\ \bibnamefont {{Siwick}}},\ }\bibfield  {title} {\emph {\bibinfo {title} {{Ultrafast phonon diffuse scattering as a tool for observing chiral phonons in monolayer hexagonal lattices}}},\ }\href {https://doi.org/10.1103/PhysRevB.107.214306} {\bibfield  {journal} {\bibinfo  {journal} {Phys. Rev. B}\ }\textbf {\bibinfo {volume} {107}},\ \bibinfo {pages} {214306} (\bibinfo {year} {2023})}\BibitemShut {NoStop}%
\bibitem [{\citenamefont {Trigo}(2018)}]{trigo2018ultrafast}%
  \BibitemOpen
  \bibfield  {author} {\bibinfo {author} {\bibfnamefont {M.}~\bibnamefont {Trigo}},\ }\bibfield  {title} {\emph {\bibinfo {title} {Ultrafast {Fourier} transform inelastic {X}-ray scattering}},\ }\href {https://doi.org/10.1557/mrs.2018.151} {\bibfield  {journal} {\bibinfo  {journal} {MRS Bull.}\ }\textbf {\bibinfo {volume} {43}},\ \bibinfo {pages} {520--526} (\bibinfo {year} {2018})}\BibitemShut {NoStop}%
\bibitem [{\citenamefont {Johnson}\ \emph {et~al.}(2009)\citenamefont {Johnson}, \citenamefont {Beaud}, \citenamefont {Vorobeva}, \citenamefont {Milne}, \citenamefont {Murray}, \citenamefont {Fahy},\ and\ \citenamefont {Ingold}}]{johnson2009directly}%
  \BibitemOpen
  \bibfield  {author} {\bibinfo {author} {\bibfnamefont {S.~L.}\ \bibnamefont {Johnson}}, \bibinfo {author} {\bibfnamefont {P.}~\bibnamefont {Beaud}}, \bibinfo {author} {\bibfnamefont {E.}~\bibnamefont {Vorobeva}}, \bibinfo {author} {\bibfnamefont {C.~J.}\ \bibnamefont {Milne}}, \bibinfo {author} {\bibfnamefont {{\'E}.~D.}\ \bibnamefont {Murray}}, \bibinfo {author} {\bibfnamefont {S.}~\bibnamefont {Fahy}},\ and\ \bibinfo {author} {\bibfnamefont {G.}~\bibnamefont {Ingold}},\ }\bibfield  {title} {\emph {\bibinfo {title} {{Directly observing squeezed phonon states with femtosecond x-ray diffraction}}},\ }\href {https://doi.org/10.1103/PhysRevLett.102.175503} {\bibfield  {journal} {\bibinfo  {journal} {Phys. Rev. Lett.}\ }\textbf {\bibinfo {volume} {102}},\ \bibinfo {pages} {175503} (\bibinfo {year} {2009})}\BibitemShut {NoStop}%
\bibitem [{\citenamefont {Trigo}\ \emph {et~al.}(2013)\citenamefont {Trigo}, \citenamefont {Fuchs}, \citenamefont {Chen}, \citenamefont {Jiang}, \citenamefont {Cammarata}, \citenamefont {Fahy}, \citenamefont {Fritz}, \citenamefont {Gaffney}, \citenamefont {Ghimire}, \citenamefont {Higginbotham}, \citenamefont {Johnson}, \citenamefont {Kozina}, \citenamefont {Larsson}, \citenamefont {Lemke}, \citenamefont {Lindenberg}, \citenamefont {Ndabashimiye}, \citenamefont {Quirin}, \citenamefont {Sokolowski-Tinten}, \citenamefont {Uher}, \citenamefont {Wang}, \citenamefont {Wark}, \citenamefont {Zhu},\ and\ \citenamefont {Reis}}]{trigo2013fourier}%
  \BibitemOpen
  \bibfield  {author} {\bibinfo {author} {\bibfnamefont {M.}~\bibnamefont {Trigo}}, \bibinfo {author} {\bibfnamefont {M.}~\bibnamefont {Fuchs}}, \bibinfo {author} {\bibfnamefont {J.}~\bibnamefont {Chen}}, \bibinfo {author} {\bibfnamefont {M.~P.}\ \bibnamefont {Jiang}}, \bibinfo {author} {\bibfnamefont {M.}~\bibnamefont {Cammarata}}, \bibinfo {author} {\bibfnamefont {S.}~\bibnamefont {Fahy}}, \bibinfo {author} {\bibfnamefont {D.~M.}\ \bibnamefont {Fritz}}, \bibinfo {author} {\bibfnamefont {K.}~\bibnamefont {Gaffney}}, \bibinfo {author} {\bibfnamefont {S.}~\bibnamefont {Ghimire}}, \bibinfo {author} {\bibfnamefont {A.}~\bibnamefont {Higginbotham}}, \bibinfo {author} {\bibfnamefont {S.~L.}\ \bibnamefont {Johnson}}, \bibinfo {author} {\bibfnamefont {M.~E.}\ \bibnamefont {Kozina}}, \bibinfo {author} {\bibfnamefont {J.}~\bibnamefont {Larsson}}, \bibinfo {author} {\bibfnamefont {H.}~\bibnamefont {Lemke}}, \bibinfo {author} {\bibfnamefont {A.~M.}\ \bibnamefont {Lindenberg}}, \bibinfo {author} {\bibfnamefont
  {G.}~\bibnamefont {Ndabashimiye}}, \bibinfo {author} {\bibfnamefont {F.}~\bibnamefont {Quirin}}, \bibinfo {author} {\bibfnamefont {K.}~\bibnamefont {Sokolowski-Tinten}}, \bibinfo {author} {\bibfnamefont {C.}~\bibnamefont {Uher}}, \bibinfo {author} {\bibfnamefont {G.}~\bibnamefont {Wang}}, \bibinfo {author} {\bibfnamefont {J.~S.}\ \bibnamefont {Wark}}, \bibinfo {author} {\bibfnamefont {D.}~\bibnamefont {Zhu}},\ and\ \bibinfo {author} {\bibfnamefont {D.~A.}\ \bibnamefont {Reis}},\ }\bibfield  {title} {\emph {\bibinfo {title} {Fourier-transform inelastic {X}-ray scattering from time- and momentum-dependent phonon–phonon correlations}},\ }\href {https://doi.org/10.1038/nphys2788} {\bibfield  {journal} {\bibinfo  {journal} {Nat. Phys.}\ }\textbf {\bibinfo {volume} {9}},\ \bibinfo {pages} {790--794} (\bibinfo {year} {2013})}\BibitemShut {NoStop}%
\bibitem [{\citenamefont {Hu}\ and\ \citenamefont {Nori}(1996)}]{hu1996squeezed}%
  \BibitemOpen
  \bibfield  {author} {\bibinfo {author} {\bibfnamefont {X.}~\bibnamefont {Hu}}\ and\ \bibinfo {author} {\bibfnamefont {F.}~\bibnamefont {Nori}},\ }\bibfield  {title} {\emph {\bibinfo {title} {{Squeezed phonon states: Modulating quantum fluctuations of atomic displacements}}},\ }\href {https://doi.org/10.1103/PhysRevLett.76.2294} {\bibfield  {journal} {\bibinfo  {journal} {Phys. Rev. Lett.}\ }\textbf {\bibinfo {volume} {76}},\ \bibinfo {pages} {2294--2297} (\bibinfo {year} {1996})}\BibitemShut {NoStop}%
\bibitem [{\citenamefont {Garrett}\ \emph {et~al.}(1997)\citenamefont {Garrett}, \citenamefont {Rojo}, \citenamefont {Sood}, \citenamefont {Whitaker},\ and\ \citenamefont {Merlin}}]{garrett1997vacuum}%
  \BibitemOpen
  \bibfield  {author} {\bibinfo {author} {\bibfnamefont {G.~A.}\ \bibnamefont {Garrett}}, \bibinfo {author} {\bibfnamefont {A.~G.}\ \bibnamefont {Rojo}}, \bibinfo {author} {\bibfnamefont {A.~K.}\ \bibnamefont {Sood}}, \bibinfo {author} {\bibfnamefont {J.~F.}\ \bibnamefont {Whitaker}},\ and\ \bibinfo {author} {\bibfnamefont {R.}~\bibnamefont {Merlin}},\ }\bibfield  {title} {\emph {\bibinfo {title} {{Vacuum squeezing of solids: Macroscopic quantum states driven by light pulses}}},\ }\href {https://doi.org/10.1126/science.275.5306.1638} {\bibfield  {journal} {\bibinfo  {journal} {Science}\ }\textbf {\bibinfo {volume} {275}},\ \bibinfo {pages} {1638--1640} (\bibinfo {year} {1997})}\BibitemShut {NoStop}%
\bibitem [{\citenamefont {Henighan}\ \emph {et~al.}(2016)\citenamefont {Henighan}, \citenamefont {Trigo}, \citenamefont {Chollet}, \citenamefont {Clark}, \citenamefont {Fahy}, \citenamefont {Glownia}, \citenamefont {Jiang}, \citenamefont {Kozina}, \citenamefont {Liu}, \citenamefont {Song}, \citenamefont {Zhu},\ and\ \citenamefont {Reis}}]{henighan2016control}%
  \BibitemOpen
  \bibfield  {author} {\bibinfo {author} {\bibfnamefont {T.}~\bibnamefont {Henighan}}, \bibinfo {author} {\bibfnamefont {M.}~\bibnamefont {Trigo}}, \bibinfo {author} {\bibfnamefont {M.}~\bibnamefont {Chollet}}, \bibinfo {author} {\bibfnamefont {J.~N.}\ \bibnamefont {Clark}}, \bibinfo {author} {\bibfnamefont {S.}~\bibnamefont {Fahy}}, \bibinfo {author} {\bibfnamefont {J.~M.}\ \bibnamefont {Glownia}}, \bibinfo {author} {\bibfnamefont {M.~P.}\ \bibnamefont {Jiang}}, \bibinfo {author} {\bibfnamefont {M.}~\bibnamefont {Kozina}}, \bibinfo {author} {\bibfnamefont {H.}~\bibnamefont {Liu}}, \bibinfo {author} {\bibfnamefont {S.}~\bibnamefont {Song}}, \bibinfo {author} {\bibfnamefont {D.}~\bibnamefont {Zhu}},\ and\ \bibinfo {author} {\bibfnamefont {D.~A.}\ \bibnamefont {Reis}},\ }\bibfield  {title} {\emph {\bibinfo {title} {Control of two-phonon correlations and the mechanism of high-wavevector phonon generation by ultrafast light pulses}},\ }\href {https://doi.org/10.1103/PhysRevB.94.020302} {\bibfield  {journal}
  {\bibinfo  {journal} {Phys. Rev. B}\ }\textbf {\bibinfo {volume} {94}},\ \bibinfo {pages} {020302} (\bibinfo {year} {2016})}\BibitemShut {NoStop}%
\bibitem [{\citenamefont {{Dean}}\ \emph {et~al.}(2016)\citenamefont {{Dean}}, \citenamefont {{Cao}}, \citenamefont {{Liu}}, \citenamefont {{Wall}}, \citenamefont {{Zhu}}, \citenamefont {{Mankowsky}}, \citenamefont {{Thampy}}, \citenamefont {{Chen}}, \citenamefont {{Vale}}, \citenamefont {{Casa}}, \citenamefont {{Kim}}, \citenamefont {{Said}}, \citenamefont {{Juhas}}, \citenamefont {{Alonso-Mori}}, \citenamefont {{Glownia}}, \citenamefont {{Robert}}, \citenamefont {{Robinson}}, \citenamefont {{Sikorski}}, \citenamefont {{Song}}, \citenamefont {{Kozina}}, \citenamefont {{Lemke}}, \citenamefont {{Patthey}}, \citenamefont {{Owada}}, \citenamefont {{Katayama}}, \citenamefont {{Yabashi}}, \citenamefont {{Tanaka}}, \citenamefont {{Togashi}}, \citenamefont {{Liu}}, \citenamefont {{Rayan Serrao}}, \citenamefont {{Kim}}, \citenamefont {{Huber}}, \citenamefont {{Chang}}, \citenamefont {{McMorrow}}, \citenamefont {{F{\"o}rst}},\ and\ \citenamefont {{Hill}}}]{dean2016ultrafast}%
  \BibitemOpen
  \bibfield  {author} {\bibinfo {author} {\bibfnamefont {M.~P.~M.}\ \bibnamefont {{Dean}}}, \bibinfo {author} {\bibfnamefont {Y.}~\bibnamefont {{Cao}}}, \bibinfo {author} {\bibfnamefont {X.}~\bibnamefont {{Liu}}}, \bibinfo {author} {\bibfnamefont {S.}~\bibnamefont {{Wall}}}, \bibinfo {author} {\bibfnamefont {D.}~\bibnamefont {{Zhu}}}, \bibinfo {author} {\bibfnamefont {R.}~\bibnamefont {{Mankowsky}}}, \bibinfo {author} {\bibfnamefont {V.}~\bibnamefont {{Thampy}}}, \bibinfo {author} {\bibfnamefont {X.~M.}\ \bibnamefont {{Chen}}}, \bibinfo {author} {\bibfnamefont {J.~G.}\ \bibnamefont {{Vale}}}, \bibinfo {author} {\bibfnamefont {D.}~\bibnamefont {{Casa}}}, \bibinfo {author} {\bibfnamefont {J.}~\bibnamefont {{Kim}}}, \bibinfo {author} {\bibfnamefont {A.~H.}\ \bibnamefont {{Said}}}, \bibinfo {author} {\bibfnamefont {P.}~\bibnamefont {{Juhas}}}, \bibinfo {author} {\bibfnamefont {R.}~\bibnamefont {{Alonso-Mori}}}, \bibinfo {author} {\bibfnamefont {J.~M.}\ \bibnamefont {{Glownia}}}, \bibinfo {author} {\bibfnamefont
  {A.}~\bibnamefont {{Robert}}}, \bibinfo {author} {\bibfnamefont {J.}~\bibnamefont {{Robinson}}}, \bibinfo {author} {\bibfnamefont {M.}~\bibnamefont {{Sikorski}}}, \bibinfo {author} {\bibfnamefont {S.}~\bibnamefont {{Song}}}, \bibinfo {author} {\bibfnamefont {M.}~\bibnamefont {{Kozina}}}, \bibinfo {author} {\bibfnamefont {H.}~\bibnamefont {{Lemke}}}, \bibinfo {author} {\bibfnamefont {L.}~\bibnamefont {{Patthey}}}, \bibinfo {author} {\bibfnamefont {S.}~\bibnamefont {{Owada}}}, \bibinfo {author} {\bibfnamefont {T.}~\bibnamefont {{Katayama}}}, \bibinfo {author} {\bibfnamefont {M.}~\bibnamefont {{Yabashi}}}, \bibinfo {author} {\bibfnamefont {Y.}~\bibnamefont {{Tanaka}}}, \bibinfo {author} {\bibfnamefont {T.}~\bibnamefont {{Togashi}}}, \bibinfo {author} {\bibfnamefont {J.}~\bibnamefont {{Liu}}}, \bibinfo {author} {\bibfnamefont {C.}~\bibnamefont {{Rayan Serrao}}}, \bibinfo {author} {\bibfnamefont {B.~J.}\ \bibnamefont {{Kim}}}, \bibinfo {author} {\bibfnamefont {L.}~\bibnamefont {{Huber}}}, \bibinfo {author}
  {\bibfnamefont {C.~L.}\ \bibnamefont {{Chang}}}, \bibinfo {author} {\bibfnamefont {D.~F.}\ \bibnamefont {{McMorrow}}}, \bibinfo {author} {\bibfnamefont {M.}~\bibnamefont {{F{\"o}rst}}},\ and\ \bibinfo {author} {\bibfnamefont {J.~P.}\ \bibnamefont {{Hill}}},\ }\bibfield  {title} {\emph {\bibinfo {title} {{{Ultrafast energy- and momentum-resolved dynamics of magnetic correlations in the photo-doped Mott insulator Sr$_{2}$IrO$_{4}$}}}},\ }\href {https://doi.org/10.1038/nmat4641} {\bibfield  {journal} {\bibinfo  {journal} {Nat. Mater.}\ }\textbf {\bibinfo {volume} {15}},\ \bibinfo {pages} {601} (\bibinfo {year} {2016})}\BibitemShut {NoStop}%
\bibitem [{\citenamefont {{Mazzone}}\ \emph {et~al.}(2021)\citenamefont {{Mazzone}}, \citenamefont {{Meyers}}, \citenamefont {{Cao}}, \citenamefont {{Vale}}, \citenamefont {{Dashwood}}, \citenamefont {{Shi}}, \citenamefont {{James}}, \citenamefont {{Robinson}}, \citenamefont {{Lin}}, \citenamefont {{Thampy}}, \citenamefont {{Tanaka}}, \citenamefont {{Johnson}}, \citenamefont {{Miao}}, \citenamefont {{Wang}}, \citenamefont {{Assefa}}, \citenamefont {{Kim}}, \citenamefont {{Casa}}, \citenamefont {{Mankowsky}}, \citenamefont {{Zhu}}, \citenamefont {{Alonso-Mori}}, \citenamefont {{Song}}, \citenamefont {{Yavas}}, \citenamefont {{Katayama}}, \citenamefont {{Yabashi}}, \citenamefont {{Kubota}}, \citenamefont {{Owada}}, \citenamefont {{Liu}}, \citenamefont {{Yang}}, \citenamefont {{Konik}}, \citenamefont {{Robinson}}, \citenamefont {{Hill}}, \citenamefont {{McMorrow}}, \citenamefont {{F{\"o}rst}}, \citenamefont {{Wall}}, \citenamefont {{Liu}},\ and\ \citenamefont {{Dean}}}]{mazzone2021laser}%
  \BibitemOpen
  \bibfield  {author} {\bibinfo {author} {\bibfnamefont {D.~G.}\ \bibnamefont {{Mazzone}}}, \bibinfo {author} {\bibfnamefont {D.}~\bibnamefont {{Meyers}}}, \bibinfo {author} {\bibfnamefont {Y.}~\bibnamefont {{Cao}}}, \bibinfo {author} {\bibfnamefont {J.~G.}\ \bibnamefont {{Vale}}}, \bibinfo {author} {\bibfnamefont {C.~D.}\ \bibnamefont {{Dashwood}}}, \bibinfo {author} {\bibfnamefont {Y.}~\bibnamefont {{Shi}}}, \bibinfo {author} {\bibfnamefont {A.~J.~A.}\ \bibnamefont {{James}}}, \bibinfo {author} {\bibfnamefont {N.~J.}\ \bibnamefont {{Robinson}}}, \bibinfo {author} {\bibfnamefont {J.}~\bibnamefont {{Lin}}}, \bibinfo {author} {\bibfnamefont {V.}~\bibnamefont {{Thampy}}}, \bibinfo {author} {\bibfnamefont {Y.}~\bibnamefont {{Tanaka}}}, \bibinfo {author} {\bibfnamefont {A.~S.}\ \bibnamefont {{Johnson}}}, \bibinfo {author} {\bibfnamefont {H.}~\bibnamefont {{Miao}}}, \bibinfo {author} {\bibfnamefont {R.}~\bibnamefont {{Wang}}}, \bibinfo {author} {\bibfnamefont {T.~A.}\ \bibnamefont {{Assefa}}}, \bibinfo {author}
  {\bibfnamefont {J.}~\bibnamefont {{Kim}}}, \bibinfo {author} {\bibfnamefont {D.}~\bibnamefont {{Casa}}}, \bibinfo {author} {\bibfnamefont {R.}~\bibnamefont {{Mankowsky}}}, \bibinfo {author} {\bibfnamefont {D.}~\bibnamefont {{Zhu}}}, \bibinfo {author} {\bibfnamefont {R.}~\bibnamefont {{Alonso-Mori}}}, \bibinfo {author} {\bibfnamefont {S.}~\bibnamefont {{Song}}}, \bibinfo {author} {\bibfnamefont {H.}~\bibnamefont {{Yavas}}}, \bibinfo {author} {\bibfnamefont {T.}~\bibnamefont {{Katayama}}}, \bibinfo {author} {\bibfnamefont {M.}~\bibnamefont {{Yabashi}}}, \bibinfo {author} {\bibfnamefont {Y.}~\bibnamefont {{Kubota}}}, \bibinfo {author} {\bibfnamefont {S.}~\bibnamefont {{Owada}}}, \bibinfo {author} {\bibfnamefont {J.}~\bibnamefont {{Liu}}}, \bibinfo {author} {\bibfnamefont {J.}~\bibnamefont {{Yang}}}, \bibinfo {author} {\bibfnamefont {R.~M.}\ \bibnamefont {{Konik}}}, \bibinfo {author} {\bibfnamefont {I.~K.}\ \bibnamefont {{Robinson}}}, \bibinfo {author} {\bibfnamefont {J.~P.}\ \bibnamefont {{Hill}}}, \bibinfo
  {author} {\bibfnamefont {D.~F.}\ \bibnamefont {{McMorrow}}}, \bibinfo {author} {\bibfnamefont {M.}~\bibnamefont {{F{\"o}rst}}}, \bibinfo {author} {\bibfnamefont {S.}~\bibnamefont {{Wall}}}, \bibinfo {author} {\bibfnamefont {X.}~\bibnamefont {{Liu}}},\ and\ \bibinfo {author} {\bibfnamefont {M.~P.~M.}\ \bibnamefont {{Dean}}},\ }\bibfield  {title} {\emph {\bibinfo {title} {{{Laser-induced transient magnons in Sr$_{3}$Ir$_{2}$O$_{7}$ throughout the Brillouin zone}}}},\ }\href {https://doi.org/10.1073/pnas.2103696118} {\bibfield  {journal} {\bibinfo  {journal} {Proc. Natl. Acad. Sci. U.S.A.}\ }\textbf {\bibinfo {volume} {118}},\ \bibinfo {pages} {e2103696118} (\bibinfo {year} {2021})}\BibitemShut {NoStop}%
\bibitem [{\citenamefont {Kim}\ \emph {et~al.}(2018)\citenamefont {Kim}, \citenamefont {Casa}, \citenamefont {Said}, \citenamefont {Krakora}, \citenamefont {Kim}, \citenamefont {Kasman}, \citenamefont {Huang},\ and\ \citenamefont {Gog}}]{kim2018quartz}%
  \BibitemOpen
  \bibfield  {author} {\bibinfo {author} {\bibfnamefont {J.}~\bibnamefont {Kim}}, \bibinfo {author} {\bibfnamefont {D.}~\bibnamefont {Casa}}, \bibinfo {author} {\bibfnamefont {A.}~\bibnamefont {Said}}, \bibinfo {author} {\bibfnamefont {R.}~\bibnamefont {Krakora}}, \bibinfo {author} {\bibfnamefont {B.~J.}\ \bibnamefont {Kim}}, \bibinfo {author} {\bibfnamefont {E.}~\bibnamefont {Kasman}}, \bibinfo {author} {\bibfnamefont {X.}~\bibnamefont {Huang}},\ and\ \bibinfo {author} {\bibfnamefont {T.}~\bibnamefont {Gog}},\ }\bibfield  {title} {\emph {\bibinfo {title} {{Quartz-based flat-crystal resonant inelastic x-ray scattering spectrometer with sub-10~{meV} energy resolution}}},\ }\href {https://doi.org/10.1038/s41598-018-20396-z} {\bibfield  {journal} {\bibinfo  {journal} {Sci. Rep.}\ }\textbf {\bibinfo {volume} {8}},\ \bibinfo {pages} {1958} (\bibinfo {year} {2018})}\BibitemShut {NoStop}%
\bibitem [{\citenamefont {Damascelli}\ \emph {et~al.}(2003)\citenamefont {Damascelli}, \citenamefont {Hussain},\ and\ \citenamefont {Shen}}]{damascelli2003angle}%
  \BibitemOpen
  \bibfield  {author} {\bibinfo {author} {\bibfnamefont {A.}~\bibnamefont {Damascelli}}, \bibinfo {author} {\bibfnamefont {Z.}~\bibnamefont {Hussain}},\ and\ \bibinfo {author} {\bibfnamefont {Z.-X.}\ \bibnamefont {Shen}},\ }\bibfield  {title} {\emph {\bibinfo {title} {Angle-resolved photoemission studies of the cuprate superconductors}},\ }\href {https://doi.org/10.1103/RevModPhys.75.473} {\bibfield  {journal} {\bibinfo  {journal} {Rev. Mod. Phys.}\ }\textbf {\bibinfo {volume} {75}},\ \bibinfo {pages} {473} (\bibinfo {year} {2003})}\BibitemShut {NoStop}%
\bibitem [{\citenamefont {Sobota}\ \emph {et~al.}(2021)\citenamefont {Sobota}, \citenamefont {He},\ and\ \citenamefont {Shen}}]{sobota2021angle}%
  \BibitemOpen
  \bibfield  {author} {\bibinfo {author} {\bibfnamefont {J.~A.}\ \bibnamefont {Sobota}}, \bibinfo {author} {\bibfnamefont {Y.}~\bibnamefont {He}},\ and\ \bibinfo {author} {\bibfnamefont {Z.-X.}\ \bibnamefont {Shen}},\ }\bibfield  {title} {\emph {\bibinfo {title} {Angle-resolved photoemission studies of quantum materials}},\ }\href {https://doi.org/10.1103/RevModPhys.93.025006} {\bibfield  {journal} {\bibinfo  {journal} {Rev. Mod. Phys.}\ }\textbf {\bibinfo {volume} {93}},\ \bibinfo {pages} {025006} (\bibinfo {year} {2021})}\BibitemShut {NoStop}%
\bibitem [{\citenamefont {{Zong}}\ \emph {et~al.}(2019{\natexlab{b}})\citenamefont {{Zong}}, \citenamefont {{Kogar}}, \citenamefont {{Bie}}, \citenamefont {{Rohwer}}, \citenamefont {{Lee}}, \citenamefont {{Baldini}}, \citenamefont {{Erge{\c{c}}en}}, \citenamefont {{Yilmaz}}, \citenamefont {{Freelon}}, \citenamefont {{Sie}}, \citenamefont {{Zhou}}, \citenamefont {{Straquadine}}, \citenamefont {{Walmsley}}, \citenamefont {{Dolgirev}}, \citenamefont {{Rozhkov}}, \citenamefont {{Fisher}}, \citenamefont {{Jarillo-Herrero}}, \citenamefont {{Fine}},\ and\ \citenamefont {{Gedik}}}]{zong2019evidence}%
  \BibitemOpen
  \bibfield  {author} {\bibinfo {author} {\bibfnamefont {A.}~\bibnamefont {{Zong}}}, \bibinfo {author} {\bibfnamefont {A.}~\bibnamefont {{Kogar}}}, \bibinfo {author} {\bibfnamefont {Y.-Q.}\ \bibnamefont {{Bie}}}, \bibinfo {author} {\bibfnamefont {T.}~\bibnamefont {{Rohwer}}}, \bibinfo {author} {\bibfnamefont {C.}~\bibnamefont {{Lee}}}, \bibinfo {author} {\bibfnamefont {E.}~\bibnamefont {{Baldini}}}, \bibinfo {author} {\bibfnamefont {E.}~\bibnamefont {{Erge{\c{c}}en}}}, \bibinfo {author} {\bibfnamefont {M.~B.}\ \bibnamefont {{Yilmaz}}}, \bibinfo {author} {\bibfnamefont {B.}~\bibnamefont {{Freelon}}}, \bibinfo {author} {\bibfnamefont {E.~J.}\ \bibnamefont {{Sie}}}, \bibinfo {author} {\bibfnamefont {H.}~\bibnamefont {{Zhou}}}, \bibinfo {author} {\bibfnamefont {J.}~\bibnamefont {{Straquadine}}}, \bibinfo {author} {\bibfnamefont {P.}~\bibnamefont {{Walmsley}}}, \bibinfo {author} {\bibfnamefont {P.~E.}\ \bibnamefont {{Dolgirev}}}, \bibinfo {author} {\bibfnamefont {A.~V.}\ \bibnamefont {{Rozhkov}}}, \bibinfo
  {author} {\bibfnamefont {I.~R.}\ \bibnamefont {{Fisher}}}, \bibinfo {author} {\bibfnamefont {P.}~\bibnamefont {{Jarillo-Herrero}}}, \bibinfo {author} {\bibfnamefont {B.~V.}\ \bibnamefont {{Fine}}},\ and\ \bibinfo {author} {\bibfnamefont {N.}~\bibnamefont {{Gedik}}},\ }\bibfield  {title} {\emph {\bibinfo {title} {{Evidence for topological defects in a photoinduced phase transition}}},\ }\href {https://doi.org/10.1038/s41567-018-0311-9} {\bibfield  {journal} {\bibinfo  {journal} {Nat. Phys.}\ }\textbf {\bibinfo {volume} {15}},\ \bibinfo {pages} {27} (\bibinfo {year} {2019}{\natexlab{b}})}\BibitemShut {NoStop}%
\bibitem [{\citenamefont {{Duan}}\ \emph {et~al.}(2023)\citenamefont {{Duan}}, \citenamefont {{Xia}}, \citenamefont {{Huang}}, \citenamefont {{Wang}}, \citenamefont {{Gu}}, \citenamefont {{Liu}}, \citenamefont {{Xiang}}, \citenamefont {{Qian}}, \citenamefont {{Guo}},\ and\ \citenamefont {{Zhang}}}]{duan2023ultrafast}%
  \BibitemOpen
  \bibfield  {author} {\bibinfo {author} {\bibfnamefont {S.}~\bibnamefont {{Duan}}}, \bibinfo {author} {\bibfnamefont {W.}~\bibnamefont {{Xia}}}, \bibinfo {author} {\bibfnamefont {C.}~\bibnamefont {{Huang}}}, \bibinfo {author} {\bibfnamefont {S.}~\bibnamefont {{Wang}}}, \bibinfo {author} {\bibfnamefont {L.}~\bibnamefont {{Gu}}}, \bibinfo {author} {\bibfnamefont {H.}~\bibnamefont {{Liu}}}, \bibinfo {author} {\bibfnamefont {D.}~\bibnamefont {{Xiang}}}, \bibinfo {author} {\bibfnamefont {D.}~\bibnamefont {{Qian}}}, \bibinfo {author} {\bibfnamefont {Y.}~\bibnamefont {{Guo}}},\ and\ \bibinfo {author} {\bibfnamefont {W.}~\bibnamefont {{Zhang}}},\ }\bibfield  {title} {\emph {\bibinfo {title} {{{Ultrafast switching from the charge density wave phase to a metastable metallic state in 1$T$-TiSe$_{2}$}}}},\ }\href {https://doi.org/10.1103/PhysRevLett.130.226501} {\bibfield  {journal} {\bibinfo  {journal} {Phys. Rev. Lett.}\ }\textbf {\bibinfo {volume} {130}},\ \bibinfo {pages} {226501} (\bibinfo {year}
  {2023})}\BibitemShut {NoStop}%
\bibitem [{\citenamefont {{Mad{\'e}o}}\ \emph {et~al.}(2020)\citenamefont {{Mad{\'e}o}}, \citenamefont {{Man}}, \citenamefont {{Sahoo}}, \citenamefont {{Campbell}}, \citenamefont {{Pareek}}, \citenamefont {{Wong}}, \citenamefont {{Al-Mahboob}}, \citenamefont {{Chan}}, \citenamefont {{Karmakar}}, \citenamefont {{Mariserla}}, \citenamefont {{Li}}, \citenamefont {{Heinz}}, \citenamefont {{Cao}},\ and\ \citenamefont {{Dani}}}]{madeo2020directly}%
  \BibitemOpen
  \bibfield  {author} {\bibinfo {author} {\bibfnamefont {J.}~\bibnamefont {{Mad{\'e}o}}}, \bibinfo {author} {\bibfnamefont {M.~K.~L.}\ \bibnamefont {{Man}}}, \bibinfo {author} {\bibfnamefont {C.}~\bibnamefont {{Sahoo}}}, \bibinfo {author} {\bibfnamefont {M.}~\bibnamefont {{Campbell}}}, \bibinfo {author} {\bibfnamefont {V.}~\bibnamefont {{Pareek}}}, \bibinfo {author} {\bibfnamefont {E.~L.}\ \bibnamefont {{Wong}}}, \bibinfo {author} {\bibfnamefont {A.}~\bibnamefont {{Al-Mahboob}}}, \bibinfo {author} {\bibfnamefont {N.~S.}\ \bibnamefont {{Chan}}}, \bibinfo {author} {\bibfnamefont {A.}~\bibnamefont {{Karmakar}}}, \bibinfo {author} {\bibfnamefont {B.~M.~K.}\ \bibnamefont {{Mariserla}}}, \bibinfo {author} {\bibfnamefont {X.}~\bibnamefont {{Li}}}, \bibinfo {author} {\bibfnamefont {T.~F.}\ \bibnamefont {{Heinz}}}, \bibinfo {author} {\bibfnamefont {T.}~\bibnamefont {{Cao}}},\ and\ \bibinfo {author} {\bibfnamefont {K.~M.}\ \bibnamefont {{Dani}}},\ }\bibfield  {title} {\emph {\bibinfo {title} {{Directly visualizing the
  momentum-forbidden dark excitons and their dynamics in atomically thin semiconductors}}},\ }\href {https://doi.org/10.1126/science.aba1029} {\bibfield  {journal} {\bibinfo  {journal} {Science}\ }\textbf {\bibinfo {volume} {370}},\ \bibinfo {pages} {1199} (\bibinfo {year} {2020})}\BibitemShut {NoStop}%
\bibitem [{\citenamefont {Karni}\ \emph {et~al.}(2022)\citenamefont {Karni}, \citenamefont {Barré}, \citenamefont {Pareek}, \citenamefont {Georgaras}, \citenamefont {Man}, \citenamefont {Sahoo}, \citenamefont {Bacon}, \citenamefont {Zhu}, \citenamefont {Ribeiro}, \citenamefont {O’Beirne}, \citenamefont {Hu}, \citenamefont {Al-Mahboob}, \citenamefont {Abdelrasoul}, \citenamefont {Chan}, \citenamefont {Karmakar}, \citenamefont {Winchester}, \citenamefont {Kim}, \citenamefont {Watanabe}, \citenamefont {Taniguchi}, \citenamefont {Barmak}, \citenamefont {Madéo}, \citenamefont {da~Jornada}, \citenamefont {Heinz},\ and\ \citenamefont {Dani}}]{karni2022structure}%
  \BibitemOpen
  \bibfield  {author} {\bibinfo {author} {\bibfnamefont {O.}~\bibnamefont {Karni}}, \bibinfo {author} {\bibfnamefont {E.}~\bibnamefont {Barré}}, \bibinfo {author} {\bibfnamefont {V.}~\bibnamefont {Pareek}}, \bibinfo {author} {\bibfnamefont {J.~D.}\ \bibnamefont {Georgaras}}, \bibinfo {author} {\bibfnamefont {M.~K.~L.}\ \bibnamefont {Man}}, \bibinfo {author} {\bibfnamefont {C.}~\bibnamefont {Sahoo}}, \bibinfo {author} {\bibfnamefont {D.~R.}\ \bibnamefont {Bacon}}, \bibinfo {author} {\bibfnamefont {X.}~\bibnamefont {Zhu}}, \bibinfo {author} {\bibfnamefont {H.~B.}\ \bibnamefont {Ribeiro}}, \bibinfo {author} {\bibfnamefont {A.~L.}\ \bibnamefont {O’Beirne}}, \bibinfo {author} {\bibfnamefont {J.}~\bibnamefont {Hu}}, \bibinfo {author} {\bibfnamefont {A.}~\bibnamefont {Al-Mahboob}}, \bibinfo {author} {\bibfnamefont {M.~M.~M.}\ \bibnamefont {Abdelrasoul}}, \bibinfo {author} {\bibfnamefont {N.~S.}\ \bibnamefont {Chan}}, \bibinfo {author} {\bibfnamefont {A.}~\bibnamefont {Karmakar}}, \bibinfo {author} {\bibfnamefont
  {A.~J.}\ \bibnamefont {Winchester}}, \bibinfo {author} {\bibfnamefont {B.}~\bibnamefont {Kim}}, \bibinfo {author} {\bibfnamefont {K.}~\bibnamefont {Watanabe}}, \bibinfo {author} {\bibfnamefont {T.}~\bibnamefont {Taniguchi}}, \bibinfo {author} {\bibfnamefont {K.}~\bibnamefont {Barmak}}, \bibinfo {author} {\bibfnamefont {J.}~\bibnamefont {Madéo}}, \bibinfo {author} {\bibfnamefont {F.~H.}\ \bibnamefont {da~Jornada}}, \bibinfo {author} {\bibfnamefont {T.~F.}\ \bibnamefont {Heinz}},\ and\ \bibinfo {author} {\bibfnamefont {K.~M.}\ \bibnamefont {Dani}},\ }\bibfield  {title} {\emph {\bibinfo {title} {{Structure of the moir{\'e} exciton captured by imaging its electron and hole}}},\ }\href {https://doi.org/10.1038/s41586-021-04360-y} {\bibfield  {journal} {\bibinfo  {journal} {Nature}\ }\textbf {\bibinfo {volume} {603}},\ \bibinfo {pages} {247--252} (\bibinfo {year} {2022})}\BibitemShut {NoStop}%
\bibitem [{\citenamefont {{Perfetto}}\ \emph {et~al.}(2016)\citenamefont {{Perfetto}}, \citenamefont {{Sangalli}}, \citenamefont {{Marini}},\ and\ \citenamefont {{Stefanucci}}}]{prefetto2016first}%
  \BibitemOpen
  \bibfield  {author} {\bibinfo {author} {\bibfnamefont {E.}~\bibnamefont {{Perfetto}}}, \bibinfo {author} {\bibfnamefont {D.}~\bibnamefont {{Sangalli}}}, \bibinfo {author} {\bibfnamefont {A.}~\bibnamefont {{Marini}}},\ and\ \bibinfo {author} {\bibfnamefont {G.}~\bibnamefont {{Stefanucci}}},\ }\bibfield  {title} {\emph {\bibinfo {title} {{First-principles approach to excitons in time-resolved and angle-resolved photoemission spectra}}},\ }\href {https://doi.org/10.1103/PhysRevB.94.245303} {\bibfield  {journal} {\bibinfo  {journal} {Phys. Rev. B}\ }\textbf {\bibinfo {volume} {94}},\ \bibinfo {pages} {245303} (\bibinfo {year} {2016})}\BibitemShut {NoStop}%
\bibitem [{\citenamefont {{Steinhoff}}\ \emph {et~al.}(2017)\citenamefont {{Steinhoff}}, \citenamefont {{Florian}}, \citenamefont {{R{\"o}sner}}, \citenamefont {{Sch{\"o}nhoff}}, \citenamefont {{Wehling}},\ and\ \citenamefont {{Jahnke}}}]{steinhoff2017exciton}%
  \BibitemOpen
  \bibfield  {author} {\bibinfo {author} {\bibfnamefont {A.}~\bibnamefont {{Steinhoff}}}, \bibinfo {author} {\bibfnamefont {M.}~\bibnamefont {{Florian}}}, \bibinfo {author} {\bibfnamefont {M.}~\bibnamefont {{R{\"o}sner}}}, \bibinfo {author} {\bibfnamefont {G.}~\bibnamefont {{Sch{\"o}nhoff}}}, \bibinfo {author} {\bibfnamefont {T.~O.}\ \bibnamefont {{Wehling}}},\ and\ \bibinfo {author} {\bibfnamefont {F.}~\bibnamefont {{Jahnke}}},\ }\bibfield  {title} {\emph {\bibinfo {title} {{Exciton fission in monolayer transition metal dichalcogenide semiconductors}}},\ }\href {https://doi.org/10.1038/s41467-017-01298-6} {\bibfield  {journal} {\bibinfo  {journal} {Nat. Commun.}\ }\textbf {\bibinfo {volume} {8}},\ \bibinfo {pages} {1166} (\bibinfo {year} {2017})}\BibitemShut {NoStop}%
\bibitem [{\citenamefont {Rustagi}\ and\ \citenamefont {Kemper}(2018)}]{rustagi2018photoemission}%
  \BibitemOpen
  \bibfield  {author} {\bibinfo {author} {\bibfnamefont {A.}~\bibnamefont {Rustagi}}\ and\ \bibinfo {author} {\bibfnamefont {A.~F.}\ \bibnamefont {Kemper}},\ }\bibfield  {title} {\emph {\bibinfo {title} {Photoemission signature of excitons}},\ }\href {https://doi.org/10.1103/PhysRevB.97.235310} {\bibfield  {journal} {\bibinfo  {journal} {Phys. Rev. B}\ }\textbf {\bibinfo {volume} {97}},\ \bibinfo {pages} {235310} (\bibinfo {year} {2018})}\BibitemShut {NoStop}%
\bibitem [{\citenamefont {{Rustagi}}\ and\ \citenamefont {{Kemper}}(2019)}]{rustagi2019coherent}%
  \BibitemOpen
  \bibfield  {author} {\bibinfo {author} {\bibfnamefont {A.}~\bibnamefont {{Rustagi}}}\ and\ \bibinfo {author} {\bibfnamefont {A.~F.}\ \bibnamefont {{Kemper}}},\ }\bibfield  {title} {\emph {\bibinfo {title} {{Coherent excitonic quantum beats in time-resolved photoemission measurements}}},\ }\href {https://doi.org/10.1103/PhysRevB.99.125303} {\bibfield  {journal} {\bibinfo  {journal} {Phys. Rev. B}\ }\textbf {\bibinfo {volume} {99}},\ \bibinfo {pages} {125303} (\bibinfo {year} {2019})}\BibitemShut {NoStop}%
\bibitem [{\citenamefont {{Christiansen}}\ \emph {et~al.}(2019)\citenamefont {{Christiansen}}, \citenamefont {{Selig}}, \citenamefont {{Malic}}, \citenamefont {{Ernstorfer}},\ and\ \citenamefont {{Knorr}}}]{christiansen2019theory}%
  \BibitemOpen
  \bibfield  {author} {\bibinfo {author} {\bibfnamefont {D.}~\bibnamefont {{Christiansen}}}, \bibinfo {author} {\bibfnamefont {M.}~\bibnamefont {{Selig}}}, \bibinfo {author} {\bibfnamefont {E.}~\bibnamefont {{Malic}}}, \bibinfo {author} {\bibfnamefont {R.}~\bibnamefont {{Ernstorfer}}},\ and\ \bibinfo {author} {\bibfnamefont {A.}~\bibnamefont {{Knorr}}},\ }\bibfield  {title} {\emph {\bibinfo {title} {{Theory of exciton dynamics in time-resolved ARPES: Intra- and intervalley scattering in two-dimensional semiconductors}}},\ }\href {https://doi.org/10.1103/PhysRevB.100.205401} {\bibfield  {journal} {\bibinfo  {journal} {Phys. Rev. B}\ }\textbf {\bibinfo {volume} {100}},\ \bibinfo {pages} {205401} (\bibinfo {year} {2019})}\BibitemShut {NoStop}%
\bibitem [{\citenamefont {Zewail}\ and\ \citenamefont {Thomas}(2009)}]{zewail20094d}%
  \BibitemOpen
  \bibfield  {author} {\bibinfo {author} {\bibfnamefont {A.~H.}\ \bibnamefont {Zewail}}\ and\ \bibinfo {author} {\bibfnamefont {J.~M.}\ \bibnamefont {Thomas}},\ }\href {https://doi.org/10.1142/p641} {\emph {\bibinfo {title} {{4D electron microscopy: imaging in space and time}}}}\ (\bibinfo  {publisher} {World Scientific},\ \bibinfo {year} {2009})\BibitemShut {NoStop}%
\bibitem [{\citenamefont {{Zewail}}(2010)}]{zwail2010fourdimentional}%
  \BibitemOpen
  \bibfield  {author} {\bibinfo {author} {\bibfnamefont {A.~H.}\ \bibnamefont {{Zewail}}},\ }\bibfield  {title} {\emph {\bibinfo {title} {{Four-dimensional electron microscopy}}},\ }\href {https://doi.org/10.1126/science.1166135} {\bibfield  {journal} {\bibinfo  {journal} {Science}\ }\textbf {\bibinfo {volume} {328}},\ \bibinfo {pages} {187} (\bibinfo {year} {2010})}\BibitemShut {NoStop}%
\bibitem [{\citenamefont {{Barwick}}\ \emph {et~al.}(2009)\citenamefont {{Barwick}}, \citenamefont {{Flannigan}},\ and\ \citenamefont {{Zewail}}}]{barwick2009photon}%
  \BibitemOpen
  \bibfield  {author} {\bibinfo {author} {\bibfnamefont {B.}~\bibnamefont {{Barwick}}}, \bibinfo {author} {\bibfnamefont {D.~J.}\ \bibnamefont {{Flannigan}}},\ and\ \bibinfo {author} {\bibfnamefont {A.~H.}\ \bibnamefont {{Zewail}}},\ }\bibfield  {title} {\emph {\bibinfo {title} {{Photon-induced near-field electron microscopy}}},\ }\href {https://doi.org/10.1038/nature08662} {\bibfield  {journal} {\bibinfo  {journal} {Nature}\ }\textbf {\bibinfo {volume} {462}},\ \bibinfo {pages} {902} (\bibinfo {year} {2009})}\BibitemShut {NoStop}%
\bibitem [{\citenamefont {{Liu}}\ \emph {et~al.}(2016)\citenamefont {{Liu}}, \citenamefont {{Baskin}},\ and\ \citenamefont {{Zewail}}}]{liu2016infrared}%
  \BibitemOpen
  \bibfield  {author} {\bibinfo {author} {\bibfnamefont {H.}~\bibnamefont {{Liu}}}, \bibinfo {author} {\bibfnamefont {J.~S.}\ \bibnamefont {{Baskin}}},\ and\ \bibinfo {author} {\bibfnamefont {A.~H.}\ \bibnamefont {{Zewail}}},\ }\bibfield  {title} {\emph {\bibinfo {title} {{Infrared PINEM developed by diffraction in 4D UEM}}},\ }\href {https://doi.org/10.1073/pnas.1600317113} {\bibfield  {journal} {\bibinfo  {journal} {Proc. Natl. Acad. Sci. U.S.A.}\ }\textbf {\bibinfo {volume} {113}},\ \bibinfo {pages} {2041} (\bibinfo {year} {2016})}\BibitemShut {NoStop}%
\bibitem [{\citenamefont {{B{\"u}ttner}}\ \emph {et~al.}(2021)\citenamefont {{B{\"u}ttner}}, \citenamefont {{Pfau}}, \citenamefont {{B{\"o}ttcher}}, \citenamefont {{Schneider}}, \citenamefont {{Mercurio}}, \citenamefont {{G{\"u}nther}}, \citenamefont {{Hessing}}, \citenamefont {{Klose}}, \citenamefont {{Wittmann}}, \citenamefont {{Gerlinger}}, \citenamefont {{Kern}}, \citenamefont {{Str{\"u}ber}}, \citenamefont {{von Korff Schmising}}, \citenamefont {{Fuchs}}, \citenamefont {{Engel}}, \citenamefont {{Churikova}}, \citenamefont {{Huang}}, \citenamefont {{Suzuki}}, \citenamefont {{Lemesh}}, \citenamefont {{Huang}}, \citenamefont {{Caretta}}, \citenamefont {{Weder}}, \citenamefont {{Gaida}}, \citenamefont {{M{\"o}ller}}, \citenamefont {{Harvey}}, \citenamefont {{Zayko}}, \citenamefont {{Bagschik}}, \citenamefont {{Carley}}, \citenamefont {{Mercadier}}, \citenamefont {{Schlappa}}, \citenamefont {{Yaroslavtsev}}, \citenamefont {{Le Guyarder}}, \citenamefont {{Gerasimova}}, \citenamefont {{Scherz}}, \citenamefont
  {{Deiter}}, \citenamefont {{Gort}}, \citenamefont {{Hickin}}, \citenamefont {{Zhu}}, \citenamefont {{Turcato}}, \citenamefont {{Lomidze}}, \citenamefont {{Erdinger}}, \citenamefont {{Castoldi}}, \citenamefont {{Maffessanti}}, \citenamefont {{Porro}}, \citenamefont {{Samartsev}}, \citenamefont {{Sinova}}, \citenamefont {{Ropers}}, \citenamefont {{Mentink}}, \citenamefont {{Dup{\'e}}}, \citenamefont {{Beach}},\ and\ \citenamefont {{Eisebitt}}}]{buttner2021observation}%
  \BibitemOpen
  \bibfield  {author} {\bibinfo {author} {\bibfnamefont {F.}~\bibnamefont {{B{\"u}ttner}}}, \bibinfo {author} {\bibfnamefont {B.}~\bibnamefont {{Pfau}}}, \bibinfo {author} {\bibfnamefont {M.}~\bibnamefont {{B{\"o}ttcher}}}, \bibinfo {author} {\bibfnamefont {M.}~\bibnamefont {{Schneider}}}, \bibinfo {author} {\bibfnamefont {G.}~\bibnamefont {{Mercurio}}}, \bibinfo {author} {\bibfnamefont {C.~M.}\ \bibnamefont {{G{\"u}nther}}}, \bibinfo {author} {\bibfnamefont {P.}~\bibnamefont {{Hessing}}}, \bibinfo {author} {\bibfnamefont {C.}~\bibnamefont {{Klose}}}, \bibinfo {author} {\bibfnamefont {A.}~\bibnamefont {{Wittmann}}}, \bibinfo {author} {\bibfnamefont {K.}~\bibnamefont {{Gerlinger}}}, \bibinfo {author} {\bibfnamefont {L.-M.}\ \bibnamefont {{Kern}}}, \bibinfo {author} {\bibfnamefont {C.}~\bibnamefont {{Str{\"u}ber}}}, \bibinfo {author} {\bibfnamefont {C.}~\bibnamefont {{von Korff Schmising}}}, \bibinfo {author} {\bibfnamefont {J.}~\bibnamefont {{Fuchs}}}, \bibinfo {author} {\bibfnamefont {D.}~\bibnamefont
  {{Engel}}}, \bibinfo {author} {\bibfnamefont {A.}~\bibnamefont {{Churikova}}}, \bibinfo {author} {\bibfnamefont {S.}~\bibnamefont {{Huang}}}, \bibinfo {author} {\bibfnamefont {D.}~\bibnamefont {{Suzuki}}}, \bibinfo {author} {\bibfnamefont {I.}~\bibnamefont {{Lemesh}}}, \bibinfo {author} {\bibfnamefont {M.}~\bibnamefont {{Huang}}}, \bibinfo {author} {\bibfnamefont {L.}~\bibnamefont {{Caretta}}}, \bibinfo {author} {\bibfnamefont {D.}~\bibnamefont {{Weder}}}, \bibinfo {author} {\bibfnamefont {J.~H.}\ \bibnamefont {{Gaida}}}, \bibinfo {author} {\bibfnamefont {M.}~\bibnamefont {{M{\"o}ller}}}, \bibinfo {author} {\bibfnamefont {T.~R.}\ \bibnamefont {{Harvey}}}, \bibinfo {author} {\bibfnamefont {S.}~\bibnamefont {{Zayko}}}, \bibinfo {author} {\bibfnamefont {K.}~\bibnamefont {{Bagschik}}}, \bibinfo {author} {\bibfnamefont {R.}~\bibnamefont {{Carley}}}, \bibinfo {author} {\bibfnamefont {L.}~\bibnamefont {{Mercadier}}}, \bibinfo {author} {\bibfnamefont {J.}~\bibnamefont {{Schlappa}}}, \bibinfo {author} {\bibfnamefont
  {A.}~\bibnamefont {{Yaroslavtsev}}}, \bibinfo {author} {\bibfnamefont {L.}~\bibnamefont {{Le Guyarder}}}, \bibinfo {author} {\bibfnamefont {N.}~\bibnamefont {{Gerasimova}}}, \bibinfo {author} {\bibfnamefont {A.}~\bibnamefont {{Scherz}}}, \bibinfo {author} {\bibfnamefont {C.}~\bibnamefont {{Deiter}}}, \bibinfo {author} {\bibfnamefont {R.}~\bibnamefont {{Gort}}}, \bibinfo {author} {\bibfnamefont {D.}~\bibnamefont {{Hickin}}}, \bibinfo {author} {\bibfnamefont {J.}~\bibnamefont {{Zhu}}}, \bibinfo {author} {\bibfnamefont {M.}~\bibnamefont {{Turcato}}}, \bibinfo {author} {\bibfnamefont {D.}~\bibnamefont {{Lomidze}}}, \bibinfo {author} {\bibfnamefont {F.}~\bibnamefont {{Erdinger}}}, \bibinfo {author} {\bibfnamefont {A.}~\bibnamefont {{Castoldi}}}, \bibinfo {author} {\bibfnamefont {S.}~\bibnamefont {{Maffessanti}}}, \bibinfo {author} {\bibfnamefont {M.}~\bibnamefont {{Porro}}}, \bibinfo {author} {\bibfnamefont {A.}~\bibnamefont {{Samartsev}}}, \bibinfo {author} {\bibfnamefont {J.}~\bibnamefont {{Sinova}}}, \bibinfo
  {author} {\bibfnamefont {C.}~\bibnamefont {{Ropers}}}, \bibinfo {author} {\bibfnamefont {J.~H.}\ \bibnamefont {{Mentink}}}, \bibinfo {author} {\bibfnamefont {B.}~\bibnamefont {{Dup{\'e}}}}, \bibinfo {author} {\bibfnamefont {G.~S.~D.}\ \bibnamefont {{Beach}}},\ and\ \bibinfo {author} {\bibfnamefont {S.}~\bibnamefont {{Eisebitt}}},\ }\bibfield  {title} {\emph {\bibinfo {title} {{Observation of fluctuation-mediated picosecond nucleation of a topological phase}}},\ }\href {https://doi.org/10.1038/s41563-020-00807-1} {\bibfield  {journal} {\bibinfo  {journal} {Nat. Mater.}\ }\textbf {\bibinfo {volume} {20}},\ \bibinfo {pages} {30} (\bibinfo {year} {2021})}\BibitemShut {NoStop}%
\bibitem [{\citenamefont {{Klose}}\ \emph {et~al.}(2023)\citenamefont {{Klose}}, \citenamefont {{B{\"u}ttner}}, \citenamefont {{Hu}}, \citenamefont {{Mazzoli}}, \citenamefont {{Litzius}}, \citenamefont {{Battistelli}}, \citenamefont {{Lemesh}}, \citenamefont {{Bartell}}, \citenamefont {{Huang}}, \citenamefont {{G{\"u}nther}}, \citenamefont {{Schneider}}, \citenamefont {{Barbour}}, \citenamefont {{Wilkins}}, \citenamefont {{Beach}}, \citenamefont {{Eisebitt}},\ and\ \citenamefont {{Pfau}}}]{klose2023coherent}%
  \BibitemOpen
  \bibfield  {author} {\bibinfo {author} {\bibfnamefont {C.}~\bibnamefont {{Klose}}}, \bibinfo {author} {\bibfnamefont {F.}~\bibnamefont {{B{\"u}ttner}}}, \bibinfo {author} {\bibfnamefont {W.}~\bibnamefont {{Hu}}}, \bibinfo {author} {\bibfnamefont {C.}~\bibnamefont {{Mazzoli}}}, \bibinfo {author} {\bibfnamefont {K.}~\bibnamefont {{Litzius}}}, \bibinfo {author} {\bibfnamefont {R.}~\bibnamefont {{Battistelli}}}, \bibinfo {author} {\bibfnamefont {I.}~\bibnamefont {{Lemesh}}}, \bibinfo {author} {\bibfnamefont {J.~M.}\ \bibnamefont {{Bartell}}}, \bibinfo {author} {\bibfnamefont {M.}~\bibnamefont {{Huang}}}, \bibinfo {author} {\bibfnamefont {C.~M.}\ \bibnamefont {{G{\"u}nther}}}, \bibinfo {author} {\bibfnamefont {M.}~\bibnamefont {{Schneider}}}, \bibinfo {author} {\bibfnamefont {A.}~\bibnamefont {{Barbour}}}, \bibinfo {author} {\bibfnamefont {S.~B.}\ \bibnamefont {{Wilkins}}}, \bibinfo {author} {\bibfnamefont {G.~S.~D.}\ \bibnamefont {{Beach}}}, \bibinfo {author} {\bibfnamefont {S.}~\bibnamefont {{Eisebitt}}},\
  and\ \bibinfo {author} {\bibfnamefont {B.}~\bibnamefont {{Pfau}}},\ }\bibfield  {title} {\emph {\bibinfo {title} {{Coherent correlation imaging for resolving fluctuating states of matter}}},\ }\href {https://doi.org/10.1038/s41586-022-05537-9} {\bibfield  {journal} {\bibinfo  {journal} {Nature}\ }\textbf {\bibinfo {volume} {614}},\ \bibinfo {pages} {256} (\bibinfo {year} {2023})}\BibitemShut {NoStop}%
\bibitem [{\citenamefont {{Johnson}}\ \emph {et~al.}(2023)\citenamefont {{Johnson}}, \citenamefont {{Perez-Salinas}}, \citenamefont {{Siddiqui}}, \citenamefont {{Kim}}, \citenamefont {{Choi}}, \citenamefont {{Volckaert}}, \citenamefont {{Majchrzak}}, \citenamefont {{Ulstrup}}, \citenamefont {{Agarwal}}, \citenamefont {{Hallman}}, \citenamefont {{Haglund}}, \citenamefont {{G{\"u}nther}}, \citenamefont {{Pfau}}, \citenamefont {{Eisebitt}}, \citenamefont {{Backes}}, \citenamefont {{Maccherozzi}}, \citenamefont {{Fitzpatrick}}, \citenamefont {{Dhesi}}, \citenamefont {{Gargiani}}, \citenamefont {{Valvidares}}, \citenamefont {{Artrith}}, \citenamefont {{de Groot}}, \citenamefont {{Choi}}, \citenamefont {{Jang}}, \citenamefont {{Katoch}}, \citenamefont {{Kwon}}, \citenamefont {{Park}}, \citenamefont {{Kim}},\ and\ \citenamefont {{Wall}}}]{johnson2023ultrafast}%
  \BibitemOpen
  \bibfield  {author} {\bibinfo {author} {\bibfnamefont {A.~S.}\ \bibnamefont {{Johnson}}}, \bibinfo {author} {\bibfnamefont {D.}~\bibnamefont {{Perez-Salinas}}}, \bibinfo {author} {\bibfnamefont {K.~M.}\ \bibnamefont {{Siddiqui}}}, \bibinfo {author} {\bibfnamefont {S.}~\bibnamefont {{Kim}}}, \bibinfo {author} {\bibfnamefont {S.}~\bibnamefont {{Choi}}}, \bibinfo {author} {\bibfnamefont {K.}~\bibnamefont {{Volckaert}}}, \bibinfo {author} {\bibfnamefont {P.~E.}\ \bibnamefont {{Majchrzak}}}, \bibinfo {author} {\bibfnamefont {S.}~\bibnamefont {{Ulstrup}}}, \bibinfo {author} {\bibfnamefont {N.}~\bibnamefont {{Agarwal}}}, \bibinfo {author} {\bibfnamefont {K.}~\bibnamefont {{Hallman}}}, \bibinfo {author} {\bibfnamefont {R.~F.}\ \bibnamefont {{Haglund}}}, \bibinfo {author} {\bibfnamefont {C.~M.}\ \bibnamefont {{G{\"u}nther}}}, \bibinfo {author} {\bibfnamefont {B.}~\bibnamefont {{Pfau}}}, \bibinfo {author} {\bibfnamefont {S.}~\bibnamefont {{Eisebitt}}}, \bibinfo {author} {\bibfnamefont {D.}~\bibnamefont {{Backes}}},
  \bibinfo {author} {\bibfnamefont {F.}~\bibnamefont {{Maccherozzi}}}, \bibinfo {author} {\bibfnamefont {A.}~\bibnamefont {{Fitzpatrick}}}, \bibinfo {author} {\bibfnamefont {S.~S.}\ \bibnamefont {{Dhesi}}}, \bibinfo {author} {\bibfnamefont {P.}~\bibnamefont {{Gargiani}}}, \bibinfo {author} {\bibfnamefont {M.}~\bibnamefont {{Valvidares}}}, \bibinfo {author} {\bibfnamefont {N.}~\bibnamefont {{Artrith}}}, \bibinfo {author} {\bibfnamefont {F.}~\bibnamefont {{de Groot}}}, \bibinfo {author} {\bibfnamefont {H.}~\bibnamefont {{Choi}}}, \bibinfo {author} {\bibfnamefont {D.}~\bibnamefont {{Jang}}}, \bibinfo {author} {\bibfnamefont {A.}~\bibnamefont {{Katoch}}}, \bibinfo {author} {\bibfnamefont {S.}~\bibnamefont {{Kwon}}}, \bibinfo {author} {\bibfnamefont {S.~H.}\ \bibnamefont {{Park}}}, \bibinfo {author} {\bibfnamefont {H.}~\bibnamefont {{Kim}}},\ and\ \bibinfo {author} {\bibfnamefont {S.~E.}\ \bibnamefont {{Wall}}},\ }\bibfield  {title} {\emph {\bibinfo {title} {{{Ultrafast X-ray imaging of the light-induced phase
  transition in VO$_2$}}}},\ }\href {https://doi.org/10.1038/s41567-022-01848-w} {\bibfield  {journal} {\bibinfo  {journal} {Nat. Phys.}\ }\textbf {\bibinfo {volume} {19}},\ \bibinfo {pages} {215} (\bibinfo {year} {2023})}\BibitemShut {NoStop}%
\bibitem [{\citenamefont {Jelic}\ \emph {et~al.}(2024)\citenamefont {Jelic}, \citenamefont {Adams}, \citenamefont {Hassan}, \citenamefont {Cleland-Host}, \citenamefont {Ammerman},\ and\ \citenamefont {Cocker}}]{jelic2024atomic}%
  \BibitemOpen
  \bibfield  {author} {\bibinfo {author} {\bibfnamefont {V.}~\bibnamefont {Jelic}}, \bibinfo {author} {\bibfnamefont {S.}~\bibnamefont {Adams}}, \bibinfo {author} {\bibfnamefont {M.}~\bibnamefont {Hassan}}, \bibinfo {author} {\bibfnamefont {K.}~\bibnamefont {Cleland-Host}}, \bibinfo {author} {\bibfnamefont {S.~E.}\ \bibnamefont {Ammerman}},\ and\ \bibinfo {author} {\bibfnamefont {T.~L.}\ \bibnamefont {Cocker}},\ }\bibfield  {title} {\emph {\bibinfo {title} {{Atomic-scale terahertz time-domain spectroscopy}}},\ }\href {https://doi.org/10.1038/s41566-024-01467-2} {\bibfield  {journal} {\bibinfo  {journal} {Nat. Photonics}\ }\textbf {\bibinfo {volume} {18}},\ \bibinfo {pages} {898--904} (\bibinfo {year} {2024})}\BibitemShut {NoStop}%
\bibitem [{\citenamefont {Sheng}\ \emph {et~al.}(2024)\citenamefont {Sheng}, \citenamefont {Abdo}, \citenamefont {Rolf-Pissarczyk}, \citenamefont {Lichtenberg}, \citenamefont {Baumann}, \citenamefont {Burgess}, \citenamefont {Malavolti},\ and\ \citenamefont {Loth}}]{sheng2024terahertz}%
  \BibitemOpen
  \bibfield  {author} {\bibinfo {author} {\bibfnamefont {S.}~\bibnamefont {Sheng}}, \bibinfo {author} {\bibfnamefont {M.}~\bibnamefont {Abdo}}, \bibinfo {author} {\bibfnamefont {S.}~\bibnamefont {Rolf-Pissarczyk}}, \bibinfo {author} {\bibfnamefont {K.}~\bibnamefont {Lichtenberg}}, \bibinfo {author} {\bibfnamefont {S.}~\bibnamefont {Baumann}}, \bibinfo {author} {\bibfnamefont {J.~A.~J.}\ \bibnamefont {Burgess}}, \bibinfo {author} {\bibfnamefont {L.}~\bibnamefont {Malavolti}},\ and\ \bibinfo {author} {\bibfnamefont {S.}~\bibnamefont {Loth}},\ }\bibfield  {title} {\emph {\bibinfo {title} {{Terahertz spectroscopy of collective charge density wave dynamics at the atomic scale}}},\ }\href {https://doi.org/10.1038/s41567-024-02552-7} {\bibfield  {journal} {\bibinfo  {journal} {Nat. Phys.}\ }\textbf {\bibinfo {volume} {20}},\ \bibinfo {pages} {1603--1608} (\bibinfo {year} {2024})}\BibitemShut {NoStop}%
\bibitem [{\citenamefont {Barwick}\ and\ \citenamefont {Zewail}(2015)}]{barwick2015photonics..pinem}%
  \BibitemOpen
  \bibfield  {author} {\bibinfo {author} {\bibfnamefont {B.}~\bibnamefont {Barwick}}\ and\ \bibinfo {author} {\bibfnamefont {A.~H.}\ \bibnamefont {Zewail}},\ }\bibfield  {title} {\emph {\bibinfo {title} {{Photonics and plasmonics in 4D ultrafast electron microscopy}}},\ }\href {https://doi.org/10.1021/acsphotonics.5b00427} {\bibfield  {journal} {\bibinfo  {journal} {ACS Photonics}\ }\textbf {\bibinfo {volume} {2}},\ \bibinfo {pages} {1391} (\bibinfo {year} {2015})}\BibitemShut {NoStop}%
\bibitem [{\citenamefont {{Piazza}}\ \emph {et~al.}(2015)\citenamefont {{Piazza}}, \citenamefont {{Lummen}}, \citenamefont {{Qui{\~n}onez}}, \citenamefont {{Murooka}}, \citenamefont {{Reed}}, \citenamefont {{Barwick}},\ and\ \citenamefont {{Carbone}}}]{piazza2015simultaneous}%
  \BibitemOpen
  \bibfield  {author} {\bibinfo {author} {\bibfnamefont {L.}~\bibnamefont {{Piazza}}}, \bibinfo {author} {\bibfnamefont {T.~T.~A.}\ \bibnamefont {{Lummen}}}, \bibinfo {author} {\bibfnamefont {E.}~\bibnamefont {{Qui{\~n}onez}}}, \bibinfo {author} {\bibfnamefont {Y.}~\bibnamefont {{Murooka}}}, \bibinfo {author} {\bibfnamefont {B.~W.}\ \bibnamefont {{Reed}}}, \bibinfo {author} {\bibfnamefont {B.}~\bibnamefont {{Barwick}}},\ and\ \bibinfo {author} {\bibfnamefont {F.}~\bibnamefont {{Carbone}}},\ }\bibfield  {title} {\emph {\bibinfo {title} {{Simultaneous observation of the quantization and the interference pattern of a plasmonic near-field}}},\ }\href {https://doi.org/10.1038/ncomms7407} {\bibfield  {journal} {\bibinfo  {journal} {Nat. Commun.}\ }\textbf {\bibinfo {volume} {6}},\ \bibinfo {pages} {6407} (\bibinfo {year} {2015})}\BibitemShut {NoStop}%
\bibitem [{\citenamefont {{Kurman}}\ \emph {et~al.}(2021)\citenamefont {{Kurman}}, \citenamefont {{Dahan}}, \citenamefont {{Sheinfux}}, \citenamefont {{Wang}}, \citenamefont {{Yannai}}, \citenamefont {{Adiv}}, \citenamefont {{Reinhardt}}, \citenamefont {{Tizei}}, \citenamefont {{Woo}}, \citenamefont {{Li}}, \citenamefont {{Edgar}}, \citenamefont {{Kociak}}, \citenamefont {{Koppens}},\ and\ \citenamefont {{Kaminer}}}]{kurman2021spatiotemporal}%
  \BibitemOpen
  \bibfield  {author} {\bibinfo {author} {\bibfnamefont {Y.}~\bibnamefont {{Kurman}}}, \bibinfo {author} {\bibfnamefont {R.}~\bibnamefont {{Dahan}}}, \bibinfo {author} {\bibfnamefont {H.~H.}\ \bibnamefont {{Sheinfux}}}, \bibinfo {author} {\bibfnamefont {K.}~\bibnamefont {{Wang}}}, \bibinfo {author} {\bibfnamefont {M.}~\bibnamefont {{Yannai}}}, \bibinfo {author} {\bibfnamefont {Y.}~\bibnamefont {{Adiv}}}, \bibinfo {author} {\bibfnamefont {O.}~\bibnamefont {{Reinhardt}}}, \bibinfo {author} {\bibfnamefont {L.~H.~G.}\ \bibnamefont {{Tizei}}}, \bibinfo {author} {\bibfnamefont {S.~Y.}\ \bibnamefont {{Woo}}}, \bibinfo {author} {\bibfnamefont {J.}~\bibnamefont {{Li}}}, \bibinfo {author} {\bibfnamefont {J.~H.}\ \bibnamefont {{Edgar}}}, \bibinfo {author} {\bibfnamefont {M.}~\bibnamefont {{Kociak}}}, \bibinfo {author} {\bibfnamefont {F.~H.~L.}\ \bibnamefont {{Koppens}}},\ and\ \bibinfo {author} {\bibfnamefont {I.}~\bibnamefont {{Kaminer}}},\ }\bibfield  {title} {\emph {\bibinfo {title} {{Spatiotemporal imaging of 2D
  polariton wave packet dynamics using free electrons}}},\ }\href {https://doi.org/10.1126/science.abg9015} {\bibfield  {journal} {\bibinfo  {journal} {Science}\ }\textbf {\bibinfo {volume} {372}},\ \bibinfo {pages} {1181} (\bibinfo {year} {2021})}\BibitemShut {NoStop}%
\bibitem [{\citenamefont {{Yurtsever}}\ \emph {et~al.}(2012)\citenamefont {{Yurtsever}}, \citenamefont {{van der Veen}},\ and\ \citenamefont {{Zewail}}}]{yurtsever2012subparticle}%
  \BibitemOpen
  \bibfield  {author} {\bibinfo {author} {\bibfnamefont {A.}~\bibnamefont {{Yurtsever}}}, \bibinfo {author} {\bibfnamefont {R.~M.}\ \bibnamefont {{van der Veen}}},\ and\ \bibinfo {author} {\bibfnamefont {A.~H.}\ \bibnamefont {{Zewail}}},\ }\bibfield  {title} {\emph {\bibinfo {title} {{Subparticle ultrafast spectrum imaging in 4D electron microscopy}}},\ }\href {https://doi.org/10.1126/science.1213504} {\bibfield  {journal} {\bibinfo  {journal} {Science}\ }\textbf {\bibinfo {volume} {335}},\ \bibinfo {pages} {59} (\bibinfo {year} {2012})}\BibitemShut {NoStop}%
\bibitem [{\citenamefont {{Bakker}}\ \emph {et~al.}(1994)\citenamefont {{Bakker}}, \citenamefont {{Hunsche}},\ and\ \citenamefont {{Kurz}}}]{bakker1994investigation}%
  \BibitemOpen
  \bibfield  {author} {\bibinfo {author} {\bibfnamefont {H.~J.}\ \bibnamefont {{Bakker}}}, \bibinfo {author} {\bibfnamefont {S.}~\bibnamefont {{Hunsche}}},\ and\ \bibinfo {author} {\bibfnamefont {H.}~\bibnamefont {{Kurz}}},\ }\bibfield  {title} {\emph {\bibinfo {title} {{Investigation of anharmonic lattice vibrations with coherent phonon polaritons}}},\ }\href {https://doi.org/10.1103/PhysRevB.50.914} {\bibfield  {journal} {\bibinfo  {journal} {Phys. Rev. B}\ }\textbf {\bibinfo {volume} {50}},\ \bibinfo {pages} {914} (\bibinfo {year} {1994})}\BibitemShut {NoStop}%
\bibitem [{\citenamefont {{Luo}}\ \emph {et~al.}(2024)\citenamefont {{Luo}}, \citenamefont {{Ilyas}}, \citenamefont {{Hoegen}}, \citenamefont {{Lee}}, \citenamefont {{Park}}, \citenamefont {{Park}},\ and\ \citenamefont {{Gedik}}}]{luo2024time}%
  \BibitemOpen
  \bibfield  {author} {\bibinfo {author} {\bibfnamefont {T.}~\bibnamefont {{Luo}}}, \bibinfo {author} {\bibfnamefont {B.}~\bibnamefont {{Ilyas}}}, \bibinfo {author} {\bibfnamefont {A.~v.}\ \bibnamefont {{Hoegen}}}, \bibinfo {author} {\bibfnamefont {Y.}~\bibnamefont {{Lee}}}, \bibinfo {author} {\bibfnamefont {J.}~\bibnamefont {{Park}}}, \bibinfo {author} {\bibfnamefont {J.-G.}\ \bibnamefont {{Park}}},\ and\ \bibinfo {author} {\bibfnamefont {N.}~\bibnamefont {{Gedik}}},\ }\bibfield  {title} {\emph {\bibinfo {title} {{Time-of-flight detection of terahertz phonon-polariton}}},\ }\href {https://doi.org/10.1038/s41467-024-46515-1} {\bibfield  {journal} {\bibinfo  {journal} {Nat. Commun.}\ }\textbf {\bibinfo {volume} {15}},\ \bibinfo {pages} {2276} (\bibinfo {year} {2024})}\BibitemShut {NoStop}%
\bibitem [{\citenamefont {{Feurer}}\ \emph {et~al.}(2007)\citenamefont {{Feurer}}, \citenamefont {{Stoyanov}}, \citenamefont {{Ward}}, \citenamefont {{Vaughan}}, \citenamefont {{Statz}},\ and\ \citenamefont {{Nelson}}}]{Feurer2007Polaritonics}%
  \BibitemOpen
  \bibfield  {author} {\bibinfo {author} {\bibfnamefont {T.}~\bibnamefont {{Feurer}}}, \bibinfo {author} {\bibfnamefont {N.~S.}\ \bibnamefont {{Stoyanov}}}, \bibinfo {author} {\bibfnamefont {D.~W.}\ \bibnamefont {{Ward}}}, \bibinfo {author} {\bibfnamefont {J.~C.}\ \bibnamefont {{Vaughan}}}, \bibinfo {author} {\bibfnamefont {E.~R.}\ \bibnamefont {{Statz}}},\ and\ \bibinfo {author} {\bibfnamefont {K.~A.}\ \bibnamefont {{Nelson}}},\ }\bibfield  {title} {\emph {\bibinfo {title} {{Terahertz polaritonics}}},\ }\href {https://doi.org/10.1146/annurev.matsci.37.052506.084327} {\bibfield  {journal} {\bibinfo  {journal} {Annu. Rev. Mater. Res.}\ }\textbf {\bibinfo {volume} {37}},\ \bibinfo {pages} {317} (\bibinfo {year} {2007})}\BibitemShut {NoStop}%
\bibitem [{\citenamefont {{Bakker}}\ \emph {et~al.}(1998)\citenamefont {{Bakker}}, \citenamefont {{Hunsche}},\ and\ \citenamefont {{Kurz}}}]{bakker1998coherent}%
  \BibitemOpen
  \bibfield  {author} {\bibinfo {author} {\bibfnamefont {H.~J.}\ \bibnamefont {{Bakker}}}, \bibinfo {author} {\bibfnamefont {S.}~\bibnamefont {{Hunsche}}},\ and\ \bibinfo {author} {\bibfnamefont {H.}~\bibnamefont {{Kurz}}},\ }\bibfield  {title} {\emph {\bibinfo {title} {{Coherent phonon polaritons as probes of anharmonic phonons in ferroelectrics}}},\ }\href {https://doi.org/10.1103/RevModPhys.70.523} {\bibfield  {journal} {\bibinfo  {journal} {Rev. Mod. Phys.}\ }\textbf {\bibinfo {volume} {70}},\ \bibinfo {pages} {523} (\bibinfo {year} {1998})}\BibitemShut {NoStop}%
\bibitem [{\citenamefont {{Henry}}\ and\ \citenamefont {{Hopfield}}(1965)}]{henry1965raman}%
  \BibitemOpen
  \bibfield  {author} {\bibinfo {author} {\bibfnamefont {C.~H.}\ \bibnamefont {{Henry}}}\ and\ \bibinfo {author} {\bibfnamefont {J.~J.}\ \bibnamefont {{Hopfield}}},\ }\bibfield  {title} {\emph {\bibinfo {title} {{Raman scattering by polaritons}}},\ }\href {https://doi.org/10.1103/PhysRevLett.15.964} {\bibfield  {journal} {\bibinfo  {journal} {Phys. Rev. Lett.}\ }\textbf {\bibinfo {volume} {15}},\ \bibinfo {pages} {964} (\bibinfo {year} {1965})}\BibitemShut {NoStop}%
\bibitem [{\citenamefont {{Cheung}}\ and\ \citenamefont {{Auston}}(1985)}]{cheung1985excitation}%
  \BibitemOpen
  \bibfield  {author} {\bibinfo {author} {\bibfnamefont {K.~P.}\ \bibnamefont {{Cheung}}}\ and\ \bibinfo {author} {\bibfnamefont {D.~H.}\ \bibnamefont {{Auston}}},\ }\bibfield  {title} {\emph {\bibinfo {title} {{Excitation of coherent phonon polaritons with femtosecond optical pulses}}},\ }\href {https://doi.org/10.1103/PhysRevLett.55.2152} {\bibfield  {journal} {\bibinfo  {journal} {Phys. Rev. Lett.}\ }\textbf {\bibinfo {volume} {55}},\ \bibinfo {pages} {2152} (\bibinfo {year} {1985})}\BibitemShut {NoStop}%
\bibitem [{\citenamefont {{Yan}}\ \emph {et~al.}(1985)\citenamefont {{Yan}}, \citenamefont {{Gamble}},\ and\ \citenamefont {{Nelson}}}]{yan1985impulsive}%
  \BibitemOpen
  \bibfield  {author} {\bibinfo {author} {\bibfnamefont {Y.~X.}\ \bibnamefont {{Yan}}}, \bibinfo {author} {\bibfnamefont {J.}~\bibnamefont {{Gamble}}, \bibfnamefont {E.~B.}},\ and\ \bibinfo {author} {\bibfnamefont {K.~A.}\ \bibnamefont {{Nelson}}},\ }\bibfield  {title} {\emph {\bibinfo {title} {{Impulsive stimulated scattering: General importance in femtosecond laser pulse interactions with matter, and spectroscopic applications}}},\ }\href {https://doi.org/10.1063/1.449708} {\bibfield  {journal} {\bibinfo  {journal} {J. Chem. Phys.}\ }\textbf {\bibinfo {volume} {83}},\ \bibinfo {pages} {5391} (\bibinfo {year} {1985})}\BibitemShut {NoStop}%
\bibitem [{\citenamefont {{Planken}}\ \emph {et~al.}(1992)\citenamefont {{Planken}}, \citenamefont {{Noordam}}, \citenamefont {{Kennis}},\ and\ \citenamefont {{Lagendijk}}}]{planken1992femtosecond}%
  \BibitemOpen
  \bibfield  {author} {\bibinfo {author} {\bibfnamefont {P.~C.~M.}\ \bibnamefont {{Planken}}}, \bibinfo {author} {\bibfnamefont {L.~D.}\ \bibnamefont {{Noordam}}}, \bibinfo {author} {\bibfnamefont {J.~T.~M.}\ \bibnamefont {{Kennis}}},\ and\ \bibinfo {author} {\bibfnamefont {A.}~\bibnamefont {{Lagendijk}}},\ }\bibfield  {title} {\emph {\bibinfo {title} {{Femtosecond time-resolved study of the generation and propagation of phonon polaritons in LiNbO$_{3}$}}},\ }\href {https://doi.org/10.1103/PhysRevB.45.7106} {\bibfield  {journal} {\bibinfo  {journal} {Phys. Rev. B}\ }\textbf {\bibinfo {volume} {45}},\ \bibinfo {pages} {7106} (\bibinfo {year} {1992})}\BibitemShut {NoStop}%
\bibitem [{\citenamefont {{Koehl}}\ \emph {et~al.}(1999{\natexlab{a}})\citenamefont {{Koehl}}, \citenamefont {{Adachi}},\ and\ \citenamefont {{Nelson}}}]{koehl1999direct}%
  \BibitemOpen
  \bibfield  {author} {\bibinfo {author} {\bibfnamefont {R.~M.}\ \bibnamefont {{Koehl}}}, \bibinfo {author} {\bibfnamefont {S.}~\bibnamefont {{Adachi}}},\ and\ \bibinfo {author} {\bibfnamefont {K.~A.}\ \bibnamefont {{Nelson}}},\ }\bibfield  {title} {\emph {\bibinfo {title} {{Direct visualization of collective wavepacket dynamics}}},\ }\href {https://doi.org/10.1021/jp9922007} {\bibfield  {journal} {\bibinfo  {journal} {J. Phys. Chem. A}\ }\textbf {\bibinfo {volume} {103}},\ \bibinfo {pages} {10260} (\bibinfo {year} {1999}{\natexlab{a}})}\BibitemShut {NoStop}%
\bibitem [{\citenamefont {{Koehl}}\ \emph {et~al.}(1999{\natexlab{b}})\citenamefont {{Koehl}}, \citenamefont {{Adachi}},\ and\ \citenamefont {{Nelson}}}]{koehl1999real}%
  \BibitemOpen
  \bibfield  {author} {\bibinfo {author} {\bibfnamefont {R.~M.}\ \bibnamefont {{Koehl}}}, \bibinfo {author} {\bibfnamefont {S.}~\bibnamefont {{Adachi}}},\ and\ \bibinfo {author} {\bibfnamefont {K.~A.}\ \bibnamefont {{Nelson}}},\ }\bibfield  {title} {\emph {\bibinfo {title} {{Real-space polariton wave packet imaging}}},\ }\href {https://doi.org/10.1063/1.478007} {\bibfield  {journal} {\bibinfo  {journal} {J. Chem. Phys.}\ }\textbf {\bibinfo {volume} {110}},\ \bibinfo {pages} {1317--1320} (\bibinfo {year} {1999}{\natexlab{b}})}\BibitemShut {NoStop}%
\bibitem [{\citenamefont {Kojima}\ \emph {et~al.}(2003)\citenamefont {Kojima}, \citenamefont {Tsumura}, \citenamefont {Takeda},\ and\ \citenamefont {Nishizawa}}]{kojima2003far}%
  \BibitemOpen
  \bibfield  {author} {\bibinfo {author} {\bibfnamefont {S.}~\bibnamefont {Kojima}}, \bibinfo {author} {\bibfnamefont {N.}~\bibnamefont {Tsumura}}, \bibinfo {author} {\bibfnamefont {M.~W.}\ \bibnamefont {Takeda}},\ and\ \bibinfo {author} {\bibfnamefont {S.}~\bibnamefont {Nishizawa}},\ }\bibfield  {title} {\emph {\bibinfo {title} {{Far-infrared phonon-polariton dispersion probed by terahertz time-domain spectroscopy}}},\ }\href {https://doi.org/10.1103/PhysRevB.67.035102} {\bibfield  {journal} {\bibinfo  {journal} {Phys. Rev. B}\ }\textbf {\bibinfo {volume} {67}},\ \bibinfo {pages} {035102} (\bibinfo {year} {2003})}\BibitemShut {NoStop}%
\bibitem [{\citenamefont {{Alfaro-Mozaz}}\ \emph {et~al.}(2019)\citenamefont {{Alfaro-Mozaz}}, \citenamefont {{Rodrigo}}, \citenamefont {{Alonso-Gonz{\'a}lez}}, \citenamefont {{V{\'e}lez}}, \citenamefont {{Dolado}}, \citenamefont {{Casanova}}, \citenamefont {{Hueso}}, \citenamefont {{Mart{\'\i}n-Moreno}}, \citenamefont {{Hillenbrand}},\ and\ \citenamefont {{Nikitin}}}]{alfaro2019deeply}%
  \BibitemOpen
  \bibfield  {author} {\bibinfo {author} {\bibfnamefont {F.~J.}\ \bibnamefont {{Alfaro-Mozaz}}}, \bibinfo {author} {\bibfnamefont {S.~G.}\ \bibnamefont {{Rodrigo}}}, \bibinfo {author} {\bibfnamefont {P.}~\bibnamefont {{Alonso-Gonz{\'a}lez}}}, \bibinfo {author} {\bibfnamefont {S.}~\bibnamefont {{V{\'e}lez}}}, \bibinfo {author} {\bibfnamefont {I.}~\bibnamefont {{Dolado}}}, \bibinfo {author} {\bibfnamefont {F.}~\bibnamefont {{Casanova}}}, \bibinfo {author} {\bibfnamefont {L.~E.}\ \bibnamefont {{Hueso}}}, \bibinfo {author} {\bibfnamefont {L.}~\bibnamefont {{Mart{\'\i}n-Moreno}}}, \bibinfo {author} {\bibfnamefont {R.}~\bibnamefont {{Hillenbrand}}},\ and\ \bibinfo {author} {\bibfnamefont {A.~Y.}\ \bibnamefont {{Nikitin}}},\ }\bibfield  {title} {\emph {\bibinfo {title} {{Deeply subwavelength phonon-polaritonic crystal made of a van der Waals material}}},\ }\href {https://doi.org/10.1038/s41467-018-07795-6} {\bibfield  {journal} {\bibinfo  {journal} {Nat. Commun.}\ }\textbf {\bibinfo {volume} {10}},\ \bibinfo {pages}
  {42} (\bibinfo {year} {2019})}\BibitemShut {NoStop}%
\bibitem [{\citenamefont {{Foteinopoulou}}\ \emph {et~al.}(2019)\citenamefont {{Foteinopoulou}}, \citenamefont {{Devarapu}}, \citenamefont {{Subramania}}, \citenamefont {{Krishna}},\ and\ \citenamefont {{Wasserman}}}]{fotei2019phonon}%
  \BibitemOpen
  \bibfield  {author} {\bibinfo {author} {\bibfnamefont {S.}~\bibnamefont {{Foteinopoulou}}}, \bibinfo {author} {\bibfnamefont {G.~C.~R.}\ \bibnamefont {{Devarapu}}}, \bibinfo {author} {\bibfnamefont {G.~S.}\ \bibnamefont {{Subramania}}}, \bibinfo {author} {\bibfnamefont {S.}~\bibnamefont {{Krishna}}},\ and\ \bibinfo {author} {\bibfnamefont {D.}~\bibnamefont {{Wasserman}}},\ }\bibfield  {title} {\emph {\bibinfo {title} {{Phonon-polaritonics: enabling powerful capabilities for infrared photonics}}},\ }\href {https://doi.org/10.1515/nanoph-2019-0232} {\bibfield  {journal} {\bibinfo  {journal} {Nanophotonics}\ }\textbf {\bibinfo {volume} {8}},\ \bibinfo {pages} {232} (\bibinfo {year} {2019})}\BibitemShut {NoStop}%
\bibitem [{\citenamefont {Stuart}(2004)}]{stuart2004infrared}%
  \BibitemOpen
  \bibfield  {author} {\bibinfo {author} {\bibfnamefont {B.~H.}\ \bibnamefont {Stuart}},\ }\href {https://doi.org/10.1002/0470011149} {\emph {\bibinfo {title} {{Infrared spectroscopy: fundamentals and applications}}}}\ (\bibinfo  {publisher} {John Wiley \& Sons, Ltd},\ \bibinfo {year} {2004})\BibitemShut {NoStop}%
\bibitem [{\citenamefont {Gardiner}(1989)}]{gardiner1989introduction}%
  \BibitemOpen
  \bibfield  {author} {\bibinfo {author} {\bibfnamefont {D.~J.}\ \bibnamefont {Gardiner}},\ }in\ \href {https://doi.org/10.1007/978-3-642-74040-4_1} {\emph {\bibinfo {booktitle} {{Practical Raman Spectroscopy}}}}\ (\bibinfo  {publisher} {Springer},\ \bibinfo {year} {1989})\ p.~\bibinfo {pages} {1}\BibitemShut {NoStop}%
\bibitem [{\citenamefont {Burkel}(2000)}]{burkel2000phonon}%
  \BibitemOpen
  \bibfield  {author} {\bibinfo {author} {\bibfnamefont {E.}~\bibnamefont {Burkel}},\ }\bibfield  {title} {\emph {\bibinfo {title} {{Phonon spectroscopy by inelastic x-ray scattering}}},\ }\href {https://doi.org/10.1088/0034-4885/63/2/203} {\bibfield  {journal} {\bibinfo  {journal} {Rep. Prog. Phys.}\ }\textbf {\bibinfo {volume} {63}},\ \bibinfo {pages} {171} (\bibinfo {year} {2000})}\BibitemShut {NoStop}%
\bibitem [{\citenamefont {{F{\"o}rst}}\ \emph {et~al.}(2011)\citenamefont {{F{\"o}rst}}, \citenamefont {{Manzoni}}, \citenamefont {{Kaiser}}, \citenamefont {{Tomioka}}, \citenamefont {{Tokura}}, \citenamefont {{Merlin}},\ and\ \citenamefont {{Cavalleri}}}]{forst2011nonlinearphononics}%
  \BibitemOpen
  \bibfield  {author} {\bibinfo {author} {\bibfnamefont {M.}~\bibnamefont {{F{\"o}rst}}}, \bibinfo {author} {\bibfnamefont {C.}~\bibnamefont {{Manzoni}}}, \bibinfo {author} {\bibfnamefont {S.}~\bibnamefont {{Kaiser}}}, \bibinfo {author} {\bibfnamefont {Y.}~\bibnamefont {{Tomioka}}}, \bibinfo {author} {\bibfnamefont {Y.}~\bibnamefont {{Tokura}}}, \bibinfo {author} {\bibfnamefont {R.}~\bibnamefont {{Merlin}}},\ and\ \bibinfo {author} {\bibfnamefont {A.}~\bibnamefont {{Cavalleri}}},\ }\bibfield  {title} {\emph {\bibinfo {title} {{Nonlinear phononics as an ultrafast route to lattice control}}},\ }\href {https://doi.org/10.1038/nphys2055} {\bibfield  {journal} {\bibinfo  {journal} {Nat. Phys.}\ }\textbf {\bibinfo {volume} {7}},\ \bibinfo {pages} {854} (\bibinfo {year} {2011})}\BibitemShut {NoStop}%
\bibitem [{\citenamefont {Radaelli}(2018)}]{radaelli2018breaking}%
  \BibitemOpen
  \bibfield  {author} {\bibinfo {author} {\bibfnamefont {P.~G.}\ \bibnamefont {Radaelli}},\ }\bibfield  {title} {\emph {\bibinfo {title} {Breaking symmetry with light: {Ultrafast} ferroelectricity and magnetism from three-phonon coupling}},\ }\href {https://doi.org/10.1103/PhysRevB.97.085145} {\bibfield  {journal} {\bibinfo  {journal} {Phys. Rev. B}\ }\textbf {\bibinfo {volume} {97}},\ \bibinfo {pages} {085145} (\bibinfo {year} {2018})}\BibitemShut {NoStop}%
\bibitem [{\citenamefont {{Hortensius}}\ \emph {et~al.}(2020)\citenamefont {{Hortensius}}, \citenamefont {{Afanasiev}}, \citenamefont {{Sasani}}, \citenamefont {{Bousquet}},\ and\ \citenamefont {{Caviglia}}}]{hortensius2020ultrafast}%
  \BibitemOpen
  \bibfield  {author} {\bibinfo {author} {\bibfnamefont {J.~R.}\ \bibnamefont {{Hortensius}}}, \bibinfo {author} {\bibfnamefont {D.}~\bibnamefont {{Afanasiev}}}, \bibinfo {author} {\bibfnamefont {A.}~\bibnamefont {{Sasani}}}, \bibinfo {author} {\bibfnamefont {E.}~\bibnamefont {{Bousquet}}},\ and\ \bibinfo {author} {\bibfnamefont {A.~D.}\ \bibnamefont {{Caviglia}}},\ }\bibfield  {title} {\emph {\bibinfo {title} {{Ultrafast strain engineering and coherent structural dynamics from resonantly driven optical phonons in LaAlO$_{3}$}}},\ }\href {https://doi.org/10.1038/s41535-020-00297-z} {\bibfield  {journal} {\bibinfo  {journal} {npj Quantum Mater.}\ }\textbf {\bibinfo {volume} {5}},\ \bibinfo {pages} {95} (\bibinfo {year} {2020})}\BibitemShut {NoStop}%
\bibitem [{\citenamefont {{Mankowsky}}\ \emph {et~al.}(2017)\citenamefont {{Mankowsky}}, \citenamefont {{von Hoegen}}, \citenamefont {{F{\"o}rst}},\ and\ \citenamefont {{Cavalleri}}}]{mankowsky2017ultrafast}%
  \BibitemOpen
  \bibfield  {author} {\bibinfo {author} {\bibfnamefont {R.}~\bibnamefont {{Mankowsky}}}, \bibinfo {author} {\bibfnamefont {A.}~\bibnamefont {{von Hoegen}}}, \bibinfo {author} {\bibfnamefont {M.}~\bibnamefont {{F{\"o}rst}}},\ and\ \bibinfo {author} {\bibfnamefont {A.}~\bibnamefont {{Cavalleri}}},\ }\bibfield  {title} {\emph {\bibinfo {title} {{Ultrafast reversal of the ferroelectric polarization}}},\ }\href {https://doi.org/10.1103/PhysRevLett.118.197601} {\bibfield  {journal} {\bibinfo  {journal} {Phys. Rev. Lett.}\ }\textbf {\bibinfo {volume} {118}},\ \bibinfo {pages} {197601} (\bibinfo {year} {2017})}\BibitemShut {NoStop}%
\bibitem [{\citenamefont {{Henstridge}}\ \emph {et~al.}(2022)\citenamefont {{Henstridge}}, \citenamefont {{F{\"o}rst}}, \citenamefont {{Rowe}}, \citenamefont {{Fechner}},\ and\ \citenamefont {{Cavalleri}}}]{henstridge2022nonlocal}%
  \BibitemOpen
  \bibfield  {author} {\bibinfo {author} {\bibfnamefont {M.}~\bibnamefont {{Henstridge}}}, \bibinfo {author} {\bibfnamefont {M.}~\bibnamefont {{F{\"o}rst}}}, \bibinfo {author} {\bibfnamefont {E.}~\bibnamefont {{Rowe}}}, \bibinfo {author} {\bibfnamefont {M.}~\bibnamefont {{Fechner}}},\ and\ \bibinfo {author} {\bibfnamefont {A.}~\bibnamefont {{Cavalleri}}},\ }\bibfield  {title} {\emph {\bibinfo {title} {{Nonlocal nonlinear phononics}}},\ }\href {https://doi.org/10.1038/s41567-022-01512-3} {\bibfield  {journal} {\bibinfo  {journal} {Nat. Phys.}\ }\textbf {\bibinfo {volume} {18}},\ \bibinfo {pages} {457} (\bibinfo {year} {2022})}\BibitemShut {NoStop}%
\bibitem [{\citenamefont {{Li}}\ \emph {et~al.}(2019)\citenamefont {{Li}}, \citenamefont {{Qiu}}, \citenamefont {{Zhang}}, \citenamefont {{Baldini}}, \citenamefont {{Lu}}, \citenamefont {{Rappe}},\ and\ \citenamefont {{Nelson}}}]{li2019terahertz}%
  \BibitemOpen
  \bibfield  {author} {\bibinfo {author} {\bibfnamefont {X.}~\bibnamefont {{Li}}}, \bibinfo {author} {\bibfnamefont {T.}~\bibnamefont {{Qiu}}}, \bibinfo {author} {\bibfnamefont {J.}~\bibnamefont {{Zhang}}}, \bibinfo {author} {\bibfnamefont {E.}~\bibnamefont {{Baldini}}}, \bibinfo {author} {\bibfnamefont {J.}~\bibnamefont {{Lu}}}, \bibinfo {author} {\bibfnamefont {A.~M.}\ \bibnamefont {{Rappe}}},\ and\ \bibinfo {author} {\bibfnamefont {K.~A.}\ \bibnamefont {{Nelson}}},\ }\bibfield  {title} {\emph {\bibinfo {title} {{Terahertz field-induced ferroelectricity in quantum paraelectric SrTiO$_{3}$}}},\ }\href {https://doi.org/10.1126/science.aaw4913} {\bibfield  {journal} {\bibinfo  {journal} {Science}\ }\textbf {\bibinfo {volume} {364}},\ \bibinfo {pages} {1079} (\bibinfo {year} {2019})}\BibitemShut {NoStop}%
\bibitem [{\citenamefont {{Nova}}\ \emph {et~al.}(2019)\citenamefont {{Nova}}, \citenamefont {{Disa}}, \citenamefont {{Fechner}},\ and\ \citenamefont {{Cavalleri}}}]{nova2019metastable}%
  \BibitemOpen
  \bibfield  {author} {\bibinfo {author} {\bibfnamefont {T.~F.}\ \bibnamefont {{Nova}}}, \bibinfo {author} {\bibfnamefont {A.~S.}\ \bibnamefont {{Disa}}}, \bibinfo {author} {\bibfnamefont {M.}~\bibnamefont {{Fechner}}},\ and\ \bibinfo {author} {\bibfnamefont {A.}~\bibnamefont {{Cavalleri}}},\ }\bibfield  {title} {\emph {\bibinfo {title} {{Metastable ferroelectricity in optically strained SrTiO$_3$}}},\ }\href {https://doi.org/10.1126/science.aaw4911} {\bibfield  {journal} {\bibinfo  {journal} {Science}\ }\textbf {\bibinfo {volume} {364}},\ \bibinfo {pages} {1075} (\bibinfo {year} {2019})}\BibitemShut {NoStop}%
\bibitem [{\citenamefont {{Fechner}}\ \emph {et~al.}(2024)\citenamefont {{Fechner}}, \citenamefont {{F{\"o}rst}}, \citenamefont {{Orenstein}}, \citenamefont {{Krapivin}}, \citenamefont {{Disa}}, \citenamefont {{Buzzi}}, \citenamefont {{von Hoegen}}, \citenamefont {{de la Pena}}, \citenamefont {{Nguyen}}, \citenamefont {{Mankowsky}}, \citenamefont {{Sander}}, \citenamefont {{Lemke}}, \citenamefont {{Deng}}, \citenamefont {{Trigo}},\ and\ \citenamefont {{Cavalleri}}}]{fechner2024quenchedlattice}%
  \BibitemOpen
  \bibfield  {author} {\bibinfo {author} {\bibfnamefont {M.}~\bibnamefont {{Fechner}}}, \bibinfo {author} {\bibfnamefont {M.}~\bibnamefont {{F{\"o}rst}}}, \bibinfo {author} {\bibfnamefont {G.}~\bibnamefont {{Orenstein}}}, \bibinfo {author} {\bibfnamefont {V.}~\bibnamefont {{Krapivin}}}, \bibinfo {author} {\bibfnamefont {A.~S.}\ \bibnamefont {{Disa}}}, \bibinfo {author} {\bibfnamefont {M.}~\bibnamefont {{Buzzi}}}, \bibinfo {author} {\bibfnamefont {A.}~\bibnamefont {{von Hoegen}}}, \bibinfo {author} {\bibfnamefont {G.}~\bibnamefont {{de la Pena}}}, \bibinfo {author} {\bibfnamefont {Q.~L.}\ \bibnamefont {{Nguyen}}}, \bibinfo {author} {\bibfnamefont {R.}~\bibnamefont {{Mankowsky}}}, \bibinfo {author} {\bibfnamefont {M.}~\bibnamefont {{Sander}}}, \bibinfo {author} {\bibfnamefont {H.}~\bibnamefont {{Lemke}}}, \bibinfo {author} {\bibfnamefont {Y.}~\bibnamefont {{Deng}}}, \bibinfo {author} {\bibfnamefont {M.}~\bibnamefont {{Trigo}}},\ and\ \bibinfo {author} {\bibfnamefont {A.}~\bibnamefont {{Cavalleri}}},\ }\bibfield
   {title} {\emph {\bibinfo {title} {{Quenched lattice fluctuations in optically driven SrTiO$_{3}$}}},\ }\href {https://doi.org/10.1038/s41563-023-01791-y} {\bibfield  {journal} {\bibinfo  {journal} {Nat. Mater.}\ }\textbf {\bibinfo {volume} {23}},\ \bibinfo {pages} {363} (\bibinfo {year} {2024})}\BibitemShut {NoStop}%
\bibitem [{\citenamefont {{Bustamante Lopez}}\ \emph {et~al.}(2023)\citenamefont {{Bustamante Lopez}}, \citenamefont {{Juraschek}}, \citenamefont {{Fechner}}, \citenamefont {{Xu}}, \citenamefont {{Cheong}},\ and\ \citenamefont {{Hu}}}]{lopez2023ultrafast}%
  \BibitemOpen
  \bibfield  {author} {\bibinfo {author} {\bibfnamefont {D.~A.}\ \bibnamefont {{Bustamante Lopez}}}, \bibinfo {author} {\bibfnamefont {D.~M.}\ \bibnamefont {{Juraschek}}}, \bibinfo {author} {\bibfnamefont {M.}~\bibnamefont {{Fechner}}}, \bibinfo {author} {\bibfnamefont {X.}~\bibnamefont {{Xu}}}, \bibinfo {author} {\bibfnamefont {S.-W.}\ \bibnamefont {{Cheong}}},\ and\ \bibinfo {author} {\bibfnamefont {W.}~\bibnamefont {{Hu}}},\ }\href {https://doi.org/10.48550/arXiv.2305.08250} {\bibinfo {title} {{\textit{Ultrafast simultaneous manipulation of multiple ferroic orders through nonlinear phonon excitation}}}} (\bibinfo {year} {2023}),\ \Eprint {https://arxiv.org/abs/2305.08250} {arXiv:2305.08250} \BibitemShut {NoStop}%
\bibitem [{\citenamefont {{Fausti}}\ \emph {et~al.}(2011)\citenamefont {{Fausti}}, \citenamefont {{Tobey}}, \citenamefont {{Dean}}, \citenamefont {{Kaiser}}, \citenamefont {{Dienst}}, \citenamefont {{Hoffmann}}, \citenamefont {{Pyon}}, \citenamefont {{Takayama}}, \citenamefont {{Takagi}},\ and\ \citenamefont {{Cavalleri}}}]{fausti2011light}%
  \BibitemOpen
  \bibfield  {author} {\bibinfo {author} {\bibfnamefont {D.}~\bibnamefont {{Fausti}}}, \bibinfo {author} {\bibfnamefont {R.~I.}\ \bibnamefont {{Tobey}}}, \bibinfo {author} {\bibfnamefont {N.}~\bibnamefont {{Dean}}}, \bibinfo {author} {\bibfnamefont {S.}~\bibnamefont {{Kaiser}}}, \bibinfo {author} {\bibfnamefont {A.}~\bibnamefont {{Dienst}}}, \bibinfo {author} {\bibfnamefont {M.~C.}\ \bibnamefont {{Hoffmann}}}, \bibinfo {author} {\bibfnamefont {S.}~\bibnamefont {{Pyon}}}, \bibinfo {author} {\bibfnamefont {T.}~\bibnamefont {{Takayama}}}, \bibinfo {author} {\bibfnamefont {H.}~\bibnamefont {{Takagi}}},\ and\ \bibinfo {author} {\bibfnamefont {A.}~\bibnamefont {{Cavalleri}}},\ }\bibfield  {title} {\emph {\bibinfo {title} {{Light-induced superconductivity in a stripe-ordered cuprate}}},\ }\href {https://doi.org/10.1126/science.1197294} {\bibfield  {journal} {\bibinfo  {journal} {Science}\ }\textbf {\bibinfo {volume} {331}},\ \bibinfo {pages} {189} (\bibinfo {year} {2011})}\BibitemShut {NoStop}%
\bibitem [{\citenamefont {Mankowsky}\ \emph {et~al.}(2014)\citenamefont {Mankowsky}, \citenamefont {Subedi}, \citenamefont {Först}, \citenamefont {Mariager}, \citenamefont {Chollet}, \citenamefont {Lemke}, \citenamefont {Robinson}, \citenamefont {Glownia}, \citenamefont {Minitti}, \citenamefont {Frano}, \citenamefont {Fechner}, \citenamefont {Spaldin}, \citenamefont {Loew}, \citenamefont {Keimer}, \citenamefont {Georges},\ and\ \citenamefont {Cavalleri}}]{mankowsky2014nonlinear}%
  \BibitemOpen
  \bibfield  {author} {\bibinfo {author} {\bibfnamefont {R.}~\bibnamefont {Mankowsky}}, \bibinfo {author} {\bibfnamefont {A.}~\bibnamefont {Subedi}}, \bibinfo {author} {\bibfnamefont {M.}~\bibnamefont {Först}}, \bibinfo {author} {\bibfnamefont {S.~O.}\ \bibnamefont {Mariager}}, \bibinfo {author} {\bibfnamefont {M.}~\bibnamefont {Chollet}}, \bibinfo {author} {\bibfnamefont {H.~T.}\ \bibnamefont {Lemke}}, \bibinfo {author} {\bibfnamefont {J.~S.}\ \bibnamefont {Robinson}}, \bibinfo {author} {\bibfnamefont {J.~M.}\ \bibnamefont {Glownia}}, \bibinfo {author} {\bibfnamefont {M.~P.}\ \bibnamefont {Minitti}}, \bibinfo {author} {\bibfnamefont {A.}~\bibnamefont {Frano}}, \bibinfo {author} {\bibfnamefont {M.}~\bibnamefont {Fechner}}, \bibinfo {author} {\bibfnamefont {N.~A.}\ \bibnamefont {Spaldin}}, \bibinfo {author} {\bibfnamefont {T.}~\bibnamefont {Loew}}, \bibinfo {author} {\bibfnamefont {B.}~\bibnamefont {Keimer}}, \bibinfo {author} {\bibfnamefont {A.}~\bibnamefont {Georges}},\ and\ \bibinfo {author} {\bibfnamefont
  {A.}~\bibnamefont {Cavalleri}},\ }\bibfield  {title} {\emph {\bibinfo {title} {Nonlinear lattice dynamics as a basis for enhanced superconductivity in {YBa$_2$Cu$_3$O$_{6.5}$}}},\ }\href {https://doi.org/10.1038/nature13875} {\bibfield  {journal} {\bibinfo  {journal} {Nature}\ }\textbf {\bibinfo {volume} {516}},\ \bibinfo {pages} {71--73} (\bibinfo {year} {2014})}\BibitemShut {NoStop}%
\bibitem [{\citenamefont {Mankowsky}\ \emph {et~al.}(2017)\citenamefont {Mankowsky}, \citenamefont {Fechner}, \citenamefont {F\"orst}, \citenamefont {von Hoegen}, \citenamefont {Porras}, \citenamefont {Loew}, \citenamefont {Dakovski}, \citenamefont {Seaberg}, \citenamefont {Möller}, \citenamefont {Coslovich}, \citenamefont {Keimer}, \citenamefont {Dhesi},\ and\ \citenamefont {Cavalleri}}]{mankowsky2017optically}%
  \BibitemOpen
  \bibfield  {author} {\bibinfo {author} {\bibfnamefont {R.}~\bibnamefont {Mankowsky}}, \bibinfo {author} {\bibfnamefont {M.}~\bibnamefont {Fechner}}, \bibinfo {author} {\bibfnamefont {M.}~\bibnamefont {F\"orst}}, \bibinfo {author} {\bibfnamefont {A.}~\bibnamefont {von Hoegen}}, \bibinfo {author} {\bibfnamefont {J.}~\bibnamefont {Porras}}, \bibinfo {author} {\bibfnamefont {T.}~\bibnamefont {Loew}}, \bibinfo {author} {\bibfnamefont {G.~L.}\ \bibnamefont {Dakovski}}, \bibinfo {author} {\bibfnamefont {M.}~\bibnamefont {Seaberg}}, \bibinfo {author} {\bibfnamefont {S.}~\bibnamefont {Möller}}, \bibinfo {author} {\bibfnamefont {G.}~\bibnamefont {Coslovich}}, \bibinfo {author} {\bibfnamefont {B.}~\bibnamefont {Keimer}}, \bibinfo {author} {\bibfnamefont {S.~S.}\ \bibnamefont {Dhesi}},\ and\ \bibinfo {author} {\bibfnamefont {A.}~\bibnamefont {Cavalleri}},\ }\bibfield  {title} {\emph {\bibinfo {title} {Optically induced lattice deformations, electronic structure changes, and enhanced superconductivity in
  {YBa$_2$Cu$_3$O$_{6.48}$}}},\ }\href {https://doi.org/10.1063/1.4977672} {\bibfield  {journal} {\bibinfo  {journal} {Struct. Dyn.}\ }\textbf {\bibinfo {volume} {4}},\ \bibinfo {pages} {044007} (\bibinfo {year} {2017})}\BibitemShut {NoStop}%
\bibitem [{\citenamefont {Kaiser}(2017)}]{kaiser2017light}%
  \BibitemOpen
  \bibfield  {author} {\bibinfo {author} {\bibfnamefont {S.}~\bibnamefont {Kaiser}},\ }\bibfield  {title} {\emph {\bibinfo {title} {Light-induced superconductivity in high-{$T_c$} cuprates}},\ }\href {https://doi.org/10.1088/1402-4896/aa8201} {\bibfield  {journal} {\bibinfo  {journal} {Phys. Scr.}\ }\textbf {\bibinfo {volume} {92}},\ \bibinfo {pages} {103001} (\bibinfo {year} {2017})}\BibitemShut {NoStop}%
\bibitem [{\citenamefont {Liu}\ \emph {et~al.}(2020)\citenamefont {Liu}, \citenamefont {F\"orst}, \citenamefont {Fechner}, \citenamefont {Nicoletti}, \citenamefont {Porras}, \citenamefont {Loew}, \citenamefont {Keimer},\ and\ \citenamefont {Cavalleri}}]{liu2020pump}%
  \BibitemOpen
  \bibfield  {author} {\bibinfo {author} {\bibfnamefont {B.}~\bibnamefont {Liu}}, \bibinfo {author} {\bibfnamefont {M.}~\bibnamefont {F\"orst}}, \bibinfo {author} {\bibfnamefont {M.}~\bibnamefont {Fechner}}, \bibinfo {author} {\bibfnamefont {D.}~\bibnamefont {Nicoletti}}, \bibinfo {author} {\bibfnamefont {J.}~\bibnamefont {Porras}}, \bibinfo {author} {\bibfnamefont {T.}~\bibnamefont {Loew}}, \bibinfo {author} {\bibfnamefont {B.}~\bibnamefont {Keimer}},\ and\ \bibinfo {author} {\bibfnamefont {A.}~\bibnamefont {Cavalleri}},\ }\bibfield  {title} {\emph {\bibinfo {title} {Pump frequency resonances for light-induced incipient superconductivity in {YBa$_2$Cu$_3$O$_{6.5}$}}},\ }\href {https://doi.org/10.1103/PhysRevX.10.011053} {\bibfield  {journal} {\bibinfo  {journal} {Phys. Rev. X}\ }\textbf {\bibinfo {volume} {10}},\ \bibinfo {pages} {011053} (\bibinfo {year} {2020})}\BibitemShut {NoStop}%
\bibitem [{\citenamefont {{Fava}}\ \emph {et~al.}(2024)\citenamefont {{Fava}}, \citenamefont {{De Vecchi}}, \citenamefont {{Jotzu}}, \citenamefont {{Buzzi}}, \citenamefont {{Gebert}}, \citenamefont {{Liu}}, \citenamefont {{Keimer}},\ and\ \citenamefont {{Cavalleri}}}]{fava2024magnetic}%
  \BibitemOpen
  \bibfield  {author} {\bibinfo {author} {\bibfnamefont {S.}~\bibnamefont {{Fava}}}, \bibinfo {author} {\bibfnamefont {G.}~\bibnamefont {{De Vecchi}}}, \bibinfo {author} {\bibfnamefont {G.}~\bibnamefont {{Jotzu}}}, \bibinfo {author} {\bibfnamefont {M.}~\bibnamefont {{Buzzi}}}, \bibinfo {author} {\bibfnamefont {T.}~\bibnamefont {{Gebert}}}, \bibinfo {author} {\bibfnamefont {Y.}~\bibnamefont {{Liu}}}, \bibinfo {author} {\bibfnamefont {B.}~\bibnamefont {{Keimer}}},\ and\ \bibinfo {author} {\bibfnamefont {A.}~\bibnamefont {{Cavalleri}}},\ }\bibfield  {title} {\emph {\bibinfo {title} {{Magnetic field expulsion in optically driven YBa$_{2}$Cu$_{3}$O$_{6.48}$}}},\ }\href {https://doi.org/10.1038/s41586-024-07635-2} {\bibfield  {journal} {\bibinfo  {journal} {Nature}\ }\textbf {\bibinfo {volume} {632}},\ \bibinfo {pages} {75} (\bibinfo {year} {2024})}\BibitemShut {NoStop}%
\bibitem [{\citenamefont {{Zhang}}\ \emph {et~al.}(2020)\citenamefont {{Zhang}}, \citenamefont {{Wang}}, \citenamefont {{Xiang}}, \citenamefont {{Yao}}, \citenamefont {{Liu}}, \citenamefont {{Shi}}, \citenamefont {{Lin}}, \citenamefont {{Dong}}, \citenamefont {{Wu}},\ and\ \citenamefont {{Wang}}}]{zhang2020photoinduced}%
  \BibitemOpen
  \bibfield  {author} {\bibinfo {author} {\bibfnamefont {S.~J.}\ \bibnamefont {{Zhang}}}, \bibinfo {author} {\bibfnamefont {Z.~X.}\ \bibnamefont {{Wang}}}, \bibinfo {author} {\bibfnamefont {H.}~\bibnamefont {{Xiang}}}, \bibinfo {author} {\bibfnamefont {X.}~\bibnamefont {{Yao}}}, \bibinfo {author} {\bibfnamefont {Q.~M.}\ \bibnamefont {{Liu}}}, \bibinfo {author} {\bibfnamefont {L.~Y.}\ \bibnamefont {{Shi}}}, \bibinfo {author} {\bibfnamefont {T.}~\bibnamefont {{Lin}}}, \bibinfo {author} {\bibfnamefont {T.}~\bibnamefont {{Dong}}}, \bibinfo {author} {\bibfnamefont {D.}~\bibnamefont {{Wu}}},\ and\ \bibinfo {author} {\bibfnamefont {N.~L.}\ \bibnamefont {{Wang}}},\ }\bibfield  {title} {\emph {\bibinfo {title} {{Photoinduced nonequilibrium response in underdoped YBa$_{2}$Cu$_{3}$O$_{6 +x}$ probed by time-resolved terahertz spectroscopy}}},\ }\href {https://doi.org/10.1103/PhysRevX.10.011056} {\bibfield  {journal} {\bibinfo  {journal} {Phys. Rev. X}\ }\textbf {\bibinfo {volume} {10}},\ \bibinfo {pages} {011056}
  (\bibinfo {year} {2020})}\BibitemShut {NoStop}%
\bibitem [{\citenamefont {{Zhang}}\ \emph {et~al.}(2024{\natexlab{a}})\citenamefont {{Zhang}}, \citenamefont {{Zhou}}, \citenamefont {{Xu}}, \citenamefont {{Wu}}, \citenamefont {{Yue}}, \citenamefont {{Liu}}, \citenamefont {{Hu}}, \citenamefont {{Li}}, \citenamefont {{Yuan}}, \citenamefont {{Homes}}, \citenamefont {{Gu}}, \citenamefont {{Dong}},\ and\ \citenamefont {{Wang}}}]{zhang2024light}%
  \BibitemOpen
  \bibfield  {author} {\bibinfo {author} {\bibfnamefont {S.~J.}\ \bibnamefont {{Zhang}}}, \bibinfo {author} {\bibfnamefont {X.~Y.}\ \bibnamefont {{Zhou}}}, \bibinfo {author} {\bibfnamefont {S.~X.}\ \bibnamefont {{Xu}}}, \bibinfo {author} {\bibfnamefont {Q.}~\bibnamefont {{Wu}}}, \bibinfo {author} {\bibfnamefont {L.}~\bibnamefont {{Yue}}}, \bibinfo {author} {\bibfnamefont {Q.~M.}\ \bibnamefont {{Liu}}}, \bibinfo {author} {\bibfnamefont {T.~C.}\ \bibnamefont {{Hu}}}, \bibinfo {author} {\bibfnamefont {R.~S.}\ \bibnamefont {{Li}}}, \bibinfo {author} {\bibfnamefont {J.~Y.}\ \bibnamefont {{Yuan}}}, \bibinfo {author} {\bibfnamefont {C.~C.}\ \bibnamefont {{Homes}}}, \bibinfo {author} {\bibfnamefont {G.~D.}\ \bibnamefont {{Gu}}}, \bibinfo {author} {\bibfnamefont {T.}~\bibnamefont {{Dong}}},\ and\ \bibinfo {author} {\bibfnamefont {N.~L.}\ \bibnamefont {{Wang}}},\ }\bibfield  {title} {\emph {\bibinfo {title} {{Light-induced melting of competing stripe orders without introducing superconductivity in La$_{2
  -x}$Ba$_{x}$CuO$_{4}$}}},\ }\href {https://doi.org/10.1103/PhysRevX.14.011036} {\bibfield  {journal} {\bibinfo  {journal} {Phys. Rev. X}\ }\textbf {\bibinfo {volume} {14}},\ \bibinfo {pages} {011036} (\bibinfo {year} {2024}{\natexlab{a}})}\BibitemShut {NoStop}%
\bibitem [{\citenamefont {{Dodge}}\ \emph {et~al.}(2023{\natexlab{a}})\citenamefont {{Dodge}}, \citenamefont {{Lopez}},\ and\ \citenamefont {{Sahota}}}]{dodge2023optical}%
  \BibitemOpen
  \bibfield  {author} {\bibinfo {author} {\bibfnamefont {J.~S.}\ \bibnamefont {{Dodge}}}, \bibinfo {author} {\bibfnamefont {L.}~\bibnamefont {{Lopez}}},\ and\ \bibinfo {author} {\bibfnamefont {D.~G.}\ \bibnamefont {{Sahota}}},\ }\bibfield  {title} {\emph {\bibinfo {title} {{Optical saturation produces spurious evidence for photoinduced superconductivity in K$_{3}$C$_{60}$}}},\ }\href {https://doi.org/10.1103/PhysRevLett.130.146002} {\bibfield  {journal} {\bibinfo  {journal} {Phys. Rev. Lett.}\ }\textbf {\bibinfo {volume} {130}},\ \bibinfo {pages} {146002} (\bibinfo {year} {2023}{\natexlab{a}})}\BibitemShut {NoStop}%
\bibitem [{\citenamefont {{Dodge}}\ \emph {et~al.}(2023{\natexlab{b}})\citenamefont {{Dodge}}, \citenamefont {{Lopez}},\ and\ \citenamefont {{Sahota}}}]{dodge2023status}%
  \BibitemOpen
  \bibfield  {author} {\bibinfo {author} {\bibfnamefont {J.~S.}\ \bibnamefont {{Dodge}}}, \bibinfo {author} {\bibfnamefont {L.}~\bibnamefont {{Lopez}}},\ and\ \bibinfo {author} {\bibfnamefont {D.~G.}\ \bibnamefont {{Sahota}}},\ }in\ \href {https://doi.org/10.1109/IRMMW-THz57677.2023.10299137} {\emph {\bibinfo {booktitle} {{48th International Conference on Infrared, Millimeter, and Terahertz Waves (IRMMW-THz)}}}}\ (\bibinfo {organization} {IEEE},\ \bibinfo {year} {2023})\ p.~\bibinfo {pages} {1}\BibitemShut {NoStop}%
\bibitem [{\citenamefont {{Buzzi}}\ \emph {et~al.}(2023)\citenamefont {{Buzzi}}, \citenamefont {{Nicoletti}}, \citenamefont {{Rowe}}, \citenamefont {{Wang}},\ and\ \citenamefont {{Cavalleri}}}]{buzzi2023comment}%
  \BibitemOpen
  \bibfield  {author} {\bibinfo {author} {\bibfnamefont {M.}~\bibnamefont {{Buzzi}}}, \bibinfo {author} {\bibfnamefont {D.}~\bibnamefont {{Nicoletti}}}, \bibinfo {author} {\bibfnamefont {E.}~\bibnamefont {{Rowe}}}, \bibinfo {author} {\bibfnamefont {E.}~\bibnamefont {{Wang}}},\ and\ \bibinfo {author} {\bibfnamefont {A.}~\bibnamefont {{Cavalleri}}},\ }\href {https://doi.org/10.48550/arXiv.2303.10169} {\bibinfo {title} {{\textit{Comment on arXiv:2210.01114: Optical Saturation Produces Spurious Evidence for Photoinduced Superconductivity in K$_3$C$_{60}$}}}} (\bibinfo {year} {2023}),\ \Eprint {https://arxiv.org/abs/2303.10169} {arXiv:2303.10169} \BibitemShut {NoStop}%
\bibitem [{\citenamefont {{Kozina}}\ \emph {et~al.}(2019)\citenamefont {{Kozina}}, \citenamefont {{Fechner}}, \citenamefont {{Marsik}}, \citenamefont {{van Driel}}, \citenamefont {{Glownia}}, \citenamefont {{Bernhard}}, \citenamefont {{Radovic}}, \citenamefont {{Zhu}}, \citenamefont {{Bonetti}}, \citenamefont {{Staub}},\ and\ \citenamefont {{Hoffmann}}}]{kozina2019terahertz}%
  \BibitemOpen
  \bibfield  {author} {\bibinfo {author} {\bibfnamefont {M.}~\bibnamefont {{Kozina}}}, \bibinfo {author} {\bibfnamefont {M.}~\bibnamefont {{Fechner}}}, \bibinfo {author} {\bibfnamefont {P.}~\bibnamefont {{Marsik}}}, \bibinfo {author} {\bibfnamefont {T.}~\bibnamefont {{van Driel}}}, \bibinfo {author} {\bibfnamefont {J.~M.}\ \bibnamefont {{Glownia}}}, \bibinfo {author} {\bibfnamefont {C.}~\bibnamefont {{Bernhard}}}, \bibinfo {author} {\bibfnamefont {M.}~\bibnamefont {{Radovic}}}, \bibinfo {author} {\bibfnamefont {D.}~\bibnamefont {{Zhu}}}, \bibinfo {author} {\bibfnamefont {S.}~\bibnamefont {{Bonetti}}}, \bibinfo {author} {\bibfnamefont {U.}~\bibnamefont {{Staub}}},\ and\ \bibinfo {author} {\bibfnamefont {M.~C.}\ \bibnamefont {{Hoffmann}}},\ }\bibfield  {title} {\emph {\bibinfo {title} {{Terahertz-driven phonon upconversion in SrTiO$_{3}$}}},\ }\href {https://doi.org/10.1038/s41567-018-0408-1} {\bibfield  {journal} {\bibinfo  {journal} {Nat. Phys.}\ }\textbf {\bibinfo {volume} {15}},\ \bibinfo {pages} {387}
  (\bibinfo {year} {2019})}\BibitemShut {NoStop}%
\bibitem [{\citenamefont {{Blank}}\ \emph {et~al.}(2023)\citenamefont {{Blank}}, \citenamefont {{Grishunin}}, \citenamefont {{Zvezdin}}, \citenamefont {{Hai}}, \citenamefont {{Wu}}, \citenamefont {{Su}}, \citenamefont {{Huang}}, \citenamefont {{Zvezdin}},\ and\ \citenamefont {{Kimel}}}]{blank2023twodimensionaltera}%
  \BibitemOpen
  \bibfield  {author} {\bibinfo {author} {\bibfnamefont {T.~G.~H.}\ \bibnamefont {{Blank}}}, \bibinfo {author} {\bibfnamefont {K.~A.}\ \bibnamefont {{Grishunin}}}, \bibinfo {author} {\bibfnamefont {K.~A.}\ \bibnamefont {{Zvezdin}}}, \bibinfo {author} {\bibfnamefont {N.~T.}\ \bibnamefont {{Hai}}}, \bibinfo {author} {\bibfnamefont {J.~C.}\ \bibnamefont {{Wu}}}, \bibinfo {author} {\bibfnamefont {S.~H.}\ \bibnamefont {{Su}}}, \bibinfo {author} {\bibfnamefont {J.~C.~A.}\ \bibnamefont {{Huang}}}, \bibinfo {author} {\bibfnamefont {A.~K.}\ \bibnamefont {{Zvezdin}}},\ and\ \bibinfo {author} {\bibfnamefont {A.~V.}\ \bibnamefont {{Kimel}}},\ }\bibfield  {title} {\emph {\bibinfo {title} {{Two-dimensional terahertz spectroscopy of nonlinear phononics in the topological Insulator MnBi$_{2}$Te$_{4}$}}},\ }\href {https://doi.org/10.1103/PhysRevLett.131.026902} {\bibfield  {journal} {\bibinfo  {journal} {Phys. Rev. Lett.}\ }\textbf {\bibinfo {volume} {131}},\ \bibinfo {pages} {026902} (\bibinfo {year} {2023})}\BibitemShut
  {NoStop}%
\bibitem [{\citenamefont {{Cheng}}\ \emph {et~al.}(2023)\citenamefont {{Cheng}}, \citenamefont {{Kramer}}, \citenamefont {{Shen}},\ and\ \citenamefont {{Hoffmann}}}]{cheng2023terahertz}%
  \BibitemOpen
  \bibfield  {author} {\bibinfo {author} {\bibfnamefont {B.}~\bibnamefont {{Cheng}}}, \bibinfo {author} {\bibfnamefont {P.~L.}\ \bibnamefont {{Kramer}}}, \bibinfo {author} {\bibfnamefont {Z.-X.}\ \bibnamefont {{Shen}}},\ and\ \bibinfo {author} {\bibfnamefont {M.~C.}\ \bibnamefont {{Hoffmann}}},\ }\bibfield  {title} {\emph {\bibinfo {title} {{Terahertz-driven local dipolar correlation in a quantum paraelectric}}},\ }\href {https://doi.org/10.1103/PhysRevLett.130.126902} {\bibfield  {journal} {\bibinfo  {journal} {Phys. Rev. Lett.}\ }\textbf {\bibinfo {volume} {130}},\ \bibinfo {pages} {126902} (\bibinfo {year} {2023})}\BibitemShut {NoStop}%
\bibitem [{\citenamefont {{Cui}}\ \emph {et~al.}(2023)\citenamefont {{Cui}}, \citenamefont {{Bostr{\"o}m}}, \citenamefont {{Ozerov}}, \citenamefont {{Wu}}, \citenamefont {{Jiang}}, \citenamefont {{Chu}}, \citenamefont {{Li}}, \citenamefont {{Liu}}, \citenamefont {{Xu}}, \citenamefont {{Rubio}},\ and\ \citenamefont {{Zhang}}}]{cui2023chirality}%
  \BibitemOpen
  \bibfield  {author} {\bibinfo {author} {\bibfnamefont {J.}~\bibnamefont {{Cui}}}, \bibinfo {author} {\bibfnamefont {E.~V.}\ \bibnamefont {{Bostr{\"o}m}}}, \bibinfo {author} {\bibfnamefont {M.}~\bibnamefont {{Ozerov}}}, \bibinfo {author} {\bibfnamefont {F.}~\bibnamefont {{Wu}}}, \bibinfo {author} {\bibfnamefont {Q.}~\bibnamefont {{Jiang}}}, \bibinfo {author} {\bibfnamefont {J.-H.}\ \bibnamefont {{Chu}}}, \bibinfo {author} {\bibfnamefont {C.}~\bibnamefont {{Li}}}, \bibinfo {author} {\bibfnamefont {F.}~\bibnamefont {{Liu}}}, \bibinfo {author} {\bibfnamefont {X.}~\bibnamefont {{Xu}}}, \bibinfo {author} {\bibfnamefont {A.}~\bibnamefont {{Rubio}}},\ and\ \bibinfo {author} {\bibfnamefont {Q.}~\bibnamefont {{Zhang}}},\ }\bibfield  {title} {\emph {\bibinfo {title} {{Chirality selective magnon-phonon hybridization and magnon-induced chiral phonons in a layered zigzag antiferromagnet}}},\ }\href {https://doi.org/10.1038/s41467-023-39123-y} {\bibfield  {journal} {\bibinfo  {journal} {Nat. Commun.}\ }\textbf {\bibinfo
  {volume} {14}},\ \bibinfo {pages} {3396} (\bibinfo {year} {2023})}\BibitemShut {NoStop}%
\bibitem [{\citenamefont {{Mertens}}\ \emph {et~al.}(2023)\citenamefont {{Mertens}}, \citenamefont {{M{\"o}nkeb{\"u}scher}}, \citenamefont {{Parlak}}, \citenamefont {{Boix-Constant}}, \citenamefont {{Ma{\~n}as-Valero}}, \citenamefont {{Matzer}}, \citenamefont {{Adhikari}}, \citenamefont {{Bonanni}}, \citenamefont {{Coronado}}, \citenamefont {{Kalashnikova}}, \citenamefont {{Bossini}},\ and\ \citenamefont {{Cinchetti}}}]{mertens2023ultrafastcoherent}%
  \BibitemOpen
  \bibfield  {author} {\bibinfo {author} {\bibfnamefont {F.}~\bibnamefont {{Mertens}}}, \bibinfo {author} {\bibfnamefont {D.}~\bibnamefont {{M{\"o}nkeb{\"u}scher}}}, \bibinfo {author} {\bibfnamefont {U.}~\bibnamefont {{Parlak}}}, \bibinfo {author} {\bibfnamefont {C.}~\bibnamefont {{Boix-Constant}}}, \bibinfo {author} {\bibfnamefont {S.}~\bibnamefont {{Ma{\~n}as-Valero}}}, \bibinfo {author} {\bibfnamefont {M.}~\bibnamefont {{Matzer}}}, \bibinfo {author} {\bibfnamefont {R.}~\bibnamefont {{Adhikari}}}, \bibinfo {author} {\bibfnamefont {A.}~\bibnamefont {{Bonanni}}}, \bibinfo {author} {\bibfnamefont {E.}~\bibnamefont {{Coronado}}}, \bibinfo {author} {\bibfnamefont {A.~M.}\ \bibnamefont {{Kalashnikova}}}, \bibinfo {author} {\bibfnamefont {D.}~\bibnamefont {{Bossini}}},\ and\ \bibinfo {author} {\bibfnamefont {M.}~\bibnamefont {{Cinchetti}}},\ }\bibfield  {title} {\emph {\bibinfo {title} {{Ultrafast coherent THz lattice dynamics coupled to spins in the van der Waals antiferromagnet FePS$_3$}}},\ }\href
  {https://doi.org/10.1002/adma.202208355} {\bibfield  {journal} {\bibinfo  {journal} {Adv. Mater.}\ }\textbf {\bibinfo {volume} {35}},\ \bibinfo {pages} {2208355} (\bibinfo {year} {2023})}\BibitemShut {NoStop}%
\bibitem [{\citenamefont {{Nova}}\ \emph {et~al.}(2017)\citenamefont {{Nova}}, \citenamefont {{Cartella}}, \citenamefont {{Cantaluppi}}, \citenamefont {{F{\"o}rst}}, \citenamefont {{Bossini}}, \citenamefont {{Mikhaylovskiy}}, \citenamefont {{Kimel}}, \citenamefont {{Merlin}},\ and\ \citenamefont {{Cavalleri}}}]{nova2017aneffective}%
  \BibitemOpen
  \bibfield  {author} {\bibinfo {author} {\bibfnamefont {T.~F.}\ \bibnamefont {{Nova}}}, \bibinfo {author} {\bibfnamefont {A.}~\bibnamefont {{Cartella}}}, \bibinfo {author} {\bibfnamefont {A.}~\bibnamefont {{Cantaluppi}}}, \bibinfo {author} {\bibfnamefont {M.}~\bibnamefont {{F{\"o}rst}}}, \bibinfo {author} {\bibfnamefont {D.}~\bibnamefont {{Bossini}}}, \bibinfo {author} {\bibfnamefont {R.~V.}\ \bibnamefont {{Mikhaylovskiy}}}, \bibinfo {author} {\bibfnamefont {A.~V.}\ \bibnamefont {{Kimel}}}, \bibinfo {author} {\bibfnamefont {R.}~\bibnamefont {{Merlin}}},\ and\ \bibinfo {author} {\bibfnamefont {A.}~\bibnamefont {{Cavalleri}}},\ }\bibfield  {title} {\emph {\bibinfo {title} {{An effective magnetic field from optically driven phonons}}},\ }\href {https://doi.org/10.1038/nphys3925} {\bibfield  {journal} {\bibinfo  {journal} {Nat. Phys.}\ }\textbf {\bibinfo {volume} {13}},\ \bibinfo {pages} {132} (\bibinfo {year} {2017})}\BibitemShut {NoStop}%
\bibitem [{\citenamefont {{Afanasiev}}\ \emph {et~al.}(2021)\citenamefont {{Afanasiev}}, \citenamefont {{Hortensius}}, \citenamefont {{Ivanov}}, \citenamefont {{Sasani}}, \citenamefont {{Bousquet}}, \citenamefont {{Blanter}}, \citenamefont {{Mikhaylovskiy}}, \citenamefont {{Kimel}},\ and\ \citenamefont {{Caviglia}}}]{afanasiev2021ultrafastcontrol}%
  \BibitemOpen
  \bibfield  {author} {\bibinfo {author} {\bibfnamefont {D.}~\bibnamefont {{Afanasiev}}}, \bibinfo {author} {\bibfnamefont {J.~R.}\ \bibnamefont {{Hortensius}}}, \bibinfo {author} {\bibfnamefont {B.~A.}\ \bibnamefont {{Ivanov}}}, \bibinfo {author} {\bibfnamefont {A.}~\bibnamefont {{Sasani}}}, \bibinfo {author} {\bibfnamefont {E.}~\bibnamefont {{Bousquet}}}, \bibinfo {author} {\bibfnamefont {Y.~M.}\ \bibnamefont {{Blanter}}}, \bibinfo {author} {\bibfnamefont {R.~V.}\ \bibnamefont {{Mikhaylovskiy}}}, \bibinfo {author} {\bibfnamefont {A.~V.}\ \bibnamefont {{Kimel}}},\ and\ \bibinfo {author} {\bibfnamefont {A.~D.}\ \bibnamefont {{Caviglia}}},\ }\bibfield  {title} {\emph {\bibinfo {title} {{Ultrafast control of magnetic interactions via light-driven phonons}}},\ }\href {https://doi.org/10.1038/s41563-021-00922-7} {\bibfield  {journal} {\bibinfo  {journal} {Nat. Mater.}\ }\textbf {\bibinfo {volume} {20}},\ \bibinfo {pages} {607} (\bibinfo {year} {2021})}\BibitemShut {NoStop}%
\bibitem [{\citenamefont {{Frenkel'}}(1979)}]{frenkel1979history}%
  \BibitemOpen
  \bibfield  {author} {\bibinfo {author} {\bibfnamefont {V.~Y.}\ \bibnamefont {{Frenkel'}}},\ }\bibfield  {title} {\emph {\bibinfo {title} {{On the history of the Einstein-de Haas effect}}},\ }\href {https://doi.org/10.1070/PU1979v022n07ABEH005587} {\bibfield  {journal} {\bibinfo  {journal} {Soviet Physics Uspekhi}\ }\textbf {\bibinfo {volume} {22}},\ \bibinfo {pages} {580} (\bibinfo {year} {1979})}\BibitemShut {NoStop}%
\bibitem [{\citenamefont {{Barnett}}(1915)}]{barnett1915magnetization}%
  \BibitemOpen
  \bibfield  {author} {\bibinfo {author} {\bibfnamefont {S.~J.}\ \bibnamefont {{Barnett}}},\ }\bibfield  {title} {\emph {\bibinfo {title} {{Magnetization by rotation}}},\ }\href {https://doi.org/10.1103/PhysRev.6.239} {\bibfield  {journal} {\bibinfo  {journal} {Phys. Rev.}\ }\textbf {\bibinfo {volume} {6}},\ \bibinfo {pages} {239} (\bibinfo {year} {1915})}\BibitemShut {NoStop}%
\bibitem [{\citenamefont {Barnett}(1935)}]{barnett1935gyromagnetic}%
  \BibitemOpen
  \bibfield  {author} {\bibinfo {author} {\bibfnamefont {S.~J.}\ \bibnamefont {Barnett}},\ }\bibfield  {title} {\emph {\bibinfo {title} {{Gyromagnetic and electron-inertia effects}}},\ }\href {https://doi.org/10.1103/RevModPhys.7.129} {\bibfield  {journal} {\bibinfo  {journal} {Rev. Mod. Phys}\ }\textbf {\bibinfo {volume} {7}},\ \bibinfo {pages} {129} (\bibinfo {year} {1935})}\BibitemShut {NoStop}%
\bibitem [{\citenamefont {{Lee}}(1955)}]{lee1955magnetostriction}%
  \BibitemOpen
  \bibfield  {author} {\bibinfo {author} {\bibfnamefont {E.~W.}\ \bibnamefont {{Lee}}},\ }\bibfield  {title} {\emph {\bibinfo {title} {{Magnetostriction and magnetomechanical effects}}},\ }\href {https://doi.org/10.1088/0034-4885/18/1/305} {\bibfield  {journal} {\bibinfo  {journal} {Rep. Prog. Phys.}\ }\textbf {\bibinfo {volume} {18}},\ \bibinfo {pages} {184} (\bibinfo {year} {1955})}\BibitemShut {NoStop}%
\bibitem [{\citenamefont {{Dzialoshinskii}}(1958)}]{dzialoshinskii1958problem}%
  \BibitemOpen
  \bibfield  {author} {\bibinfo {author} {\bibfnamefont {I.~E.}\ \bibnamefont {{Dzialoshinskii}}},\ }\bibfield  {title} {\emph {\bibinfo {title} {{The problem of piezomagnetism}}},\ }\href@noop {} {\bibfield  {journal} {\bibinfo  {journal} {Soviet Journal of Experimental and Theoretical Physics}\ }\textbf {\bibinfo {volume} {6}},\ \bibinfo {pages} {621} (\bibinfo {year} {1958})}\BibitemShut {NoStop}%
\bibitem [{\citenamefont {{Borovik-romanov}}(1994)}]{borovik1994piezomagnetism}%
  \BibitemOpen
  \bibfield  {author} {\bibinfo {author} {\bibfnamefont {A.~S.}\ \bibnamefont {{Borovik-romanov}}},\ }\bibfield  {title} {\emph {\bibinfo {title} {{Piezomagnetism, linear magnetostriction and magnetooptic effect}}},\ }\href {https://doi.org/10.1080/00150199408245101} {\bibfield  {journal} {\bibinfo  {journal} {Ferroelectrics}\ }\textbf {\bibinfo {volume} {162}},\ \bibinfo {pages} {153} (\bibinfo {year} {1994})}\BibitemShut {NoStop}%
\bibitem [{\citenamefont {{Broholm}}\ \emph {et~al.}(2020)\citenamefont {{Broholm}}, \citenamefont {{Cava}}, \citenamefont {{Kivelson}}, \citenamefont {{Nocera}}, \citenamefont {{Norman}},\ and\ \citenamefont {{Senthil}}}]{broholm2020quantum}%
  \BibitemOpen
  \bibfield  {author} {\bibinfo {author} {\bibfnamefont {C.}~\bibnamefont {{Broholm}}}, \bibinfo {author} {\bibfnamefont {R.~J.}\ \bibnamefont {{Cava}}}, \bibinfo {author} {\bibfnamefont {S.~A.}\ \bibnamefont {{Kivelson}}}, \bibinfo {author} {\bibfnamefont {D.~G.}\ \bibnamefont {{Nocera}}}, \bibinfo {author} {\bibfnamefont {M.~R.}\ \bibnamefont {{Norman}}},\ and\ \bibinfo {author} {\bibfnamefont {T.}~\bibnamefont {{Senthil}}},\ }\bibfield  {title} {\emph {\bibinfo {title} {{Quantum spin liquids}}},\ }\href {https://doi.org/10.1126/science.aay0668} {\bibfield  {journal} {\bibinfo  {journal} {Science}\ }\textbf {\bibinfo {volume} {367}},\ \bibinfo {pages} {eaay0668} (\bibinfo {year} {2020})}\BibitemShut {NoStop}%
\bibitem [{\citenamefont {Hu}\ \emph {et~al.}(2023)\citenamefont {Hu}, \citenamefont {Du}, \citenamefont {Chen}, \citenamefont {Zhai}, \citenamefont {Wang},\ and\ \citenamefont {Xiong}}]{hu2023spin}%
  \BibitemOpen
  \bibfield  {author} {\bibinfo {author} {\bibfnamefont {L.}~\bibnamefont {Hu}}, \bibinfo {author} {\bibfnamefont {K.-z.}\ \bibnamefont {Du}}, \bibinfo {author} {\bibfnamefont {Y.}~\bibnamefont {Chen}}, \bibinfo {author} {\bibfnamefont {Y.}~\bibnamefont {Zhai}}, \bibinfo {author} {\bibfnamefont {X.}~\bibnamefont {Wang}},\ and\ \bibinfo {author} {\bibfnamefont {Q.}~\bibnamefont {Xiong}},\ }\bibfield  {title} {\emph {\bibinfo {title} {{Spin-phonon coupling in two-dimensional magnetic materials}}},\ }\href {https://doi.org/10.1360/nso/20230002} {\bibfield  {journal} {\bibinfo  {journal} {Natl. Sci. Open}\ }\textbf {\bibinfo {volume} {2}},\ \bibinfo {pages} {20230002} (\bibinfo {year} {2023})}\BibitemShut {NoStop}%
\bibitem [{\citenamefont {Beaurepaire}\ \emph {et~al.}(1996)\citenamefont {Beaurepaire}, \citenamefont {Merle}, \citenamefont {Daunois},\ and\ \citenamefont {Bigot}}]{beaurepaire1996ultrafast}%
  \BibitemOpen
  \bibfield  {author} {\bibinfo {author} {\bibfnamefont {E.}~\bibnamefont {Beaurepaire}}, \bibinfo {author} {\bibfnamefont {J.-C.}\ \bibnamefont {Merle}}, \bibinfo {author} {\bibfnamefont {A.}~\bibnamefont {Daunois}},\ and\ \bibinfo {author} {\bibfnamefont {J.-Y.}\ \bibnamefont {Bigot}},\ }\bibfield  {title} {\emph {\bibinfo {title} {{Ultrafast spin dynamics in ferromagnetic nickel}}},\ }\href {https://doi.org/10.1103/PhysRevLett.76.4250} {\bibfield  {journal} {\bibinfo  {journal} {Phys. Rev. Lett.}\ }\textbf {\bibinfo {volume} {76}},\ \bibinfo {pages} {4250--4253} (\bibinfo {year} {1996})}\BibitemShut {NoStop}%
\bibitem [{\citenamefont {{Dornes}}\ \emph {et~al.}(2019)\citenamefont {{Dornes}}, \citenamefont {{Acremann}}, \citenamefont {{Savoini}}, \citenamefont {{Kubli}}, \citenamefont {{Neugebauer}}, \citenamefont {{Abreu}}, \citenamefont {{Huber}}, \citenamefont {{Lantz}}, \citenamefont {{Vaz}}, \citenamefont {{Lemke}}, \citenamefont {{Bothschafter}}, \citenamefont {{Porer}}, \citenamefont {{Esposito}}, \citenamefont {{Rettig}}, \citenamefont {{Buzzi}}, \citenamefont {{Alberca}}, \citenamefont {{Windsor}}, \citenamefont {{Beaud}}, \citenamefont {{Staub}}, \citenamefont {{Zhu}}, \citenamefont {{Song}}, \citenamefont {{Glownia}},\ and\ \citenamefont {{Johnson}}}]{dornes2019ultrafast}%
  \BibitemOpen
  \bibfield  {author} {\bibinfo {author} {\bibfnamefont {C.}~\bibnamefont {{Dornes}}}, \bibinfo {author} {\bibfnamefont {Y.}~\bibnamefont {{Acremann}}}, \bibinfo {author} {\bibfnamefont {M.}~\bibnamefont {{Savoini}}}, \bibinfo {author} {\bibfnamefont {M.}~\bibnamefont {{Kubli}}}, \bibinfo {author} {\bibfnamefont {M.~J.}\ \bibnamefont {{Neugebauer}}}, \bibinfo {author} {\bibfnamefont {E.}~\bibnamefont {{Abreu}}}, \bibinfo {author} {\bibfnamefont {L.}~\bibnamefont {{Huber}}}, \bibinfo {author} {\bibfnamefont {G.}~\bibnamefont {{Lantz}}}, \bibinfo {author} {\bibfnamefont {C.~A.~F.}\ \bibnamefont {{Vaz}}}, \bibinfo {author} {\bibfnamefont {H.}~\bibnamefont {{Lemke}}}, \bibinfo {author} {\bibfnamefont {E.~M.}\ \bibnamefont {{Bothschafter}}}, \bibinfo {author} {\bibfnamefont {M.}~\bibnamefont {{Porer}}}, \bibinfo {author} {\bibfnamefont {V.}~\bibnamefont {{Esposito}}}, \bibinfo {author} {\bibfnamefont {L.}~\bibnamefont {{Rettig}}}, \bibinfo {author} {\bibfnamefont {M.}~\bibnamefont {{Buzzi}}}, \bibinfo {author}
  {\bibfnamefont {A.}~\bibnamefont {{Alberca}}}, \bibinfo {author} {\bibfnamefont {Y.~W.}\ \bibnamefont {{Windsor}}}, \bibinfo {author} {\bibfnamefont {P.}~\bibnamefont {{Beaud}}}, \bibinfo {author} {\bibfnamefont {U.}~\bibnamefont {{Staub}}}, \bibinfo {author} {\bibfnamefont {D.}~\bibnamefont {{Zhu}}}, \bibinfo {author} {\bibfnamefont {S.}~\bibnamefont {{Song}}}, \bibinfo {author} {\bibfnamefont {J.~M.}\ \bibnamefont {{Glownia}}},\ and\ \bibinfo {author} {\bibfnamefont {S.~L.}\ \bibnamefont {{Johnson}}},\ }\bibfield  {title} {\emph {\bibinfo {title} {{The ultrafast Einstein-de Haas effect}}},\ }\href {https://doi.org/10.1038/s41586-018-0822-7} {\bibfield  {journal} {\bibinfo  {journal} {Nature}\ }\textbf {\bibinfo {volume} {565}},\ \bibinfo {pages} {209} (\bibinfo {year} {2019})}\BibitemShut {NoStop}%
\bibitem [{\citenamefont {{Tauchert}}\ \emph {et~al.}(2022)\citenamefont {{Tauchert}}, \citenamefont {{Volkov}}, \citenamefont {{Ehberger}}, \citenamefont {{Kazenwadel}}, \citenamefont {{Evers}}, \citenamefont {{Lange}}, \citenamefont {{Donges}}, \citenamefont {{Book}}, \citenamefont {{Kreuzpaintner}}, \citenamefont {{Nowak}},\ and\ \citenamefont {{Baum}}}]{tauchert2022polarized}%
  \BibitemOpen
  \bibfield  {author} {\bibinfo {author} {\bibfnamefont {S.~R.}\ \bibnamefont {{Tauchert}}}, \bibinfo {author} {\bibfnamefont {M.}~\bibnamefont {{Volkov}}}, \bibinfo {author} {\bibfnamefont {D.}~\bibnamefont {{Ehberger}}}, \bibinfo {author} {\bibfnamefont {D.}~\bibnamefont {{Kazenwadel}}}, \bibinfo {author} {\bibfnamefont {M.}~\bibnamefont {{Evers}}}, \bibinfo {author} {\bibfnamefont {H.}~\bibnamefont {{Lange}}}, \bibinfo {author} {\bibfnamefont {A.}~\bibnamefont {{Donges}}}, \bibinfo {author} {\bibfnamefont {A.}~\bibnamefont {{Book}}}, \bibinfo {author} {\bibfnamefont {W.}~\bibnamefont {{Kreuzpaintner}}}, \bibinfo {author} {\bibfnamefont {U.}~\bibnamefont {{Nowak}}},\ and\ \bibinfo {author} {\bibfnamefont {P.}~\bibnamefont {{Baum}}},\ }\bibfield  {title} {\emph {\bibinfo {title} {{Polarized phonons carry angular momentum in ultrafast demagnetization}}},\ }\href {https://doi.org/10.1038/s41586-021-04306-4} {\bibfield  {journal} {\bibinfo  {journal} {Nature}\ }\textbf {\bibinfo {volume} {602}},\ \bibinfo
  {pages} {73} (\bibinfo {year} {2022})}\BibitemShut {NoStop}%
\bibitem [{\citenamefont {Zhou}\ \emph {et~al.}(2022)\citenamefont {Zhou}, \citenamefont {Hwangbo}, \citenamefont {Zhang}, \citenamefont {Wang}, \citenamefont {Shen}, \citenamefont {Zhang}, \citenamefont {Jiang}, \citenamefont {Zong}, \citenamefont {Su}, \citenamefont {Zajac}, \citenamefont {Ahn}, \citenamefont {Walko}, \citenamefont {Schaller}, \citenamefont {Chu}, \citenamefont {Gedik}, \citenamefont {Xu}, \citenamefont {Xiao},\ and\ \citenamefont {Wen}}]{zhou2022dynamical}%
  \BibitemOpen
  \bibfield  {author} {\bibinfo {author} {\bibfnamefont {F.}~\bibnamefont {Zhou}}, \bibinfo {author} {\bibfnamefont {K.}~\bibnamefont {Hwangbo}}, \bibinfo {author} {\bibfnamefont {Q.}~\bibnamefont {Zhang}}, \bibinfo {author} {\bibfnamefont {C.}~\bibnamefont {Wang}}, \bibinfo {author} {\bibfnamefont {L.}~\bibnamefont {Shen}}, \bibinfo {author} {\bibfnamefont {J.}~\bibnamefont {Zhang}}, \bibinfo {author} {\bibfnamefont {Q.}~\bibnamefont {Jiang}}, \bibinfo {author} {\bibfnamefont {A.}~\bibnamefont {Zong}}, \bibinfo {author} {\bibfnamefont {Y.}~\bibnamefont {Su}}, \bibinfo {author} {\bibfnamefont {M.}~\bibnamefont {Zajac}}, \bibinfo {author} {\bibfnamefont {Y.}~\bibnamefont {Ahn}}, \bibinfo {author} {\bibfnamefont {D.~A.}\ \bibnamefont {Walko}}, \bibinfo {author} {\bibfnamefont {R.~D.}\ \bibnamefont {Schaller}}, \bibinfo {author} {\bibfnamefont {J.-h.}\ \bibnamefont {Chu}}, \bibinfo {author} {\bibfnamefont {N.}~\bibnamefont {Gedik}}, \bibinfo {author} {\bibfnamefont {X.}~\bibnamefont {Xu}}, \bibinfo {author}
  {\bibfnamefont {D.}~\bibnamefont {Xiao}},\ and\ \bibinfo {author} {\bibfnamefont {H.}~\bibnamefont {Wen}},\ }\bibfield  {title} {\emph {\bibinfo {title} {{Dynamical criticality of spin-shear coupling in van der Waals antiferromagnets}}},\ }\href {https://doi.org/10.1038/s41467-022-34376-5} {\bibfield  {journal} {\bibinfo  {journal} {Nat. Commun.}\ }\textbf {\bibinfo {volume} {13}},\ \bibinfo {pages} {6598} (\bibinfo {year} {2022})}\BibitemShut {NoStop}%
\bibitem [{\citenamefont {Zhou}\ \emph {et~al.}(2023)\citenamefont {Zhou}, \citenamefont {Liu}, \citenamefont {Zajac}, \citenamefont {Hwangbo}, \citenamefont {Jiang}, \citenamefont {Chu}, \citenamefont {Xu}, \citenamefont {Arslan}, \citenamefont {Gage},\ and\ \citenamefont {Wen}}]{zhou2023ultrafast}%
  \BibitemOpen
  \bibfield  {author} {\bibinfo {author} {\bibfnamefont {F.}~\bibnamefont {Zhou}}, \bibinfo {author} {\bibfnamefont {H.}~\bibnamefont {Liu}}, \bibinfo {author} {\bibfnamefont {M.}~\bibnamefont {Zajac}}, \bibinfo {author} {\bibfnamefont {K.}~\bibnamefont {Hwangbo}}, \bibinfo {author} {\bibfnamefont {Q.}~\bibnamefont {Jiang}}, \bibinfo {author} {\bibfnamefont {J.-h.}\ \bibnamefont {Chu}}, \bibinfo {author} {\bibfnamefont {X.}~\bibnamefont {Xu}}, \bibinfo {author} {\bibfnamefont {I.}~\bibnamefont {Arslan}}, \bibinfo {author} {\bibfnamefont {T.~E.}\ \bibnamefont {Gage}},\ and\ \bibinfo {author} {\bibfnamefont {H.}~\bibnamefont {Wen}},\ }\bibfield  {title} {\emph {\bibinfo {title} {{Ultrafast nanoimaging of spin-mediated shear waves in an acoustic cavity}}},\ }\href {https://doi.org/10.1021/acs.nanolett.3c02747} {\bibfield  {journal} {\bibinfo  {journal} {Nano Lett.}\ }\textbf {\bibinfo {volume} {23}},\ \bibinfo {pages} {10213--10220} (\bibinfo {year} {2023})}\BibitemShut {NoStop}%
\bibitem [{\citenamefont {Liu}\ \emph {et~al.}(2021)\citenamefont {Liu}, \citenamefont {Granados~del Águila}, \citenamefont {Bhowmick}, \citenamefont {Gan}, \citenamefont {Thu Ha~Do}, \citenamefont {Prosnikov}, \citenamefont {Sedmidubský}, \citenamefont {Sofer}, \citenamefont {Christianen}, \citenamefont {Sengupta},\ and\ \citenamefont {Xiong}}]{liu2021direct}%
  \BibitemOpen
  \bibfield  {author} {\bibinfo {author} {\bibfnamefont {S.}~\bibnamefont {Liu}}, \bibinfo {author} {\bibfnamefont {A.}~\bibnamefont {Granados~del Águila}}, \bibinfo {author} {\bibfnamefont {D.}~\bibnamefont {Bhowmick}}, \bibinfo {author} {\bibfnamefont {C.~K.}\ \bibnamefont {Gan}}, \bibinfo {author} {\bibfnamefont {T.}~\bibnamefont {Thu Ha~Do}}, \bibinfo {author} {\bibfnamefont {M.~A.}\ \bibnamefont {Prosnikov}}, \bibinfo {author} {\bibfnamefont {D.}~\bibnamefont {Sedmidubský}}, \bibinfo {author} {\bibfnamefont {Z.}~\bibnamefont {Sofer}}, \bibinfo {author} {\bibfnamefont {P.~C.~M.}\ \bibnamefont {Christianen}}, \bibinfo {author} {\bibfnamefont {P.}~\bibnamefont {Sengupta}},\ and\ \bibinfo {author} {\bibfnamefont {Q.}~\bibnamefont {Xiong}},\ }\bibfield  {title} {\emph {\bibinfo {title} {Direct observation of magnon-phonon strong coupling in two-dimensional antiferromagnet at high magnetic fields}},\ }\href {https://doi.org/10.1103/PhysRevLett.127.097401} {\bibfield  {journal} {\bibinfo  {journal} {Phys.
  Rev. Lett.}\ }\textbf {\bibinfo {volume} {127}},\ \bibinfo {pages} {097401} (\bibinfo {year} {2021})}\BibitemShut {NoStop}%
\bibitem [{\citenamefont {Vaclavkova}\ \emph {et~al.}(2021)\citenamefont {Vaclavkova}, \citenamefont {Palit}, \citenamefont {Wyzula}, \citenamefont {Ghosh}, \citenamefont {Delhomme}, \citenamefont {Maity}, \citenamefont {Kapuscinski}, \citenamefont {Ghosh}, \citenamefont {Veis}, \citenamefont {Grzeszczyk}, \citenamefont {Faugeras}, \citenamefont {Orlita}, \citenamefont {Datta},\ and\ \citenamefont {Potemski}}]{vaclavkova2021magnon}%
  \BibitemOpen
  \bibfield  {author} {\bibinfo {author} {\bibfnamefont {D.}~\bibnamefont {Vaclavkova}}, \bibinfo {author} {\bibfnamefont {M.}~\bibnamefont {Palit}}, \bibinfo {author} {\bibfnamefont {J.}~\bibnamefont {Wyzula}}, \bibinfo {author} {\bibfnamefont {S.}~\bibnamefont {Ghosh}}, \bibinfo {author} {\bibfnamefont {A.}~\bibnamefont {Delhomme}}, \bibinfo {author} {\bibfnamefont {S.}~\bibnamefont {Maity}}, \bibinfo {author} {\bibfnamefont {P.}~\bibnamefont {Kapuscinski}}, \bibinfo {author} {\bibfnamefont {A.}~\bibnamefont {Ghosh}}, \bibinfo {author} {\bibfnamefont {M.}~\bibnamefont {Veis}}, \bibinfo {author} {\bibfnamefont {M.}~\bibnamefont {Grzeszczyk}}, \bibinfo {author} {\bibfnamefont {C.}~\bibnamefont {Faugeras}}, \bibinfo {author} {\bibfnamefont {M.}~\bibnamefont {Orlita}}, \bibinfo {author} {\bibfnamefont {S.}~\bibnamefont {Datta}},\ and\ \bibinfo {author} {\bibfnamefont {M.}~\bibnamefont {Potemski}},\ }\bibfield  {title} {\emph {\bibinfo {title} {Magnon polarons in the van der {Waals} antiferromagnet
  {FePS$_3$}}},\ }\href {https://doi.org/10.1103/PhysRevB.104.134437} {\bibfield  {journal} {\bibinfo  {journal} {Phys. Rev. B}\ }\textbf {\bibinfo {volume} {104}},\ \bibinfo {pages} {134437} (\bibinfo {year} {2021})}\BibitemShut {NoStop}%
\bibitem [{\citenamefont {Zhang}\ \emph {et~al.}(2021{\natexlab{b}})\citenamefont {Zhang}, \citenamefont {Ozerov}, \citenamefont {Boström}, \citenamefont {Cui}, \citenamefont {Suri}, \citenamefont {Jiang}, \citenamefont {Wang}, \citenamefont {Wu}, \citenamefont {Hwangbo}, \citenamefont {Chu}, \citenamefont {Xiao}, \citenamefont {Rubio},\ and\ \citenamefont {Xu}}]{zhang2021coherent}%
  \BibitemOpen
  \bibfield  {author} {\bibinfo {author} {\bibfnamefont {Q.}~\bibnamefont {Zhang}}, \bibinfo {author} {\bibfnamefont {M.}~\bibnamefont {Ozerov}}, \bibinfo {author} {\bibfnamefont {E.~V.}\ \bibnamefont {Boström}}, \bibinfo {author} {\bibfnamefont {J.}~\bibnamefont {Cui}}, \bibinfo {author} {\bibfnamefont {N.}~\bibnamefont {Suri}}, \bibinfo {author} {\bibfnamefont {Q.}~\bibnamefont {Jiang}}, \bibinfo {author} {\bibfnamefont {C.}~\bibnamefont {Wang}}, \bibinfo {author} {\bibfnamefont {F.}~\bibnamefont {Wu}}, \bibinfo {author} {\bibfnamefont {K.}~\bibnamefont {Hwangbo}}, \bibinfo {author} {\bibfnamefont {J.-H.}\ \bibnamefont {Chu}}, \bibinfo {author} {\bibfnamefont {D.}~\bibnamefont {Xiao}}, \bibinfo {author} {\bibfnamefont {A.}~\bibnamefont {Rubio}},\ and\ \bibinfo {author} {\bibfnamefont {X.}~\bibnamefont {Xu}},\ }\href {https://doi.org/10.48550/arXiv.2108.11619} {\bibinfo {title} {\textit{Coherent strong-coupling of terahertz magnons and phonons in a {Van} der {Waals} antiferromagnetic insulator}}} (\bibinfo
  {year} {2021}{\natexlab{b}}),\ \Eprint {https://arxiv.org/abs/2108.11619} {arXiv:2108.11619} \BibitemShut {NoStop}%
\bibitem [{\citenamefont {{Mashkovich}}\ \emph {et~al.}(2021)\citenamefont {{Mashkovich}}, \citenamefont {{Grishunin}}, \citenamefont {{Dubrovin}}, \citenamefont {{Zvezdin}}, \citenamefont {{Pisarev}},\ and\ \citenamefont {{Kimel}}}]{mashkovich2021terahertz}%
  \BibitemOpen
  \bibfield  {author} {\bibinfo {author} {\bibfnamefont {E.~A.}\ \bibnamefont {{Mashkovich}}}, \bibinfo {author} {\bibfnamefont {K.~A.}\ \bibnamefont {{Grishunin}}}, \bibinfo {author} {\bibfnamefont {R.~M.}\ \bibnamefont {{Dubrovin}}}, \bibinfo {author} {\bibfnamefont {A.~K.}\ \bibnamefont {{Zvezdin}}}, \bibinfo {author} {\bibfnamefont {R.~V.}\ \bibnamefont {{Pisarev}}},\ and\ \bibinfo {author} {\bibfnamefont {A.~V.}\ \bibnamefont {{Kimel}}},\ }\bibfield  {title} {\emph {\bibinfo {title} {{Terahertz light-driven coupling of antiferromagnetic spins to lattice}}},\ }\href {https://doi.org/10.1126/science.abk1121} {\bibfield  {journal} {\bibinfo  {journal} {Science}\ }\textbf {\bibinfo {volume} {374}},\ \bibinfo {pages} {1608} (\bibinfo {year} {2021})}\BibitemShut {NoStop}%
\bibitem [{\citenamefont {Zhang}\ and\ \citenamefont {Niu}(2015)}]{zhang2015chiral}%
  \BibitemOpen
  \bibfield  {author} {\bibinfo {author} {\bibfnamefont {L.}~\bibnamefont {Zhang}}\ and\ \bibinfo {author} {\bibfnamefont {Q.}~\bibnamefont {Niu}},\ }\bibfield  {title} {\emph {\bibinfo {title} {{Chiral phonons at high-symmetry points in monolayer hexagonal lattices}}},\ }\href {https://doi.org/10.1103/PhysRevLett.115.115502} {\bibfield  {journal} {\bibinfo  {journal} {Phys. Rev. Lett.}\ }\textbf {\bibinfo {volume} {115}},\ \bibinfo {pages} {115502} (\bibinfo {year} {2015})}\BibitemShut {NoStop}%
\bibitem [{\citenamefont {{Suri}}\ \emph {et~al.}(2021)\citenamefont {{Suri}}, \citenamefont {{Wang}}, \citenamefont {{Zhang}},\ and\ \citenamefont {{Xiao}}}]{suri2021chiral}%
  \BibitemOpen
  \bibfield  {author} {\bibinfo {author} {\bibfnamefont {N.}~\bibnamefont {{Suri}}}, \bibinfo {author} {\bibfnamefont {C.}~\bibnamefont {{Wang}}}, \bibinfo {author} {\bibfnamefont {Y.}~\bibnamefont {{Zhang}}},\ and\ \bibinfo {author} {\bibfnamefont {D.}~\bibnamefont {{Xiao}}},\ }\bibfield  {title} {\emph {\bibinfo {title} {{Chiral phonons in Moir{\'e} superlattices}}},\ }\href {https://doi.org/10.1021/acs.nanolett.1c03692} {\bibfield  {journal} {\bibinfo  {journal} {Nano Lett.}\ }\textbf {\bibinfo {volume} {21}},\ \bibinfo {pages} {10026} (\bibinfo {year} {2021})}\BibitemShut {NoStop}%
\bibitem [{\citenamefont {{Zhu}}\ \emph {et~al.}(2018)\citenamefont {{Zhu}}, \citenamefont {{Yi}}, \citenamefont {{Li}}, \citenamefont {{Xiao}}, \citenamefont {{Zhang}}, \citenamefont {{Yang}}, \citenamefont {{Kaindl}}, \citenamefont {{Li}}, \citenamefont {{Wang}},\ and\ \citenamefont {{Zhang}}}]{zhu2018observation}%
  \BibitemOpen
  \bibfield  {author} {\bibinfo {author} {\bibfnamefont {H.}~\bibnamefont {{Zhu}}}, \bibinfo {author} {\bibfnamefont {J.}~\bibnamefont {{Yi}}}, \bibinfo {author} {\bibfnamefont {M.-Y.}\ \bibnamefont {{Li}}}, \bibinfo {author} {\bibfnamefont {J.}~\bibnamefont {{Xiao}}}, \bibinfo {author} {\bibfnamefont {L.}~\bibnamefont {{Zhang}}}, \bibinfo {author} {\bibfnamefont {C.-W.}\ \bibnamefont {{Yang}}}, \bibinfo {author} {\bibfnamefont {R.~A.}\ \bibnamefont {{Kaindl}}}, \bibinfo {author} {\bibfnamefont {L.-J.}\ \bibnamefont {{Li}}}, \bibinfo {author} {\bibfnamefont {Y.}~\bibnamefont {{Wang}}},\ and\ \bibinfo {author} {\bibfnamefont {X.}~\bibnamefont {{Zhang}}},\ }\bibfield  {title} {\emph {\bibinfo {title} {{Observation of chiral phonons}}},\ }\href {https://doi.org/10.1126/science.aar2711} {\bibfield  {journal} {\bibinfo  {journal} {Science}\ }\textbf {\bibinfo {volume} {359}},\ \bibinfo {pages} {579} (\bibinfo {year} {2018})}\BibitemShut {NoStop}%
\bibitem [{\citenamefont {Kalashnikova}\ \emph {et~al.}(2007)\citenamefont {Kalashnikova}, \citenamefont {Kimel}, \citenamefont {Pisarev}, \citenamefont {Gridnev}, \citenamefont {Kirilyuk},\ and\ \citenamefont {Rasing}}]{kalashnikova2007impulsive}%
  \BibitemOpen
  \bibfield  {author} {\bibinfo {author} {\bibfnamefont {A.~M.}\ \bibnamefont {Kalashnikova}}, \bibinfo {author} {\bibfnamefont {A.~V.}\ \bibnamefont {Kimel}}, \bibinfo {author} {\bibfnamefont {R.~V.}\ \bibnamefont {Pisarev}}, \bibinfo {author} {\bibfnamefont {V.~N.}\ \bibnamefont {Gridnev}}, \bibinfo {author} {\bibfnamefont {A.}~\bibnamefont {Kirilyuk}},\ and\ \bibinfo {author} {\bibfnamefont {T.}~\bibnamefont {Rasing}},\ }\bibfield  {title} {\emph {\bibinfo {title} {{Impulsive generation of coherent magnons by linearly polarized light in the easy-plane antiferromagnet ${\mathrm{FeBO}}_{3}$}}},\ }\href {https://doi.org/10.1103/PhysRevLett.99.167205} {\bibfield  {journal} {\bibinfo  {journal} {Phys. Rev. Lett.}\ }\textbf {\bibinfo {volume} {99}},\ \bibinfo {pages} {167205} (\bibinfo {year} {2007})}\BibitemShut {NoStop}%
\bibitem [{\citenamefont {{Baldini}}\ \emph {et~al.}(2019)\citenamefont {{Baldini}}, \citenamefont {{Dominguez}}, \citenamefont {{Palmieri}}, \citenamefont {{Cannelli}}, \citenamefont {{Rubio}}, \citenamefont {{Ruello}},\ and\ \citenamefont {{Chergui}}}]{baldini2019exciton}%
  \BibitemOpen
  \bibfield  {author} {\bibinfo {author} {\bibfnamefont {E.}~\bibnamefont {{Baldini}}}, \bibinfo {author} {\bibfnamefont {A.}~\bibnamefont {{Dominguez}}}, \bibinfo {author} {\bibfnamefont {T.}~\bibnamefont {{Palmieri}}}, \bibinfo {author} {\bibfnamefont {O.}~\bibnamefont {{Cannelli}}}, \bibinfo {author} {\bibfnamefont {A.}~\bibnamefont {{Rubio}}}, \bibinfo {author} {\bibfnamefont {P.}~\bibnamefont {{Ruello}}},\ and\ \bibinfo {author} {\bibfnamefont {M.}~\bibnamefont {{Chergui}}},\ }\bibfield  {title} {\emph {\bibinfo {title} {{Exciton control in a room temperature bulk semiconductor with coherent strain pulses}}},\ }\href {https://doi.org/10.1126/sciadv.aax2937} {\bibfield  {journal} {\bibinfo  {journal} {Sci. Adv.}\ }\textbf {\bibinfo {volume} {5}},\ \bibinfo {pages} {eaax2937} (\bibinfo {year} {2019})}\BibitemShut {NoStop}%
\bibitem [{\citenamefont {{Thouin}}\ \emph {et~al.}(2019)\citenamefont {{Thouin}}, \citenamefont {{Valverde-Ch{\'a}vez}}, \citenamefont {{Quarti}}, \citenamefont {{Cortecchia}}, \citenamefont {{Bargigia}}, \citenamefont {{Beljonne}}, \citenamefont {{Petrozza}}, \citenamefont {{Silva}},\ and\ \citenamefont {{Srimath Kandada}}}]{thouin2019phonon}%
  \BibitemOpen
  \bibfield  {author} {\bibinfo {author} {\bibfnamefont {F.}~\bibnamefont {{Thouin}}}, \bibinfo {author} {\bibfnamefont {D.~A.}\ \bibnamefont {{Valverde-Ch{\'a}vez}}}, \bibinfo {author} {\bibfnamefont {C.}~\bibnamefont {{Quarti}}}, \bibinfo {author} {\bibfnamefont {D.}~\bibnamefont {{Cortecchia}}}, \bibinfo {author} {\bibfnamefont {I.}~\bibnamefont {{Bargigia}}}, \bibinfo {author} {\bibfnamefont {D.}~\bibnamefont {{Beljonne}}}, \bibinfo {author} {\bibfnamefont {A.}~\bibnamefont {{Petrozza}}}, \bibinfo {author} {\bibfnamefont {C.}~\bibnamefont {{Silva}}},\ and\ \bibinfo {author} {\bibfnamefont {A.~R.}\ \bibnamefont {{Srimath Kandada}}},\ }\bibfield  {title} {\emph {\bibinfo {title} {{Phonon coherences reveal the polaronic character of excitons in two-dimensional lead halide perovskites}}},\ }\href {https://doi.org/10.1038/s41563-018-0262-7} {\bibfield  {journal} {\bibinfo  {journal} {Nat. Mater.}\ }\textbf {\bibinfo {volume} {18}},\ \bibinfo {pages} {349} (\bibinfo {year} {2019})}\BibitemShut {NoStop}%
\bibitem [{\citenamefont {{Trovatello}}\ \emph {et~al.}(2020{\natexlab{a}})\citenamefont {{Trovatello}}, \citenamefont {{Miranda}}, \citenamefont {{Molina-S{\'a}nchez}}, \citenamefont {{Borrego Varillas}}, \citenamefont {{Manzoni}}, \citenamefont {{Moretti}}, \citenamefont {{Ganzer}}, \citenamefont {{Maiuri}}, \citenamefont {{Wang}}, \citenamefont {{Dumcenco}}, \citenamefont {{Kis}}, \citenamefont {{Wirtz}}, \citenamefont {{Marini}}, \citenamefont {{Soavi}}, \citenamefont {{Ferrari}}, \citenamefont {{Cerullo}}, \citenamefont {{Sangalli}},\ and\ \citenamefont {{Dal Conte}}}]{trovatello2020strongly}%
  \BibitemOpen
  \bibfield  {author} {\bibinfo {author} {\bibfnamefont {C.}~\bibnamefont {{Trovatello}}}, \bibinfo {author} {\bibfnamefont {H.~P.~C.}\ \bibnamefont {{Miranda}}}, \bibinfo {author} {\bibfnamefont {A.}~\bibnamefont {{Molina-S{\'a}nchez}}}, \bibinfo {author} {\bibfnamefont {R.}~\bibnamefont {{Borrego Varillas}}}, \bibinfo {author} {\bibfnamefont {C.}~\bibnamefont {{Manzoni}}}, \bibinfo {author} {\bibfnamefont {L.}~\bibnamefont {{Moretti}}}, \bibinfo {author} {\bibfnamefont {L.}~\bibnamefont {{Ganzer}}}, \bibinfo {author} {\bibfnamefont {M.}~\bibnamefont {{Maiuri}}}, \bibinfo {author} {\bibfnamefont {J.}~\bibnamefont {{Wang}}}, \bibinfo {author} {\bibfnamefont {D.}~\bibnamefont {{Dumcenco}}}, \bibinfo {author} {\bibfnamefont {A.}~\bibnamefont {{Kis}}}, \bibinfo {author} {\bibfnamefont {L.}~\bibnamefont {{Wirtz}}}, \bibinfo {author} {\bibfnamefont {A.}~\bibnamefont {{Marini}}}, \bibinfo {author} {\bibfnamefont {G.}~\bibnamefont {{Soavi}}}, \bibinfo {author} {\bibfnamefont {A.~C.}\ \bibnamefont {{Ferrari}}},
  \bibinfo {author} {\bibfnamefont {G.}~\bibnamefont {{Cerullo}}}, \bibinfo {author} {\bibfnamefont {D.}~\bibnamefont {{Sangalli}}},\ and\ \bibinfo {author} {\bibfnamefont {S.}~\bibnamefont {{Dal Conte}}},\ }\bibfield  {title} {\emph {\bibinfo {title} {{Strongly coupled coherent phonons in single-layer MoS$_2$}}},\ }\href {https://doi.org/10.1021/acsnano.0c00309} {\bibfield  {journal} {\bibinfo  {journal} {ACS nano}\ }\textbf {\bibinfo {volume} {14}},\ \bibinfo {pages} {5700} (\bibinfo {year} {2020}{\natexlab{a}})}\BibitemShut {NoStop}%
\bibitem [{\citenamefont {{Trovatello}}\ \emph {et~al.}(2020{\natexlab{b}})\citenamefont {{Trovatello}}, \citenamefont {{Katsch}}, \citenamefont {{Borys}}, \citenamefont {{Selig}}, \citenamefont {{Yao}}, \citenamefont {{Borrego-Varillas}}, \citenamefont {{Scotognella}}, \citenamefont {{Kriegel}}, \citenamefont {{Yan}}, \citenamefont {{Zettl}}, \citenamefont {{Schuck}}, \citenamefont {{Knorr}}, \citenamefont {{Cerullo}},\ and\ \citenamefont {{Conte}}}]{Trovatello2020theultrafast}%
  \BibitemOpen
  \bibfield  {author} {\bibinfo {author} {\bibfnamefont {C.}~\bibnamefont {{Trovatello}}}, \bibinfo {author} {\bibfnamefont {F.}~\bibnamefont {{Katsch}}}, \bibinfo {author} {\bibfnamefont {N.~J.}\ \bibnamefont {{Borys}}}, \bibinfo {author} {\bibfnamefont {M.}~\bibnamefont {{Selig}}}, \bibinfo {author} {\bibfnamefont {K.}~\bibnamefont {{Yao}}}, \bibinfo {author} {\bibfnamefont {R.}~\bibnamefont {{Borrego-Varillas}}}, \bibinfo {author} {\bibfnamefont {F.}~\bibnamefont {{Scotognella}}}, \bibinfo {author} {\bibfnamefont {I.}~\bibnamefont {{Kriegel}}}, \bibinfo {author} {\bibfnamefont {A.}~\bibnamefont {{Yan}}}, \bibinfo {author} {\bibfnamefont {A.}~\bibnamefont {{Zettl}}}, \bibinfo {author} {\bibfnamefont {P.~J.}\ \bibnamefont {{Schuck}}}, \bibinfo {author} {\bibfnamefont {A.}~\bibnamefont {{Knorr}}}, \bibinfo {author} {\bibfnamefont {G.}~\bibnamefont {{Cerullo}}},\ and\ \bibinfo {author} {\bibfnamefont {S.~D.}\ \bibnamefont {{Conte}}},\ }\bibfield  {title} {\emph {\bibinfo {title} {{The ultrafast onset of
  exciton formation in 2D semiconductors}}},\ }\href {https://doi.org/10.1038/s41467-020-18835-5} {\bibfield  {journal} {\bibinfo  {journal} {Nat. Commun.}\ ,\ \bibinfo {pages} {5277}} (\bibinfo {year} {2020}{\natexlab{b}})}\BibitemShut {NoStop}%
\bibitem [{\citenamefont {{Peng}}\ \emph {et~al.}(2022)\citenamefont {{Peng}}, \citenamefont {{Ripin}}, \citenamefont {{Ye}}, \citenamefont {{Zhu}}, \citenamefont {{Wu}}, \citenamefont {{Lee}}, \citenamefont {{Li}}, \citenamefont {{Taniguchi}}, \citenamefont {{Watanabe}}, \citenamefont {{Cao}}, \citenamefont {{Xu}},\ and\ \citenamefont {{Li}}}]{peng2022long}%
  \BibitemOpen
  \bibfield  {author} {\bibinfo {author} {\bibfnamefont {R.}~\bibnamefont {{Peng}}}, \bibinfo {author} {\bibfnamefont {A.}~\bibnamefont {{Ripin}}}, \bibinfo {author} {\bibfnamefont {Y.}~\bibnamefont {{Ye}}}, \bibinfo {author} {\bibfnamefont {J.}~\bibnamefont {{Zhu}}}, \bibinfo {author} {\bibfnamefont {C.}~\bibnamefont {{Wu}}}, \bibinfo {author} {\bibfnamefont {S.}~\bibnamefont {{Lee}}}, \bibinfo {author} {\bibfnamefont {H.}~\bibnamefont {{Li}}}, \bibinfo {author} {\bibfnamefont {T.}~\bibnamefont {{Taniguchi}}}, \bibinfo {author} {\bibfnamefont {K.}~\bibnamefont {{Watanabe}}}, \bibinfo {author} {\bibfnamefont {T.}~\bibnamefont {{Cao}}}, \bibinfo {author} {\bibfnamefont {X.}~\bibnamefont {{Xu}}},\ and\ \bibinfo {author} {\bibfnamefont {M.}~\bibnamefont {{Li}}},\ }\bibfield  {title} {\emph {\bibinfo {title} {{Long-range transport of 2D excitons with acoustic waves}}},\ }\href {https://doi.org/10.1038/s41467-022-29042-9} {\bibfield  {journal} {\bibinfo  {journal} {Nat. Commun.}\ }\textbf {\bibinfo {volume}
  {13}},\ \bibinfo {pages} {1334} (\bibinfo {year} {2022})}\BibitemShut {NoStop}%
\bibitem [{\citenamefont {{Morello}}\ \emph {et~al.}(2007)\citenamefont {{Morello}}, \citenamefont {{De Giorgi}}, \citenamefont {{Kudera}}, \citenamefont {{Manna}}, \citenamefont {{Cingolani}},\ and\ \citenamefont {{Anni}}}]{morello2007temperature}%
  \BibitemOpen
  \bibfield  {author} {\bibinfo {author} {\bibfnamefont {G.}~\bibnamefont {{Morello}}}, \bibinfo {author} {\bibfnamefont {M.}~\bibnamefont {{De Giorgi}}}, \bibinfo {author} {\bibfnamefont {S.}~\bibnamefont {{Kudera}}}, \bibinfo {author} {\bibfnamefont {L.}~\bibnamefont {{Manna}}}, \bibinfo {author} {\bibfnamefont {R.}~\bibnamefont {{Cingolani}}},\ and\ \bibinfo {author} {\bibfnamefont {M.}~\bibnamefont {{Anni}}},\ }\bibfield  {title} {\emph {\bibinfo {title} {{Temperature and size dependence of nonradiative relaxation and exciton- phonon coupling in colloidal CdTe quantum dots}}},\ }\href {https://doi.org/10.1021/jp068307t} {\bibfield  {journal} {\bibinfo  {journal} {J. Phys. Chem. C}\ }\textbf {\bibinfo {volume} {111}},\ \bibinfo {pages} {5846} (\bibinfo {year} {2007})}\BibitemShut {NoStop}%
\bibitem [{\citenamefont {{Sagar}}\ \emph {et~al.}(2008)\citenamefont {{Sagar}}, \citenamefont {{Cooney}}, \citenamefont {{Sewall}}, \citenamefont {{Dias}}, \citenamefont {{Barsan}}, \citenamefont {{Butler}},\ and\ \citenamefont {{Kambhampati}}}]{sagar2008size}%
  \BibitemOpen
  \bibfield  {author} {\bibinfo {author} {\bibfnamefont {D.~M.}\ \bibnamefont {{Sagar}}}, \bibinfo {author} {\bibfnamefont {R.~R.}\ \bibnamefont {{Cooney}}}, \bibinfo {author} {\bibfnamefont {S.~L.}\ \bibnamefont {{Sewall}}}, \bibinfo {author} {\bibfnamefont {E.~A.}\ \bibnamefont {{Dias}}}, \bibinfo {author} {\bibfnamefont {M.~M.}\ \bibnamefont {{Barsan}}}, \bibinfo {author} {\bibfnamefont {I.~S.}\ \bibnamefont {{Butler}}},\ and\ \bibinfo {author} {\bibfnamefont {P.}~\bibnamefont {{Kambhampati}}},\ }\bibfield  {title} {\emph {\bibinfo {title} {{Size dependent, state-resolved studies of exciton-phonon couplings in strongly confined semiconductor quantum dots}}},\ }\href {https://doi.org/10.1103/PhysRevB.77.235321} {\bibfield  {journal} {\bibinfo  {journal} {Phys. Rev. B}\ }\textbf {\bibinfo {volume} {77}},\ \bibinfo {pages} {235321} (\bibinfo {year} {2008})}\BibitemShut {NoStop}%
\bibitem [{\citenamefont {{Nazir}}\ and\ \citenamefont {{McCutcheon}}(2016)}]{Nazir2016modelling}%
  \BibitemOpen
  \bibfield  {author} {\bibinfo {author} {\bibfnamefont {A.}~\bibnamefont {{Nazir}}}\ and\ \bibinfo {author} {\bibfnamefont {D.~P.~S.}\ \bibnamefont {{McCutcheon}}},\ }\bibfield  {title} {\emph {\bibinfo {title} {{Modelling exciton-phonon interactions in optically driven quantum dots}}},\ }\href {https://doi.org/10.1088/0953-8984/28/10/103002} {\bibfield  {journal} {\bibinfo  {journal} {J. Phys. Condens. Matter}\ }\textbf {\bibinfo {volume} {28}},\ \bibinfo {pages} {103002} (\bibinfo {year} {2016})}\BibitemShut {NoStop}%
\bibitem [{\citenamefont {{Zhang}}\ \emph {et~al.}(2024{\natexlab{b}})\citenamefont {{Zhang}}, \citenamefont {{Gao}}, \citenamefont {{Curtis}}, \citenamefont {{Liu}}, \citenamefont {{Chien}}, \citenamefont {{von Hoegen}}, \citenamefont {{Wong}}, \citenamefont {{Kurihara}}, \citenamefont {{Suemoto}}, \citenamefont {{Narang}}, \citenamefont {{Baldini}},\ and\ \citenamefont {{Nelson}}}]{zhang2024terahertzDFG}%
  \BibitemOpen
  \bibfield  {author} {\bibinfo {author} {\bibfnamefont {Z.}~\bibnamefont {{Zhang}}}, \bibinfo {author} {\bibfnamefont {F.~Y.}\ \bibnamefont {{Gao}}}, \bibinfo {author} {\bibfnamefont {J.~B.}\ \bibnamefont {{Curtis}}}, \bibinfo {author} {\bibfnamefont {Z.-J.}\ \bibnamefont {{Liu}}}, \bibinfo {author} {\bibfnamefont {Y.-C.}\ \bibnamefont {{Chien}}}, \bibinfo {author} {\bibfnamefont {A.}~\bibnamefont {{von Hoegen}}}, \bibinfo {author} {\bibfnamefont {M.~T.}\ \bibnamefont {{Wong}}}, \bibinfo {author} {\bibfnamefont {T.}~\bibnamefont {{Kurihara}}}, \bibinfo {author} {\bibfnamefont {T.}~\bibnamefont {{Suemoto}}}, \bibinfo {author} {\bibfnamefont {P.}~\bibnamefont {{Narang}}}, \bibinfo {author} {\bibfnamefont {E.}~\bibnamefont {{Baldini}}},\ and\ \bibinfo {author} {\bibfnamefont {K.~A.}\ \bibnamefont {{Nelson}}},\ }\bibfield  {title} {\emph {\bibinfo {title} {{Terahertz field-induced nonlinear coupling of two magnon modes in an antiferromagnet}}},\ }\href {https://doi.org/10.1038/s41567-024-02386-3} {\bibfield
  {journal} {\bibinfo  {journal} {Nat. Phys.}\ }\textbf {\bibinfo {volume} {20}},\ \bibinfo {pages} {801} (\bibinfo {year} {2024}{\natexlab{b}})}\BibitemShut {NoStop}%
\bibitem [{\citenamefont {{Zhang}}\ \emph {et~al.}(2024{\natexlab{c}})\citenamefont {{Zhang}}, \citenamefont {{Gao}}, \citenamefont {{Chien}}, \citenamefont {{Liu}}, \citenamefont {{Curtis}}, \citenamefont {{Sung}}, \citenamefont {{Ma}}, \citenamefont {{Ren}}, \citenamefont {{Cao}}, \citenamefont {{Narang}}, \citenamefont {{von Hoegen}}, \citenamefont {{Baldini}},\ and\ \citenamefont {{Nelson}}}]{zhang2024terahertzup}%
  \BibitemOpen
  \bibfield  {author} {\bibinfo {author} {\bibfnamefont {Z.}~\bibnamefont {{Zhang}}}, \bibinfo {author} {\bibfnamefont {F.~Y.}\ \bibnamefont {{Gao}}}, \bibinfo {author} {\bibfnamefont {Y.-C.}\ \bibnamefont {{Chien}}}, \bibinfo {author} {\bibfnamefont {Z.-J.}\ \bibnamefont {{Liu}}}, \bibinfo {author} {\bibfnamefont {J.~B.}\ \bibnamefont {{Curtis}}}, \bibinfo {author} {\bibfnamefont {E.~R.}\ \bibnamefont {{Sung}}}, \bibinfo {author} {\bibfnamefont {X.}~\bibnamefont {{Ma}}}, \bibinfo {author} {\bibfnamefont {W.}~\bibnamefont {{Ren}}}, \bibinfo {author} {\bibfnamefont {S.}~\bibnamefont {{Cao}}}, \bibinfo {author} {\bibfnamefont {P.}~\bibnamefont {{Narang}}}, \bibinfo {author} {\bibfnamefont {A.}~\bibnamefont {{von Hoegen}}}, \bibinfo {author} {\bibfnamefont {E.}~\bibnamefont {{Baldini}}},\ and\ \bibinfo {author} {\bibfnamefont {K.~A.}\ \bibnamefont {{Nelson}}},\ }\bibfield  {title} {\emph {\bibinfo {title} {{Terahertz-field-driven magnon upconversion in an antiferromagnet}}},\ }\href
  {https://doi.org/10.1038/s41567-023-02350-7} {\bibfield  {journal} {\bibinfo  {journal} {Nat. Phys.}\ }\textbf {\bibinfo {volume} {20}},\ \bibinfo {pages} {788} (\bibinfo {year} {2024}{\natexlab{c}})}\BibitemShut {NoStop}%
\bibitem [{\citenamefont {Blank}\ \emph {et~al.}(2023)\citenamefont {Blank}, \citenamefont {Grishunin}, \citenamefont {Ivanov}, \citenamefont {Mashkovich}, \citenamefont {Afanasiev},\ and\ \citenamefont {Kimel}}]{Blank2023empowering}%
  \BibitemOpen
  \bibfield  {author} {\bibinfo {author} {\bibfnamefont {T.~G.~H.}\ \bibnamefont {Blank}}, \bibinfo {author} {\bibfnamefont {K.~A.}\ \bibnamefont {Grishunin}}, \bibinfo {author} {\bibfnamefont {B.~A.}\ \bibnamefont {Ivanov}}, \bibinfo {author} {\bibfnamefont {E.~A.}\ \bibnamefont {Mashkovich}}, \bibinfo {author} {\bibfnamefont {D.}~\bibnamefont {Afanasiev}},\ and\ \bibinfo {author} {\bibfnamefont {A.~V.}\ \bibnamefont {Kimel}},\ }\bibfield  {title} {\emph {\bibinfo {title} {{Empowering control of antiferromagnets by THz-induced spin coherence}}},\ }\href {https://doi.org/10.1103/PhysRevLett.131.096701} {\bibfield  {journal} {\bibinfo  {journal} {Phys. Rev. Lett.}\ }\textbf {\bibinfo {volume} {131}},\ \bibinfo {pages} {096701} (\bibinfo {year} {2023})}\BibitemShut {NoStop}%
\bibitem [{\citenamefont {Leenders}\ \emph {et~al.}(2024)\citenamefont {Leenders}, \citenamefont {Afanasiev}, \citenamefont {Kimel},\ and\ \citenamefont {Mikhaylovskiy}}]{leenders2024canted}%
  \BibitemOpen
  \bibfield  {author} {\bibinfo {author} {\bibfnamefont {R.~A.}\ \bibnamefont {Leenders}}, \bibinfo {author} {\bibfnamefont {D.}~\bibnamefont {Afanasiev}}, \bibinfo {author} {\bibfnamefont {A.~V.}\ \bibnamefont {Kimel}},\ and\ \bibinfo {author} {\bibfnamefont {R.~V.}\ \bibnamefont {Mikhaylovskiy}},\ }\bibfield  {title} {\emph {\bibinfo {title} {Canted spin order as a platform for ultrafast conversion of magnons}},\ }\href {https://doi.org/10.1038/s41586-024-07448-3} {\bibfield  {journal} {\bibinfo  {journal} {Nature}\ }\textbf {\bibinfo {volume} {630}},\ \bibinfo {pages} {335--339} (\bibinfo {year} {2024})}\BibitemShut {NoStop}%
\bibitem [{\citenamefont {{Zhang}}\ \emph {et~al.}(2024{\natexlab{d}})\citenamefont {{Zhang}}, \citenamefont {{Chien}}, \citenamefont {{Tou Wong}}, \citenamefont {{Gao}}, \citenamefont {{Liu}}, \citenamefont {{Ma}}, \citenamefont {{Cao}}, \citenamefont {{Baldini}},\ and\ \citenamefont {{Nelson}}}]{zhang2024terahertzstimu}%
  \BibitemOpen
  \bibfield  {author} {\bibinfo {author} {\bibfnamefont {Z.}~\bibnamefont {{Zhang}}}, \bibinfo {author} {\bibfnamefont {Y.-C.}\ \bibnamefont {{Chien}}}, \bibinfo {author} {\bibfnamefont {M.}~\bibnamefont {{Tou Wong}}}, \bibinfo {author} {\bibfnamefont {F.~Y.}\ \bibnamefont {{Gao}}}, \bibinfo {author} {\bibfnamefont {Z.-J.}\ \bibnamefont {{Liu}}}, \bibinfo {author} {\bibfnamefont {X.}~\bibnamefont {{Ma}}}, \bibinfo {author} {\bibfnamefont {S.}~\bibnamefont {{Cao}}}, \bibinfo {author} {\bibfnamefont {E.}~\bibnamefont {{Baldini}}},\ and\ \bibinfo {author} {\bibfnamefont {K.~A.}\ \bibnamefont {{Nelson}}},\ }\href {https://doi.org/10.48550/arXiv.2412.01989} {\bibinfo {title} {{Terahertz stimulated parametric downconversion of a magnon mode in an antiferromagnet}}} (\bibinfo {year} {2024}{\natexlab{d}}),\ \Eprint {https://arxiv.org/abs/2412.01989} {arXiv:2412.01989} \BibitemShut {NoStop}%
\bibitem [{\citenamefont {{Bae}}\ \emph {et~al.}(2022)\citenamefont {{Bae}}, \citenamefont {{Wang}}, \citenamefont {{Scheie}}, \citenamefont {{Xu}}, \citenamefont {{Chica}}, \citenamefont {{Diederich}}, \citenamefont {{Cenker}}, \citenamefont {{Ziebel}}, \citenamefont {{Bai}}, \citenamefont {{Ren}}, \citenamefont {{Dean}}, \citenamefont {{Delor}}, \citenamefont {{Xu}}, \citenamefont {{Roy}}, \citenamefont {{Kent}},\ and\ \citenamefont {{Zhu}}}]{bae2022exciton}%
  \BibitemOpen
  \bibfield  {author} {\bibinfo {author} {\bibfnamefont {Y.~J.}\ \bibnamefont {{Bae}}}, \bibinfo {author} {\bibfnamefont {J.}~\bibnamefont {{Wang}}}, \bibinfo {author} {\bibfnamefont {A.}~\bibnamefont {{Scheie}}}, \bibinfo {author} {\bibfnamefont {J.}~\bibnamefont {{Xu}}}, \bibinfo {author} {\bibfnamefont {D.~G.}\ \bibnamefont {{Chica}}}, \bibinfo {author} {\bibfnamefont {G.~M.}\ \bibnamefont {{Diederich}}}, \bibinfo {author} {\bibfnamefont {J.}~\bibnamefont {{Cenker}}}, \bibinfo {author} {\bibfnamefont {M.~E.}\ \bibnamefont {{Ziebel}}}, \bibinfo {author} {\bibfnamefont {Y.}~\bibnamefont {{Bai}}}, \bibinfo {author} {\bibfnamefont {H.}~\bibnamefont {{Ren}}}, \bibinfo {author} {\bibfnamefont {C.~R.}\ \bibnamefont {{Dean}}}, \bibinfo {author} {\bibfnamefont {M.}~\bibnamefont {{Delor}}}, \bibinfo {author} {\bibfnamefont {X.}~\bibnamefont {{Xu}}}, \bibinfo {author} {\bibfnamefont {X.}~\bibnamefont {{Roy}}}, \bibinfo {author} {\bibfnamefont {A.~D.}\ \bibnamefont {{Kent}}},\ and\ \bibinfo {author} {\bibfnamefont
  {X.}~\bibnamefont {{Zhu}}},\ }\bibfield  {title} {\emph {\bibinfo {title} {{Exciton-coupled coherent magnons in a 2D semiconductor}}},\ }\href {https://doi.org/10.1038/s41586-022-05024-1} {\bibfield  {journal} {\bibinfo  {journal} {Nature}\ }\textbf {\bibinfo {volume} {609}},\ \bibinfo {pages} {282} (\bibinfo {year} {2022})}\BibitemShut {NoStop}%
\bibitem [{\citenamefont {{Zayats}}\ \emph {et~al.}(2005)\citenamefont {{Zayats}}, \citenamefont {{Smolyaninov}},\ and\ \citenamefont {{Maradudin}}}]{zayats2005nano}%
  \BibitemOpen
  \bibfield  {author} {\bibinfo {author} {\bibfnamefont {A.~V.}\ \bibnamefont {{Zayats}}}, \bibinfo {author} {\bibfnamefont {I.~I.}\ \bibnamefont {{Smolyaninov}}},\ and\ \bibinfo {author} {\bibfnamefont {A.~A.}\ \bibnamefont {{Maradudin}}},\ }\bibfield  {title} {\emph {\bibinfo {title} {{Nano-optics of surface plasmon polaritons}}},\ }\href {https://doi.org/10.1016/j.physrep.2004.11.001} {\bibfield  {journal} {\bibinfo  {journal} {Phys. Rep.}\ }\textbf {\bibinfo {volume} {408}},\ \bibinfo {pages} {131} (\bibinfo {year} {2005})}\BibitemShut {NoStop}%
\bibitem [{\citenamefont {{Shen}}\ and\ \citenamefont {{Bauer}}(2018)}]{shen2018theory}%
  \BibitemOpen
  \bibfield  {author} {\bibinfo {author} {\bibfnamefont {K.}~\bibnamefont {{Shen}}}\ and\ \bibinfo {author} {\bibfnamefont {G.~E.~W.}\ \bibnamefont {{Bauer}}},\ }\bibfield  {title} {\emph {\bibinfo {title} {{Theory of spin and lattice wave dynamics excited by focused laser pulses}}},\ }\href {https://doi.org/10.1088/1361-6463/aabd68} {\bibfield  {journal} {\bibinfo  {journal} {J. Phys. D Appl. Phys.}\ }\textbf {\bibinfo {volume} {51}},\ \bibinfo {pages} {224008} (\bibinfo {year} {2018})}\BibitemShut {NoStop}%
\bibitem [{\citenamefont {Ogawa}\ \emph {et~al.}(2015)\citenamefont {Ogawa}, \citenamefont {Koshibae}, \citenamefont {Beekman}, \citenamefont {Nagaosa}, \citenamefont {Kubota}, \citenamefont {Kawasaki},\ and\ \citenamefont {Tokura}}]{ogawa2015photodrive}%
  \BibitemOpen
  \bibfield  {author} {\bibinfo {author} {\bibfnamefont {N.}~\bibnamefont {Ogawa}}, \bibinfo {author} {\bibfnamefont {W.}~\bibnamefont {Koshibae}}, \bibinfo {author} {\bibfnamefont {A.~J.}\ \bibnamefont {Beekman}}, \bibinfo {author} {\bibfnamefont {N.}~\bibnamefont {Nagaosa}}, \bibinfo {author} {\bibfnamefont {M.}~\bibnamefont {Kubota}}, \bibinfo {author} {\bibfnamefont {M.}~\bibnamefont {Kawasaki}},\ and\ \bibinfo {author} {\bibfnamefont {Y.}~\bibnamefont {Tokura}},\ }\bibfield  {title} {\emph {\bibinfo {title} {{Photodrive of magnetic bubbles via magnetoelastic waves}}},\ }\href {https://doi.org/10.1073/pnas.1504064112} {\bibfield  {journal} {\bibinfo  {journal} {Proc. Natl. Acad. Sci. U.S.A.}\ }\textbf {\bibinfo {volume} {112}},\ \bibinfo {pages} {8977} (\bibinfo {year} {2015})}\BibitemShut {NoStop}%
\bibitem [{\citenamefont {{Wang}}\ \emph {et~al.}(2023)\citenamefont {{Wang}}, \citenamefont {{Yang}}, \citenamefont {{Chen}}, \citenamefont {{Wang}}, \citenamefont {{Jia}}, \citenamefont {{Chen}}, \citenamefont {{Zhang}}, \citenamefont {{Wan}}, \citenamefont {{Liu}}, \citenamefont {{Yu}}, \citenamefont {{Han}}, \citenamefont {{Ansermet}}, \citenamefont {{Zhang}},\ and\ \citenamefont {{Yu}}}]{wang2023long}%
  \BibitemOpen
  \bibfield  {author} {\bibinfo {author} {\bibfnamefont {H.}~\bibnamefont {{Wang}}}, \bibinfo {author} {\bibfnamefont {Y.}~\bibnamefont {{Yang}}}, \bibinfo {author} {\bibfnamefont {J.}~\bibnamefont {{Chen}}}, \bibinfo {author} {\bibfnamefont {J.}~\bibnamefont {{Wang}}}, \bibinfo {author} {\bibfnamefont {H.}~\bibnamefont {{Jia}}}, \bibinfo {author} {\bibfnamefont {P.}~\bibnamefont {{Chen}}}, \bibinfo {author} {\bibfnamefont {Y.}~\bibnamefont {{Zhang}}}, \bibinfo {author} {\bibfnamefont {C.}~\bibnamefont {{Wan}}}, \bibinfo {author} {\bibfnamefont {S.}~\bibnamefont {{Liu}}}, \bibinfo {author} {\bibfnamefont {D.}~\bibnamefont {{Yu}}}, \bibinfo {author} {\bibfnamefont {X.}~\bibnamefont {{Han}}}, \bibinfo {author} {\bibfnamefont {J.-P.}\ \bibnamefont {{Ansermet}}}, \bibinfo {author} {\bibfnamefont {J.}~\bibnamefont {{Zhang}}},\ and\ \bibinfo {author} {\bibfnamefont {H.}~\bibnamefont {{Yu}}},\ }\bibfield  {title} {\emph {\bibinfo {title} {{Long-distance coherent propagation of magnon polarons in a
  ferroelectric-ferromagnetic heterostructure}}},\ }\href {https://doi.org/10.1103/PhysRevB.108.144425} {\bibfield  {journal} {\bibinfo  {journal} {Phys. Rev. B}\ }\textbf {\bibinfo {volume} {108}},\ \bibinfo {pages} {144425} (\bibinfo {year} {2023})}\BibitemShut {NoStop}%
\bibitem [{\citenamefont {Bae}\ \emph {et~al.}(2024)\citenamefont {Bae}, \citenamefont {Handa}, \citenamefont {Dai}, \citenamefont {Wang}, \citenamefont {Liu}, \citenamefont {Scheie}, \citenamefont {Chica}, \citenamefont {Ziebel}, \citenamefont {Kent}, \citenamefont {Xu}, \citenamefont {Shen}, \citenamefont {Roy},\ and\ \citenamefont {Zhu}}]{bae2024transient}%
  \BibitemOpen
  \bibfield  {author} {\bibinfo {author} {\bibfnamefont {Y.~J.}\ \bibnamefont {Bae}}, \bibinfo {author} {\bibfnamefont {T.}~\bibnamefont {Handa}}, \bibinfo {author} {\bibfnamefont {Y.}~\bibnamefont {Dai}}, \bibinfo {author} {\bibfnamefont {J.}~\bibnamefont {Wang}}, \bibinfo {author} {\bibfnamefont {H.}~\bibnamefont {Liu}}, \bibinfo {author} {\bibfnamefont {A.}~\bibnamefont {Scheie}}, \bibinfo {author} {\bibfnamefont {D.~G.}\ \bibnamefont {Chica}}, \bibinfo {author} {\bibfnamefont {M.~E.}\ \bibnamefont {Ziebel}}, \bibinfo {author} {\bibfnamefont {A.~D.}\ \bibnamefont {Kent}}, \bibinfo {author} {\bibfnamefont {X.}~\bibnamefont {Xu}}, \bibinfo {author} {\bibfnamefont {K.}~\bibnamefont {Shen}}, \bibinfo {author} {\bibfnamefont {X.}~\bibnamefont {Roy}},\ and\ \bibinfo {author} {\bibfnamefont {X.}~\bibnamefont {Zhu}},\ }\bibfield  {title} {\emph {\bibinfo {title} {{Transient magnetoelastic coupling in CrSBr}}},\ }\href {https://doi.org/10.1103/PhysRevB.109.104401} {\bibfield  {journal} {\bibinfo  {journal} {Phys.
  Rev. B}\ }\textbf {\bibinfo {volume} {109}},\ \bibinfo {pages} {104401} (\bibinfo {year} {2024})}\BibitemShut {NoStop}%
\bibitem [{\citenamefont {Sun}\ \emph {et~al.}(2024)\citenamefont {Sun}, \citenamefont {Meng}, \citenamefont {Lee}, \citenamefont {Soll}, \citenamefont {Zhang}, \citenamefont {Ramesh}, \citenamefont {Yao}, \citenamefont {Sofer},\ and\ \citenamefont {Orenstein}}]{sun2024dipolar}%
  \BibitemOpen
  \bibfield  {author} {\bibinfo {author} {\bibfnamefont {Y.}~\bibnamefont {Sun}}, \bibinfo {author} {\bibfnamefont {F.}~\bibnamefont {Meng}}, \bibinfo {author} {\bibfnamefont {C.}~\bibnamefont {Lee}}, \bibinfo {author} {\bibfnamefont {A.}~\bibnamefont {Soll}}, \bibinfo {author} {\bibfnamefont {H.}~\bibnamefont {Zhang}}, \bibinfo {author} {\bibfnamefont {R.}~\bibnamefont {Ramesh}}, \bibinfo {author} {\bibfnamefont {J.}~\bibnamefont {Yao}}, \bibinfo {author} {\bibfnamefont {Z.}~\bibnamefont {Sofer}},\ and\ \bibinfo {author} {\bibfnamefont {J.}~\bibnamefont {Orenstein}},\ }\bibfield  {title} {\emph {\bibinfo {title} {{Dipolar spin wave packet transport in a van der Waals antiferromagnet}}},\ }\href {https://doi.org/10.1038/s41567-024-02387-2} {\bibfield  {journal} {\bibinfo  {journal} {Nat. Phys.}\ }\textbf {\bibinfo {volume} {20}},\ \bibinfo {pages} {794} (\bibinfo {year} {2024})}\BibitemShut {NoStop}%
\bibitem [{\citenamefont {Violante}\ \emph {et~al.}(2014)\citenamefont {Violante}, \citenamefont {Cohen}, \citenamefont {Lazi{\'c}}, \citenamefont {Hey}, \citenamefont {Rapaport},\ and\ \citenamefont {Santos}}]{violante2014dynamics}%
  \BibitemOpen
  \bibfield  {author} {\bibinfo {author} {\bibfnamefont {A.}~\bibnamefont {Violante}}, \bibinfo {author} {\bibfnamefont {K.}~\bibnamefont {Cohen}}, \bibinfo {author} {\bibfnamefont {S.}~\bibnamefont {Lazi{\'c}}}, \bibinfo {author} {\bibfnamefont {R.}~\bibnamefont {Hey}}, \bibinfo {author} {\bibfnamefont {R.}~\bibnamefont {Rapaport}},\ and\ \bibinfo {author} {\bibfnamefont {P.~V.}\ \bibnamefont {Santos}},\ }\bibfield  {title} {\emph {\bibinfo {title} {{Dynamics of indirect exciton transport by moving acoustic fields}}},\ }\href {https://doi.org/10.1088/1367-2630/16/3/033035} {\bibfield  {journal} {\bibinfo  {journal} {New J. Phys.}\ }\textbf {\bibinfo {volume} {16}},\ \bibinfo {pages} {033035} (\bibinfo {year} {2014})}\BibitemShut {NoStop}%
\bibitem [{\citenamefont {Grasselli}\ \emph {et~al.}(2018)\citenamefont {Grasselli}, \citenamefont {Bertoni},\ and\ \citenamefont {Goldoni}}]{grasselli2018classical}%
  \BibitemOpen
  \bibfield  {author} {\bibinfo {author} {\bibfnamefont {F.}~\bibnamefont {Grasselli}}, \bibinfo {author} {\bibfnamefont {A.}~\bibnamefont {Bertoni}},\ and\ \bibinfo {author} {\bibfnamefont {G.}~\bibnamefont {Goldoni}},\ }\bibfield  {title} {\emph {\bibinfo {title} {{Classical and quantum dynamics of indirect excitons driven by surface acoustic waves}}},\ }\href {https://doi.org/10.1103/PhysRevB.98.165407} {\bibfield  {journal} {\bibinfo  {journal} {Phys. Rev. B}\ }\textbf {\bibinfo {volume} {98}},\ \bibinfo {pages} {165407} (\bibinfo {year} {2018})}\BibitemShut {NoStop}%
\bibitem [{\citenamefont {{Zhang}}\ \emph {et~al.}(2012)\citenamefont {{Zhang}}, \citenamefont {{Zhang}},\ and\ \citenamefont {{Xu}}}]{zhang2012surface}%
  \BibitemOpen
  \bibfield  {author} {\bibinfo {author} {\bibfnamefont {J.}~\bibnamefont {{Zhang}}}, \bibinfo {author} {\bibfnamefont {L.}~\bibnamefont {{Zhang}}},\ and\ \bibinfo {author} {\bibfnamefont {W.}~\bibnamefont {{Xu}}},\ }\bibfield  {title} {\emph {\bibinfo {title} {{Surface plasmon polaritons: physics and applications}}},\ }\href {https://doi.org/10.1088/0022-3727/45/11/113001} {\bibfield  {journal} {\bibinfo  {journal} {J. Phys. D}\ }\textbf {\bibinfo {volume} {45}},\ \bibinfo {pages} {113001} (\bibinfo {year} {2012})}\BibitemShut {NoStop}%
\bibitem [{\citenamefont {{Latini}}\ \emph {et~al.}(2021)\citenamefont {{Latini}}, \citenamefont {{De Giovannini}}, \citenamefont {{Sie}}, \citenamefont {{Gedik}}, \citenamefont {{H{\"u}bener}},\ and\ \citenamefont {{Rubio}}}]{latini2021phonoritons}%
  \BibitemOpen
  \bibfield  {author} {\bibinfo {author} {\bibfnamefont {S.}~\bibnamefont {{Latini}}}, \bibinfo {author} {\bibfnamefont {U.}~\bibnamefont {{De Giovannini}}}, \bibinfo {author} {\bibfnamefont {E.~J.}\ \bibnamefont {{Sie}}}, \bibinfo {author} {\bibfnamefont {N.}~\bibnamefont {{Gedik}}}, \bibinfo {author} {\bibfnamefont {H.}~\bibnamefont {{H{\"u}bener}}},\ and\ \bibinfo {author} {\bibfnamefont {A.}~\bibnamefont {{Rubio}}},\ }\bibfield  {title} {\emph {\bibinfo {title} {{{Phonoritons as hybridized exciton-photon-phonon excitations in a monolayer h-BN optical cavity}}}},\ }\href {https://doi.org/10.1103/PhysRevLett.126.227401} {\bibfield  {journal} {\bibinfo  {journal} {Phys. Rev. Lett.}\ }\textbf {\bibinfo {volume} {126}},\ \bibinfo {pages} {227401} (\bibinfo {year} {2021})}\BibitemShut {NoStop}%
\bibitem [{\citenamefont {Zhang}\ \emph {et~al.}(2020)\citenamefont {Zhang}, \citenamefont {Abhiraman}, \citenamefont {Zhang}, \citenamefont {Miao}, \citenamefont {Jo}, \citenamefont {Roccasecca}, \citenamefont {Knight}, \citenamefont {Davoyan},\ and\ \citenamefont {Jariwala}}]{zhang2020hybrid}%
  \BibitemOpen
  \bibfield  {author} {\bibinfo {author} {\bibfnamefont {H.}~\bibnamefont {Zhang}}, \bibinfo {author} {\bibfnamefont {B.}~\bibnamefont {Abhiraman}}, \bibinfo {author} {\bibfnamefont {Q.}~\bibnamefont {Zhang}}, \bibinfo {author} {\bibfnamefont {J.}~\bibnamefont {Miao}}, \bibinfo {author} {\bibfnamefont {K.}~\bibnamefont {Jo}}, \bibinfo {author} {\bibfnamefont {S.}~\bibnamefont {Roccasecca}}, \bibinfo {author} {\bibfnamefont {M.~W.}\ \bibnamefont {Knight}}, \bibinfo {author} {\bibfnamefont {A.~R.}\ \bibnamefont {Davoyan}},\ and\ \bibinfo {author} {\bibfnamefont {D.}~\bibnamefont {Jariwala}},\ }\bibfield  {title} {\emph {\bibinfo {title} {{Hybrid exciton-plasmon-polaritons in van der {Waals} semiconductor gratings}}},\ }\href {https://doi.org/10.1038/s41467-020-17313-2} {\bibfield  {journal} {\bibinfo  {journal} {Nat. Commun.}\ }\textbf {\bibinfo {volume} {11}},\ \bibinfo {pages} {3552} (\bibinfo {year} {2020})}\BibitemShut {NoStop}%
\bibitem [{\citenamefont {Keldysh}(2024)}]{keldysh2024coulomb}%
  \BibitemOpen
  \bibfield  {author} {\bibinfo {author} {\bibfnamefont {L.~V.}\ \bibnamefont {Keldysh}},\ }in\ \href {https://doi.org/10.1142/13503} {\emph {\bibinfo {booktitle} {{Selected Papers of Leonid V Keldysh}}}}\ (\bibinfo  {publisher} {World Scientific},\ \bibinfo {year} {2024})\ p.\ \bibinfo {pages} {155}\BibitemShut {NoStop}%
\bibitem [{\citenamefont {Tang}\ \emph {et~al.}(2022)\citenamefont {Tang}, \citenamefont {Zhang}, \citenamefont {Liu}, \citenamefont {Wei}, \citenamefont {Cheng}, \citenamefont {Shi},\ and\ \citenamefont {Jiang}}]{tang2022interacting}%
  \BibitemOpen
  \bibfield  {author} {\bibinfo {author} {\bibfnamefont {Y.}~\bibnamefont {Tang}}, \bibinfo {author} {\bibfnamefont {Y.}~\bibnamefont {Zhang}}, \bibinfo {author} {\bibfnamefont {Q.}~\bibnamefont {Liu}}, \bibinfo {author} {\bibfnamefont {K.}~\bibnamefont {Wei}}, \bibinfo {author} {\bibfnamefont {X.}~\bibnamefont {Cheng}}, \bibinfo {author} {\bibfnamefont {L.}~\bibnamefont {Shi}},\ and\ \bibinfo {author} {\bibfnamefont {T.}~\bibnamefont {Jiang}},\ }\bibfield  {title} {\emph {\bibinfo {title} {{Interacting plexcitons for designed ultrafast optical nonlinearity in a monolayer semiconductor}}},\ }\href {https://doi.org/10.1038/s41377-022-00754-3} {\bibfield  {journal} {\bibinfo  {journal} {Light Sci. Appl.}\ }\textbf {\bibinfo {volume} {11}},\ \bibinfo {pages} {94} (\bibinfo {year} {2022})}\BibitemShut {NoStop}%
\bibitem [{\citenamefont {{Deng}}\ \emph {et~al.}(2002)\citenamefont {{Deng}}, \citenamefont {{Weihs}}, \citenamefont {{Santori}}, \citenamefont {{Bloch}},\ and\ \citenamefont {{Yamamoto}}}]{deng2002condensation}%
  \BibitemOpen
  \bibfield  {author} {\bibinfo {author} {\bibfnamefont {H.}~\bibnamefont {{Deng}}}, \bibinfo {author} {\bibfnamefont {G.}~\bibnamefont {{Weihs}}}, \bibinfo {author} {\bibfnamefont {C.}~\bibnamefont {{Santori}}}, \bibinfo {author} {\bibfnamefont {J.}~\bibnamefont {{Bloch}}},\ and\ \bibinfo {author} {\bibfnamefont {Y.}~\bibnamefont {{Yamamoto}}},\ }\bibfield  {title} {\emph {\bibinfo {title} {{Condensation of semiconductor microcavity exciton polaritons}}},\ }\href {https://doi.org/10.1126/science.1074464} {\bibfield  {journal} {\bibinfo  {journal} {Science}\ }\textbf {\bibinfo {volume} {298}},\ \bibinfo {pages} {199} (\bibinfo {year} {2002})}\BibitemShut {NoStop}%
\bibitem [{\citenamefont {Deng}\ \emph {et~al.}(2010)\citenamefont {Deng}, \citenamefont {Haug},\ and\ \citenamefont {Yamamoto}}]{deng2010exciton}%
  \BibitemOpen
  \bibfield  {author} {\bibinfo {author} {\bibfnamefont {H.}~\bibnamefont {Deng}}, \bibinfo {author} {\bibfnamefont {H.}~\bibnamefont {Haug}},\ and\ \bibinfo {author} {\bibfnamefont {Y.}~\bibnamefont {Yamamoto}},\ }\bibfield  {title} {\emph {\bibinfo {title} {{Exciton-polariton Bose-Einstein condensation}}},\ }\href {https://doi.org/10.1103/RevModPhys.82.1489} {\bibfield  {journal} {\bibinfo  {journal} {Rev. Mod. Phys.}\ }\textbf {\bibinfo {volume} {82}},\ \bibinfo {pages} {1489--1537} (\bibinfo {year} {2010})}\BibitemShut {NoStop}%
\bibitem [{\citenamefont {{Luo}}\ \emph {et~al.}(2023{\natexlab{b}})\citenamefont {{Luo}}, \citenamefont {{Zhou}}, \citenamefont {{Zhang}},\ and\ \citenamefont {{Chen}}}]{luo2023nanophotonics}%
  \BibitemOpen
  \bibfield  {author} {\bibinfo {author} {\bibfnamefont {S.}~\bibnamefont {{Luo}}}, \bibinfo {author} {\bibfnamefont {H.}~\bibnamefont {{Zhou}}}, \bibinfo {author} {\bibfnamefont {L.}~\bibnamefont {{Zhang}}},\ and\ \bibinfo {author} {\bibfnamefont {Z.}~\bibnamefont {{Chen}}},\ }\bibfield  {title} {\emph {\bibinfo {title} {{Nanophotonics of microcavity exciton-polaritons}}},\ }\href {https://doi.org/10.1063/5.0121316} {\bibfield  {journal} {\bibinfo  {journal} {Appl. Phys. Rev.}\ }\textbf {\bibinfo {volume} {10}},\ \bibinfo {pages} {011316} (\bibinfo {year} {2023}{\natexlab{b}})}\BibitemShut {NoStop}%
\bibitem [{\citenamefont {{Ashida}}\ \emph {et~al.}(2020)\citenamefont {{Ashida}}, \citenamefont {{Imamo{\v{g}}lu}}, \citenamefont {{Faist}}, \citenamefont {{Jaksch}}, \citenamefont {{Cavalleri}},\ and\ \citenamefont {{Demler}}}]{ashida2020quantum}%
  \BibitemOpen
  \bibfield  {author} {\bibinfo {author} {\bibfnamefont {Y.}~\bibnamefont {{Ashida}}}, \bibinfo {author} {\bibfnamefont {A.}~\bibnamefont {{Imamo{\v{g}}lu}}}, \bibinfo {author} {\bibfnamefont {J.}~\bibnamefont {{Faist}}}, \bibinfo {author} {\bibfnamefont {D.}~\bibnamefont {{Jaksch}}}, \bibinfo {author} {\bibfnamefont {A.}~\bibnamefont {{Cavalleri}}},\ and\ \bibinfo {author} {\bibfnamefont {E.}~\bibnamefont {{Demler}}},\ }\bibfield  {title} {\emph {\bibinfo {title} {{Quantum electrodynamic control of matter: Cavity-enhanced ferroelectric phase transition}}},\ }\href {https://doi.org/10.1103/PhysRevX.10.041027} {\bibfield  {journal} {\bibinfo  {journal} {Phys. Rev. X}\ }\textbf {\bibinfo {volume} {10}},\ \bibinfo {pages} {041027} (\bibinfo {year} {2020})}\BibitemShut {NoStop}%
\bibitem [{\citenamefont {{Schlawin}}\ \emph {et~al.}(2022)\citenamefont {{Schlawin}}, \citenamefont {{Kennes}},\ and\ \citenamefont {{Sentef}}}]{schlawin2022cavity}%
  \BibitemOpen
  \bibfield  {author} {\bibinfo {author} {\bibfnamefont {F.}~\bibnamefont {{Schlawin}}}, \bibinfo {author} {\bibfnamefont {D.~M.}\ \bibnamefont {{Kennes}}},\ and\ \bibinfo {author} {\bibfnamefont {M.~A.}\ \bibnamefont {{Sentef}}},\ }\bibfield  {title} {\emph {\bibinfo {title} {{Cavity quantum materials}}},\ }\href {https://doi.org/10.1063/5.0083825} {\bibfield  {journal} {\bibinfo  {journal} {Appl. Phys. Rev.}\ }\textbf {\bibinfo {volume} {9}},\ \bibinfo {pages} {011312} (\bibinfo {year} {2022})}\BibitemShut {NoStop}%
\bibitem [{\citenamefont {{Peng}}\ \emph {et~al.}(2024)\citenamefont {{Peng}}, \citenamefont {{Li}}, \citenamefont {{Sun}}, \citenamefont {{Rivero}}, \citenamefont {{Ti}}, \citenamefont {{Han}}, \citenamefont {{Ge}}, \citenamefont {{Yang}}, \citenamefont {{Zhang}},\ and\ \citenamefont {{Bao}}}]{peng2024topological}%
  \BibitemOpen
  \bibfield  {author} {\bibinfo {author} {\bibfnamefont {K.}~\bibnamefont {{Peng}}}, \bibinfo {author} {\bibfnamefont {W.}~\bibnamefont {{Li}}}, \bibinfo {author} {\bibfnamefont {M.}~\bibnamefont {{Sun}}}, \bibinfo {author} {\bibfnamefont {J.~D.~H.}\ \bibnamefont {{Rivero}}}, \bibinfo {author} {\bibfnamefont {C.}~\bibnamefont {{Ti}}}, \bibinfo {author} {\bibfnamefont {X.}~\bibnamefont {{Han}}}, \bibinfo {author} {\bibfnamefont {L.}~\bibnamefont {{Ge}}}, \bibinfo {author} {\bibfnamefont {L.}~\bibnamefont {{Yang}}}, \bibinfo {author} {\bibfnamefont {X.}~\bibnamefont {{Zhang}}},\ and\ \bibinfo {author} {\bibfnamefont {W.}~\bibnamefont {{Bao}}},\ }\bibfield  {title} {\emph {\bibinfo {title} {{Topological valley Hall polariton condensation}}},\ }\href {https://doi.org/10.1038/s41565-024-01674-6} {\bibfield  {journal} {\bibinfo  {journal} {Nat. Nanotechnol.}\ }\textbf {\bibinfo {volume} {19}},\ \bibinfo {pages} {1283--1289} (\bibinfo {year} {2024})}\BibitemShut {NoStop}%
\bibitem [{\citenamefont {Garcia-Vidal}\ \emph {et~al.}(2021)\citenamefont {Garcia-Vidal}, \citenamefont {Ciuti},\ and\ \citenamefont {Ebbesen}}]{garcia-vidal2021manipulating}%
  \BibitemOpen
  \bibfield  {author} {\bibinfo {author} {\bibfnamefont {F.~J.}\ \bibnamefont {Garcia-Vidal}}, \bibinfo {author} {\bibfnamefont {C.}~\bibnamefont {Ciuti}},\ and\ \bibinfo {author} {\bibfnamefont {T.~W.}\ \bibnamefont {Ebbesen}},\ }\bibfield  {title} {\emph {\bibinfo {title} {{Manipulating matter by strong coupling to vacuum fields}}},\ }\href {https://doi.org/10.1126/science.abd0336} {\bibfield  {journal} {\bibinfo  {journal} {Science}\ }\textbf {\bibinfo {volume} {373}},\ \bibinfo {pages} {eabd0336} (\bibinfo {year} {2021})}\BibitemShut {NoStop}%
\bibitem [{\citenamefont {Jarc}\ \emph {et~al.}(2023)\citenamefont {Jarc}, \citenamefont {Mathengattil}, \citenamefont {Montanaro}, \citenamefont {Giusti}, \citenamefont {Rigoni}, \citenamefont {Sergo}, \citenamefont {Fassioli}, \citenamefont {Winnerl}, \citenamefont {Dal~Zilio}, \citenamefont {Mihailovic}, \citenamefont {Prelovšek}, \citenamefont {Eckstein},\ and\ \citenamefont {Fausti}}]{jarc2023cavity}%
  \BibitemOpen
  \bibfield  {author} {\bibinfo {author} {\bibfnamefont {G.}~\bibnamefont {Jarc}}, \bibinfo {author} {\bibfnamefont {S.~Y.}\ \bibnamefont {Mathengattil}}, \bibinfo {author} {\bibfnamefont {A.}~\bibnamefont {Montanaro}}, \bibinfo {author} {\bibfnamefont {F.}~\bibnamefont {Giusti}}, \bibinfo {author} {\bibfnamefont {E.~M.}\ \bibnamefont {Rigoni}}, \bibinfo {author} {\bibfnamefont {R.}~\bibnamefont {Sergo}}, \bibinfo {author} {\bibfnamefont {F.}~\bibnamefont {Fassioli}}, \bibinfo {author} {\bibfnamefont {S.}~\bibnamefont {Winnerl}}, \bibinfo {author} {\bibfnamefont {S.}~\bibnamefont {Dal~Zilio}}, \bibinfo {author} {\bibfnamefont {D.}~\bibnamefont {Mihailovic}}, \bibinfo {author} {\bibfnamefont {P.}~\bibnamefont {Prelovšek}}, \bibinfo {author} {\bibfnamefont {M.}~\bibnamefont {Eckstein}},\ and\ \bibinfo {author} {\bibfnamefont {D.}~\bibnamefont {Fausti}},\ }\bibfield  {title} {\emph {\bibinfo {title} {{Cavity-mediated thermal control of metal-to-insulator transition in {1$T$}-{TaS$_2$}}}},\ }\href
  {https://doi.org/10.1038/s41586-023-06596-2} {\bibfield  {journal} {\bibinfo  {journal} {Nature}\ }\textbf {\bibinfo {volume} {622}},\ \bibinfo {pages} {487--492} (\bibinfo {year} {2023})}\BibitemShut {NoStop}%
\bibitem [{\citenamefont {Zong}\ \emph {et~al.}(2024)\citenamefont {Zong}, \citenamefont {Lin}, \citenamefont {Sato}, \citenamefont {Berger}, \citenamefont {Nebgen}, \citenamefont {Hui}, \citenamefont {Lv}, \citenamefont {Cheng}, \citenamefont {Xia}, \citenamefont {Guo}, \citenamefont {Xiang},\ and\ \citenamefont {Zuerch}}]{zong2024core}%
  \BibitemOpen
  \bibfield  {author} {\bibinfo {author} {\bibfnamefont {A.}~\bibnamefont {Zong}}, \bibinfo {author} {\bibfnamefont {S.-C.}\ \bibnamefont {Lin}}, \bibinfo {author} {\bibfnamefont {S.~A.}\ \bibnamefont {Sato}}, \bibinfo {author} {\bibfnamefont {E.}~\bibnamefont {Berger}}, \bibinfo {author} {\bibfnamefont {B.~R.}\ \bibnamefont {Nebgen}}, \bibinfo {author} {\bibfnamefont {M.}~\bibnamefont {Hui}}, \bibinfo {author} {\bibfnamefont {B.~Q.}\ \bibnamefont {Lv}}, \bibinfo {author} {\bibfnamefont {Y.}~\bibnamefont {Cheng}}, \bibinfo {author} {\bibfnamefont {W.}~\bibnamefont {Xia}}, \bibinfo {author} {\bibfnamefont {Y.}~\bibnamefont {Guo}}, \bibinfo {author} {\bibfnamefont {D.}~\bibnamefont {Xiang}},\ and\ \bibinfo {author} {\bibfnamefont {M.~W.}\ \bibnamefont {Zuerch}},\ }\href {https://doi.org/10.48550/arXiv.2407.00772} {\bibinfo {title} {\textit{{Core-level signature of long-range density-wave order and short-range excitonic correlations probed by attosecond broadband spectroscopy}}}} (\bibinfo {year} {2024}),\
  \Eprint {https://arxiv.org/abs/2407.00772} {arXiv:2407.00772} \BibitemShut {NoStop}%
\bibitem [{\citenamefont {O’Callahan}\ \emph {et~al.}(2015)\citenamefont {O’Callahan}, \citenamefont {Jones}, \citenamefont {Hyung~Park}, \citenamefont {Cobden}, \citenamefont {Atkin},\ and\ \citenamefont {Raschke}}]{o2015inhomogeneity}%
  \BibitemOpen
  \bibfield  {author} {\bibinfo {author} {\bibfnamefont {B.~T.}\ \bibnamefont {O’Callahan}}, \bibinfo {author} {\bibfnamefont {A.~C.}\ \bibnamefont {Jones}}, \bibinfo {author} {\bibfnamefont {J.}~\bibnamefont {Hyung~Park}}, \bibinfo {author} {\bibfnamefont {D.~H.}\ \bibnamefont {Cobden}}, \bibinfo {author} {\bibfnamefont {J.~M.}\ \bibnamefont {Atkin}},\ and\ \bibinfo {author} {\bibfnamefont {M.~B.}\ \bibnamefont {Raschke}},\ }\bibfield  {title} {\emph {\bibinfo {title} {{Inhomogeneity of the ultrafast insulator-to-metal transition dynamics of VO$_2$}}},\ }\href {https://doi.org/10.1038/ncomms7849} {\bibfield  {journal} {\bibinfo  {journal} {Nat. Commun.}\ }\textbf {\bibinfo {volume} {6}},\ \bibinfo {pages} {6849} (\bibinfo {year} {2015})}\BibitemShut {NoStop}%
\bibitem [{\citenamefont {{Cremons}}\ \emph {et~al.}(2016)\citenamefont {{Cremons}}, \citenamefont {{Plemmons}},\ and\ \citenamefont {{Flannigan}}}]{cremons2016femtosecond}%
  \BibitemOpen
  \bibfield  {author} {\bibinfo {author} {\bibfnamefont {D.~R.}\ \bibnamefont {{Cremons}}}, \bibinfo {author} {\bibfnamefont {D.~A.}\ \bibnamefont {{Plemmons}}},\ and\ \bibinfo {author} {\bibfnamefont {D.~J.}\ \bibnamefont {{Flannigan}}},\ }\bibfield  {title} {\emph {\bibinfo {title} {{Femtosecond electron imaging of defect-modulated phonon dynamics}}},\ }\href {https://doi.org/10.1038/ncomms11230} {\bibfield  {journal} {\bibinfo  {journal} {Nat. Commun.}\ }\textbf {\bibinfo {volume} {7}},\ \bibinfo {pages} {11230} (\bibinfo {year} {2016})}\BibitemShut {NoStop}%
\bibitem [{\citenamefont {{Perez-Salinas}}\ \emph {et~al.}(2022)\citenamefont {{Perez-Salinas}}, \citenamefont {{Johnson}}, \citenamefont {{Prabhakaran}},\ and\ \citenamefont {{Wall}}}]{perez2022multi}%
  \BibitemOpen
  \bibfield  {author} {\bibinfo {author} {\bibfnamefont {D.}~\bibnamefont {{Perez-Salinas}}}, \bibinfo {author} {\bibfnamefont {A.~S.}\ \bibnamefont {{Johnson}}}, \bibinfo {author} {\bibfnamefont {D.}~\bibnamefont {{Prabhakaran}}},\ and\ \bibinfo {author} {\bibfnamefont {S.}~\bibnamefont {{Wall}}},\ }\bibfield  {title} {\emph {\bibinfo {title} {{Multi-mode excitation drives disorder during the ultrafast melting of a C4-symmetry-broken phase}}},\ }\href {https://doi.org/10.1038/s41467-021-27819-y} {\bibfield  {journal} {\bibinfo  {journal} {Nat. Commun.}\ }\textbf {\bibinfo {volume} {13}},\ \bibinfo {pages} {238} (\bibinfo {year} {2022})}\BibitemShut {NoStop}%
\bibitem [{\citenamefont {{Carbin}}\ \emph {et~al.}(2023)\citenamefont {{Carbin}}, \citenamefont {{Zhang}}, \citenamefont {{Culver}}, \citenamefont {{Zhao}}, \citenamefont {{Zong}}, \citenamefont {{Acharya}}, \citenamefont {{Abbamonte}}, \citenamefont {{Roy}}, \citenamefont {{Cao}},\ and\ \citenamefont {{Kogar}}}]{carbin2023evidence}%
  \BibitemOpen
  \bibfield  {author} {\bibinfo {author} {\bibfnamefont {T.}~\bibnamefont {{Carbin}}}, \bibinfo {author} {\bibfnamefont {X.}~\bibnamefont {{Zhang}}}, \bibinfo {author} {\bibfnamefont {A.~B.}\ \bibnamefont {{Culver}}}, \bibinfo {author} {\bibfnamefont {H.}~\bibnamefont {{Zhao}}}, \bibinfo {author} {\bibfnamefont {A.}~\bibnamefont {{Zong}}}, \bibinfo {author} {\bibfnamefont {R.}~\bibnamefont {{Acharya}}}, \bibinfo {author} {\bibfnamefont {C.~J.}\ \bibnamefont {{Abbamonte}}}, \bibinfo {author} {\bibfnamefont {R.}~\bibnamefont {{Roy}}}, \bibinfo {author} {\bibfnamefont {G.}~\bibnamefont {{Cao}}},\ and\ \bibinfo {author} {\bibfnamefont {A.}~\bibnamefont {{Kogar}}},\ }\bibfield  {title} {\emph {\bibinfo {title} {{Evidence for bootstrap percolation dynamics in a photoinduced phase transition}}},\ }\href {https://doi.org/10.1103/PhysRevLett.130.186902} {\bibfield  {journal} {\bibinfo  {journal} {Phys. Rev. Lett.}\ }\textbf {\bibinfo {volume} {130}},\ \bibinfo {pages} {186902} (\bibinfo {year} {2023})}\BibitemShut
  {NoStop}%
\bibitem [{\citenamefont {McClellan}\ \emph {et~al.}(2024)\citenamefont {McClellan}, \citenamefont {Zong}, \citenamefont {Pham}, \citenamefont {Liu}, \citenamefont {Iton}, \citenamefont {Guzelturk}, \citenamefont {Walko}, \citenamefont {Wen}, \citenamefont {Cushing},\ and\ \citenamefont {Zuerch}}]{mcclellan2024hidden}%
  \BibitemOpen
  \bibfield  {author} {\bibinfo {author} {\bibfnamefont {J.}~\bibnamefont {McClellan}}, \bibinfo {author} {\bibfnamefont {A.}~\bibnamefont {Zong}}, \bibinfo {author} {\bibfnamefont {K.~H.}\ \bibnamefont {Pham}}, \bibinfo {author} {\bibfnamefont {H.}~\bibnamefont {Liu}}, \bibinfo {author} {\bibfnamefont {Z.~W.~B.}\ \bibnamefont {Iton}}, \bibinfo {author} {\bibfnamefont {B.}~\bibnamefont {Guzelturk}}, \bibinfo {author} {\bibfnamefont {D.~A.}\ \bibnamefont {Walko}}, \bibinfo {author} {\bibfnamefont {H.}~\bibnamefont {Wen}}, \bibinfo {author} {\bibfnamefont {S.~K.}\ \bibnamefont {Cushing}},\ and\ \bibinfo {author} {\bibfnamefont {M.~W.}\ \bibnamefont {Zuerch}},\ }\href {https://doi.org/10.48550/arXiv.2406.06832} {\bibinfo {title} {\textit{Hidden correlations in stochastic photoinduced dynamics of a solid-state electrolyte}}} (\bibinfo {year} {2024}),\ \Eprint {https://arxiv.org/abs/2406.06832} {arXiv:2406.06832} \BibitemShut {NoStop}%
\bibitem [{\citenamefont {Kim}\ \emph {et~al.}(2020)\citenamefont {Kim}, \citenamefont {Vinokurov}, \citenamefont {Baek}, \citenamefont {Oang}, \citenamefont {Kim}, \citenamefont {Kim}, \citenamefont {Jang}, \citenamefont {Lee}, \citenamefont {Park}, \citenamefont {Park}, \citenamefont {Shin}, \citenamefont {Kim}, \citenamefont {Rotermund}, \citenamefont {Cho}, \citenamefont {Feurer},\ and\ \citenamefont {Jeong}}]{kim2019towards}%
  \BibitemOpen
  \bibfield  {author} {\bibinfo {author} {\bibfnamefont {H.~W.}\ \bibnamefont {Kim}}, \bibinfo {author} {\bibfnamefont {N.~A.}\ \bibnamefont {Vinokurov}}, \bibinfo {author} {\bibfnamefont {I.~H.}\ \bibnamefont {Baek}}, \bibinfo {author} {\bibfnamefont {K.~Y.}\ \bibnamefont {Oang}}, \bibinfo {author} {\bibfnamefont {M.~H.}\ \bibnamefont {Kim}}, \bibinfo {author} {\bibfnamefont {Y.~C.}\ \bibnamefont {Kim}}, \bibinfo {author} {\bibfnamefont {K.-H.}\ \bibnamefont {Jang}}, \bibinfo {author} {\bibfnamefont {K.}~\bibnamefont {Lee}}, \bibinfo {author} {\bibfnamefont {S.~H.}\ \bibnamefont {Park}}, \bibinfo {author} {\bibfnamefont {S.}~\bibnamefont {Park}}, \bibinfo {author} {\bibfnamefont {J.}~\bibnamefont {Shin}}, \bibinfo {author} {\bibfnamefont {J.}~\bibnamefont {Kim}}, \bibinfo {author} {\bibfnamefont {F.}~\bibnamefont {Rotermund}}, \bibinfo {author} {\bibfnamefont {S.}~\bibnamefont {Cho}}, \bibinfo {author} {\bibfnamefont {T.}~\bibnamefont {Feurer}},\ and\ \bibinfo {author} {\bibfnamefont {Y.~U.}\ \bibnamefont
  {Jeong}},\ }\bibfield  {title} {\emph {\bibinfo {title} {{Towards jitter-free ultrafast electron diffraction technology}}},\ }\href {https://doi.org/10.1038/s41566-019-0566-4} {\bibfield  {journal} {\bibinfo  {journal} {Nat. Photonics}\ }\textbf {\bibinfo {volume} {14}},\ \bibinfo {pages} {245--249} (\bibinfo {year} {2020})}\BibitemShut {NoStop}%
\bibitem [{\citenamefont {Nabben}\ \emph {et~al.}(2023)\citenamefont {Nabben}, \citenamefont {Kuttruff}, \citenamefont {Stolz}, \citenamefont {Ryabov},\ and\ \citenamefont {Baum}}]{nabben2023attosecond}%
  \BibitemOpen
  \bibfield  {author} {\bibinfo {author} {\bibfnamefont {D.}~\bibnamefont {Nabben}}, \bibinfo {author} {\bibfnamefont {J.}~\bibnamefont {Kuttruff}}, \bibinfo {author} {\bibfnamefont {L.}~\bibnamefont {Stolz}}, \bibinfo {author} {\bibfnamefont {A.}~\bibnamefont {Ryabov}},\ and\ \bibinfo {author} {\bibfnamefont {P.}~\bibnamefont {Baum}},\ }\bibfield  {title} {\emph {\bibinfo {title} {Attosecond electron microscopy of sub-cycle optical dynamics}},\ }\href {https://doi.org/10.1038/s41586-023-06074-9} {\bibfield  {journal} {\bibinfo  {journal} {Nature}\ }\textbf {\bibinfo {volume} {619}},\ \bibinfo {pages} {63--67} (\bibinfo {year} {2023})}\BibitemShut {NoStop}%
\bibitem [{\citenamefont {Hui}\ \emph {et~al.}(2024)\citenamefont {Hui}, \citenamefont {Alqattan}, \citenamefont {Sennary}, \citenamefont {Golubev},\ and\ \citenamefont {Hassan}}]{hui2024attosecond}%
  \BibitemOpen
  \bibfield  {author} {\bibinfo {author} {\bibfnamefont {D.}~\bibnamefont {Hui}}, \bibinfo {author} {\bibfnamefont {H.}~\bibnamefont {Alqattan}}, \bibinfo {author} {\bibfnamefont {M.}~\bibnamefont {Sennary}}, \bibinfo {author} {\bibfnamefont {N.~V.}\ \bibnamefont {Golubev}},\ and\ \bibinfo {author} {\bibfnamefont {M.~T.}\ \bibnamefont {Hassan}},\ }\bibfield  {title} {\emph {\bibinfo {title} {Attosecond electron microscopy and diffraction}},\ }\href {https://doi.org/10.1126/sciadv.adp5805} {\bibfield  {journal} {\bibinfo  {journal} {Sci. Adv.}\ }\textbf {\bibinfo {volume} {10}},\ \bibinfo {pages} {eadp5805} (\bibinfo {year} {2024})}\BibitemShut {NoStop}%
\bibitem [{\citenamefont {{Baum}}\ and\ \citenamefont {{Ropers}}(2024)}]{2024arXiv241114518B}%
  \BibitemOpen
  \bibfield  {author} {\bibinfo {author} {\bibfnamefont {P.}~\bibnamefont {{Baum}}}\ and\ \bibinfo {author} {\bibfnamefont {C.}~\bibnamefont {{Ropers}}},\ }\href {https://doi.org/10.48550/arXiv.2411.14518} {\bibinfo {title} {{Comment on 'Attosecond electron microscopy and diffraction'}}} (\bibinfo {year} {2024}),\ \Eprint {https://arxiv.org/abs/2411.14518} {arXiv:2411.14518} \BibitemShut {NoStop}%
\bibitem [{\citenamefont {{Ren{\'e} de Cotret}}\ \emph {et~al.}(2019)\citenamefont {{Ren{\'e} de Cotret}}, \citenamefont {{P{\"o}hls}}, \citenamefont {{Stern}}, \citenamefont {{Otto}}, \citenamefont {{Sutton}},\ and\ \citenamefont {{Siwick}}}]{rene2019time}%
  \BibitemOpen
  \bibfield  {author} {\bibinfo {author} {\bibfnamefont {L.~P.}\ \bibnamefont {{Ren{\'e} de Cotret}}}, \bibinfo {author} {\bibfnamefont {J.-H.}\ \bibnamefont {{P{\"o}hls}}}, \bibinfo {author} {\bibfnamefont {M.~J.}\ \bibnamefont {{Stern}}}, \bibinfo {author} {\bibfnamefont {M.~R.}\ \bibnamefont {{Otto}}}, \bibinfo {author} {\bibfnamefont {M.}~\bibnamefont {{Sutton}}},\ and\ \bibinfo {author} {\bibfnamefont {B.~J.}\ \bibnamefont {{Siwick}}},\ }\bibfield  {title} {\emph {\bibinfo {title} {{Time- and momentum-resolved phonon population dynamics with ultrafast electron diffuse scattering}}},\ }\href {https://doi.org/10.1103/PhysRevB.100.214115} {\bibfield  {journal} {\bibinfo  {journal} {Phys. Rev. B}\ }\textbf {\bibinfo {volume} {100}},\ \bibinfo {pages} {214115} (\bibinfo {year} {2019})}\BibitemShut {NoStop}%
\bibitem [{\citenamefont {{Britt}}\ \emph {et~al.}(2022)\citenamefont {{Britt}}, \citenamefont {{Li}}, \citenamefont {{Ren{\'e} de Cotret}}, \citenamefont {{Olsen}}, \citenamefont {{Otto}}, \citenamefont {{Hassan}}, \citenamefont {{Zacharias}}, \citenamefont {{Caruso}}, \citenamefont {{Zhu}},\ and\ \citenamefont {{Siwick}}}]{britt2022direct}%
  \BibitemOpen
  \bibfield  {author} {\bibinfo {author} {\bibfnamefont {T.~L.}\ \bibnamefont {{Britt}}}, \bibinfo {author} {\bibfnamefont {Q.}~\bibnamefont {{Li}}}, \bibinfo {author} {\bibfnamefont {L.~P.}\ \bibnamefont {{Ren{\'e} de Cotret}}}, \bibinfo {author} {\bibfnamefont {N.}~\bibnamefont {{Olsen}}}, \bibinfo {author} {\bibfnamefont {M.}~\bibnamefont {{Otto}}}, \bibinfo {author} {\bibfnamefont {S.~A.}\ \bibnamefont {{Hassan}}}, \bibinfo {author} {\bibfnamefont {M.}~\bibnamefont {{Zacharias}}}, \bibinfo {author} {\bibfnamefont {F.}~\bibnamefont {{Caruso}}}, \bibinfo {author} {\bibfnamefont {X.}~\bibnamefont {{Zhu}}},\ and\ \bibinfo {author} {\bibfnamefont {B.~J.}\ \bibnamefont {{Siwick}}},\ }\bibfield  {title} {\emph {\bibinfo {title} {{Direct view of phonon dynamics in atomically thin MoS$_2$}}},\ }\href {https://doi.org/10.1021/acs.nanolett.2c00850} {\bibfield  {journal} {\bibinfo  {journal} {Nano Lett.}\ }\textbf {\bibinfo {volume} {22}},\ \bibinfo {pages} {4718} (\bibinfo {year} {2022})}\BibitemShut {NoStop}%
\bibitem [{\citenamefont {Sidiropoulos}\ \emph {et~al.}(2021)\citenamefont {Sidiropoulos}, \citenamefont {Di~Palo}, \citenamefont {Rivas}, \citenamefont {Severino}, \citenamefont {Reduzzi}, \citenamefont {Nandy}, \citenamefont {Bauerhenne}, \citenamefont {Krylow}, \citenamefont {Vasileiadis}, \citenamefont {Danz}, \citenamefont {Elliott}, \citenamefont {Sharma}, \citenamefont {Dewhurst}, \citenamefont {Ropers}, \citenamefont {Joly}, \citenamefont {Garcia}, \citenamefont {Wolf}, \citenamefont {Ernstorfer},\ and\ \citenamefont {Biegert}}]{sidiropoulos2021probing}%
  \BibitemOpen
  \bibfield  {author} {\bibinfo {author} {\bibfnamefont {T.~P.~H.}\ \bibnamefont {Sidiropoulos}}, \bibinfo {author} {\bibfnamefont {N.}~\bibnamefont {Di~Palo}}, \bibinfo {author} {\bibfnamefont {D.~E.}\ \bibnamefont {Rivas}}, \bibinfo {author} {\bibfnamefont {S.}~\bibnamefont {Severino}}, \bibinfo {author} {\bibfnamefont {M.}~\bibnamefont {Reduzzi}}, \bibinfo {author} {\bibfnamefont {B.}~\bibnamefont {Nandy}}, \bibinfo {author} {\bibfnamefont {B.}~\bibnamefont {Bauerhenne}}, \bibinfo {author} {\bibfnamefont {S.}~\bibnamefont {Krylow}}, \bibinfo {author} {\bibfnamefont {T.}~\bibnamefont {Vasileiadis}}, \bibinfo {author} {\bibfnamefont {T.}~\bibnamefont {Danz}}, \bibinfo {author} {\bibfnamefont {P.}~\bibnamefont {Elliott}}, \bibinfo {author} {\bibfnamefont {S.}~\bibnamefont {Sharma}}, \bibinfo {author} {\bibfnamefont {K.}~\bibnamefont {Dewhurst}}, \bibinfo {author} {\bibfnamefont {C.}~\bibnamefont {Ropers}}, \bibinfo {author} {\bibfnamefont {Y.}~\bibnamefont {Joly}}, \bibinfo {author} {\bibfnamefont {M.~E.}\
  \bibnamefont {Garcia}}, \bibinfo {author} {\bibfnamefont {M.}~\bibnamefont {Wolf}}, \bibinfo {author} {\bibfnamefont {R.}~\bibnamefont {Ernstorfer}},\ and\ \bibinfo {author} {\bibfnamefont {J.}~\bibnamefont {Biegert}},\ }\bibfield  {title} {\emph {\bibinfo {title} {{Probing the energy conversion pathways between light, carriers, and lattice in real time with attosecond core-level spectroscopy}}},\ }\href {https://doi.org/10.1103/PhysRevX.11.041060} {\bibfield  {journal} {\bibinfo  {journal} {Phys. Rev. X}\ }\textbf {\bibinfo {volume} {11}},\ \bibinfo {pages} {041060} (\bibinfo {year} {2021})}\BibitemShut {NoStop}%
\end{thebibliography}

%

\end{document}